\documentclass[aps,preprint,floatfix,nofootinbib,showpacs]{revtex4-1}
\pdfoutput=1
\usepackage{graphicx,color,rotating}
\usepackage{hyperref}
\usepackage{epsfig,color}

\newcommand{\lsim}{\raisebox{-0.13cm}{~\shortstack{$<$ \\[-0.07cm] $\sim$}}~}
\newcommand{\gsim}{\raisebox{-0.13cm}{~\shortstack{$>$ \\[-0.07cm] $\sim$}}~}

\newcommand{\real}{\Re {\rm e}}

\def\beq{\begin{equation}}
\def\eeq{\end{equation}}

\def\bea{\begin{eqnarray}}
\def\eea{\end{eqnarray}}
\def\bei{\begin{itemize}}
\def\eei{\end{itemize}}
\def\bmat{\begin{matrix}}
\def\emat{\end{matrix}}
\def\ble{\begin{flushleft}}
\def\ele{\end{flushleft}}
\def\={\,=\,}
\def\+{\,+\,}
\def\-{\,-\,}

\begin{document}

\def\thefootnote{\fnsymbol{footnote}}

{\small
\begin{flushright}
CNU-HEP-16-02,
IPMU16-0110
\end{flushright} }

\title{Double Higgcision: \\
125 GeV Higgs boson and a potential diphoton Resonance}
\author{
Kingman Cheung$^{1,2,3}$, P. Ko$^4$,
Jae Sik Lee$^{5,3}$, Jubin Park$^{4,5,3}$, and Po-Yan Tseng$^{6,1}$}
\affiliation{
$^1$ Department of Physics, National Tsing Hua University,
Hsinchu 300, Taiwan \\
$^2$ Division of Quantum Phases and Devices, School of Physics,
Konkuk University, Seoul 143-701, Republic of Korea \\
$^3$ Physics Division, National Center for Theoretical Sciences, Hsinchu, Taiwan \\
$^4$ School of Physics, KIAS, Seoul 130-722, Republic of Korea \\
$^5$ Department of Physics, Chonnam National University, \\
300 Yongbong-dong, Buk-gu, Gwangju, 500-757, Republic of Korea\\
$^6$ Kavli IPMU (WPI), UTIAS, University of Tokyo, Kashiwa, 277-8583, Japan
}
\date{August 16, 2016}

\begin{abstract}
Searches for diphoton resonance have been shown to be very useful in
discovering new heavy spin-0 or spin-2 particles. 
Supposing that a new heavy particle shows up in the diphoton channel
and it points to a spin-0 boson, it can be allowed to have a small mixing 
with the observed 125 GeV Higgs-like boson.  
We borrow the example
of the 750 GeV particles hinted with 3.2 fb$^{-1}$ data at the end of 
2015 (though it did not appear in the 2016 data) to perform
an analysis of ``double Higgcision''.
In this work, we perform 
a complete Higgs-signal strength analysis in the Higgs-portal type
framework, using all the existing 125 GeV Higgs boson data as well as the
diphoton signal strength of the 750 GeV scalar boson.
The best fit prefers a very tiny mixing between two scalar bosons,
which has to be accommodated in models for the 750 GeV scalar boson.
\end{abstract}

\maketitle

\section{Introduction}

The first run at $\sqrt{s}=13$ TeV at the LHC has hinted a possibility 
of observing a new particle at around 750 GeV.
Both ATLAS and CMS collaborations have reported a ``bump'' in the
diphoton invariant mass distribution around 750 GeV,
indicating a local significance of $3.9\sigma$ by ATLAS \cite{atlas} 
and about $3\sigma$ by the CMS \cite{cms}.
Such an excitement has motivated a lot of speculations in many theories.
Everyone has very high expectation for the new run coming up
in May 2016 at the LHC.

Both ATLAS and CMS updated their findings in the early 2016 during
the Moriond Conference. In particular, the CMS also included a set of 
data without the magnetic field into the analysis, and improved the
significance to about $3.2\sigma$. The summary of the diphoton data
of the 750 GeV resonance is given in Table~\ref{tab:cx}.
Although the hint is preliminary, it has stimulated a lot of
phenomenological activities, bringing in a number of models for 
interpretation.
\footnote{There has been more than 300 articles appearing on arXiv
that interpret the 750 GeV particle.  We only refer to those relevant
to our work here.
}
If the particle decays directly into a pair of photons,
it can have a spin-0 or spin-2, however, one has to entertain the
possibility that the 750 GeV particle undergoes cascade decays into collimated
photon objects (aka photon-jets) \cite{photon-jet}.

During the ICHEP 2016 conference, both ATLAS and CMS reported
their searches including
the new 2016 data totaling about 12--13 fb$^{-1}$~\cite{ichep2016}.
The ATLAS collaboration reanalyzed the 2015 data of 3.2 fb$^{-1}$ and they
reported a little bit smaller excess of $3.4\sigma$ at 730 GeV,
compared to the previous $3.9\sigma$ excess at 750 GeV.
While, in the new 2016 data of 12.2 fb$^{-1}$, 
they have not observed any significant 
excess  at all.
In the combined data of 15.4 fb$^{-1}$, 
they observed $2.3\sigma$ excess at 710 GeV for the wide width case
with $\Gamma_X/m_X$=10\,\%.
In the narrow width case, the combined data show several $\sim 2\sigma$ 
excesses with the largest one at 1.6 TeV with a $2.4\sigma$ local significance.
The CMS collaboration has observed no significant excess in proximity of 750 GeV
in the new 2016 data of 12.9 fb$^{-1}$, either. But, interestingly, it reported
the largest excess newly appeared at 620 GeV with 
$\sim 2.4-2.7\sigma$ local significance. And, like as in the ATLAS  case,
the combined $\sqrt{s}=13$ TeV data of 16.2/fb show several 
excesses at the level of $\sim 2\sigma$.
Both experiments did not further find evidence of the 750 GeV
resonance.   Nevertheless, we do not give up.  Hints of new particles can 
easily show up in the near future data.
For example, the CMS data in ICHEP 2016 showed 
a new $2\sigma$ effect at around 620--650 GeV.
The ATLAS, on the other hand, did not see any new diphoton resonance
in the new 13 TeV data but still there was 
$3.4\sigma$ effect around 730 GeV in the 2015 data.
Under this situation, there are often
possibilities that potentially heavy diphoton resonances can show up
in the near future.
We borrow the example
of the 750 GeV particles hinted with 3.2 fb$^{-1}$ data at the end of 
2015 (though it did not appear in the 2016 data) to perform
an analysis of ``double Higgcision'' -- the precision-coupling
analysis involving both Higgs bosons.

In this work, we focus on the interpretation that 
this 750 GeV particle is a scalar
boson that links the SM sector with the hidden sector through the
Higgs-portal type interactions, in which 
an $SU(2)$ isospin-singlet scalar boson mixes with the SM Higgs boson 
through an angle $\alpha$ \cite{Chpoi:2013wga}.
We assume after mixing the lighter boson is the observed SM-like
Higgs boson $H_1$ at 125 GeV while the heavier one $H_2$ is the one hinted at
750 GeV.  Thus, the 750 GeV scalar boson $H_2$ opens the window to another  
hidden world containing perhaps dark matter and other exotic particles.  

In our previous global fits to the Higgs-portal type models
with all the Higgs boson data from Run I  \cite{portal} before the hint
of the 750 GeV boson, 
we have constrained the parameter space of a few 
Higgs-portal singlet-scalar models.  In those models without non-SM
contributions to the $h \gamma\gamma$ and $hgg$ vertices,
the mixing angle is constrained to $\cos\alpha > 0.86$ at 95\% CL. 
However, in those models with vector-like leptons (quarks) the mixing 
angle can be relaxed to $\cos\alpha > 0.83\;(0.7)$ at 95\% CL.
The implication was that the 750 GeV scalar boson $H_2$ can be produced in 
$gg$ fusion as if it were a 750 GeV SM Higgs boson but with a suppression
factor $\sin^2\alpha$ if there are no vector-like quarks running in the
$H_2 gg$ vertex. Additional contributions arise when there are vector-like
quarks running in the loop. Similarly, the decays of the scalar boson 
$H_2$ can be enhanced substantially into a pair of photons and gluons
in the presence of vector-like fermions.

In an earlier attempt when the 750 GeV particle was first hinted, we 
performed such an analysis in the Higgs-portal framework that the 750 GeV 
boson $H_2$ interacts with the SM particles via the mixing angle with
the 125 GeV Higgs boson and also via vector-like fermions \cite{earlier}.
Because the vector-like quarks carry electric and color charges while the
vector-like leptons carry electric charges, the 750 GeV boson can be
produced via gluon fusion and can also decay into a pair of photons and
gluons. In \cite{earlier} we used all the 125 GeV Higgs boson signal strength
data and also the diphoton cross section of the 750 GeV boson to constrain
the couplings of the 125 GeV Higgs boson, the mixing angle, and also
on the extra loop contributions to the 750 GeV boson due to the vector-like
fermions. 

In this work, we extend the earlier analysis into a full-swing analysis,
taking into account various combinations of 125 GeV Higgs couplings, 
the 750 GeV boson couplings to vector-like fermions, and the mixing angle.
Improvements are summarized as follows.
\begin{enumerate}
\item 
We include the effects of vector-like fermions in gluon fusion production
and 

\item 
We include the non-standard decay modes for the 750 GeV boson.

\item
We separately consider the choices of narrow and wide width for the 750 GeV
boson.  While the ATLAS data prefers a wide width, the CMS data prefers
the narrow width.

\end{enumerate}

The organization is as follows. In the next section, we describe briefly
the framework of Higgs-portal models with vector-like fermions,
including the production and decays of the 125 GeV and 750 GeV bosons.
In Sec. III, we describe the Higgs boson and 750 GeV boson data that we 
use in our analysis, and 
We present the fits for various combinations of couplings and the mixing angle
in Sec. IV.   In Sec. V, we discuss the decay rates for 
the 750 GeV boson into other diboson channels ($WW, ZZ$ and $Z\gamma$) 
when the vector-like  quarks in the loop are 
weak doublets and/or weak singlets with  arbitrary $U(1)_Y$ hypercharges.  
Then we conclude in Sec. VI. 

{\it Special note:}
After we posted this preprint to arXiv, both ATLAS and CMS announced
that they did not find evidence of the 750 GeV resonance in the new
2016 data.  We, nevertheless, think this double-Higgcision study would
still be a good exercise whenever another diphoton resonance shows up
in the future data. In the following, we shall borrow the data of the
750 GeV particles recorded with 3.2 fb$^{-1}$ luminosity at the end of
2015 to perform an analysis of ``double Higgcision'' -- the
precision-coupling analysis involving both Higgs bosons.

\section{Formalism}
Interpreting the $750$ GeV diphoton resonance as a scalar resonance generically
involves at least two interaction eigenstates of $h$ and $s$:
$h$ denotes the remnant of the SM Higgs doublet $H$ and 
$s$ the singlet or the remnant of additional Higgs doublets, triplets, etc.
Then the two states $h$ and $s$ mix and result in
the two mass eigenstates $H_{1,2}$. 
In the singlet case, for example, the mixing is generated from renormalizable
 potential terms such as
\begin{eqnarray}
V \supset \mu\,s\,H^\dagger H + \frac{\lambda}{2}s^2H^\dagger H\,.
\nonumber
\end{eqnarray}
In this work, for concreteness, we concentrate on the singlet case.

\subsection{Mixing and couplings}
The mass eigenstates
are related to the states $h$ and $s$
through an $SO(2)$ rotation as follows:
\begin{equation}
H_1  =  h\,\cos\alpha - s\,\sin\alpha\,; \ \ \ \
H_2  =  h\,\sin\alpha + s\,\cos\alpha\,
\end{equation}
with $\cos\alpha$ and $\sin\alpha$ describing
the mixing between the interaction eigenstates $h$ and $s$.
In the limit of $\sin\alpha\to 0$, $H_1\,(H_2)$ becomes the pure doublet
(singlet) state.
In this work,  we are taking $H_1$ for the 125 GeV boson
discovered at the 8-TeV LHC run
and $H_2$ for the 750 GeV state hinted at the early 13-TeV LHC run.
We are taking $\cos\alpha > 0$ without loss of
generality.
For the detailed description of this class of models and also   Higgs-portal
models, we refer to  Refs.~\cite{Chpoi:2013wga,portal}.

In this class of models, the singlet field $s$ does not directly couple 
to the SM
particles, but only through the mixing with the SM Higgs field at 
renormalizable 
level. The Yukawa interactions of $h$ and $s$ are described by
\begin{equation}
-{\cal L}_Y=h\sum_{f=t,b,\tau}\frac{m_f}{v}\bar{f}{f}
+s\sum_{F=Q,L}\,g^S_{s\bar{F}F}\bar{F}F\,,
\end{equation}
with $f$ denoting the 3rd-generation SM fermions and $F$ the extra
vector-like fermions (VLFs):
vector-like quarks (VLQs) and vector-like leptons (VLLs).
Thus, the couplings of the
two mass eigenstates  $H_{1,2}$ to the SM fermions and VLFs
are given by
\begin{eqnarray}
-{\cal L}_Y&=&
H_1\,\left[\cos\alpha\,\sum_{f=t,b,\tau}\frac{m_f}{v}\bar{f}{f}
-\sin\alpha\,\sum_{F=Q,L}\,g^S_{s\bar{F}F}\bar{F}F\right]
\nonumber \\
&+&
H_2\,\left[\sin\alpha\,\sum_{f=t,b,\tau}\frac{m_f}{v}\bar{f}{f}
+\cos\alpha\,\sum_{F=Q,L}\,g^S_{s\bar{F}F}\bar{F}F\right]\,.
\end{eqnarray}
Incidentally, the couplings to massive vector bosons $V=W,Z$ are
given by
\begin{equation}
{\cal L}_{HVV}=gM_W\left(W_\mu^+ W^{-\mu}+\frac{1}{2c_W^2}Z_\mu Z^\mu\right)
\left(\cos\alpha\,H_1 \ + \ \sin\alpha\,H_2\right)\,.
\end{equation}

The couplings of $H_{1,2}$ to two gluons, following the conventions and normalizations
of Ref.~\cite{Lee:2003nta}, are given by
\begin{eqnarray}
\label{eq:hgg}
S_{H_1}^g &=& \cos\alpha\,S^{g\,({\rm SM})}_{H_1}-\sin\alpha\,S^{g\,(Q)}_{H_1}
\nonumber \\
&\equiv &
\cos\alpha\,\sum_{f=t,b}\, F_{sf}(\tau_{1f})-
\sin\alpha\,\sum_Q\,g^S_{s\bar{Q}Q}\,\frac{v}{m_Q}\, F_{sf}(\tau_{1Q})\,,
\nonumber \\
S_{H_2}^g &=& \sin\alpha\,S^{g\,({\rm SM})}_{H_2}
+\cos\alpha\,S^{g\,(Q)}_{H_2} \nonumber \\
&\equiv &
\sin\alpha\,\sum_{f=t,b}\, F_{sf}(\tau_{2f})+
\cos\alpha\,\sum_Q\,g^S_{s\bar{Q}Q}\,\frac{v}{m_Q}\, F_{sf}(\tau_{2Q})\,,
\end{eqnarray}
where $\tau_{ix}=M_{H_i}^2/4m_x^2$.
We note that
$S^{g\,({\rm SM})}_{H_1}\simeq 0.651+0.050\,i$ for $M_{H_1}=125.5$ GeV
and
$S^{g\,({\rm SM})}_{H_2}\simeq 0.291+0.744\,i$ for $M_{H_2}=750$ GeV.
In the limit $\tau\to 0$, $F_{sf}(0)=2/3$.
The mass of extra fermion $F$ may be fixed by the relation
$m_F=v_s\,g^S_{s\bar{F}F}+m^0_F$ where $v_s$ denotes the VEV of the singlet $s$
while  $m^0_F$ is generated from a different origin other than $v_s$ as in
$-{\cal L}_{\rm mass}\supset m^0_F\bar{F}F$.
We note that when $m^0_Q=0$, each contribution from a VLQ is not
suppressed by $1/m_Q$ but by the common factor $1/v_s$.

Similarly, the couplings of $H_{1,2}$ to two photons are given by
\begin{eqnarray}
\label{eq:hpp}
S_{H_1}^\gamma &=& \cos\alpha\,S^{\gamma\,({\rm SM})}_{H_1}-
\sin\alpha\,S^{\gamma\,(F)}_{H_1}
\nonumber \\
&\equiv &
\cos\alpha\left[2\sum_{f=t,b,\tau}\, N_CQ_f^2 F_{sf}(\tau_{1f})
-F_1(\tau_{1W})\right]-
\sin\alpha\left[2\sum_F\,N_CQ_F^2 g^S_{s\bar{F}F}\,\frac{v}{m_F}\,
F_{sf}(\tau_{1F})\right]\,,
\nonumber \\
S_{H_2}^\gamma &=& \sin\alpha\,S^{\gamma\,({\rm SM})}_{H_2}
+\cos\alpha\,S^{\gamma\,(F)}_{H_2} \nonumber \\
&\equiv &
\sin\alpha\left[2\sum_{f=t,b,\tau}\,N_CQ_f^2 F_{sf}(\tau_{2f})
-F_1(\tau_{2W})\right]+
\cos\alpha\left[2\sum_F\,N_CQ_F^2 g^S_{s\bar{F}F}\,\frac{v}{m_F}\,
F_{sf}(\tau_{2F})\right]\,,\nonumber\\
\end{eqnarray}
where $N_C=3$ and $1$ for quarks and leptons, respectively, and
$Q_{f,F}$ denote the electric charges of fermions in the unit of $e$.
In the limit $\tau\to 0$, $F_1(0)=7$.
We note that
$S^{\gamma\,({\rm SM})}_{H_1}\simeq -6.55+0.039\,i$ for $M_{H_1}=125.5$ GeV
and
$S^{\gamma\,({\rm SM})}_{H_2}\simeq -0.94-0.043\,i$ for $M_{H_2}=750$ GeV.

\subsection{Production and Decay}
The production cross section of $H_{1,2}$ via the gluon-fusion process is given by
\begin{equation}
\sigma(gg\to H_{1,2}) =
\frac{|S^g_{H_{1,2}}|^2}{|S^{g\,({\rm SM})}_{H_{1,2}}|^2}\,\sigma_{\rm SM}(gg\to H_{1,2})
\end{equation}
with $\sigma_{\rm SM}(gg\to H_1)$ and
$\sigma_{\rm SM}(gg\to H_2)$ denoting
the corresponding SM cross sections for 
$M_{H_1}=125.5$ GeV and $M_{H_2}=750$ GeV, respectively.
We note that $\sigma_{\rm SM}(gg\to H_2) \approx 750$ fb at $\sqrt{s}=13$ TeV.

The total decay widths of $H_{1,2}$ can be cast into the form
\begin{eqnarray}
\Gamma_{H_1}&=&
\cos^2\alpha\Gamma^{\rm SM}_{H_1} +
\Delta\Gamma_{\rm vis}^{H_1\to\gamma\gamma,gg,Z\gamma} +
\Gamma^{\rm non-SM}_{H_1} \,;
\nonumber \\
\Gamma_{H_2}&=&
\sin^2\alpha\Gamma^{\rm SM}_{H_2} +
\Delta\Gamma_{\rm vis}^{H_2\to\gamma\gamma,gg,Z\gamma} +
\Gamma^{\rm non-SM}_{H_2} \,,
\end{eqnarray}
with $\Gamma^{\rm SM}_{H_1}\simeq 4$ MeV 
and $\Gamma^{\rm SM}_{H_2}\simeq 250$ GeV for the SM-like $H_2$
with $M_{H_2}=750$ GeV
\footnote{For $M_{H_2}=750$ GeV,
$\Gamma_{\rm SM}(H_2\to WW)\simeq 145$ GeV,
$\Gamma_{\rm SM}(H_2\to ZZ)\simeq 71.9$ GeV, and
$\Gamma_{\rm SM}(H_2\to t\bar{t})\simeq 30.6$ GeV.
\cite{Dittmaier:2011ti}.}. 
And $\Gamma^{\rm non-SM}_{H_{1,2}}$ denote additional partial decay widths of 
$H_{1,2}$ into non-SM particles which could be either visible or invisible. 
If the only non-SM particles into which $H_{1,2}$ can decay are invisible, 
one may have
\begin{equation}
\Gamma^{\rm non-SM}_{H_1} = \Delta\Gamma^{H_1}_{\rm inv} \,; \ \ \
\Gamma^{\rm non-SM}_{H_2} = \Gamma(H_2\to H_1H_1)+
\Delta\Gamma^{H_2}_{\rm inv} \,. 
\end{equation}
We note that $\Gamma^{\rm non-SM}_{H_2}$ includes the $H_2$ decay
into $H_1 H_1$ by definition. As we shall show that a 
sizeable $\Delta\Gamma^{H_2}_{\rm inv}$ into invisible particles 
such as dark matters
is required to accommodate a large $\Gamma_{H_2} = {\cal O}(10)$ GeV.

The quantities $\Delta\Gamma_{\rm vis}^{H_{1,2}\to\gamma\gamma\,, gg}$
are given by
\begin{eqnarray}
\Delta\Gamma^{H_1\to\gamma\gamma}_{\rm vis} &=&
\frac{M_{H_1}^3\alpha^2}{256\pi^3v^2}
\left[\left|S_{H_1}^\gamma \right|^2
-\cos^2\alpha \left|S_{H_1}^{\gamma\,({\rm SM})} \right|^2\right]
\,, \nonumber \\[2mm]
\Delta\Gamma^{H_1\to gg}_{\rm vis} &=&
\left[1+\frac{\alpha_s}{\pi}\left(\frac{95}{4}-7\right)\right]\,
\frac{M_{H_1}^3\alpha_s^2}{32\pi^3v^2}
\left[\left|S_{H_1}^g \right|^2
-\cos^2\alpha \left|S_{H_1}^{g\,({\rm SM})} \right|^2\right]\,,
\nonumber \\[2mm]
\Delta\Gamma^{H_2\to\gamma\gamma}_{\rm vis} &=&
\frac{M_{H_2}^3\alpha^2}{256\pi^3v^2}
\left[\left|S_{H_2}^\gamma \right|^2
-\sin^2\alpha \left|S_{H_2}^{\gamma\,({\rm SM})} \right|^2\right]
\,, \nonumber \\[2mm]
\Delta\Gamma^{H_2\to gg}_{\rm vis} &=&
\left[1+\frac{\alpha_s}{\pi}\left(\frac{95}{4}-7\right)\right]\,
\frac{M_{H_2}^3\alpha_s^2}{32\pi^3v^2}
\left[\left|S_{H_2}^g \right|^2
-\sin^2\alpha \left|S_{H_2}^{g\,({\rm SM})} \right|^2\right]\,,
\end{eqnarray}
with $\alpha_s=\alpha_s(M_{H_{1\,(2)}})$
for $\Delta\Gamma^{H_{1\,(2)}\to gg}$.

Before closing this section, we comment on the loop-induced decay widths
$\Delta\Gamma^{H_2\to Z\gamma}_{\rm vis}$.
The Higgs couplings are given by
\begin{equation}
\label{eq:hzp}
S^{Z\gamma}_{H_1}= \cos \alpha S^{Z\gamma(\rm SM)}_{H_1}-\sin \alpha S^{Z\gamma(\rm
F)}_{H_1}\,;  \ \ \
S^{Z\gamma}_{H_2}= \sin \alpha S^{Z\gamma(\rm SM)}_{H_2}+\cos \alpha S^{Z\gamma(\rm
F)}_{H_2}\,,
\end{equation}
with $S^{Z\gamma(\rm SM)}_{H_1} \simeq  -11.042+0.010i$ and
$S^{Z\gamma(\rm SM)}_{H_2} \simeq  -0.0771-1.805i$. The contributions from VLFs are
\begin{equation}
S^{Z\gamma(\rm F)}_{H_{1,2}} =
2\sum_F N_C Q_F \frac{2 g_{Z\bar{F}F}}{s_W c_W} \left( g^S_{s\bar{F}F} \frac{v}{m_F}
\right) m^2_F F^{(0)}_{f}(M^2_{H_{1,2}},m^2_F) ,
\end{equation}
where $Q_F$ and the couplings $g_{Z\bar{F}F}$ are defined in the interactions
$$
-{\cal L} = Q_F\; e\bar{F}\gamma^{\mu}FA_{\mu}
+\frac{e}{s_Wc_W}g_{Z\bar{F}F}\bar{F}\gamma^{\mu}FZ_{\mu}\,,
$$
and we note $2 m^2_F F^{(0)}_{f}(M^2_{H_{1,2}},m^2_F)=F_{sf}(0)=2/3$ in
the heavy $m_F$ limit, $m_F\to\infty$. Finally, the decay widths are given by
\begin{eqnarray}
\Delta\Gamma^{H_1\rightarrow Z\gamma}_{\rm vis} &=&
\frac{M^3_{H_1}\alpha^2}{128 \pi^3 v^2}\left( 1-\frac{M^2_Z}{M^2_{H_1}} \right)^3
\left[ |S^{Z\gamma}_{H_1}|^2 - \cos^2\alpha |S^{Z\gamma({\rm {SM}})}_{H_1}|^2\right] \,,
\nonumber \\
\Delta\Gamma^{H_2\rightarrow Z\gamma}_{\rm vis} &=&
\frac{M^3_{H_2}\alpha^2}{128 \pi^3 v^2}\left( 1-\frac{M^2_Z}{M^2_{H_2}} \right)^3
\left[ |S^{Z\gamma}_{H_2}|^2 - \sin^2\alpha |S^{Z\gamma({\rm {SM}})}_{H_2}|^2\right] \,.
\end{eqnarray}
With no available independent information on the $g_{Z\bar{F}F}$ couplings, 
we neglect $\Delta\Gamma^{H_{1,2}\rightarrow Z\gamma}_{\rm vis}$ by taking 
$g_{Z\bar{F}F}=0$ when we perform global fits
\footnote{
If $SU(2)$ symmetry is imposed onto the VLFs, the couplings of 
VLFs to photon and $Z$ are correlated such that $g_{Z \bar FF}$ is given by
$g_{Z \bar F F} = I_3^F - Q_F s_W^2$.
See Section~\ref{sec:vlq} for more discussions.}.

\section{Higgs Data}

\subsection{$H_1$ Data}
For $H_1$ with $M_{H_1}=125.5$ GeV, we use the signal strength data from
Refs.~\cite{Cheung:2013kla,Cheung:2014noa}.
The theoretical signal strengths may be written  as
\begin{equation}
\widehat\mu({\cal P},{\cal D}) \simeq
\widehat\mu({\cal P})\ \widehat\mu({\cal D}) \;,
\end{equation}
where ${\cal P}={\rm ggF}, {\rm VBF}, VH_1, ttH$ denote the $H_1$
production mechanisms: gluon fusion (ggF), vector-boson fusion (VBF),
and associated productions with a $V=W/Z$ boson ($VH_1$)
and top quarks ($ttH_1$)
and ${\cal D}=\gamma\gamma $, $ZZ,$ $WW,$ $b\bar{b},$ $\tau\bar\tau$
the decay channels.
Explicitly, we are taking
\begin{eqnarray}
\widehat\mu({\rm ggF}) &=& |S^g_{H_1}|^2/|S^{g({\rm SM})}_{H_1}|^2\,,
\nonumber \\
\widehat\mu({\rm VBF}) &=& \widehat\mu(VH_1) = 
\widehat\mu(ttH_1) = \cos^2\alpha
\end{eqnarray}
with $V=Z,W$. For the decay part,
\begin{equation}
\widehat\mu({\cal D}) = \frac{B(H_1\to {\cal D})}{B(H_{\rm SM}\to {\cal D})}
\end{equation}
with
\begin{equation}
B(H_1\to \gamma\gamma)=\frac{\Gamma(H_1\to\gamma\gamma)}{\Gamma_{H_1}}
=\frac{|S^\gamma_{H_1}|^2/|S^{\gamma({\rm SM})}_{H_1}|^2\,
\Gamma(H_{\rm SM}\to\gamma\gamma)}
{\cos^2\alpha\Gamma^{\rm SM}_{H_1} +
\Delta\Gamma_{\rm vis}^{H_1\to\gamma\gamma,gg,Z\gamma}+
\Gamma^{\rm non-SM}_{H_1}}
\end{equation}
and
\begin{equation}
B(H_1\to {\cal D})=\frac{\Gamma(H_1\to{\cal D})}{\Gamma_{H_1}}
=\frac{\cos^2\alpha\Gamma(H_{\rm SM}\to{\cal D})}
{\cos^2\alpha\Gamma^{\rm SM}_{H_1} +
\Delta\Gamma_{\rm vis}^{H_1\to\gamma\gamma,gg,Z\gamma}+
\Gamma^{\rm non-SM}_{H_1}}
\end{equation}
for ${\cal D}=ZZ, WW, b\bar{b}$ and $\tau\bar\tau$.
If there are no VLF contributions to the $H_1$ couplings to photons and gluons
or $S_{H_1}^{g(Q)}=S_{H_1}^{\gamma(F)}=0$,
the signal strengths are simply given by
\begin{equation}
\widehat\mu({\cal P},{\cal D}) \simeq
\frac{\cos^4\alpha}
{\cos^2\alpha+\Gamma^{\rm non-SM}_{H_1}/\Gamma^{\rm SM}_{H_1}}\,.
\end{equation}
For more details, we refer to Ref.~\cite{Cheung:2013kla}.

\subsection{$H_2$ Data}
For $H_2$ with $M_{H_2}=750$ GeV, we adopt the following cross sections
for the diphoton process $pp \to H_2 \to\gamma\gamma$
measured at $\sqrt{s}=13$ TeV~\cite{atlas2,cms2} in 2015:
\begin{eqnarray}
\sigma^{\rm ATLAS} &\approx& 9.7 \pm 3.2\, {\rm fb}~ 
({\rm for\, broad\, width})\,, 
\nonumber \\
\sigma^{\rm ATLAS} &\approx& 6.3 \pm 2.4\, {\rm fb}~ 
({\rm for\, narrow\, width})\,, 
\nonumber \\
\sigma^{\rm CMS} &\approx& 6.3^{+4.2}_{-3.1}\, {\rm fb}\,. 
\nonumber 
\end{eqnarray}
We also include the $8$-TeV CMS data which 
correspond to the following cross section at $\sqrt{s}=13$ TeV
\begin{eqnarray}
\sigma^{\rm CMS} &\approx& 3.5^{+2.2}_{-1.8}\, {\rm fb}\,. 
\nonumber 
\end{eqnarray}
In this work, we neglect the $8$-TeV ATLAS data since they
do not give positive-definite cross section at $1$-$\sigma$ level.
We note that the ATLAS Collaboration gave the cross sections for the 
broad- and narrow-width cases separately. For definiteness we apply the
broad-width value when $\Gamma_{H_2}\geq 40\,{\rm GeV}$ and
the narrow-width value when $\Gamma_{H_2}\leq 10\,{\rm GeV}$. 
However, we take the averaged value
\begin{eqnarray}
\sigma^{\rm ATLAS} &\approx& 8.0 \pm 2.8\, {\rm fb}~
\nonumber
\end{eqnarray}
for intermediate $\Gamma_{H_2}$ with $10\,{\rm GeV}<\Gamma_{H_2} < 40\,{\rm
GeV}$.
Strictly speaking, the CMS values are applicable only for the narrow-width case
but, with no available data, we apply the same value for the 
intermediate- and broad-width cases too.
Table~\ref{tab:cx} summarizes the experimental values of the 
cross sections $\sigma(pp\to H_2\to\gamma\gamma)$
at $\sqrt{s}=13$ GeV used in this work.
\begin{table}[t!]
  \caption{\small   \label{tab:cx}
The experimental values of the
cross sections $\sigma(pp\to H_2\to\gamma\gamma)\approx \sigma(pp\to H_2)\times
B(H_2\to\gamma\gamma)$
at $\sqrt{s}=13$ GeV.  }
\smallskip
\begin{tabular}{|c|rl|r|}
\hline\hline
   & & 13 TeV Data & 8 TeV Data  \\
\hline
      &  $6.3 \pm 2.4$ fb & for $\Gamma_{H_2}\leq 10$ GeV & \\
ATLAS &  $8.0 \pm 2.8$ fb & for $10$ GeV $< \Gamma_{H_2}< 40$ GeV & \\
      &  $9.7 \pm 3.2$ fb & for $\Gamma_{H_2}\geq 40$ GeV & \\
\hline
CMS & $6.3^{+4.2}_{-3.1}$ fb & & $3.5^{+2.2}_{-1.8}$ fb\\
\hline\hline
\end{tabular}
\end{table}

In this analysis, we further take into account
the following experimental constraints on
the $H_2$ production and its subsequent decays:
\begin{itemize}
\item Diboson:
$\left.\sigma(pp\rightarrow H_2)\right|_{\sqrt{s}=13\,{\rm TeV}}
\times B(H_2\rightarrow VV) \lesssim$ 150 fb~\cite{atlas3}
\item $t\bar{t}$: 
$\left.\sigma(pp\rightarrow H_2)\right|_{\sqrt{s}=8\,{\rm TeV}}
\times B(H_2\rightarrow t\bar{t}) \lesssim$ 0.5 pb~\cite{atlas5}
\item Dijet: 
$\left.\sigma(pp\rightarrow H_2)\right|_{\sqrt{s}=8\,{\rm TeV}}
\times B(H_2\rightarrow gg)
\lesssim$ 1 pb~\cite{atlas6}.
\end{itemize}

We would like to comment
on the constraint on $\Gamma(H_2\to H_1H_1)$ from the combined 95\% upper limit on
$\sigma(gg\to H_2)\times B(H_2\to H_1H_1)\lsim 45$ fb at $\sqrt{s}=8$ TeV
\cite{Aad:2015xja}:
\begin{equation}
\label{eq:h2h1h1}
\Gamma(H_2\to H_1H_1) \lsim  15\,{\rm GeV}\,
\left(\frac{150\,{\rm fb}}{\sigma(gg\to H_2)}\right)\,
\left(\frac{\Gamma_{H_2}}{50\,{\rm GeV}}\right)
\end{equation}
where we normalize the cross section $\sigma(gg\to H_2)$ using the corresponding SM 
Higgs production cross section for $M_{H_2}=750$ GeV at $\sqrt{s}=8$ TeV or
$\left.\sigma_{\rm SM}(gg\to H_2)\right|_{\sqrt{s}=8\,{\rm TeV}} \simeq 150$ fb
which is smaller by a factor of $4.693$ compared to 
$\left.\sigma_{\rm SM}(gg\to H_2)\right|_{\sqrt{s}=13\,{\rm TeV}}$.
If we parameterize the $H_2$-$H_1$-$H_1$ coupling as follows
\begin{equation}
{\cal L}_{3H}=g_{211}\,v\,H_2H_1H_1
\end{equation}
the constraint on $\Gamma(H_2\to H_1H_1)$ can be translated on 
the constraint on the coupling $g_{211}$:
\begin{equation}
\frac{g_{211}^2}{4\pi} \lsim  0.4\,
\left(\frac{150\,{\rm fb}}{\sigma(gg\to H_2)}\right)\,
\left(\frac{\Gamma_{H_2}}{50\,{\rm GeV}}\right)
\end{equation}
using
$$
\Gamma(H_2 \rightarrow H_1 H_1)=\,\frac{v^2\,g_{211}^2}{8\,\pi\,M_{H_2}}
\,\Bigg(1-\frac{4\, M_{H_1}^2}{M_{H_2}^2}\Bigg)^{\frac{1}{2}}\,.
$$
With a model-dependent coupling $g_{211}$, we do not impose any experimental
constraint from $\sigma(gg\to H_2)\times B(H_2\to H_1H_1)$. Instead,
as we shall see, we include $\Gamma(H_2\to H_1H_1)$ as a part of the 
free parameter which parameterizes
non-SM decays of $H_2$, as in
$\Gamma_{H_2}^{\rm non-SM}=\Gamma(H_2\to H_1H_1)+\Delta\Gamma^{H_2}_{\rm inv}$,
where the second term denotes additional partial decay widths of $H_1$ into 
invisible particles.

Finally, the vector-boson fusion (VBF) contribution to $H_2$ production 
is given by
\begin{equation}
\sigma^{\rm VBF}(pp\rightarrow H_2 jj)=\sin^2 \alpha \, \sigma^{\rm VBF}_{\rm
SM}(pp\rightarrow H_2 jj),
\end{equation}
with $\sigma^{\rm VBF}_{\rm SM}(pp\rightarrow H_2 jj)\simeq$ 130 fb with SM-like $H_2$
with mass 750 GeV at $\sqrt{s}=$13 TeV~\cite{handbook}.
While the gluon fusion process gives $\sigma(gg\to H_2) \sim 1250\,
\cos^2\alpha\,|S^{g(Q)}_{H_2}|^2$ fb at $\sqrt{s}=13$ TeV.
With $\sin^2\alpha\lsim 0.2$, as will be seen, and a possibly large value of 
$|S^{g(Q)}_{H_2}| \sim {\cal O}(1)$, we can 
safely ignore the VBF production of $H_2$ in this work.

\section{Fits}
In our approach, without loss of generality, we have the following 7 
model-independent parameters:
\begin{eqnarray}
&&
\sin\alpha\,; \nonumber \\
&&
S_{H_2}^{g(Q)}\,, \ \ \
S_{H_2}^{\gamma(F)}\,, \ \ \
\Gamma^{\rm non-SM}_{H_2}\,; \nonumber \\
&&
S_{H_1}^{g(Q)}\,, \ \ \
S_{H_1}^{\gamma(F)}\,, \ \ \
\Gamma^{\rm non-SM}_{H_1}\,.
\end{eqnarray}
In our numerical analysis, we shall restrict ourselves to the case
$2m_F>M_{H_2}$ so that  $H_2\to F\bar{F}$ decays are kinematically forbidden
and  $S^{g(Q),\gamma(F)}_{H_1,H_2}$ are all real.
Furthermore, we note that
\begin{eqnarray}
S_{H_1}^{g(Q)}  &=&
\sum_Q g^S_{s\bar{Q}Q}\frac{v}{m_Q}F_{sf}(\tau_{1Q})
\simeq\frac{2}{3}\sum_Q g^S_{s\bar{Q}Q}\frac{v}{m_Q}\,,
\nonumber \\[2mm]
S_{H_1}^{\gamma(F)}  &=&
2 \sum_F N_C Q_F^2 g^S_{s\bar{F}F}\frac{v}{m_F}F_{sf}(\tau_{1F})
\simeq \frac{4}{3}\sum_F N_C Q_F^2 g^S_{s\bar{F}F}\frac{v}{m_F}\,,
\end{eqnarray}
since $F_{sf}(\tau_{1F})\simeq F_{sf}(0)=2/3$. This may imply
$S^{g(Q),\gamma(F)}_{H_1}$ are not completely independent of
$S^{g(Q),\gamma(F)}_{H_2}$.
In the heavy $m_F$ limit $m_F\to\infty$, for example,
$F_{sf}(\tau_{1F})=F_{sf}(\tau_{2F})=F_{sf}(0)=2/3$ and we have
\begin{equation}
S_{H_1}^{g(Q)}=S_{H_2}^{g(Q)}\,, \ \ \
S_{H_1}^{\gamma(F)}=S_{H_2}^{\gamma(F)}\,.
\end{equation}
On the other hand, if all the VLFs are degenerate around $m_F \sim M_{H_2}/2$
or $F_{sf}(\tau_{2F})\simeq F_{sf}(1)=1$ and we have
\begin{equation}
S_{H_1}^{g(Q)}=\frac{2}{3} S_{H_2}^{g(Q)}\,, \ \ \
S_{H_1}^{\gamma(F)}=\frac{2}{3} S_{H_2}^{\gamma(F)}\,.
\end{equation}
For convenience we introduce
the parameters $\eta^{g(Q)}$ and $\eta^{\gamma(F)}$ are defined as in
\begin{equation}
S_{H_1}^{g(Q)} \equiv \eta^{g(Q)} S_{H_2}^{g(Q)}\,, \ \ \
S_{H_1}^{\gamma(F)} \equiv \eta^{\gamma(F)} S_{H_2}^{\gamma(F)}\,.
\end{equation}
We note that $|\eta^{g(Q)}|$ and $|\eta^{\gamma(F)}|$ take on values 
between $2/3$ and $1$ if all the couplings $g_{s\bar{F}F}$ are either positive or
negative,  but in general can take on any values.

\subsection {{\bf F4} fits}
We first consider the minimal {\bf F4} fit varying the following $4$ parameters:
\begin{equation}
\sin\alpha\,; \ \ 
S_{H_2}^{g(Q)}\,,  \ \
S_{H_2}^{\gamma(F)}\,, \ \
\Gamma^{\rm non-SM}_{H_2}\,. 
\end{equation}
For the remaining parameters,
first of all, we are taking $\Gamma^{\rm non-SM}_{H_1}=0$.
For the $\eta$ parameters, 
we consider the three extreme possibilities as follows:
\begin{itemize}
\item {\bf F4-1} with $\eta^{g(Q)}=\eta^{\gamma(F)}=0$ :
The VLFs are assumed not to contribute to the $H_1$ couplings to photons and gluons.
In this case, the $H_1$ sector communicates with the $H_2$ sector only through the mixing
angle $\sin\alpha$ and, accordingly, the signal strengths become
$\widehat\mu({\cal P},{\cal D}) \simeq \cos^2\alpha$
independently of the production mechanism ${\cal P}$ and the decay mode ${\cal D}$
\item {\bf F4-2} with $\eta^{g(Q)}=\eta^{\gamma(F)}=2/3$ :
The VLFs are assumed to be almost degenerate with their masses
around $M_{H_2}/2$.
\item {\bf F4-3} $\eta^{g(Q)}=\eta^{\gamma(F)}=1$ :
All the VLFs are much heavier than $H_2$.
\end{itemize}
And, the regions of the varying {\bf F4}-fit parameters are taken as follows:
\begin{itemize}
\item $|\sin\alpha|\leq 0.5$: We consider the 95\% confidence level (CL )
limit of $\cos\alpha \gsim 0.86$
obtained from the global fits to Higgs-portal models using 
the current LHC $H_1$ data~\cite{portal}.
We shall show that 
$|\sin\alpha|$ would be more stringently constrained in 
the {\bf F4-2} and {\bf F4-3} fits with non-zero $\eta^{g(Q)}$ and
$\eta^{\gamma(F)}$.
\item $|S^{g(Q)}_{H_2}|\leq 10$: We assume that $|S^{g(Q)}_{H_2}|$ 
cannot be larger than $10$. In order to achieve the maximal value of 
$S^{g(Q)}_{H_2} \sim 10$, for example,
there should be more than $20$ VLQs with 
$m_Q \sim 500$ GeV and $g_{s\bar{Q}Q} \sim 1$.
As we shall show, the $H_2$ dijet constraint gives
$|S^{g(Q)}_{H_2}|\leq 7$.
\item $|S^{\gamma(F)}_{H_2}|\leq 100$: We consider a 10 times larger region
for $S^{\gamma(F)}_{H_2}$ compared to $S^{g(Q)}_{H_2}$ because 
of the possible enhancement factor $2N_C Q_F^2$ and additional 
contributions from VLLs to the $H_2$ couplings to photons.
\item $\Gamma_{H_2}^{\rm non-SM}\lsim 50$ GeV: We restrict to the case
in which the total width of $H_2$ does not exceed $50$ GeV
\end{itemize}

%
\begin{table}[t!]
  \caption{\small \label{tab:F4_1}
Best-fit values of {\bf F4-1} with $\eta^{\gamma(F)}=\eta^{g(Q)}=0$
and $\Gamma_{H_1}^{\rm non-SM}=0$, and 
varying: $\sin\alpha\subset [-0.5:0.5], \,
S_{H_2}^{g(Q)}\subset [-10:10], \,
S_{H_2}^{\gamma(F)}\subset [-100:100], \,
\Gamma^{\rm non-SM}_{H_2}/{\rm GeV}\subset [0:50]$.
The decay widths are in units of GeV and the cross sections in units of fb.
Other than the model parameters,
the quantities shown are:
$\Gamma_{H_2}$ for the total decay width of $H_2$,
$C_{H_1}^{g,\gamma}\equiv |S_{H_1}^{g,\gamma}|/|S_{H_1}^{g({\rm SM}),\gamma ({\rm SM})}|$,
$\sigma(H_2)\equiv \sigma(gg\rightarrow H_2)$ at $\sqrt{s}=13$ TeV,  and
$\sigma(XX)\equiv\sigma(gg\rightarrow H_2\to XX)=
\sigma(gg\rightarrow H_2)\times B(H_2 \rightarrow XX)$ at 
the same value of $\sqrt{s}$.
The first line shows the best-fit values for
the global minimum over the full range of $\Gamma_{H_2}$,
whereas  the second line shows the results for the broad-width case
under the assumption of $\Gamma_{H_2}\geq$ 40 GeV.
  }
\begin{ruledtabular}
\begin{tabular}{ l|ccc|ccccccc}
Fits  & $\chi^2_{\rm tot}$ & $\chi^2_{H_1}$ & $\chi^2_{H_2}$ &
\multicolumn{7}{c}{Best-fit values} \\
& & & & $\sin \alpha$ & $S_{H_2}^{\gamma(F)}$ & $S_{H_2}^{g(Q)}$ &
$\Gamma^{\rm non-SM}_{H_1}$ &
$\Gamma^{\rm non-SM}_{H_2}$ & $\eta^{\gamma(F)}$ & $\eta^{g(Q)}$ \\
\hline\hline
{\bf F4-1}
& $17.692$ & $16.764$ & $0.929$ & $-4.6\times 10^{-4}$ &
 $55.583$ & $-0.299$ & $0$ & $3.012$ & $0$ & $0$ \\
 & $19.363$ & $16.764$ & $2.600$ & $4.4\times 10^{-4}$ &
 $-45.897$ & $1.466$ & $0$ & $46.191$ & $0$ & $0$
\end{tabular}
\begin{tabular}{lcccccccc}
 \multicolumn{8}{c}{Best-fit values} \\
$\Gamma_{H_2}$ & $C^{\gamma}_{H_1}$ & $C^g_{H_1}$ & $\sigma(H_2)$ &
$\sigma(\gamma\gamma)$ &
        $\sigma(WW)$  & $\sigma(ZZ)$ &
        $\sigma(t\bar{t})$ &
        $\sigma(gg)$   \\
\hline\hline
$3.164$ & $1.00$ & $1.00$ & $111.9$ & $5.105$ & $0.00$ & $0.00$ & $0.00$ & $0.28$  \\
$46.480$ & $1.00$ & $1.00$ & $2693$ & $5.696$ & $0.00$ & $0.00$ & $0.00$ & $11.05$
\end{tabular}
\end{ruledtabular}
\end{table}
\begin{table}[t!]
  \caption{\small \label{tab:F4_2}
The same as in TABLE~\ref{tab:F4_1} but
with $\eta^{\gamma(F)}=\eta^{g(Q)}=2/3$.
}
\begin{ruledtabular}
\begin{tabular}{l|ccc|ccccccc}
Fits  & $\chi^2_{\rm tot}$ & $\chi^2_{H_1}$ & $\chi^2_{H_2}$ &
\multicolumn{7}{c}{Best-fit values} \\
& & & & $\sin \alpha$ & $S_{H_2}^{\gamma(F)}$ & $S_{H_2}^{g(Q)}$ &
$\Gamma^{\rm non-SM}_{H_1}$ &
$\Gamma^{\rm non-SM}_{H_2}$ & $\eta^{\gamma(F)}$ & $\eta^{g(Q)}$ \\
\hline\hline
{\bf F4-2}
& $16.736$ & $15.807$ & $0.929$ & $3.570\times 10^{-2}$ &
 $29.400$ & $0.329$ & $0$ & $0.777$ & $2/3$ & $2/3$ \\
& $18.412$ & $15.812$ & $2.600$ & $1.447\times 10^{-2}$ &
 $74.395$ & $0.893$ & $0$ & $45.55$ & $2/3$ & $2/3$
\end{tabular}
\begin{tabular}{lcccccccc}
\multicolumn{8}{c}{Best-fit values} \\
$\Gamma_{H_2}$ & $C^{\gamma}_{H_1}$ & $C^g_{H_1}$ & $\sigma(H_2)$ &
$\sigma(\gamma\gamma)$ &
        $\sigma(WW)$  & $\sigma(ZZ)$ &
        $\sigma(t\bar{t})$ &
        $\sigma(gg)$   \\
\hline\hline
$1.146$ & $1.105$ & $0.987$ & $144.9$ & $5.085$ & $20.83$ & $10.21$ & $4.84$ & $1.30$
\\
$45.93$ & $1.108$ & $0.987$ & $1009$ & $5.673$ & $0.59$ & $0.29$ & $0.14$ & $1.57$
\end{tabular}
\end{ruledtabular}
\end{table}
\begin{table}[th!]
  \caption{\small \label{tab:F4_3}
The same as in TABLE~\ref{tab:F4_1} but
with $\eta^{\gamma(F)}=\eta^{g(Q)}=1$.
  }
\begin{ruledtabular}
\begin{tabular}{l|ccc|ccccccc}
Fits  & $\chi^2_{\rm tot}$ & $\chi^2_{H_1}$ & $\chi^2_{H_2}$ &
\multicolumn{7}{c}{Best-fit values} \\
& & & & $\sin \alpha$ & $S_{H_2}^{\gamma(F)}$ & $S_{H_2}^{g(Q)}$ &
$\Gamma^{\rm non-SM}_{H_1}$ &
$\Gamma^{\rm non-SM}_{H_2}$ & $\eta^{\gamma(F)}$ & $\eta^{g(Q)}$ \\
\hline\hline
{\bf F4-3}
& $16.740$ & $15.810$ & $0.929$ & $-2.193\times 10^{-2}$ &
 $-32.203$ & $-0.396$ & $0$ & $1.753$ & $1$ & $1$ \\
& $18.413$ & $15.813$ & $2.600$ & $-0.933\times 10^{-2}$ &
 $-75.915$ & $-0.857$ & $0$ & $43.50$ & $1$ & $1$
\end{tabular}
\begin{tabular}{lcccccccc}
\multicolumn{8}{c}{Best-fit values} \\
$\Gamma_{H_2}$ & $C^{\gamma}_{H_1}$ & $C^g_{H_1}$ & $\sigma(H_2)$ &
$\sigma(\gamma\gamma)$ &
        $\sigma(WW)$  & $\sigma(ZZ)$ &
        $\sigma(t\bar{t})$ &
        $\sigma(gg)$   \\
\hline\hline
$1.936$ & $1.106$ & $0.986$ & $203.5$ & $5.082$ & $6.53$ & $3.20$ & $1.52$ & $1.51$  \\
$43.85$ & $1.107$ & $0.988$ & $926.5$ & $5.686$ & $0.24$ & $0.12$ & $0.06$ & $1.39$
\end{tabular}
\end{ruledtabular}
\end{table}

In Tables~\ref{tab:F4_1}, \ref{tab:F4_2}, and \ref{tab:F4_3}, we show the 
best-fit
values for the model parameters and miscellaneous quantities for
the {\bf F4-1}, {\bf F4-2}, and {\bf F4-3} fit, respectively.
We first note that the global minima occur for the small value of $\Gamma_{H_2}=1$-$3$
GeV though the larger widths are less preferred only by a small 
$\Delta\chi^2< 2$.
The broad-width minima under the assumption of $\Gamma_{H_2}>40$ GeV give
$\Gamma_{H_2}\sim 45$ GeV.
The best-fit values of $\sin\alpha$ are either small or 
vanishingly small, independent of $\eta^{g(Q),\gamma(F)}$. 

The best-fit values for the cross section $\sigma(gg\to H_2\to\gamma\gamma)$ 
are $5.1$ fb and $5.7$ fb, again independent of $\eta^{g(Q),\gamma(F)}$, 
for the global and broad-width minima, respectively. We find that
\begin{equation}
\label{eq:cx_aa}
\sigma(gg\to H_2\to\gamma\gamma) \simeq 
0.06\,\frac{\left(S^{g(Q)}_{H_2}S^{\gamma(F)}_{H_2}\right)^2}
{\Gamma_{H_2}/{\rm GeV}}\,\ {\rm fb}\,.
\end{equation}
Incidentally, we find
\begin{equation}
\label{eq:cx_h2}
\sigma(gg\to H_2) \simeq 1250\,\left(S^{g(Q)}_{H_2}\right)^2\, \ {\rm fb}\,,
\end{equation}
and
\begin{equation}
\label{eq:cx_gg}
\sigma(gg\to H_2\to gg) \simeq 
110\,\frac{\left(S^{g(Q)}_{H_2}\right)^4}
{\Gamma_{H_2}/{\rm GeV}}\,\ {\rm fb}\,.
\end{equation}
For the {\bf F4-2} and {\bf F4-3} fits
\begin{equation}
\label{eq:cx_ww}
\sigma(gg\to H_2\to WW) \simeq 
16.5\times 10^{4}\,\frac{\left(S^{g(Q)}_{H_2}\right)^2\,\sin^2\alpha}
{\Gamma_{H_2}/{\rm GeV}}\,\ {\rm fb}\,,
\end{equation}
and $\sigma(gg\to H_2\to ZZ) \sim \sigma(gg\to H_2\to WW)/2$ and
and $\sigma(gg\to H_2\to t\bar{t}) \sim \sigma(gg\to H_2\to WW)/4$.
\footnote{ 
Here we have assumed that the VLFs are singlet and thus do not couple
directly to $W$ bosons. However, if the VLFs are arranged into $SU(2)$
doublets, the VLFs can couple directly to $W$ bosons and thus contributing
to the $H_2 \to WW$ decay via loops.  See Section \ref{sec:vlq} for more discussions.
} 
Finally,
\begin{eqnarray}
C_{H_1}^\gamma &=& \left|\frac{S_{H_1}^{\gamma}}{S_{H_1}^{\gamma ({\rm SM})}}\right|
\simeq
\left| 1 +\sin\alpha\frac{\eta^{\gamma(F)}S^{\gamma(F)}_{H_2}}{6.55}
\right| \,, \nonumber \\[2mm]
C_{H_1}^g &=& \left|\frac{S_{H_1}^{g}}{S_{H_1}^{g ({\rm SM})}}\right|
\simeq
\left| 1 -\sin\alpha\frac{\eta^{g(Q)}S^{g(Q)}_{H_2}}{0.653}
\right| \,.
\end{eqnarray}

\begin{figure}[th!]
\centering
\includegraphics[height=2.0in,angle=270]{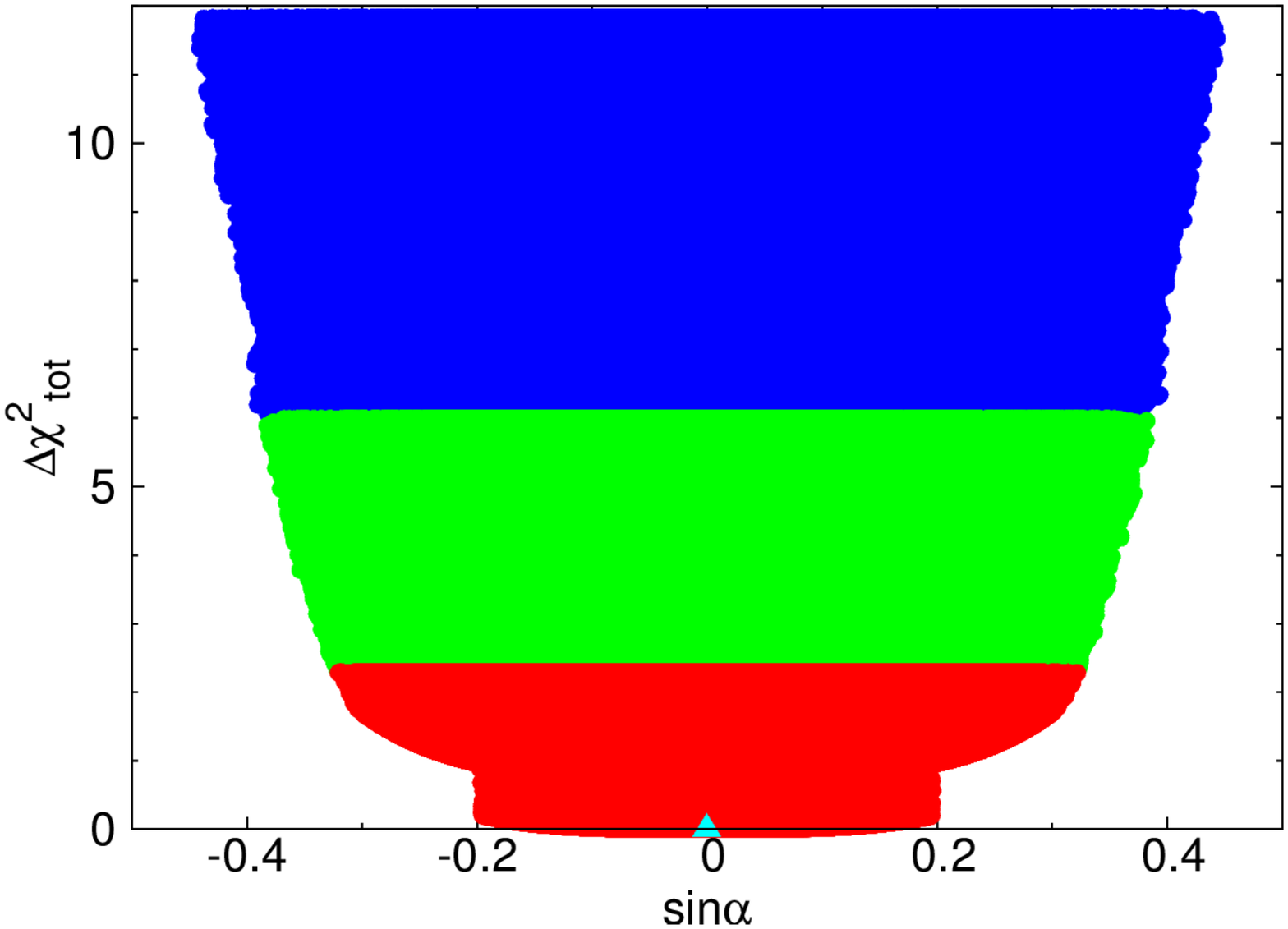}
\includegraphics[height=2.0in,angle=270]{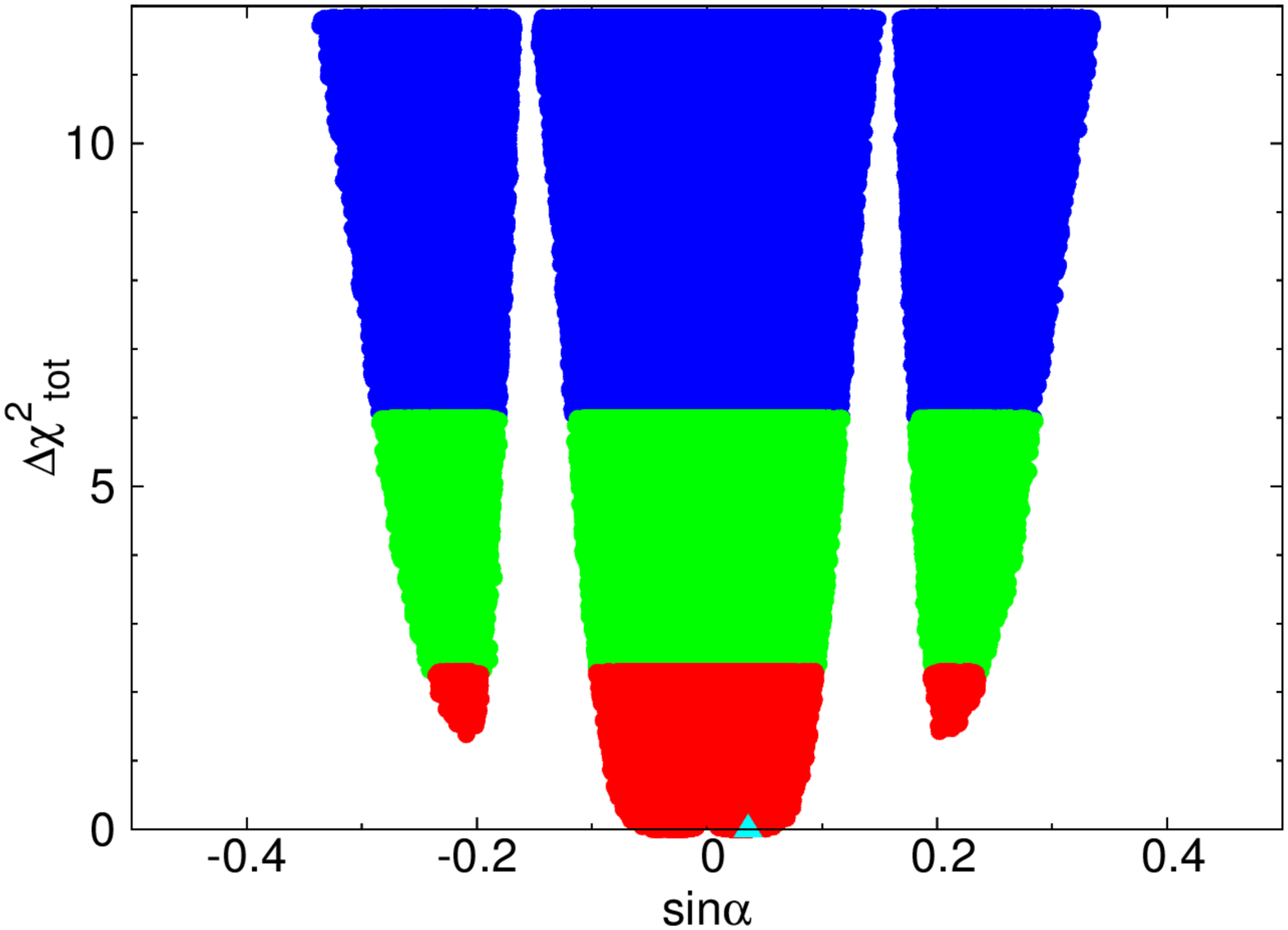}
\includegraphics[height=2.0in,angle=270]{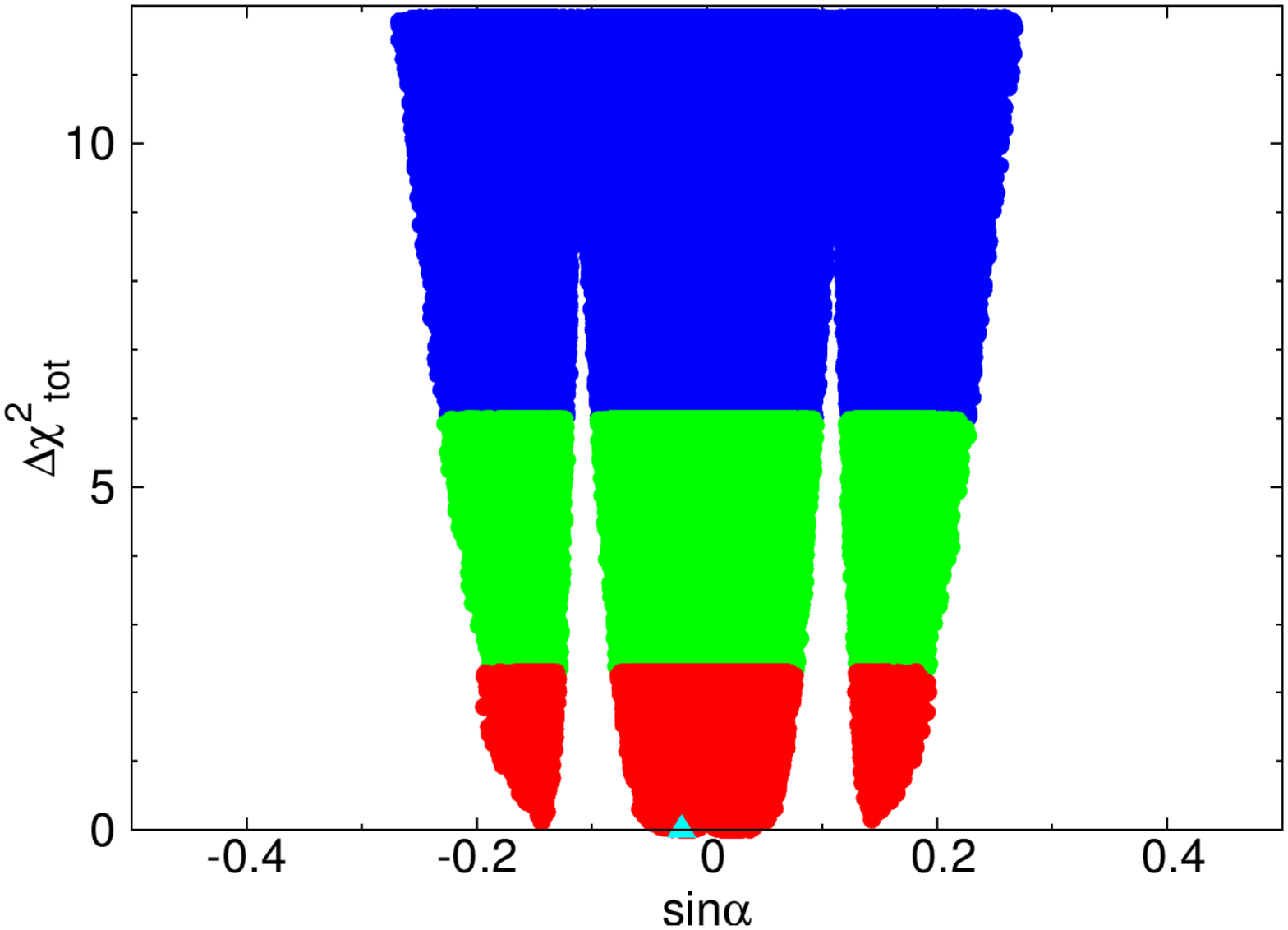}
\includegraphics[height=2.0in,angle=270]{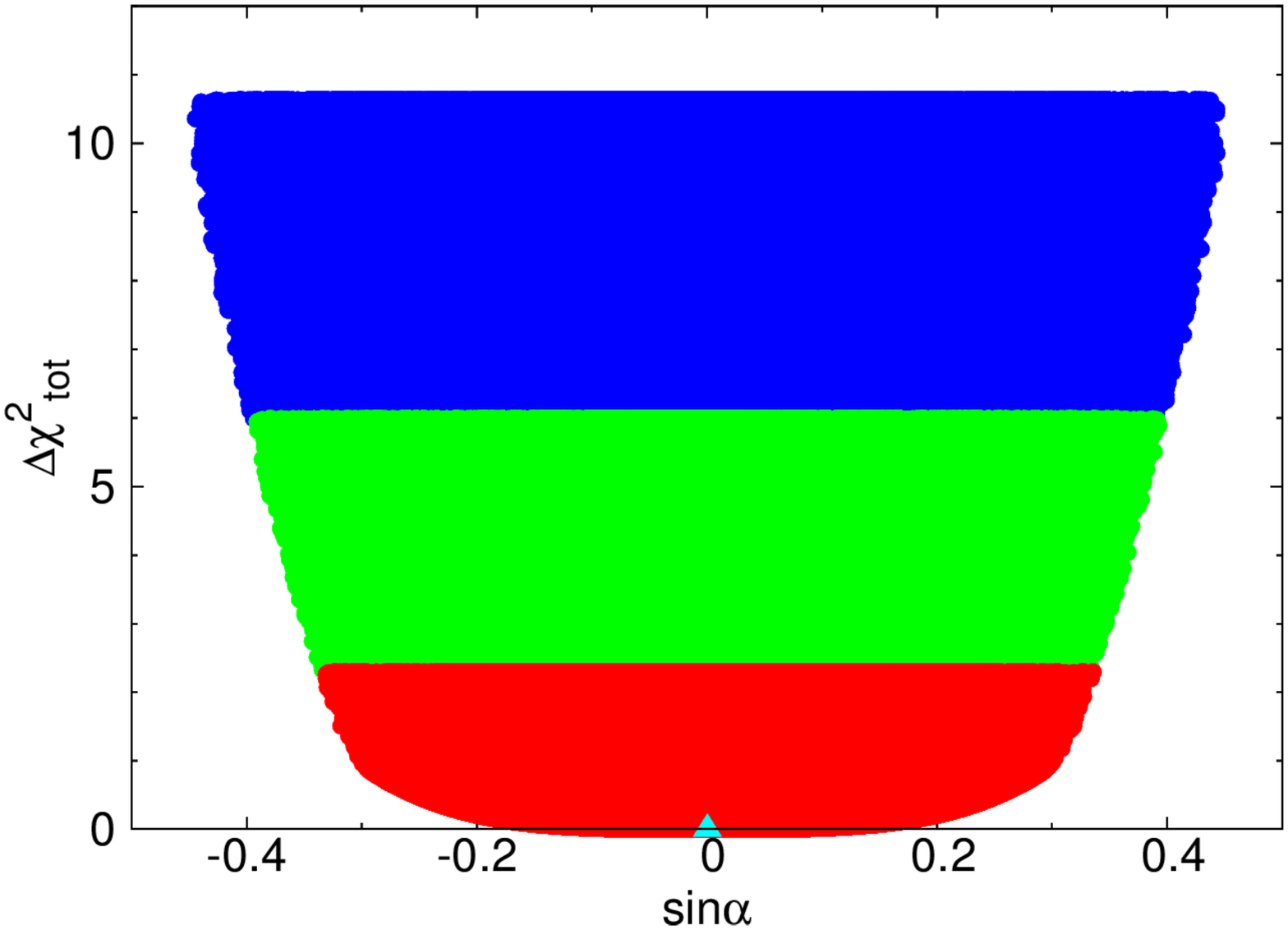}
\includegraphics[height=2.0in,angle=270]{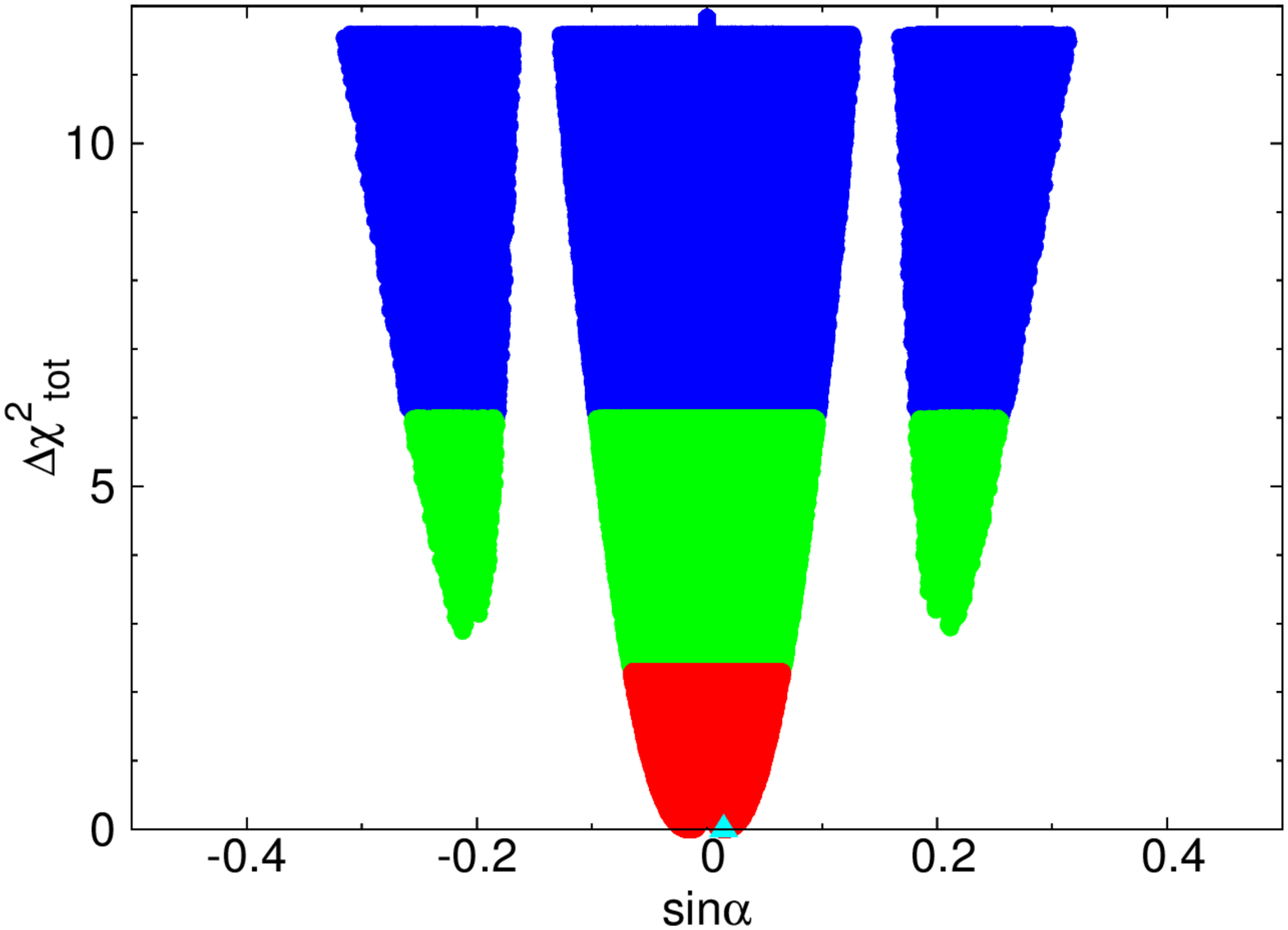}
\includegraphics[height=2.0in,angle=270]{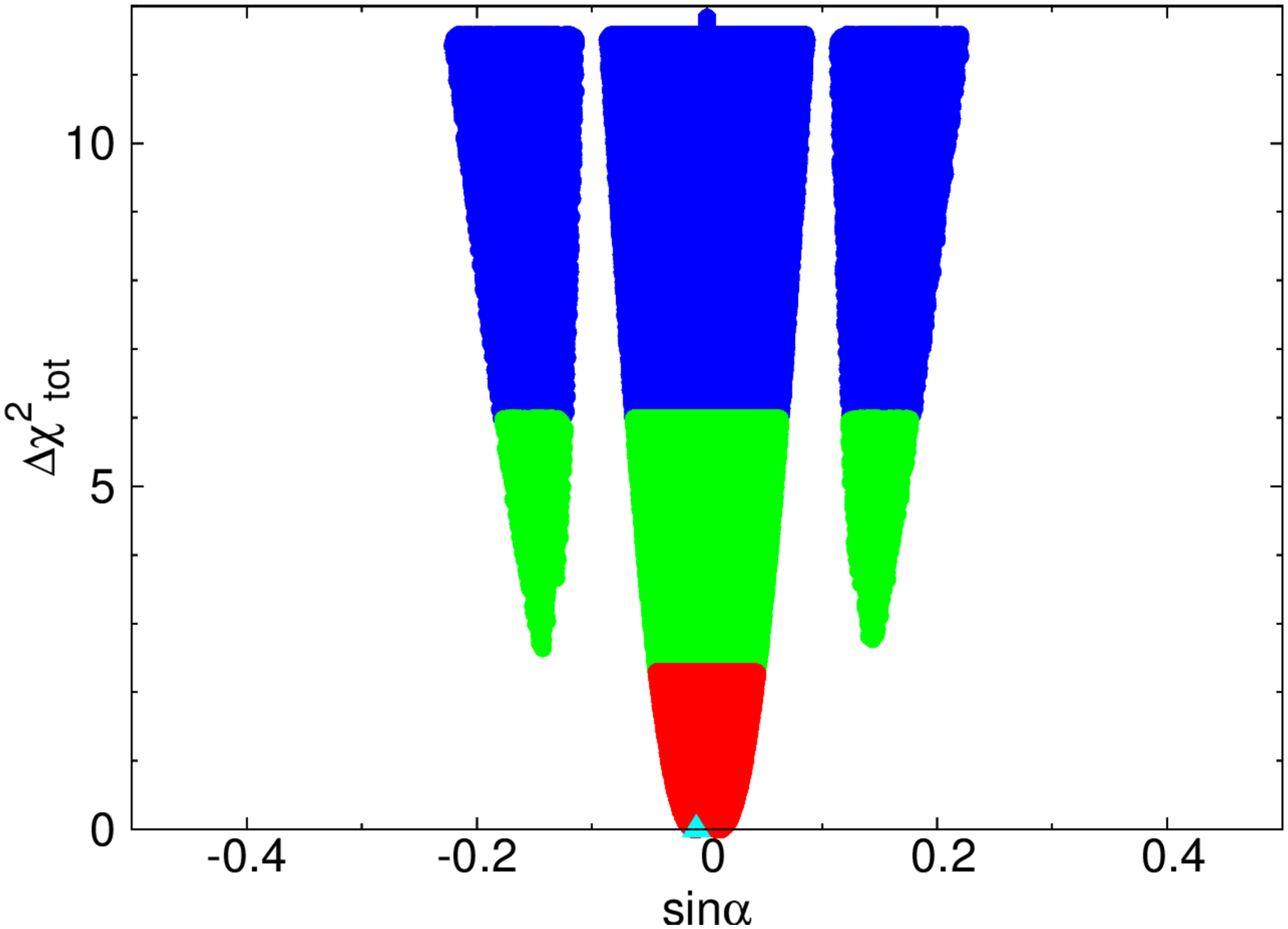}
\caption{\small \label{F4:chi_sin}
{\bf{F4 fits}}: 
Plots of $\Delta\chi^2$ vs $\sin\alpha$
for the {\bf F4-1} (left), {\bf F4-2} (middle),
and {\bf F4-3} (right) fits.
In the upper row, we consider the full range of $\Gamma_{H_2}$ while,
in the lower row, we consider the wide-width case requiring
$\Gamma_{H_2}\geq$ 40 GeV.
The regions shown are for
$\Delta \chi^2 = 2.3$ (red), $5.99$ (green), and $11.83$ (blue)
above the minimum.
The triangles denote the corresponding minima.
}
\end{figure}

Figure~\ref{F4:chi_sin} shows
$\Delta\chi^2$ vs $\sin\alpha$
for the {\bf F4-1} (left), {\bf F4-2} (middle),
and {\bf F4-3} (right) fits.
The upper frames show the results over the full range of $\Gamma_{H_2}$,
while the lower frames show the results for the broad-width case
under the assumption of $\Gamma_{H_2}\geq$ 40 GeV.
The regions shown are for
$\Delta \chi^2 = 2.3$ (red), $5.99$ (green), and $11.83$ (blue)
above the minimum and
the triangles denote the corresponding minima.
First of all, we note that the minima occur at
$\sin\alpha\sim 0$ in all the cases.
Also, since $\Gamma_{H_2}\geq\sin^2\alpha\Gamma_{H_2}^{\rm SM}$,
we observe 
\begin{equation}
\sin^2\alpha \leq \frac{\Gamma_{H_2}}{\Gamma_{H_2}^{\rm SM}}
\end{equation}
which implies, for example,
$|\sin\alpha| \lsim 0.2\,(0.4)$  when
$\Gamma_{H_2} \lsim 10\,(40)$ GeV and
$|\sin\alpha|$ cannot exceed $\sim 0.45$ if $\Gamma_{H_2}\leq 50$ GeV.
In the {\bf F4-1} fits, as shown in Table~\ref{tab:F4_1},
the minimum for the full range of $\Gamma_{H_2}$ 
is deeper than that for the broad-width case. 
From the upper-left frame of Fig.~\ref{F4:chi_sin},
$|\sin\alpha| \lsim 0.2$ in the $\Delta\chi^2 \lsim 1$ region and
we find $\Gamma_{H_2} \lsim 10$ GeV there.
In the {\bf F4-2} and {\bf F4-3} fits, in addition to the global minima
at $\sin\alpha\sim 0$, two more local minima are developed 
at non-zero $\sin\alpha$.
The local minima are developed at
$|\sin\alpha|\simeq 0.20\,(0.13)$
for the {\bf  F4-2} ({\bf  F4-3}) fits
when $S_{H_1}^\gamma = - S^{\gamma\,({\rm SM})}_{H_1}$ and
$|S^{\gamma\,(F)}_{H_1}|\simeq 100$, as we shall show soon.

\begin{figure}[th!]
\centering
\includegraphics[height=2.0in,angle=270]{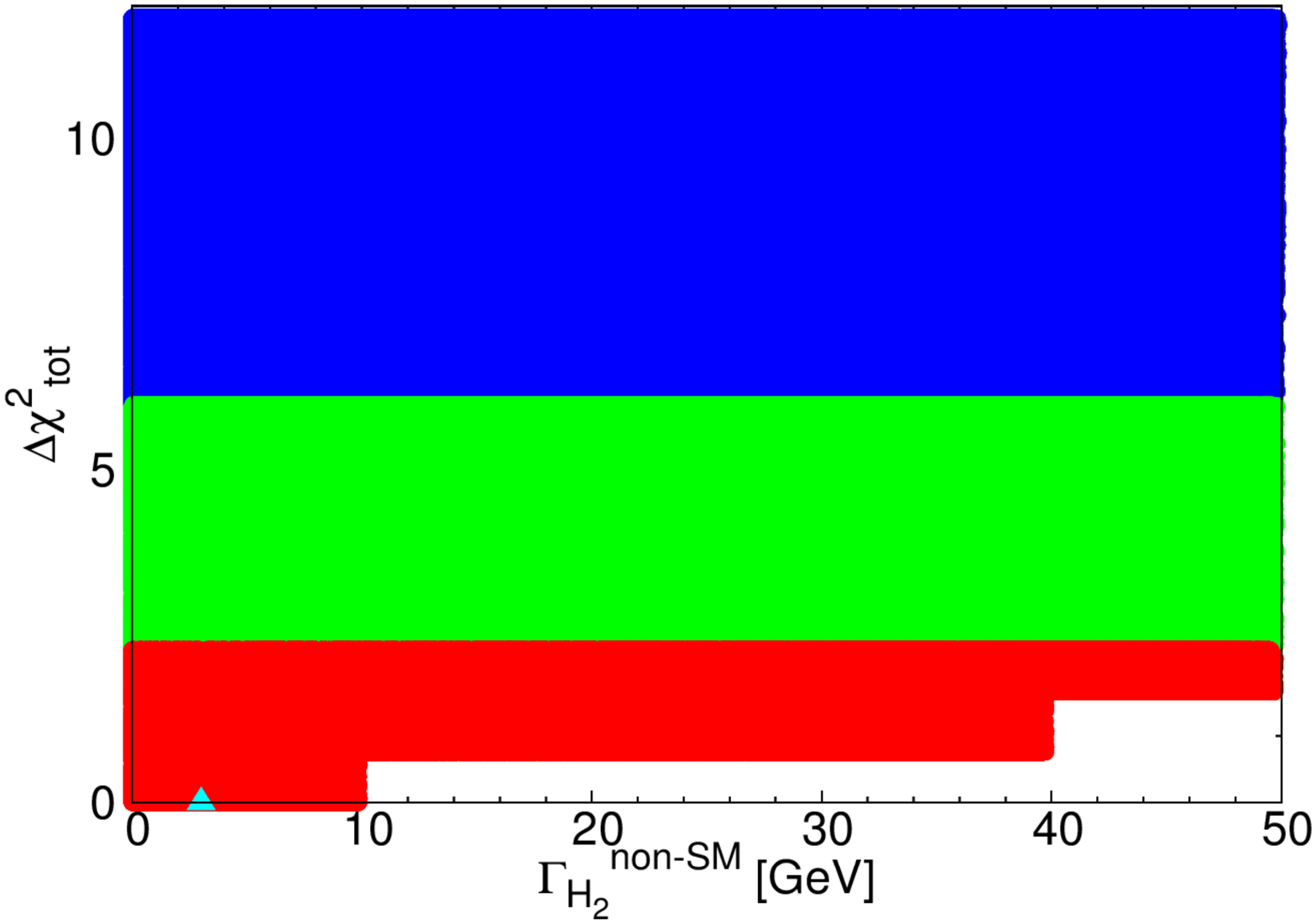}
\includegraphics[height=2.0in,angle=270]{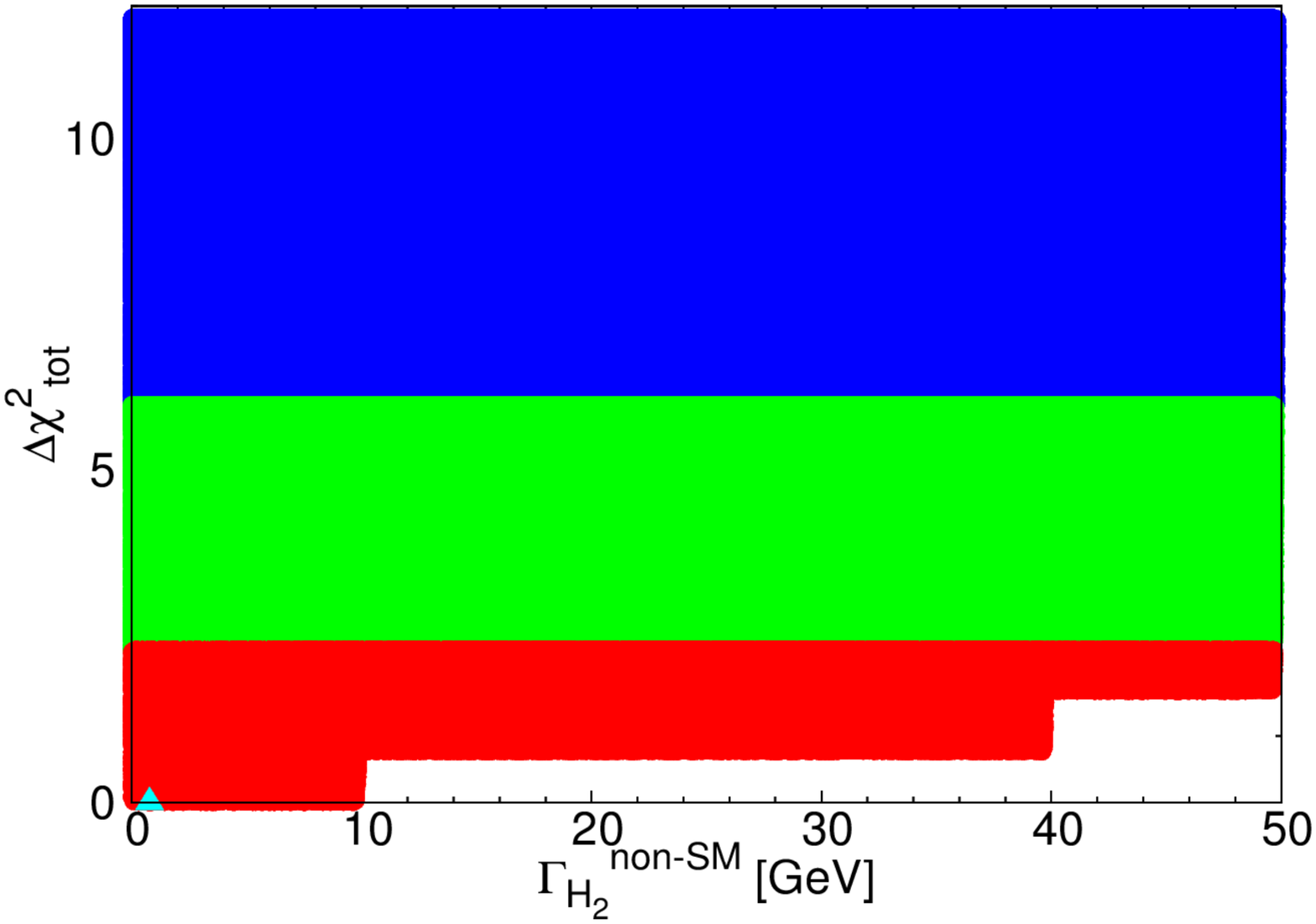}
\includegraphics[height=2.0in,angle=270]{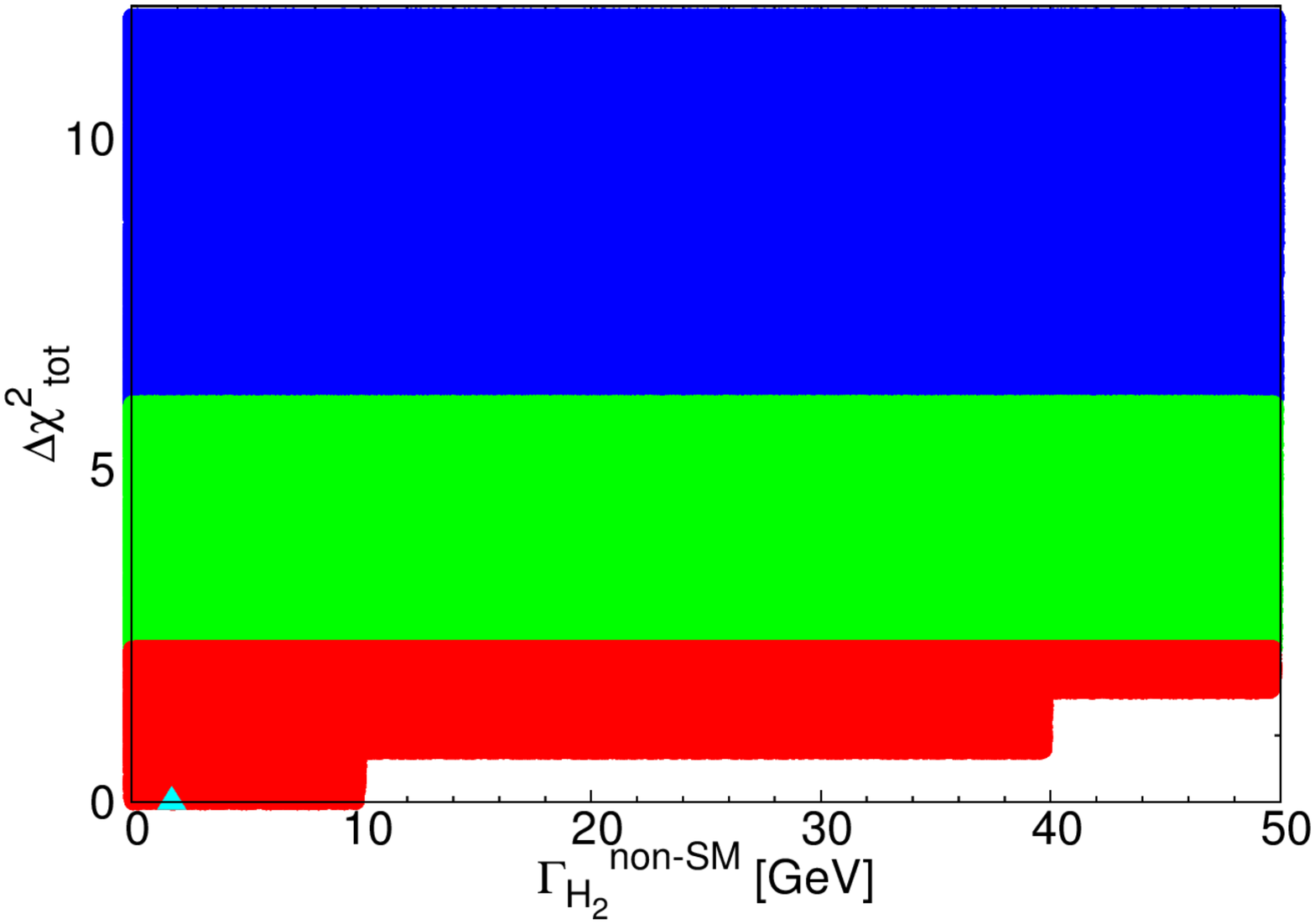}
\includegraphics[height=2.0in,angle=270]{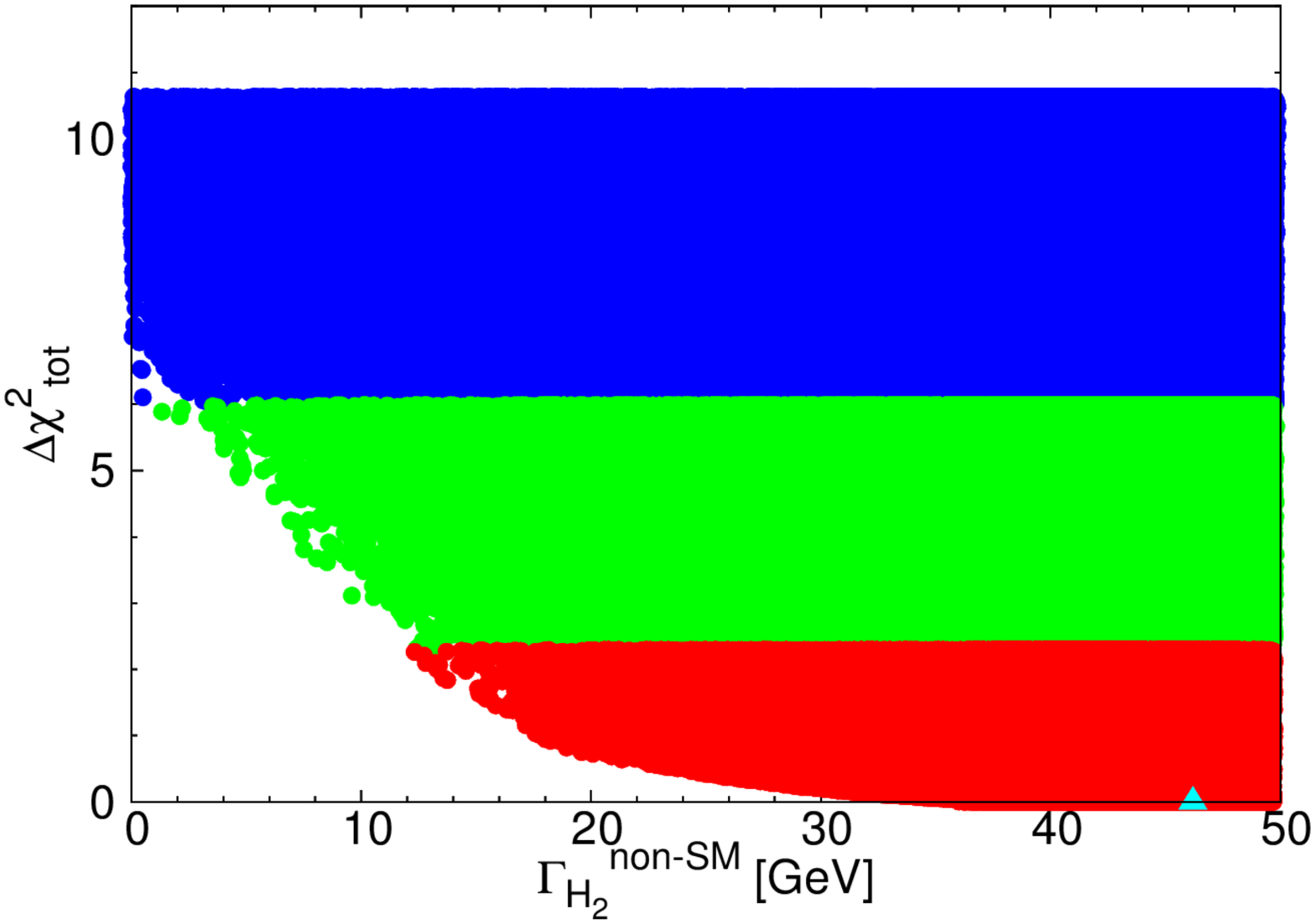}
\includegraphics[height=2.0in,angle=270]{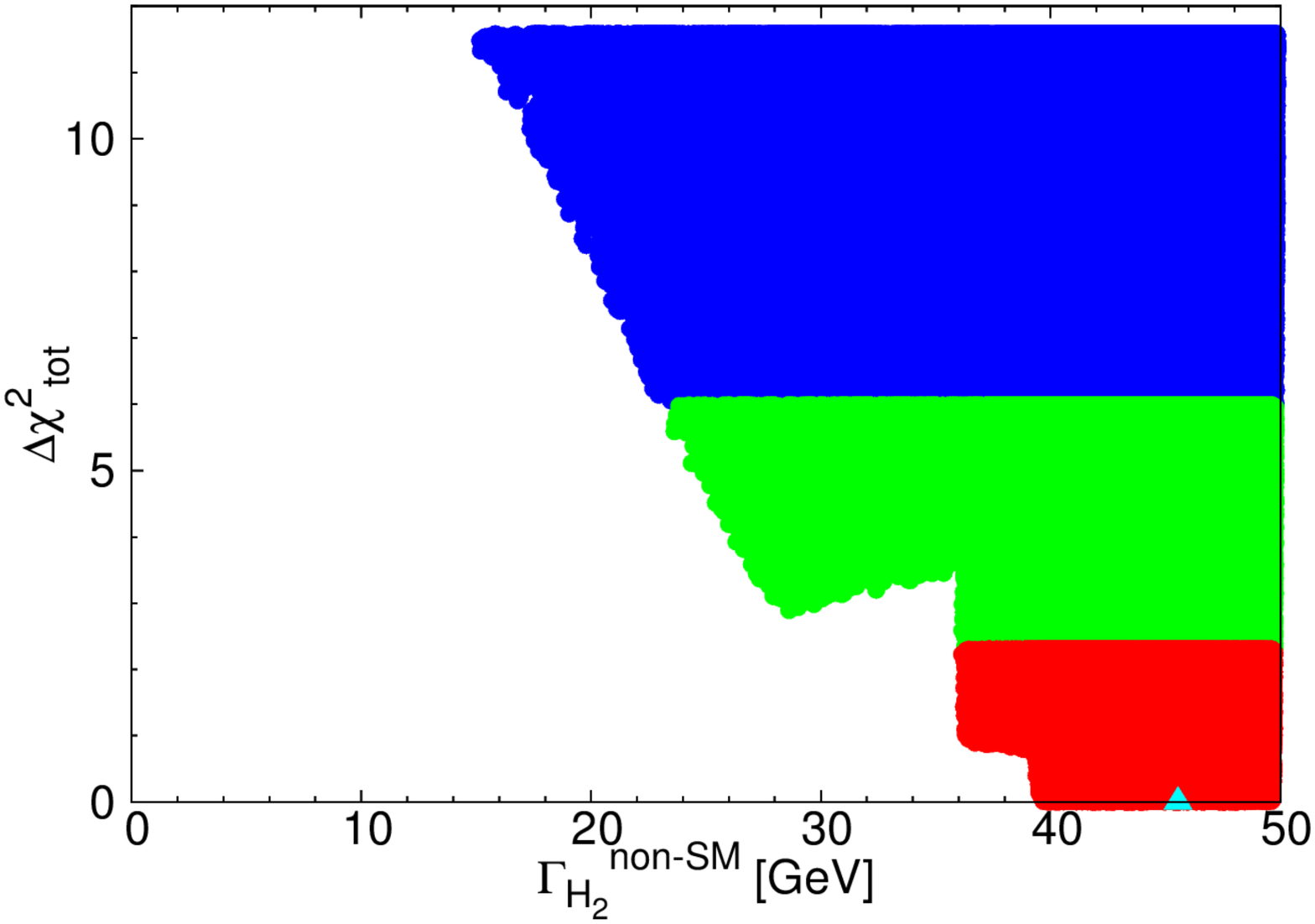}
\includegraphics[height=2.0in,angle=270]{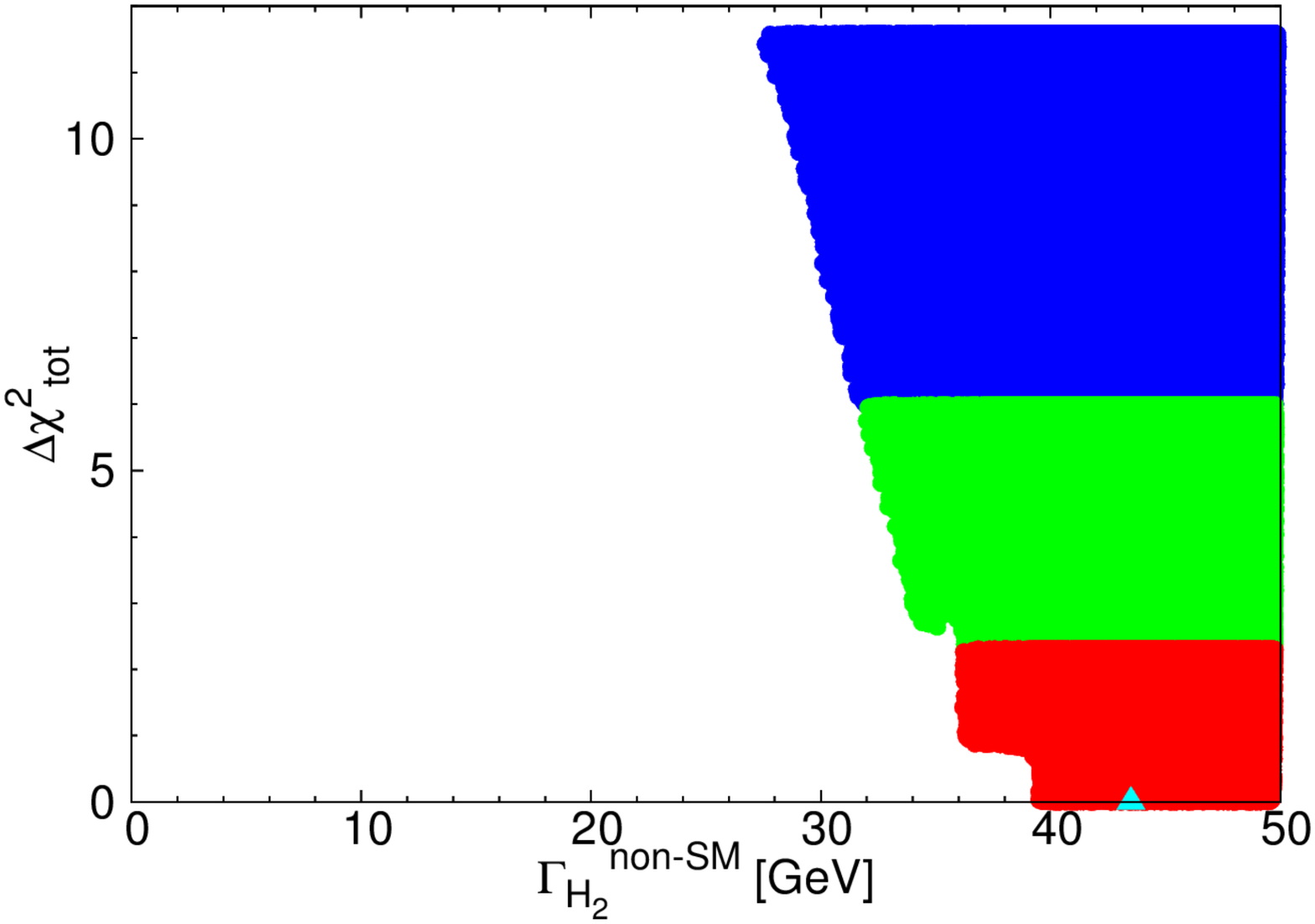}
\caption{\small \label{F4:chi_dgam}
{\bf{F4 fits}}: 
The same as in FIG.~\ref{F4:chi_sin} but for 
$\Delta\chi^2$ vs $\Gamma_{H_2}^{\rm non-SM}$.
}
\end{figure}

In Fig.~\ref{F4:chi_dgam} we show $\Delta\chi^2$ vs $\Gamma_{H_2}^{\rm non-SM}$ 
for the {\bf F4-1} (left), {\bf F4-2} (middle),
and {\bf F4-3} (right) fits.
Again, the upper frames are for the full range of $\Gamma_{H_2}$
while the lower ones for the broad-width case with $\Gamma_{H_2}\geq 40$ GeV.
We do not see any dependence on $\eta^{g(Q)\,,\gamma(F)}$ in the upper frames
since the $H_2$ width does not depend on them.  
The narrow width values are slightly preferred and $\Delta\chi^2=0$ 
is possible only when
$\Gamma_{H_2}^{\rm non-SM}\lsim 10$ GeV. This is because the ATLAS 
data on $\sigma(pp\to H_2\to \gamma\gamma)$ are closer to the CMS data
when $\Gamma_{H_2}\leq 10$ GeV, see Table~\ref{tab:cx}.
In the lower frames, we observe that,
in the $\Delta\chi^2\leq 2.3$ region (red),
$\Gamma_{H_2}^{\rm non-SM}\gsim 12$ GeV, $36$ GeV, and $36$ GeV to
achieve $\Gamma_{H_2}\geq 40$ GeV for
$\eta^{g(Q)\,,\gamma(F)}=0$ ({\bf F4-1}, left),
$\eta^{g(Q)\,,\gamma(F)}=2/3$ ({\bf F4-2}, middle), and
$\eta^{g(Q)\,,\gamma(F)}=1$ ({\bf F4-3}, right), respectively.

\begin{figure}[th!]
\centering
\includegraphics[height=2.0in,angle=270]{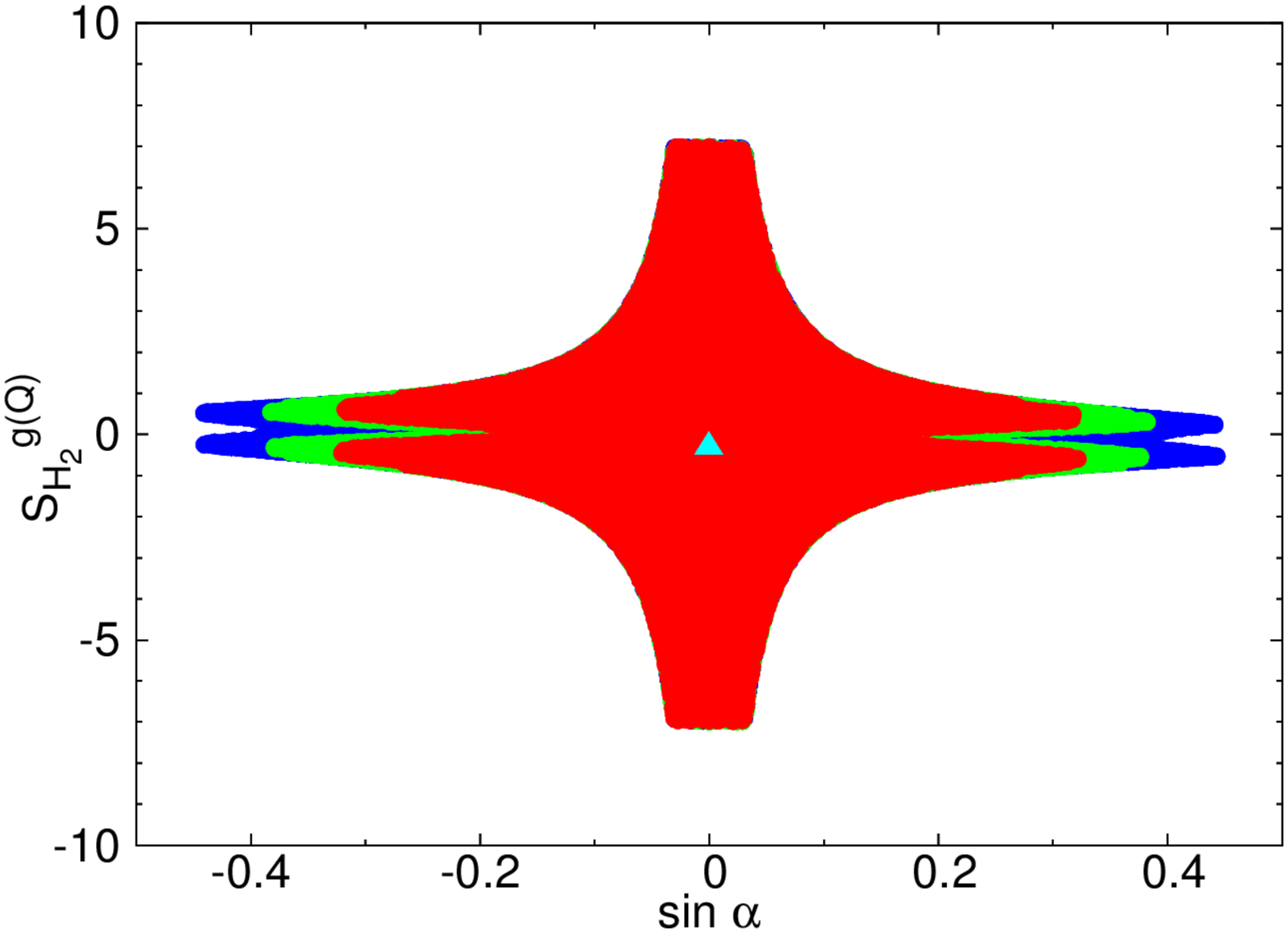}
\includegraphics[height=2.0in,angle=270]{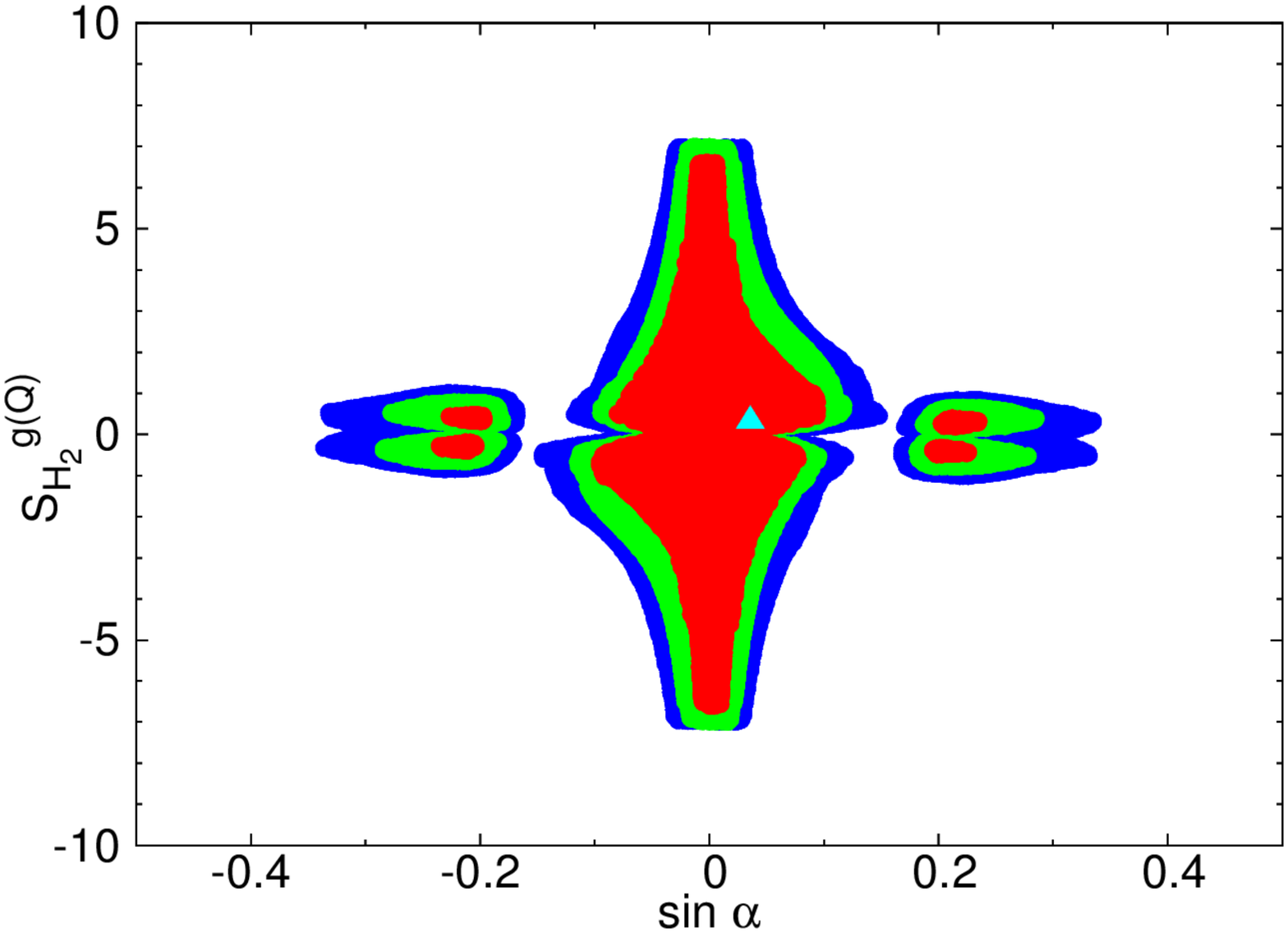}
\includegraphics[height=2.0in,angle=270]{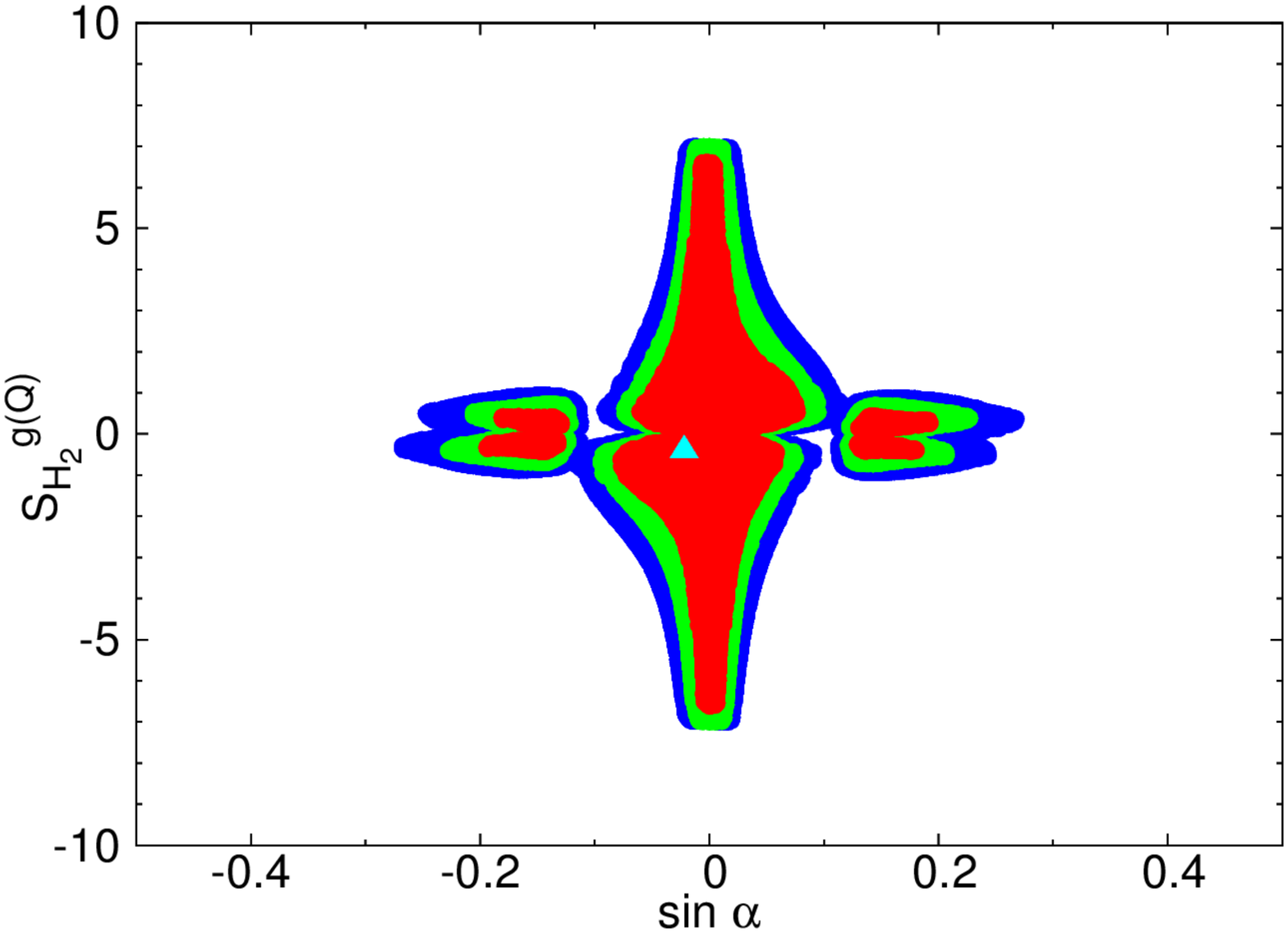}
\includegraphics[height=2.0in,angle=270]{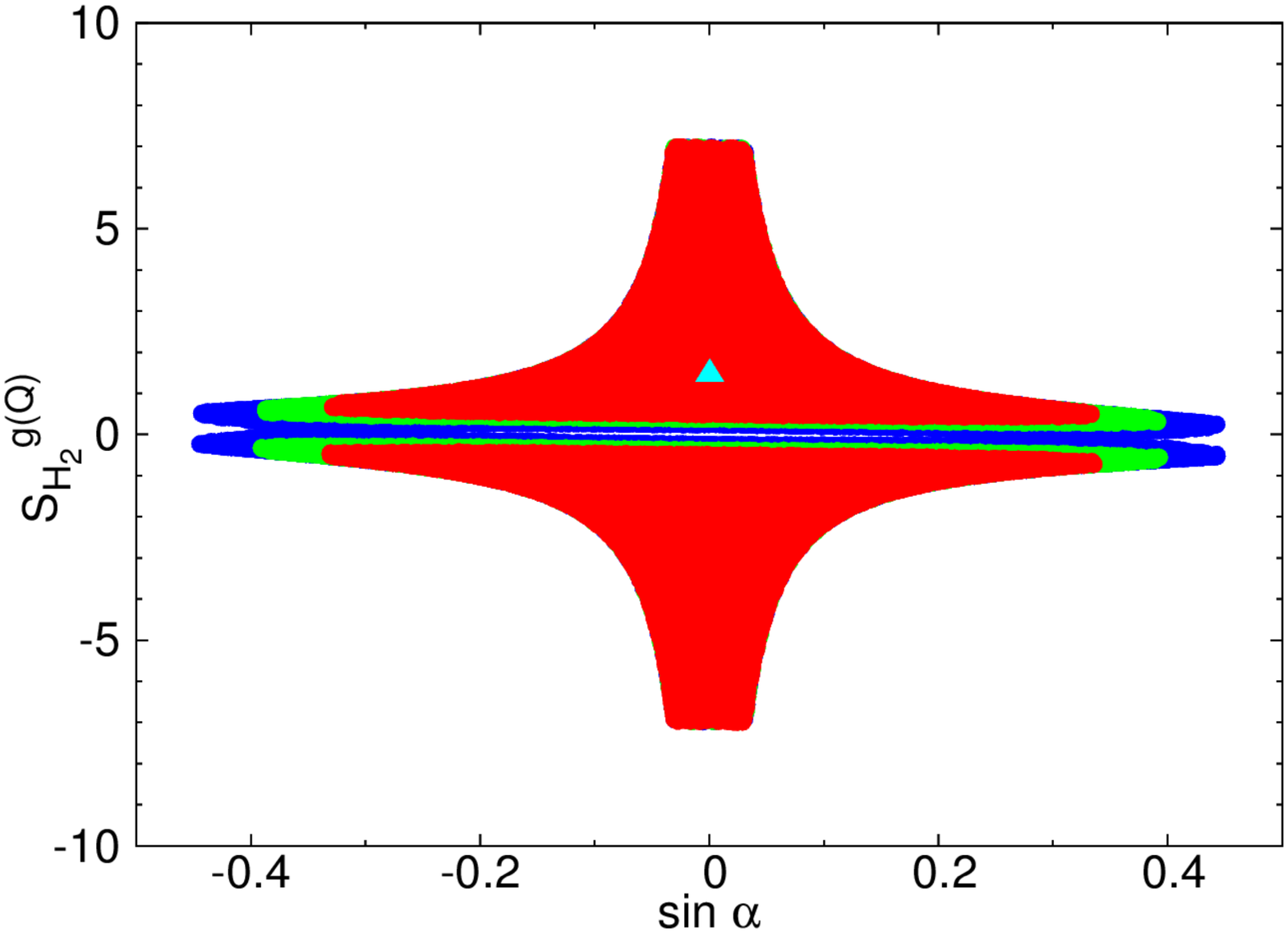}
\includegraphics[height=2.0in,angle=270]{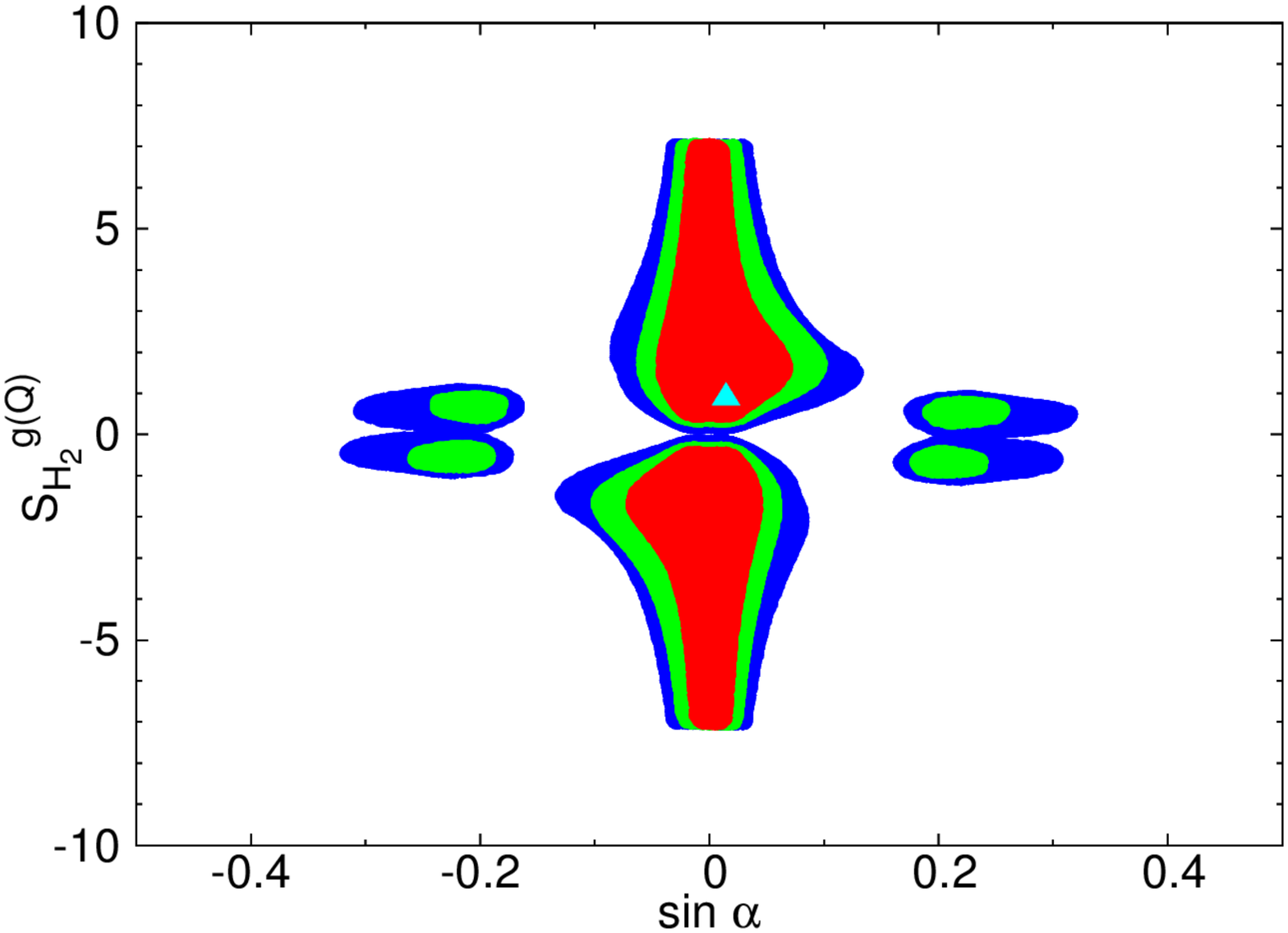}
\includegraphics[height=2.0in,angle=270]{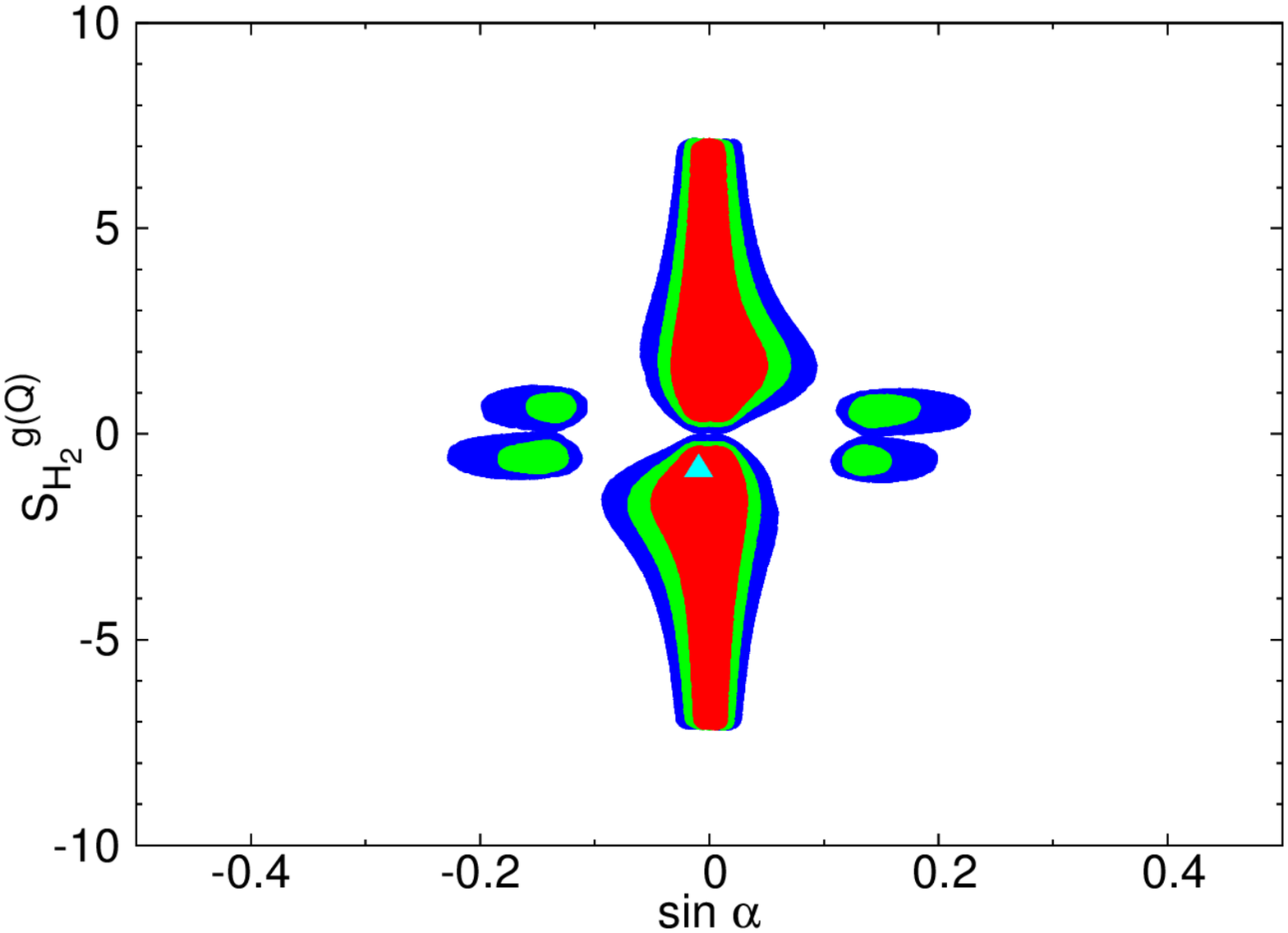}
\caption{\small \label{F4:sin_sg}
{\bf{F4 fits}}: 
The CL regions for the {\bf F4-1} (left), {\bf F4-2} (middle),
and {\bf F4-3} (right) fits in the
$(\sin\alpha,S^{g(Q)}_{H_2})$ plane.
In the upper row, we consider the full range of $\Gamma_{H_2}$ while,
in the lower row, we consider the wide width case requiring
$\Gamma_{H_2}\geq$ 40 GeV.
In all the frames, we impose the diboson, $t\bar{t}$, and dijet constraints:
$\sigma(pp\rightarrow H_2)\times B(H_2\rightarrow VV) \lesssim$ 150 fb at 13 TeV,
$\sigma(pp\rightarrow H_2)\times B(H_2\rightarrow t\bar{t}) \lesssim$ 0.5 pb at 8 TeV,
 and $\sigma(pp\rightarrow H_2)\times B(H_2\rightarrow gg) \lesssim$ 1 pb at 8 TeV.
The contour regions shown are for
$\Delta \chi^2 = 2.3$ (red), $5.99$ (green), and $11.83$ (blue)
above the minimum, which
correspond to confidence levels of
$68.3\%$, $95\%$, and $99.7\%$, respectively.
The triangles in the upper row denote the minima
over the full range of $\Gamma_{H_2}$,
while those in the lower row the minima obtained
under the assumption of $\Gamma_{H_2}\geq$ 40 GeV.
}
\end{figure}

Figure~\ref{F4:sin_sg}  shows the CL regions in the 
$(\sin\alpha,S^{g(Q)}_{H_2})$ plane.
When $\sin\alpha\sim 0$, $|S_{H_2}^{g(Q)}|$ is mostly constrained by
$\left.\sigma(pp\to H_2)\right|_{\sqrt{s}=8\,{\rm TeV}}\times B(H_2\to gg)\lsim 1$ pb.
Using Eq.~(\ref{eq:cx_gg}) and
$\left.\sigma(gg\to H_2)\right|_{\sqrt{s}=8\,{\rm TeV}}
=\left.\sigma(gg\to H_2)\right|_{\sqrt{s}=13\,{\rm TeV}}/4.7$, one may have
$|S_{H_2}^{g(Q)}|\lsim 7\,(\Gamma_{H_2}/50\,{\rm GeV})^{1/4}$.
As $\sin\alpha$ deviates from $0$, $|S_{H_2}^{g(Q)}|$ becomes constrained by
$\left.\sigma(pp\to H_2)\right|_{\sqrt{s}=13\,{\rm TeV}}\times B(H_2\to VV)\lsim 150$ fb.
Using Eq.~(\ref{eq:cx_ww}), one may have
$|S_{H_2}^{g(Q)}\,\sin\alpha|\lsim 0.2\,(\Gamma_{H_2}/50\,{\rm GeV})^{1/2}$.
These two observations mainly explain the shape of the CL regions in the left frames
for {\bf F4-1} with $\eta^{g(Q)\,,\gamma(F)}=0$.
For {\bf F4-2} and {\bf F4-3} with 
$\eta^{g(Q)\,,\gamma(F)}\neq 1$, 
the $H_1$ data provide additional constraints 
basically coming from $C_{H_1}^{g,\gamma}\sim 1$.
Definitely, the CL regions populate along the $\sin\alpha=0$ line
satisfying the $H_1$ constraints in the case
$(S^\gamma_{H_1}\,,S^g_{H_1})\sim 
(+S^{\gamma({\rm SM})}_{H_1}\,,+S^{g({\rm SM})}_{H_1})$.
On the other hand, the four islands around the points 
$\left(|\sin\alpha| = 0.20\,,|S_{H_2}^{g(Q)}| = 0.5 \right)$ and
$\left(|\sin\alpha| = 0.13\,,|S_{H_2}^{g(Q)}| = 0.5 \right)$ 
for the {\bf F4-2} and {\bf F4-3} fits, respectively,
satisfy the $H_1$ constraints in the case
$(S^\gamma_{H_1}\,,S^g_{H_1})\sim 
(-S^{\gamma({\rm SM})}_{H_1}\,,+S^{g({\rm SM})}_{H_1})$.
We find that the cases with $S^g_{H_1}\sim -S^{g({\rm SM})}_{H_1}$  
cannot satisfy the $H_1$ constraints because it requires a too large value
of $|S^{g(Q)}_{H_2}|\sim 10$~\footnote{Note that the relation
$S^g_{H_1}\sim -S^{g({\rm SM})}_{H_1}$ leads to
$
\sin\alpha\,S_{H_1}^{g(Q)} =
\sin\alpha\,\eta^{g(Q)}\,S_{H_2}^{g(Q)} \sim (1+\cos\alpha)\,S_{H_1}^{g({\rm SM})}
\simeq 1.3\,.
$
For $|\sin\alpha|=0.20\,(0.13)$ and $\eta^{g(Q)}=2/3\,(1)$, one may have
$|S_{H_2}^{g(Q)}|\sim 10$.}
which is incompatible with the $H_2$ dijet constraint
$|S^{g(Q)}_{H_2}|\lsim 7$ discussed before.
Some numerical results on 68\% CL regions
are summarized in Table~\ref{tab:68}.
\begin{table}[th!]
  \caption{\small \label{tab:68}
The 68\% CL regions of 
$\sigma(\gamma\gamma)=\sigma(gg\to H_2)\times B(H_2\to \gamma\gamma)$,
$\sin\alpha$, $\Gamma_{H_2}^{\rm non-SM}/{\rm GeV}$,
$S^{g(Q)}_{H_2}$, and $S^{\gamma(F)}_{H_2}$
in the {\bf F4} fits.}
\begin{ruledtabular}
\begin{tabular}{ l|c|c||c|c|c|c|c}
Fits  & $\eta^{g(Q),\gamma(F)}$ & $\Gamma_{H_2}/$GeV & $\sigma(\gamma\gamma)/$fb &
$|\sin\alpha|$ & $\Gamma^{\rm non-SM}_{H_2}$ & $|S^{g(Q)}_{H_2}|$ &
$|S^{\gamma(F)}_{H_2}|$ \\
\hline
{\bf F4-1}& 0 & 0$\sim$50 & 2.9$\sim$7.4  & 0$\sim$0.33  & 0$\sim$50 &
                0.05$\sim$7.0  & 2.2$\sim$100 \\
          & 0 & 40$\sim$50 & 3.3$\sim$8.2  & 0$\sim$0.34  & 12$\sim$50  &
                0.4$\sim$7.0 & 7.2$\sim$100 \\
          \hline
{\bf F4-2}& 2/3 & 0$\sim$50 & 2.9$\sim$7.4  & 0$\sim$0.095 or 0.20$\sim$0.24 &
0$\sim$50 &
                  0.05$\sim$6.6  & 2.4$\sim$100 \\
          & 2/3 & 40$\sim$50 & 3.3$\sim$8.2  & 0$\sim$0.066  & 36$\sim$50  &
                  0.5$\sim$7.0  & 7.9$\sim$100 \\
          \hline
{\bf F4-3}& 1 & 0$\sim$50 & 2.9$\sim$7.4 & 0$\sim$0.077 or 0.13$\sim$0.20  &
0$\sim$50 &
                0.05$\sim$6.6 & 2.4$\sim$100 \\
          & 1 & 40$\sim$50 & 3.3$\sim$8.2 & 0$\sim$0.044 & 36$\sim$50 &
                0.5$\sim$7.0 & 7.9$\sim$100 \\
\end{tabular}
\end{ruledtabular}
\end{table}

\begin{figure}[th!]
\centering
\includegraphics[height=2.0in,angle=270]{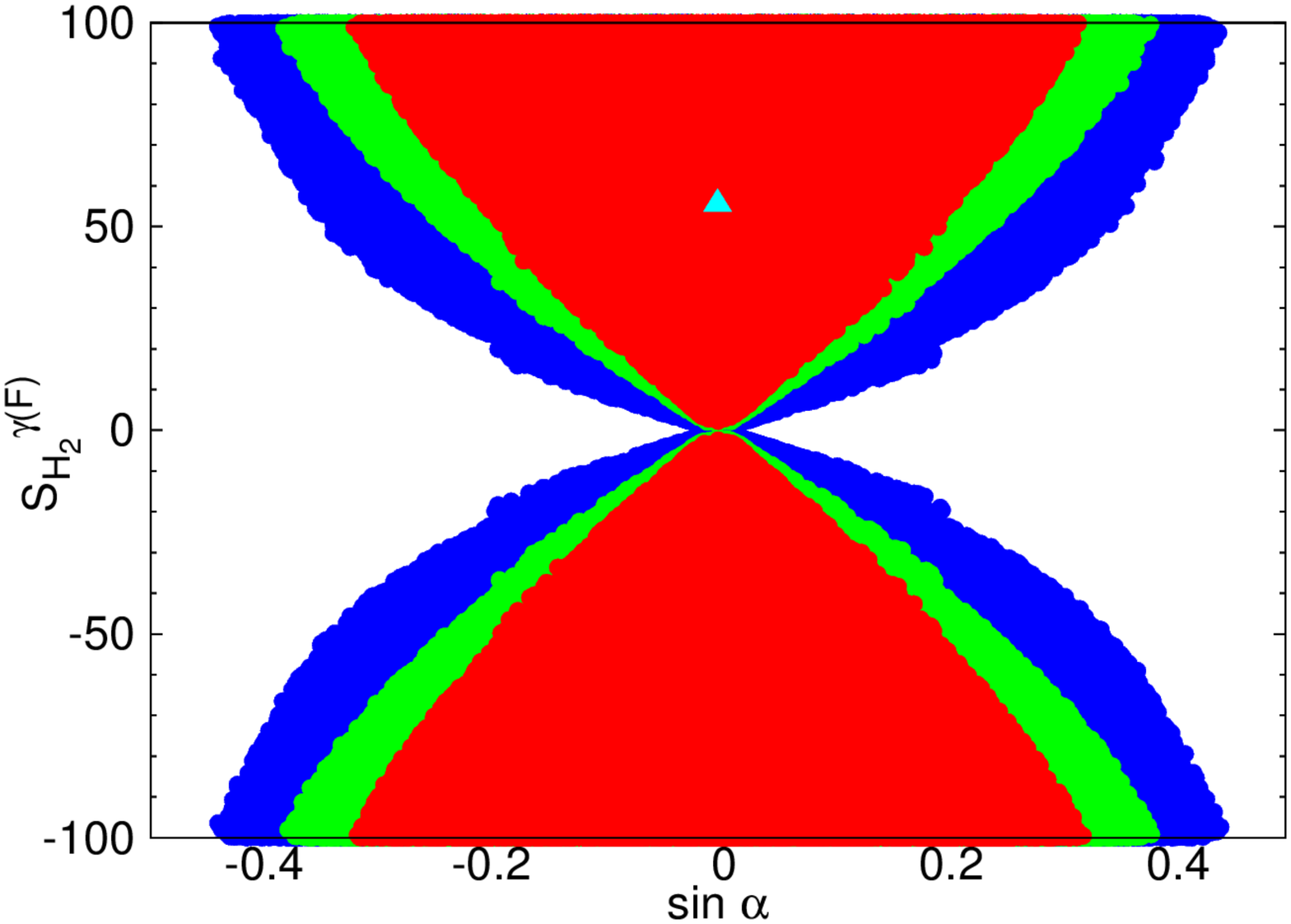}
\includegraphics[height=2.0in,angle=270]{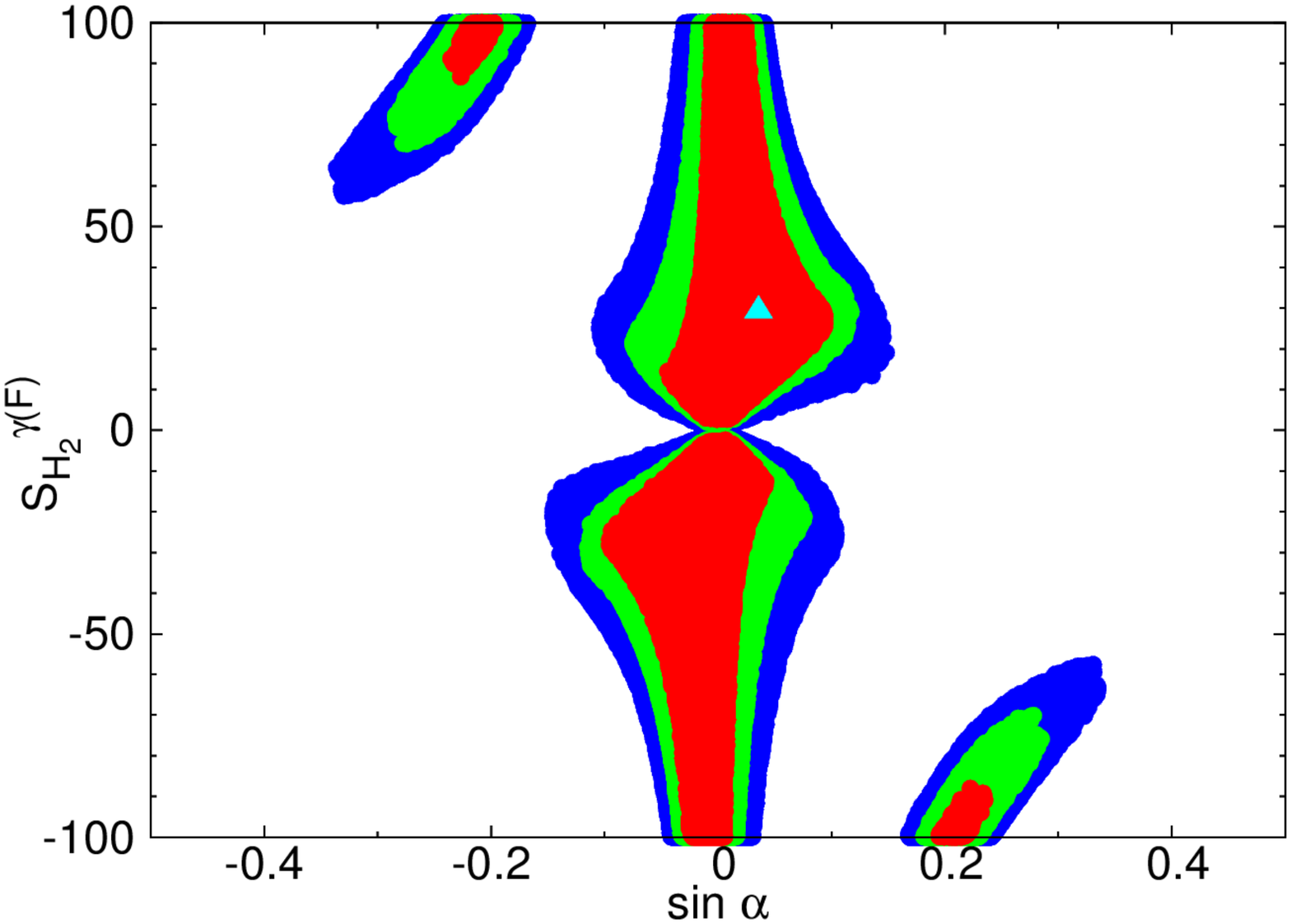}
\includegraphics[height=2.0in,angle=270]{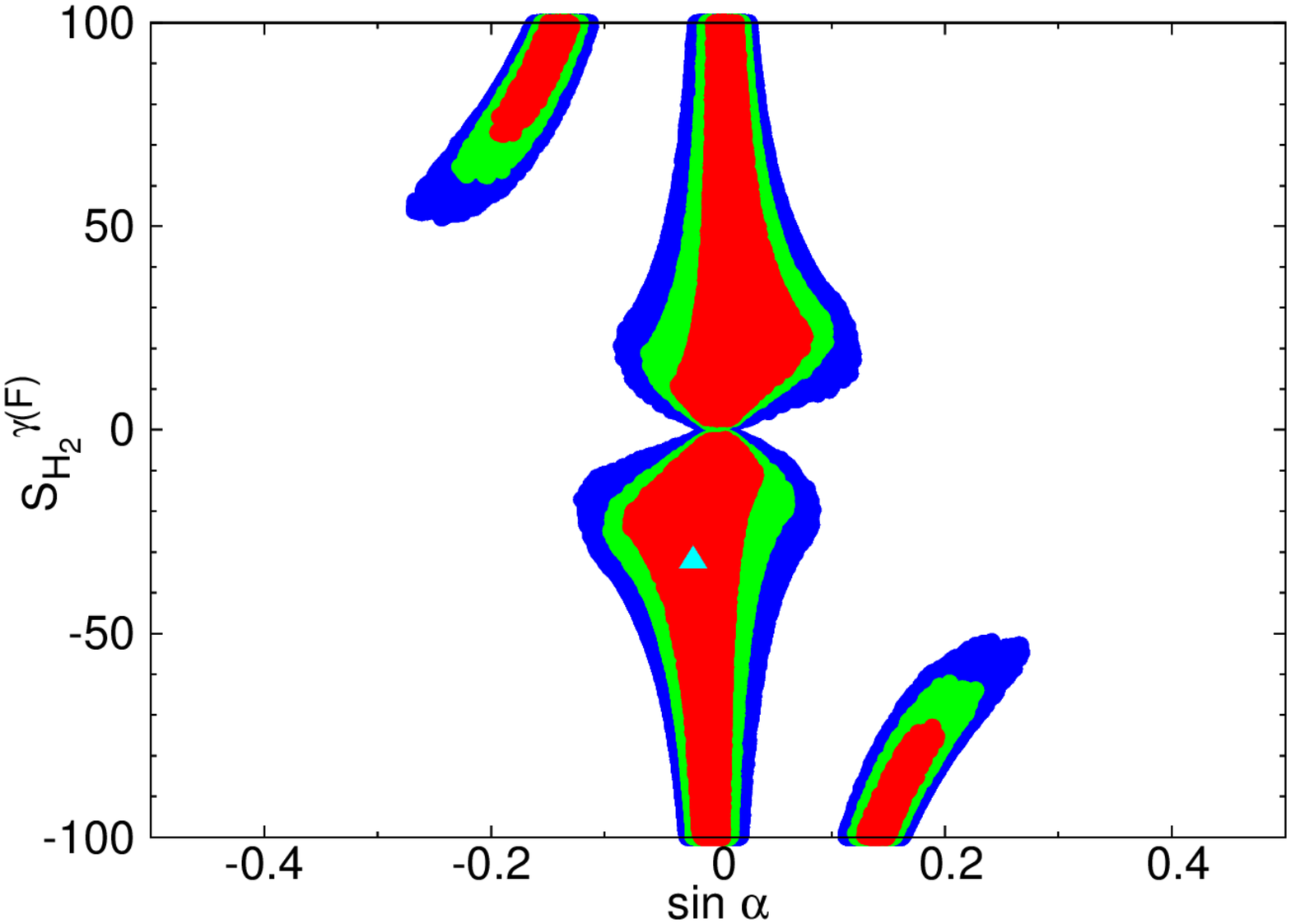}
\includegraphics[height=2.0in,angle=270]{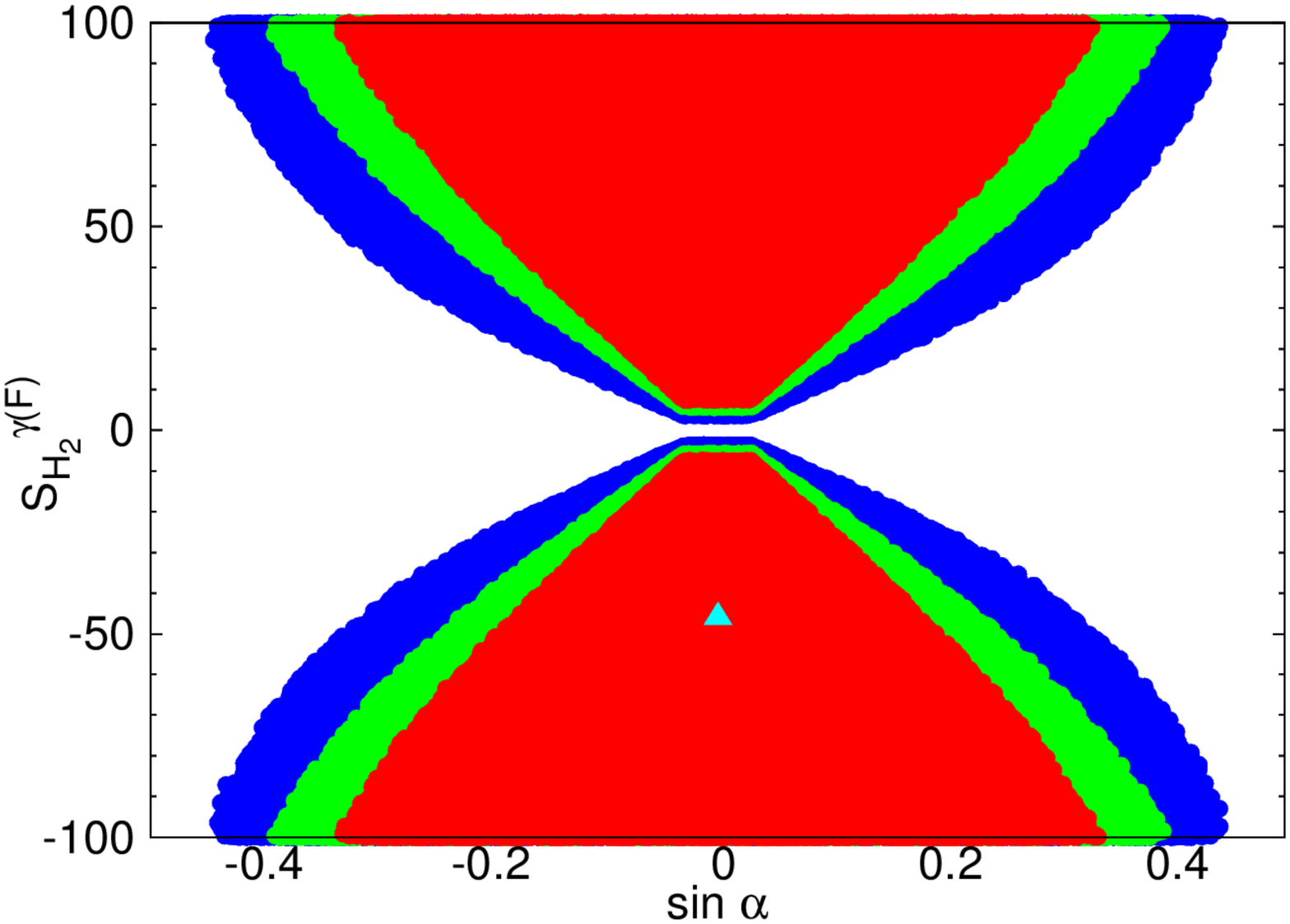}
\includegraphics[height=2.0in,angle=270]{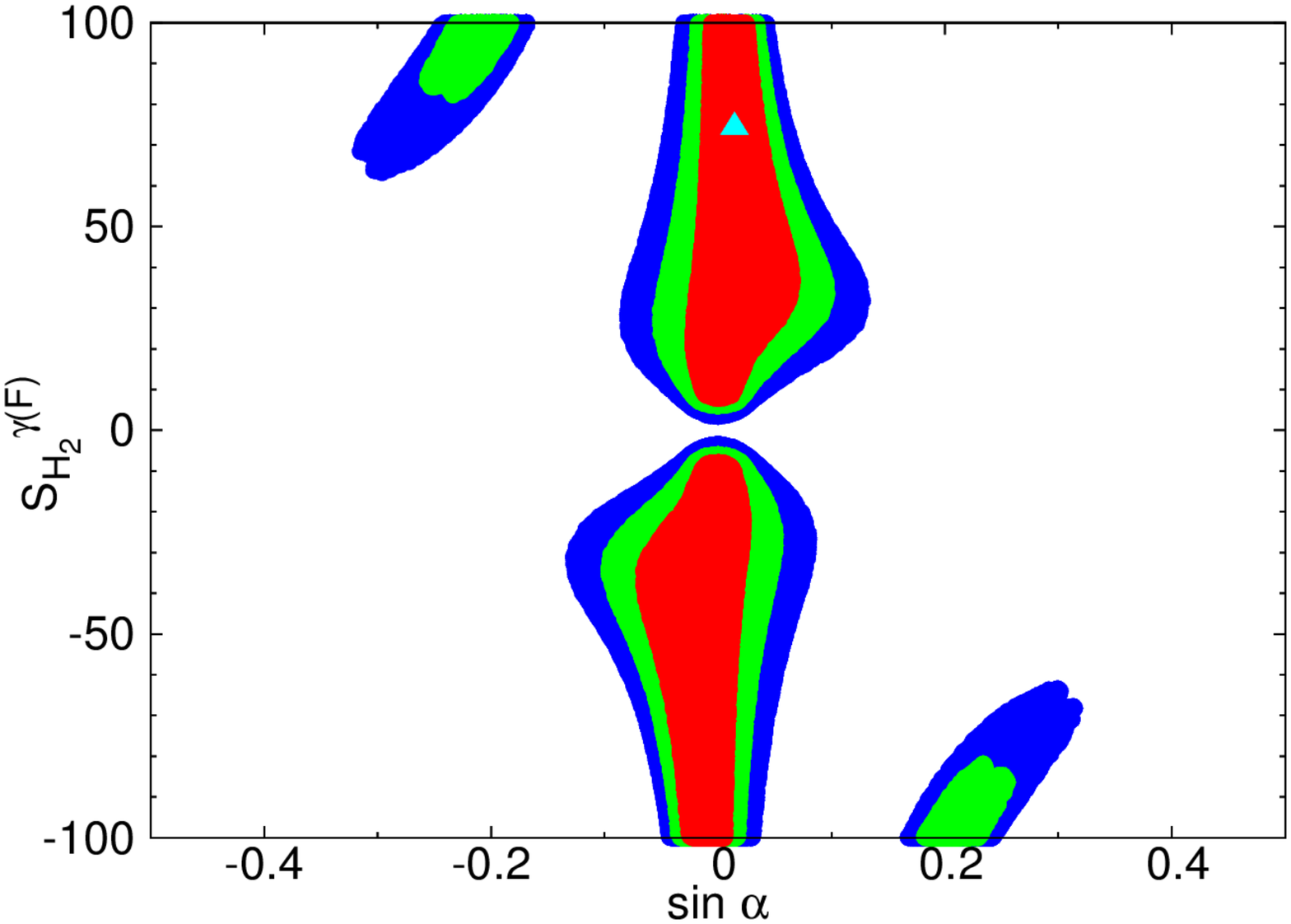}
\includegraphics[height=2.0in,angle=270]{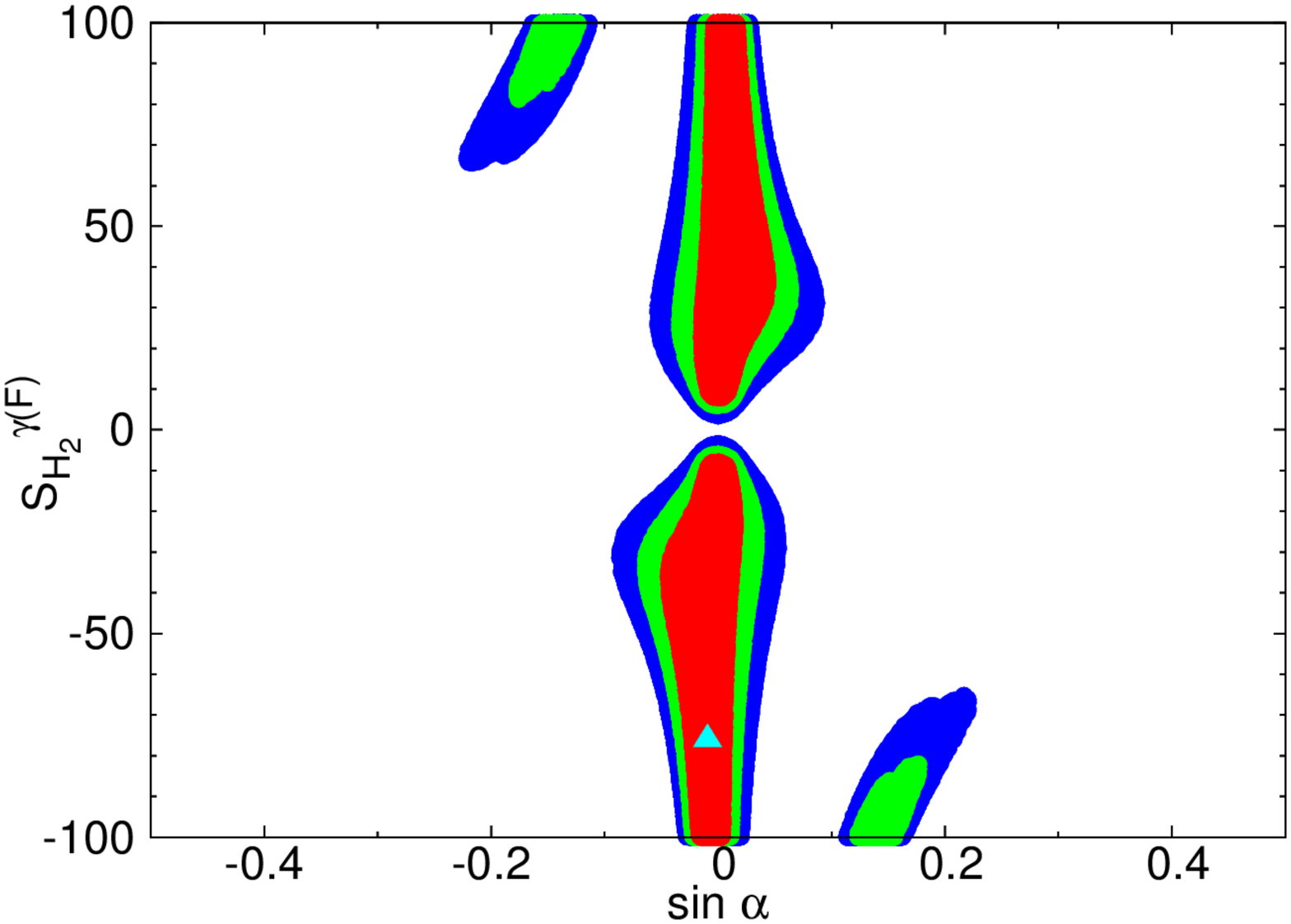}
\caption{\small \label{F4:sin_sa}
{\bf{F4 fits}}: 
The same as in FIG.~\ref{F4:sin_sg} but for the CL regions in the
$(\sin\alpha,S^{\gamma(F)}_{H_2})$ plane.
}
\end{figure}

\begin{figure}[th!]
\centering
\includegraphics[height=2.0in,angle=270]{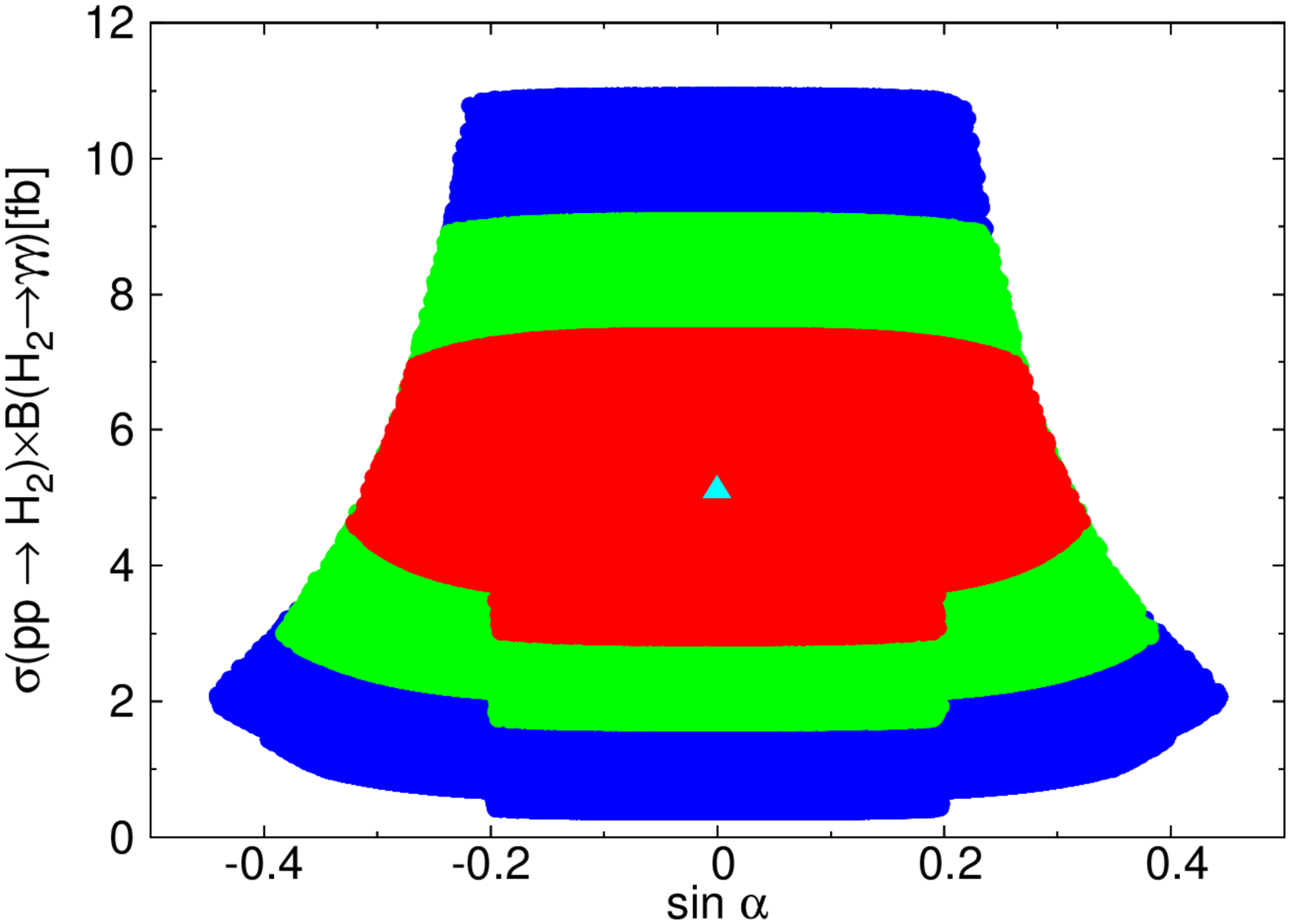}
\includegraphics[height=2.0in,angle=270]{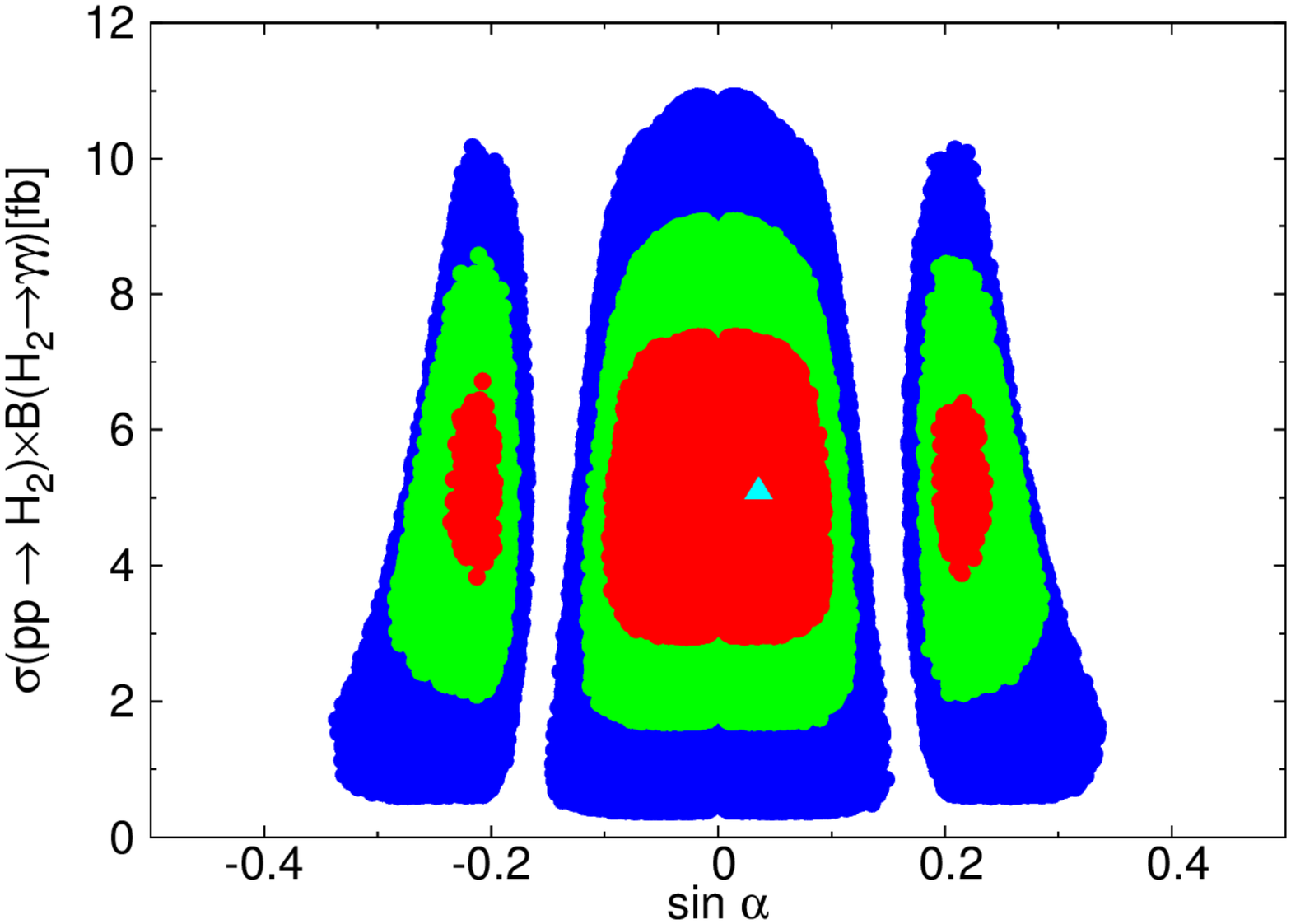}
\includegraphics[height=2.0in,angle=270]{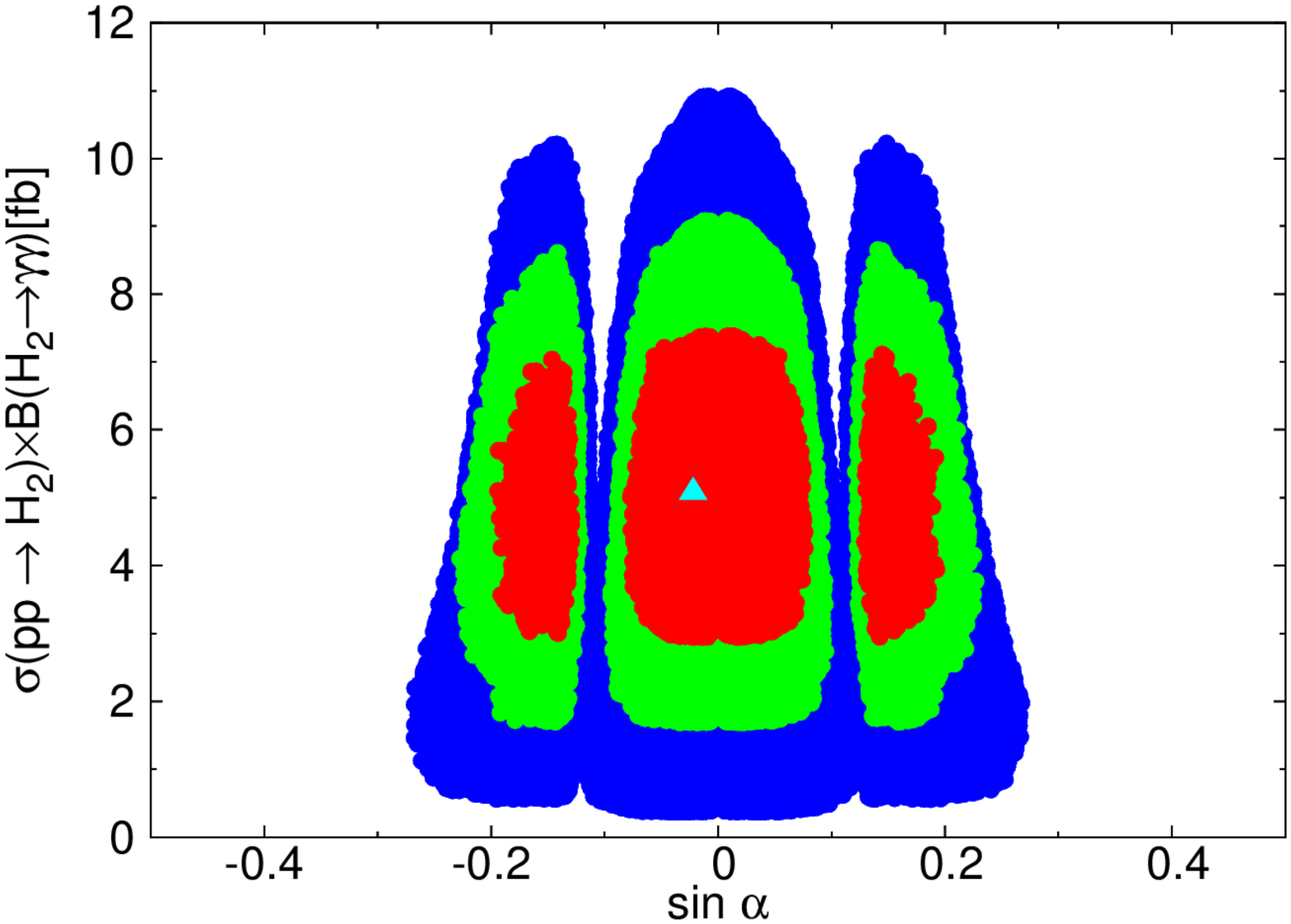}
\includegraphics[height=2.0in,angle=270]{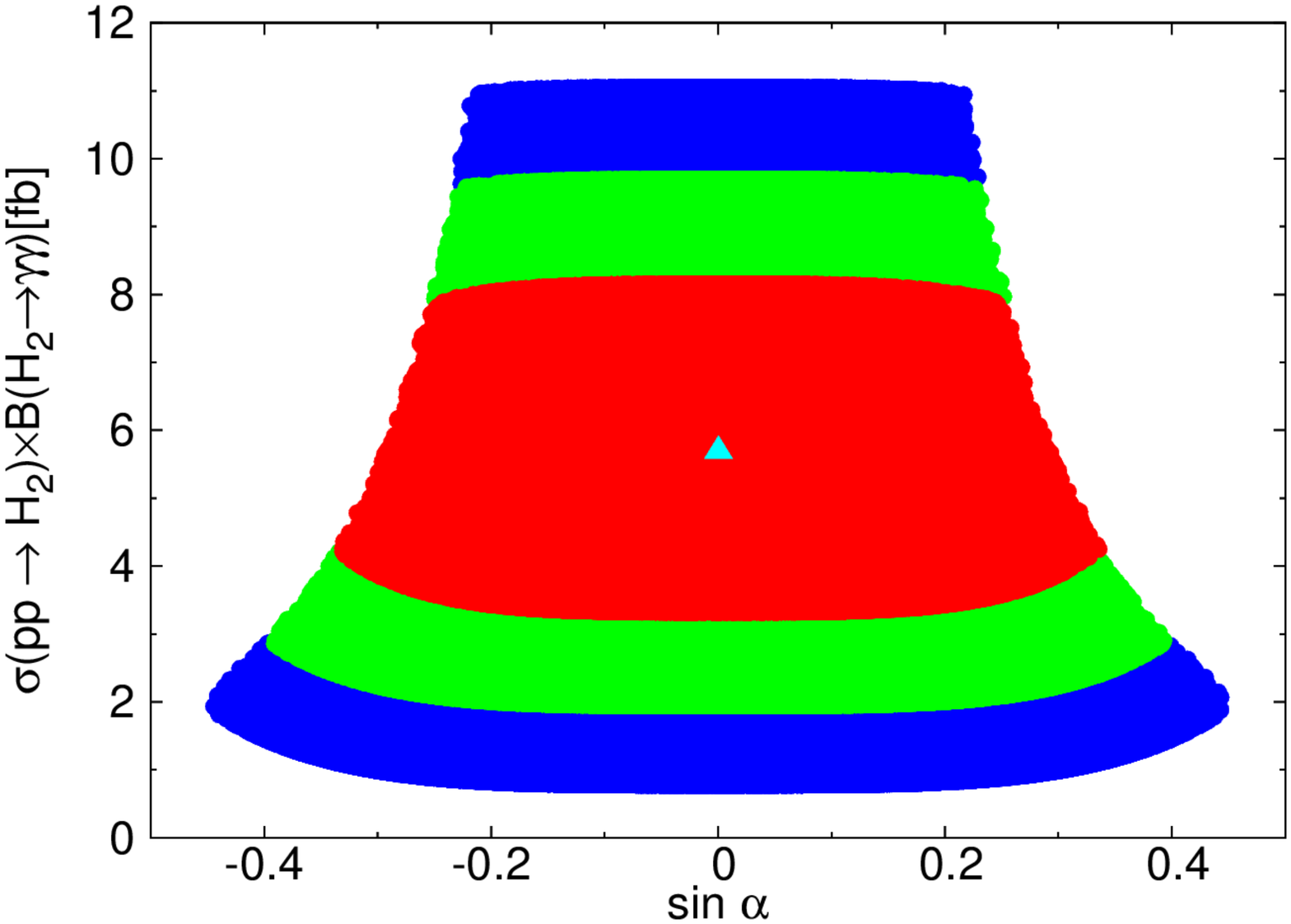}
\includegraphics[height=2.0in,angle=270]{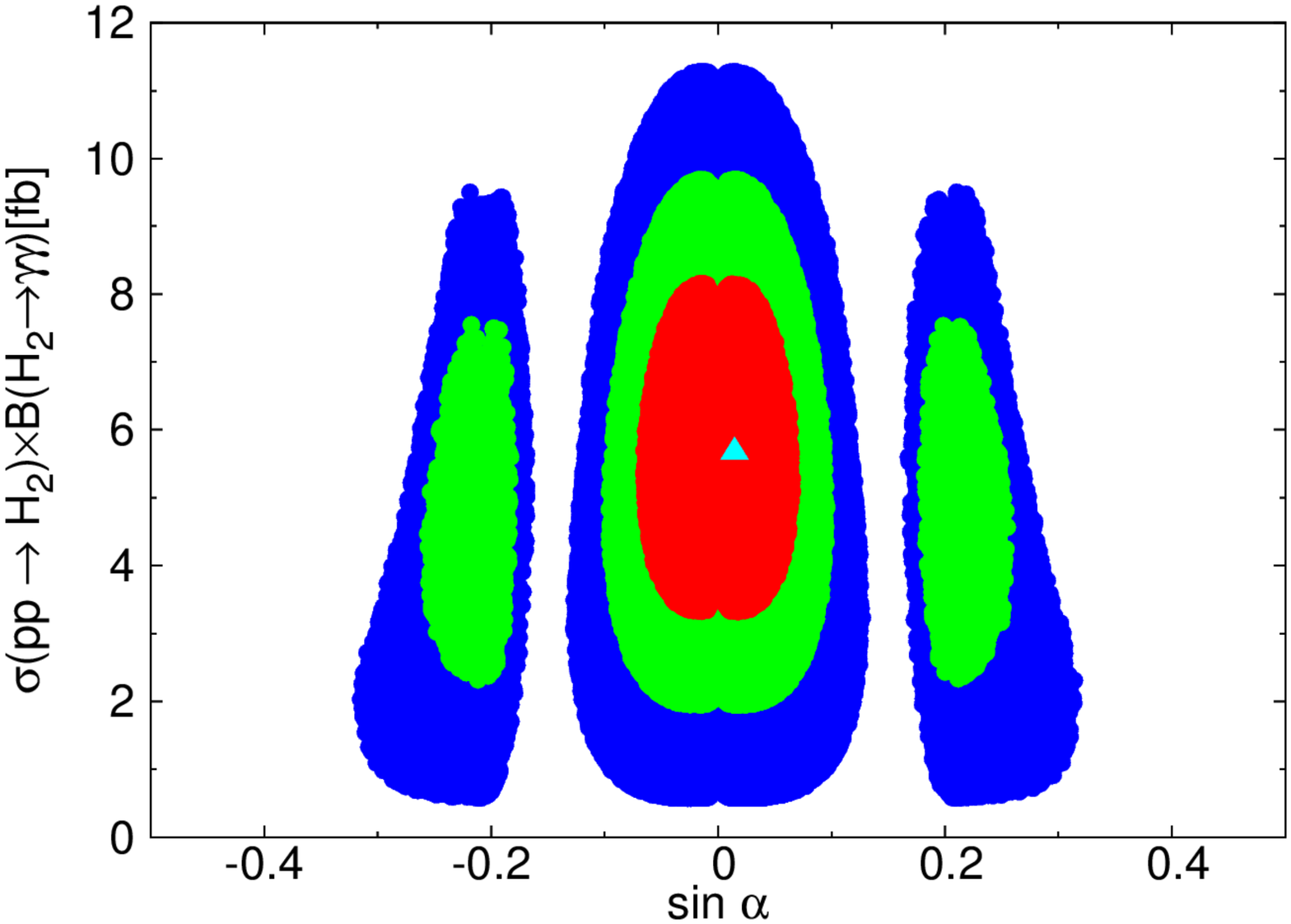}
\includegraphics[height=2.0in,angle=270]{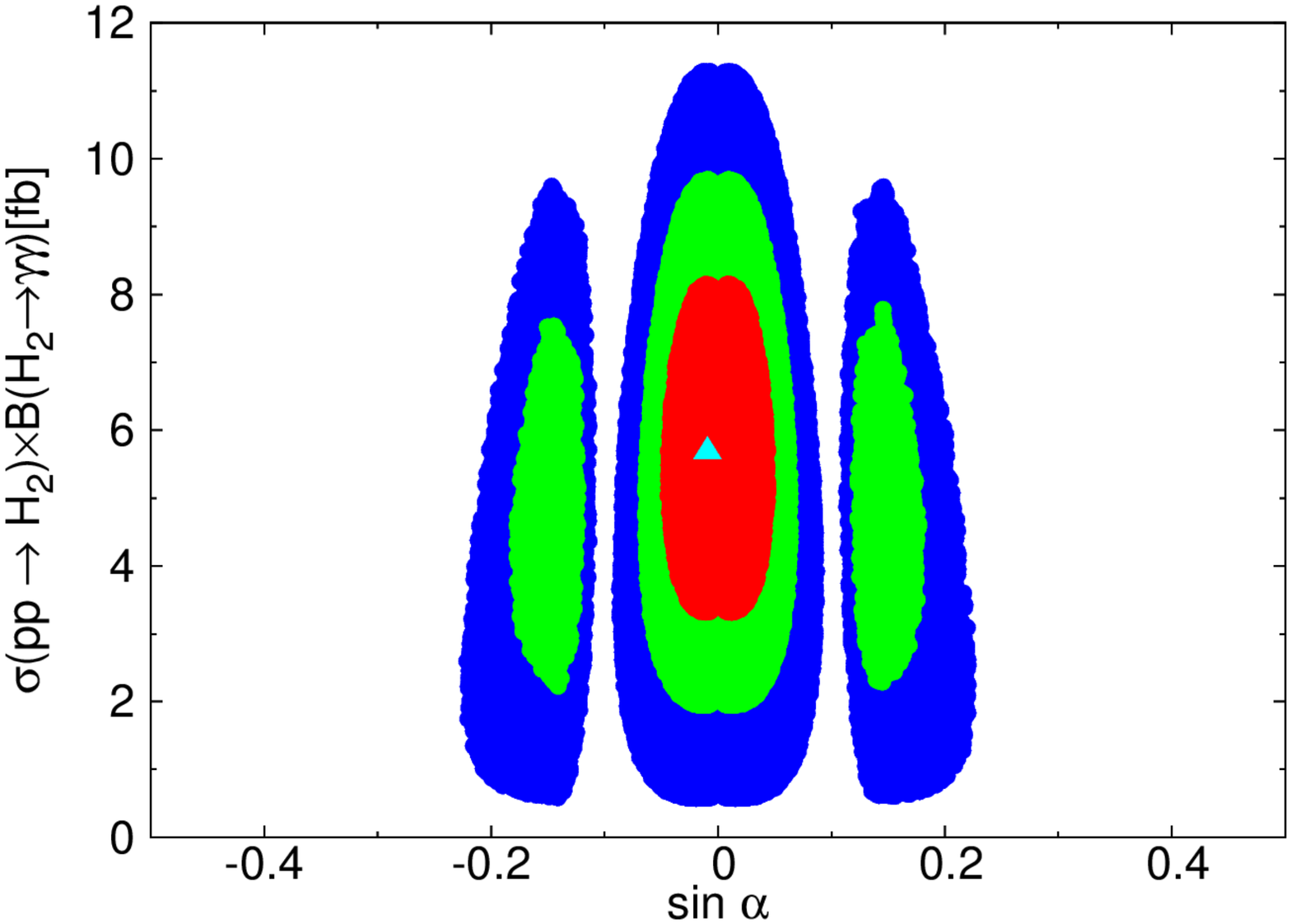}
\caption{\small \label{F4:sin_aa}
{\bf{F4 fits}}:
The same as in FIG.~\ref{F4:sin_sg} but for the CL regions in the
$(\sin\alpha,\sigma(gg\to H_2)\times B(H_2\to\gamma\gamma)$ plane.
}
\end{figure}

Figure~\ref{F4:sin_sa}  shows the CL regions in the 
$(\sin\alpha,S^{\gamma(F)}_{H_2})$ plane.
For {\bf F4-1} with $\eta^{g(Q)\,,\gamma(F)}=0$, the parameter space is constrained
basically by the lower limit on $\sigma(gg\to H_2\to  \gamma\gamma)$.
The lower limit $\sigma_{\rm min}\sim 3$ fb in at 68\% CL, see Table~\ref{tab:68}.
Then, using Eq.~(\ref{eq:cx_aa}), we have
$|S^{g(Q)}_{H_2}S^{\gamma(F)}_{H_2}|\gsim 50\,
(\Gamma_{H_2}/50\,{\rm GeV})^{1/2}\,
(\sigma_{\rm min}/3{\rm fb})^{1/2}$.
Combining this with the the $H_2$ diboson constraint
$|S_{H_2}^{g(Q)}\,\sin\alpha|\lsim 0.2\,(\Gamma_{H_2}/50\,{\rm GeV})^{1/2}$,
we obtain $|S^{\gamma(F)}_{H_2}|\gsim 250\,|\sin\alpha|\,
(\sigma_{\rm min}/3{\rm fb})^{1/2}$.
This observation basically explains the shape of CL regions in the left frames together
with the fact that the lower limit $\sigma_{\rm min}$ increases 
a little bit as $|\sin\alpha|$ deviates from $0$, see Fig.~\ref{F4:sin_aa}.
For {\bf F4-2} and {\bf 4-3} with $\eta^{g(Q)\,,\gamma(F)}\neq 0$, on the other hand,
the $H_1$ data gives further constraints like as in the
$(\sin\alpha,S^{g(Q)}_{H_2})$ case.
In the CL regions along the $\sin\alpha=0$ line, 
$(S^\gamma_{H_1}\,,S^g_{H_1})\sim 
(+S^{\gamma({\rm SM})}_{H_1}\,,+S^{g({\rm SM})}_{H_1})$. While,
on the two islands at non-zero $\sin\alpha$ and 
for large values of $|S^{\gamma(F)}_{H_2}|$,
$(S^\gamma_{H_1}\,,S^g_{H_1})\sim 
(-S^{\gamma({\rm SM})}_{H_1}\,,+S^{g({\rm SM})}_{H_1})$.
When $S_{H_1}^\gamma =- S^{\gamma\,({\rm SM})}_{H_1}$, we have
\begin{equation}
\sin\alpha\,S^{\gamma\,(F)}_{H_1} = 
\sin\alpha\,\eta^{\gamma (F)}\,S^{\gamma\,(F)}_{H_2} =
(1+\cos\alpha) S^{\gamma\,({\rm SM})}_{H_1}
\simeq -13\,,
\end{equation}
which implies that the local minima appear at
$\sin\alpha \simeq -13/(\eta^{\gamma (F)}\,S^{\gamma\,(F)}_{H_2})$.
When $S^{\gamma\,(F)}_{H_2}=\pm 100$, for example, 
the local minima may occur at $\sin\alpha\simeq \mp 0.20$ and $\mp 0.13$  
for the {\bf  F4-2} ($\eta^{\gamma (F)}=2/3$) and
the {\bf  F4-3} ($\eta^{\gamma (F)}=1$), respectively.
This finally explains why the local minima are developed at
$|\sin\alpha|\simeq 0.20\,(0.13)$ in the middle (right) frames
of Fig.~\ref{F4:chi_sin}.

Figure~\ref{F4:sin_aa}  shows the CL regions in the 
$(\sin\alpha,\sigma(gg\to H_2)\times B(H_2\to \gamma\gamma))$ plane.
We observe the cross sections are centered around 5 fb.

\begin{figure}[th!]
\centering
\includegraphics[height=2.0in,angle=270]{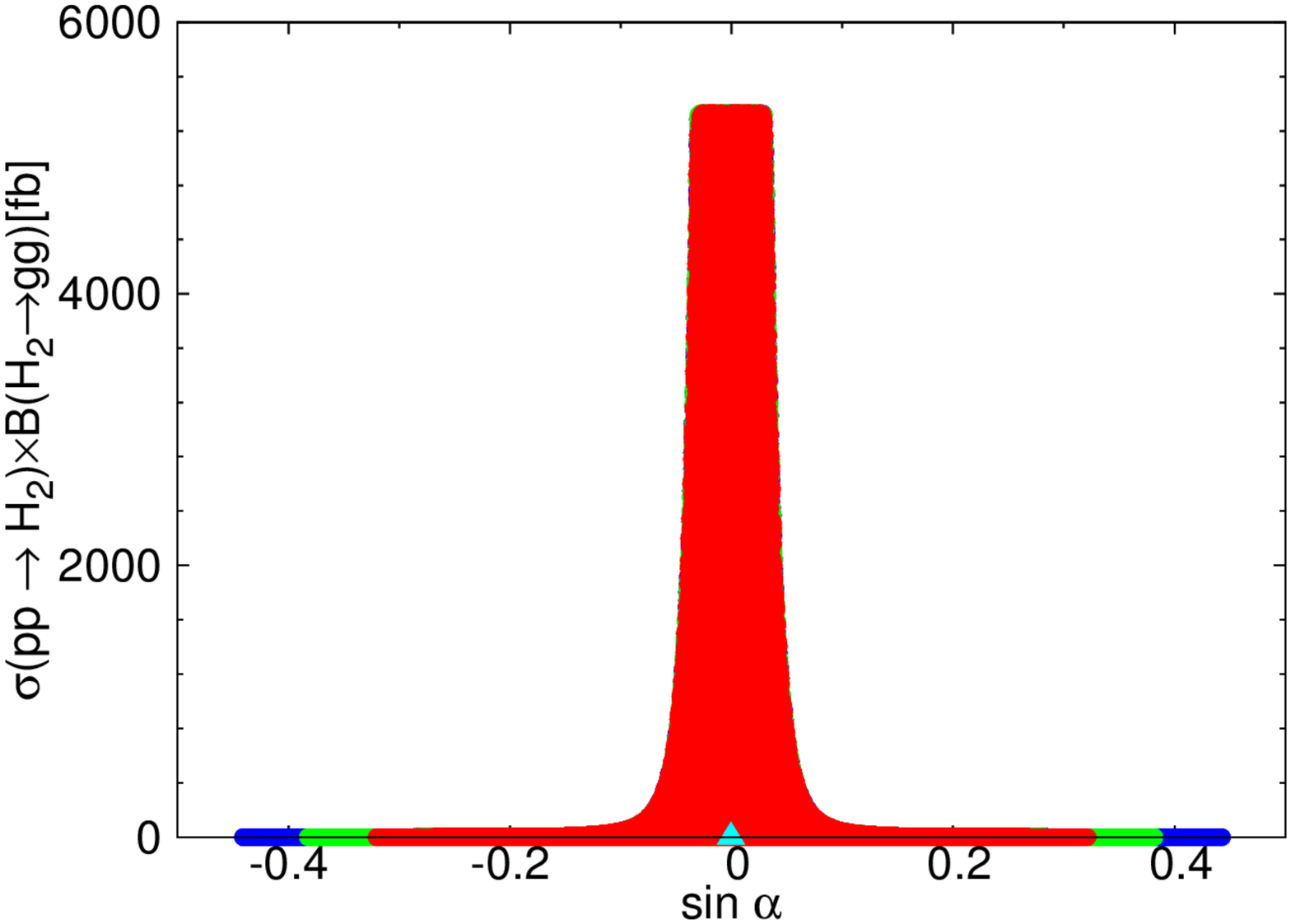}
\includegraphics[height=2.0in,angle=270]{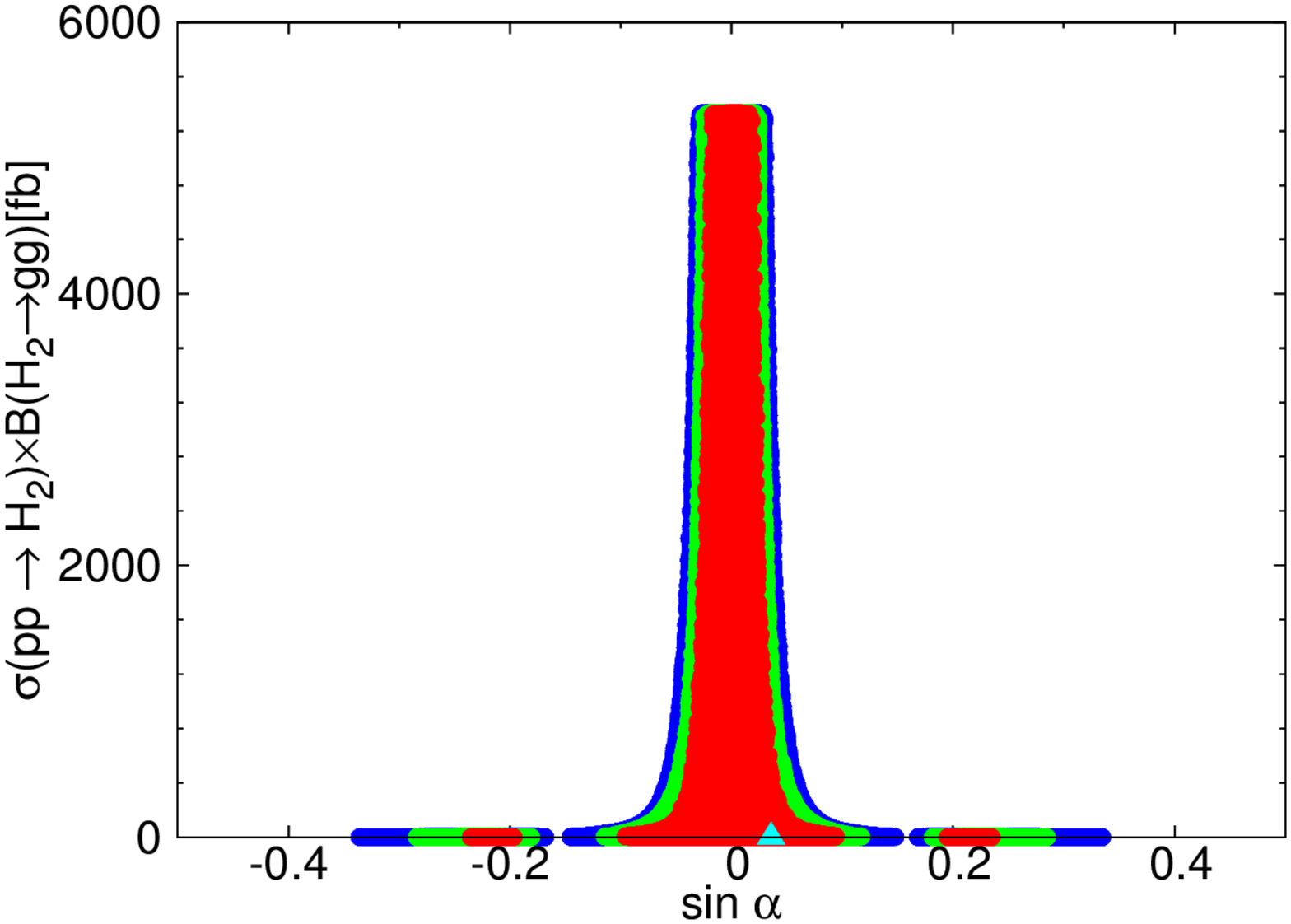}
\includegraphics[height=2.0in,angle=270]{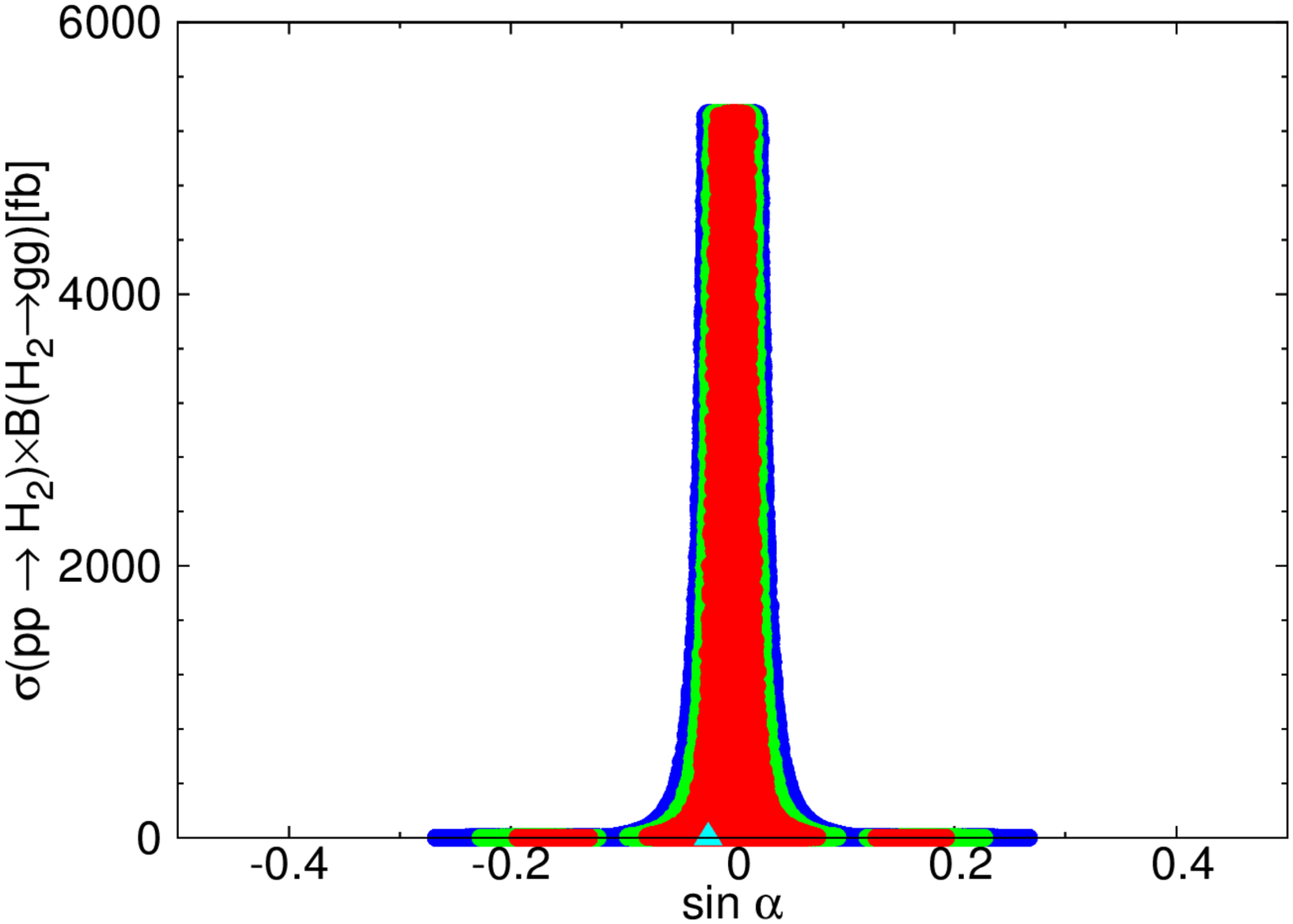}
\includegraphics[height=2.0in,angle=270]{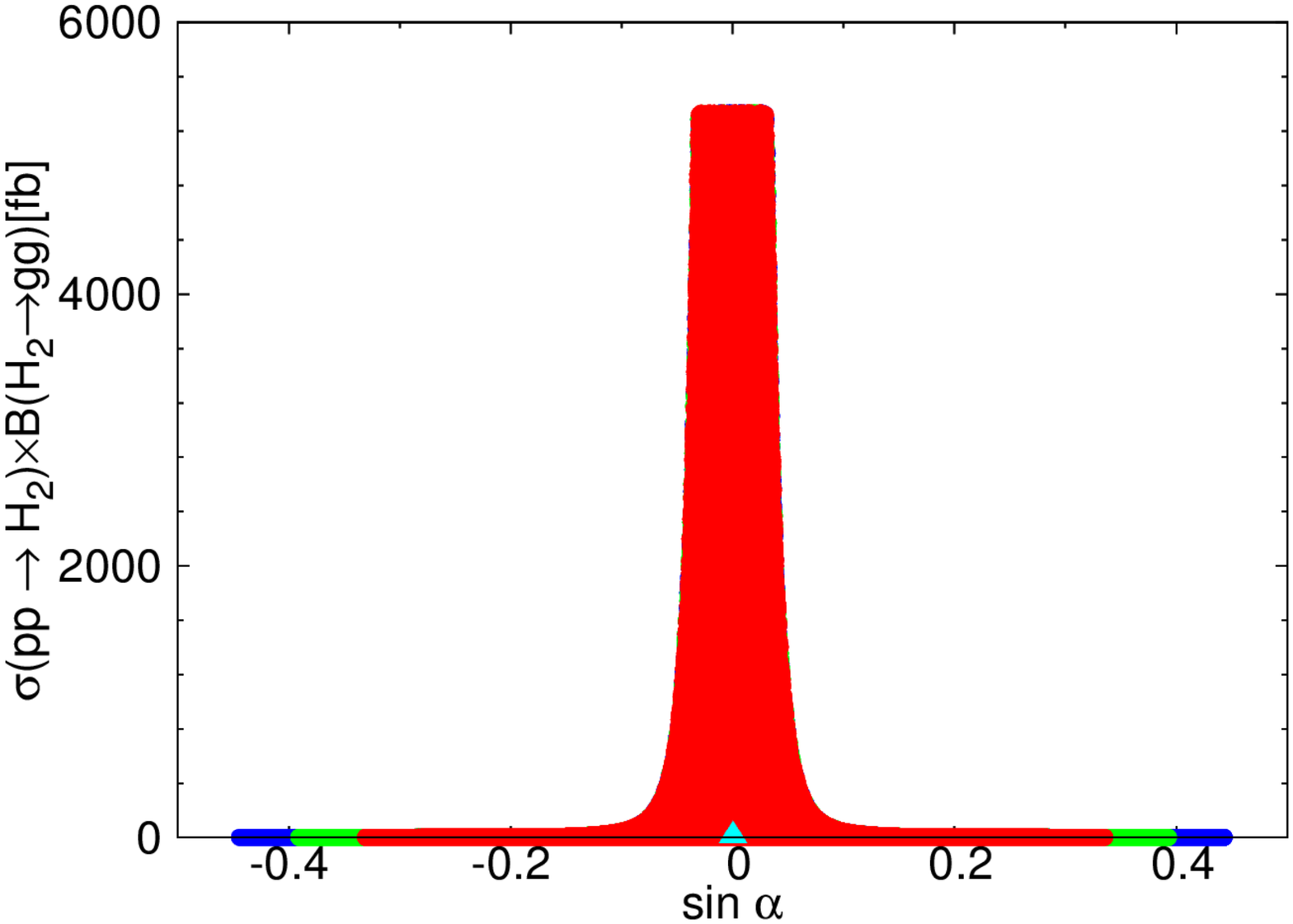}
\includegraphics[height=2.0in,angle=270]{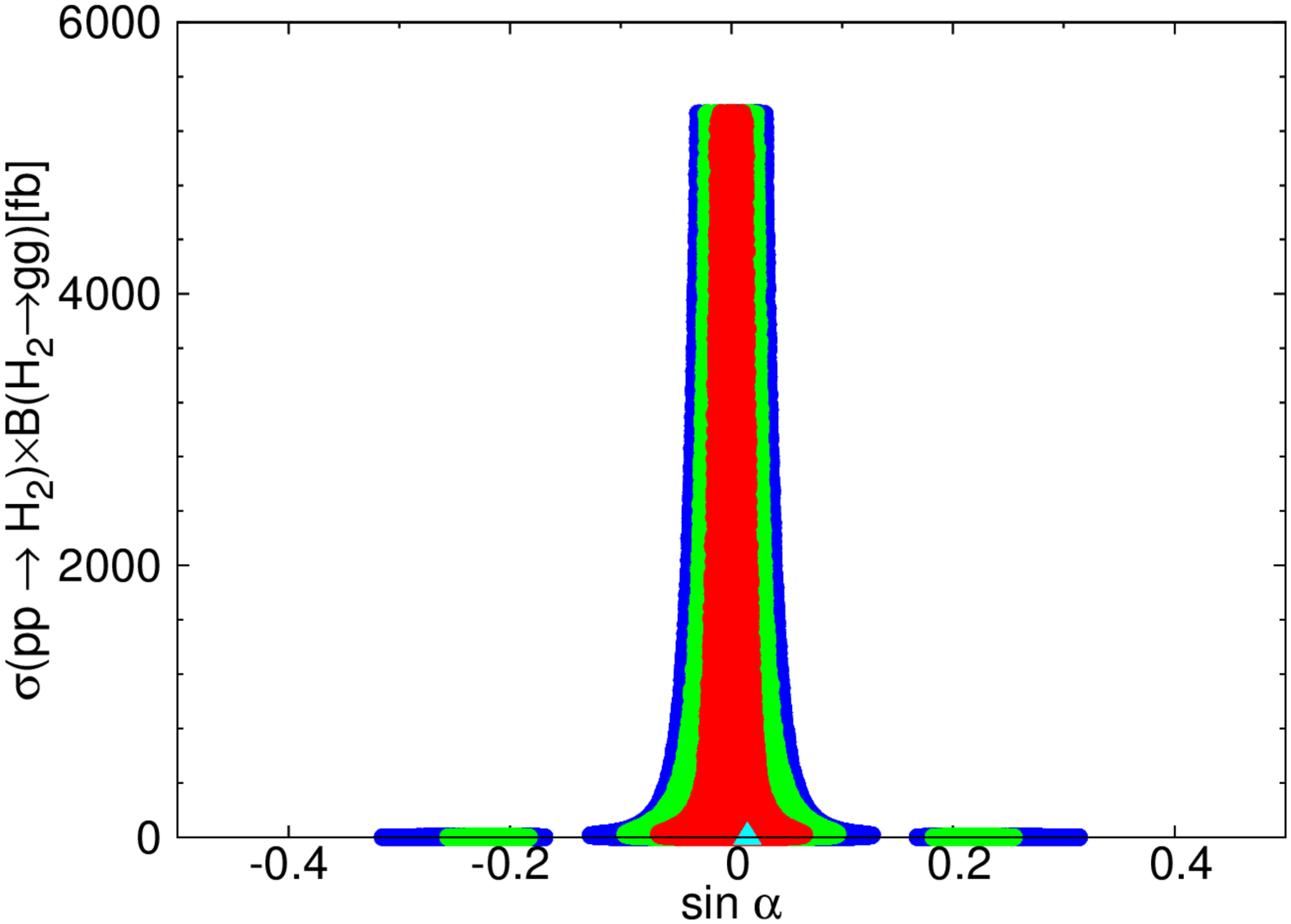}
\includegraphics[height=2.0in,angle=270]{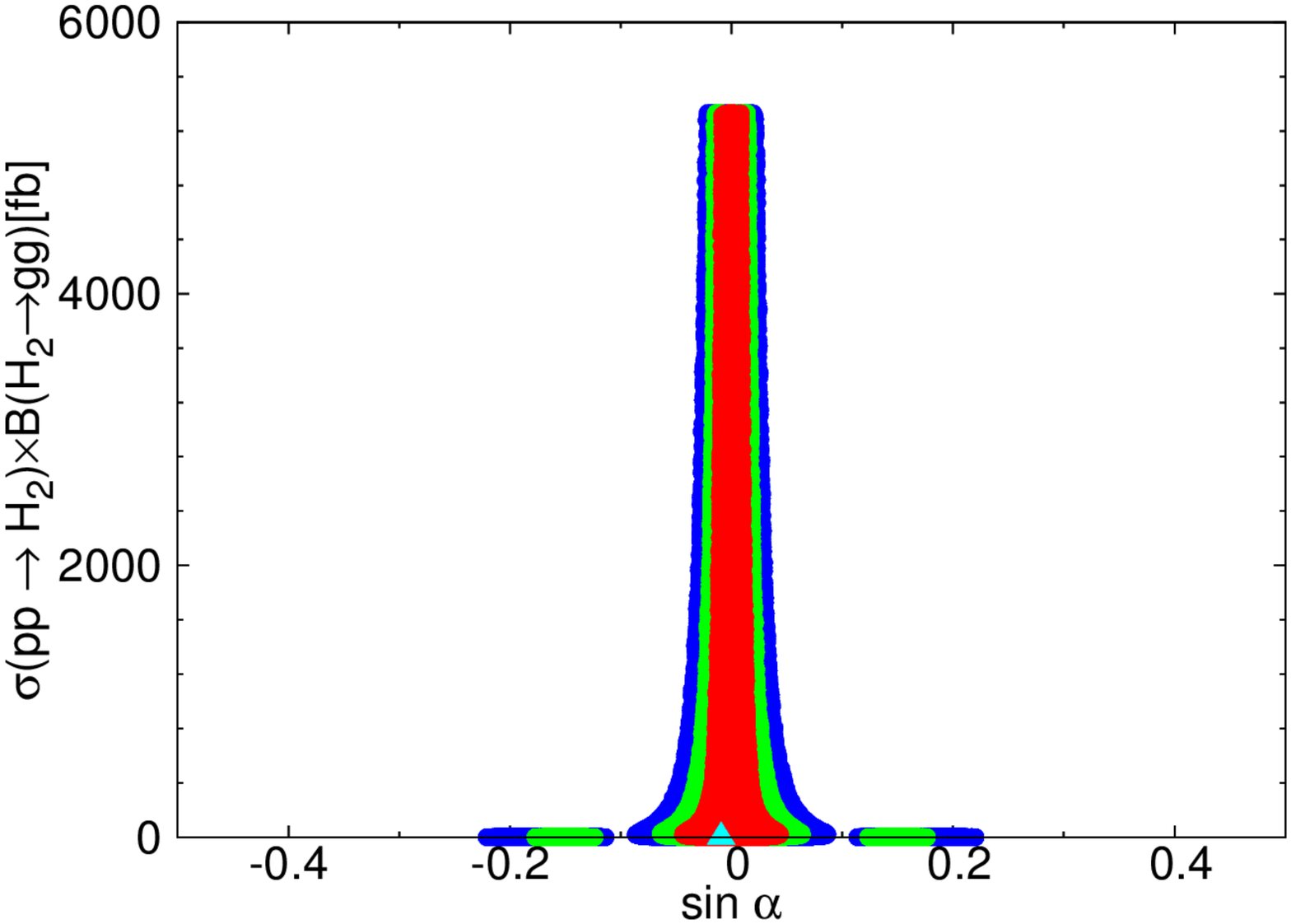}
\caption{\small \label{F4:sin_gg}
{\bf{F4 fits}}:
The same as in FIG.~\ref{F4:sin_sg} but for the CL regions in the
$(\sin\alpha,\sigma(gg\to H_2)\times B(H_2\to gg)$ plane.
}
\end{figure}

Figure~\ref{F4:sin_gg}  shows the CL regions in the 
$(\sin\alpha,\sigma(gg\to H_2)\times B(H_2\to gg))$ plane.
The cross section can be as large as up to $5$ pb
around $\sin\alpha =0$,
limited by the current $H_2$ dijet constraint.

\begin{figure}[th!]
\centering
\includegraphics[height=2.0in,angle=270]{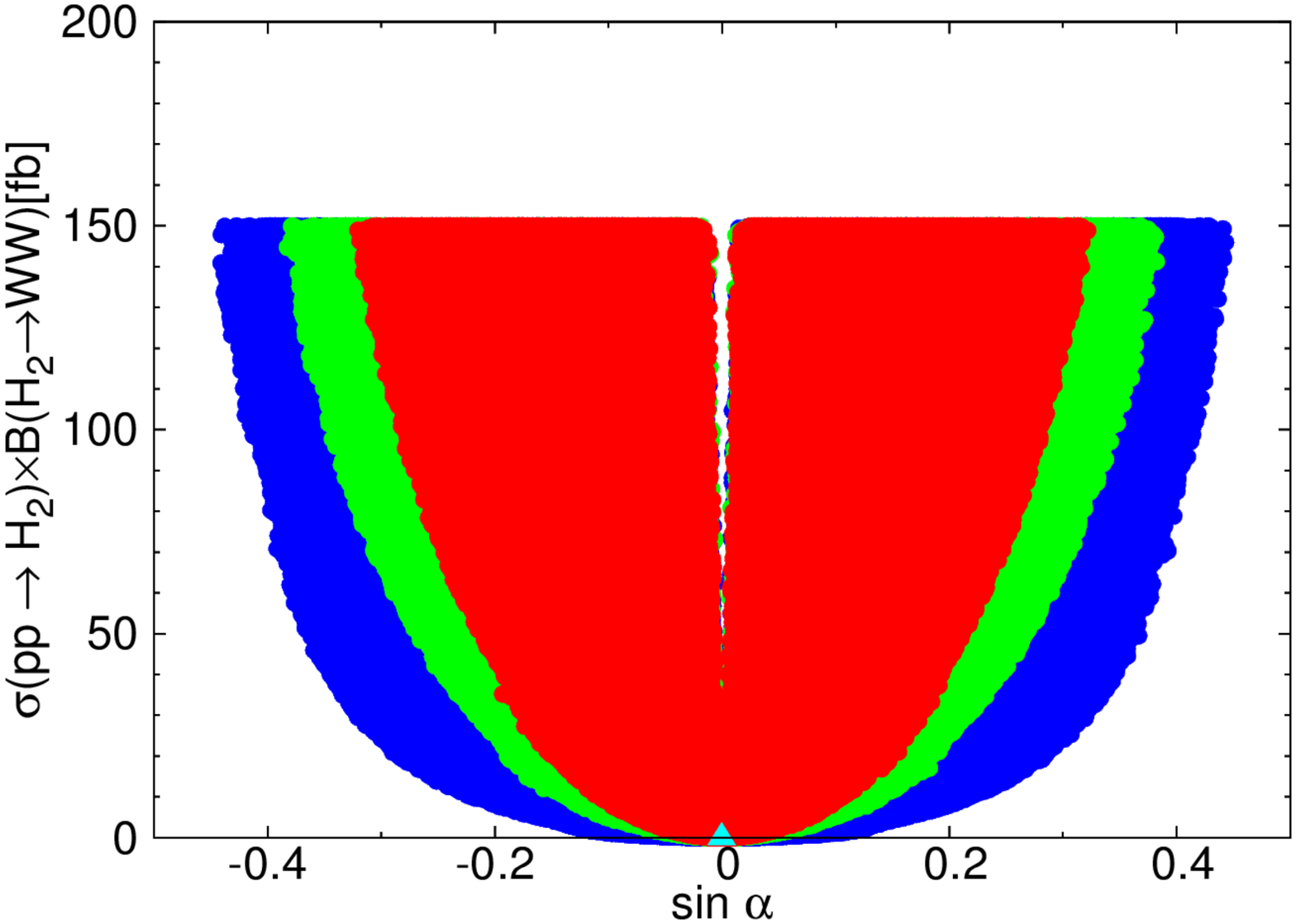}
\includegraphics[height=2.0in,angle=270]{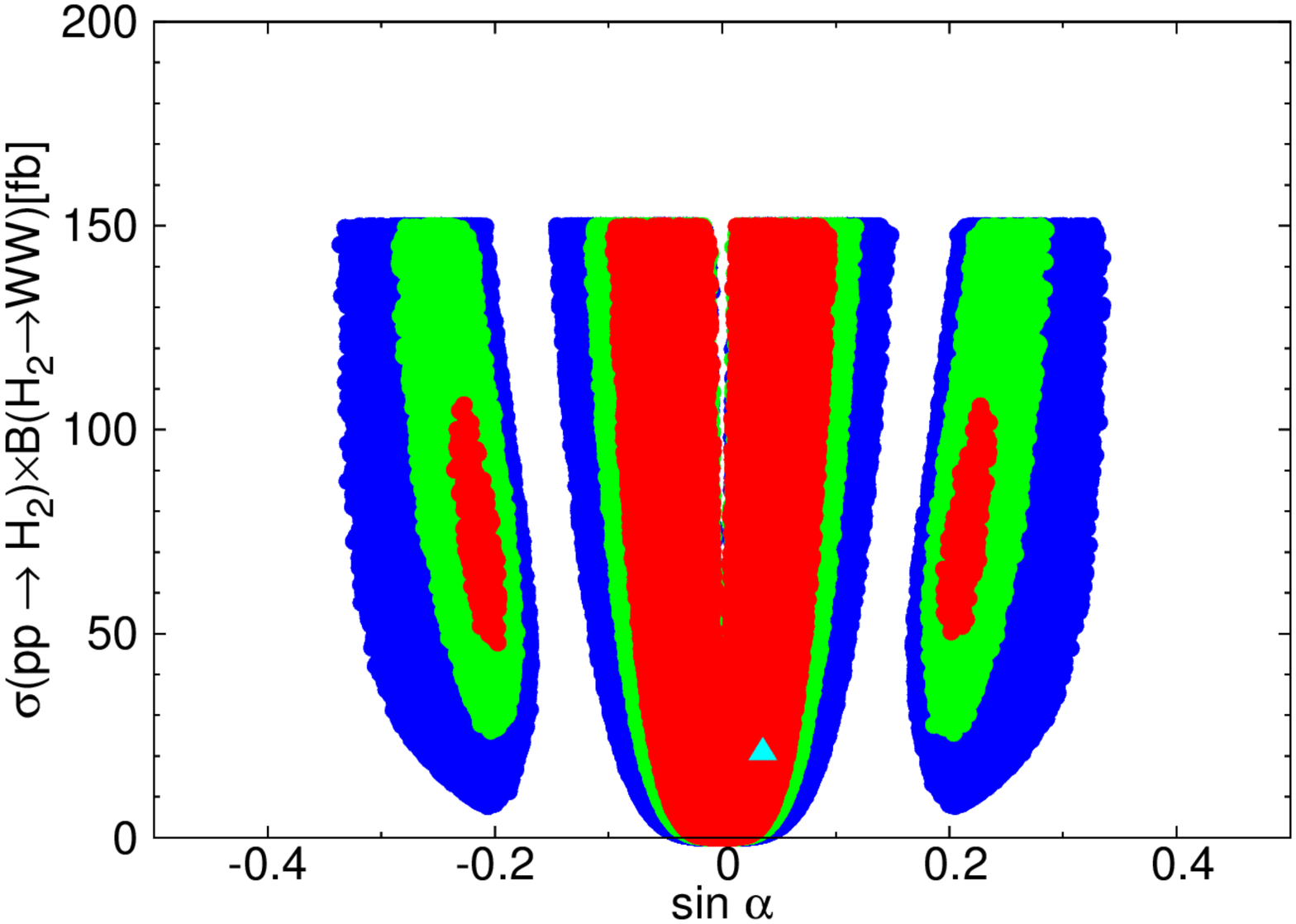}
\includegraphics[height=2.0in,angle=270]{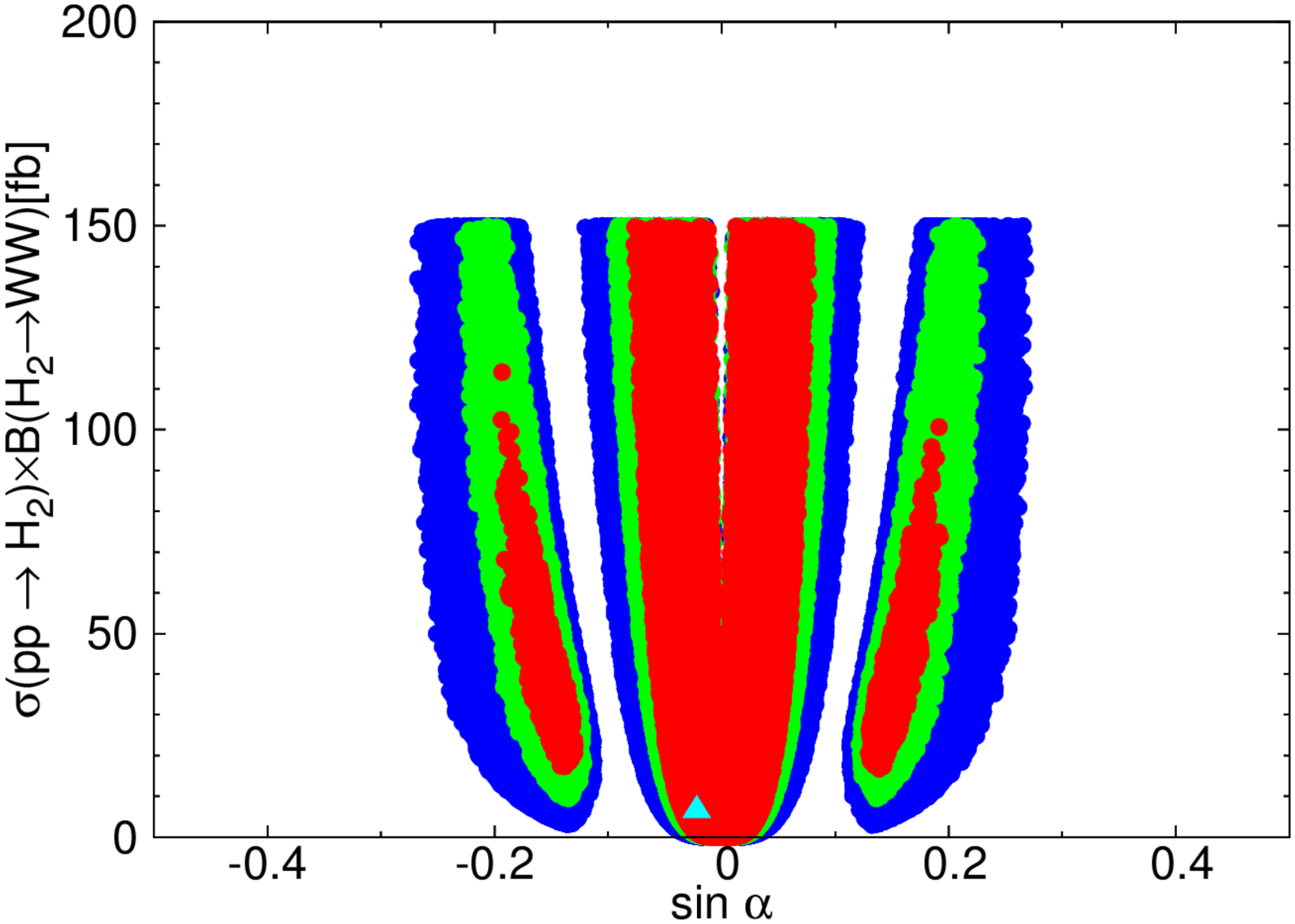}
\includegraphics[height=2.0in,angle=270]{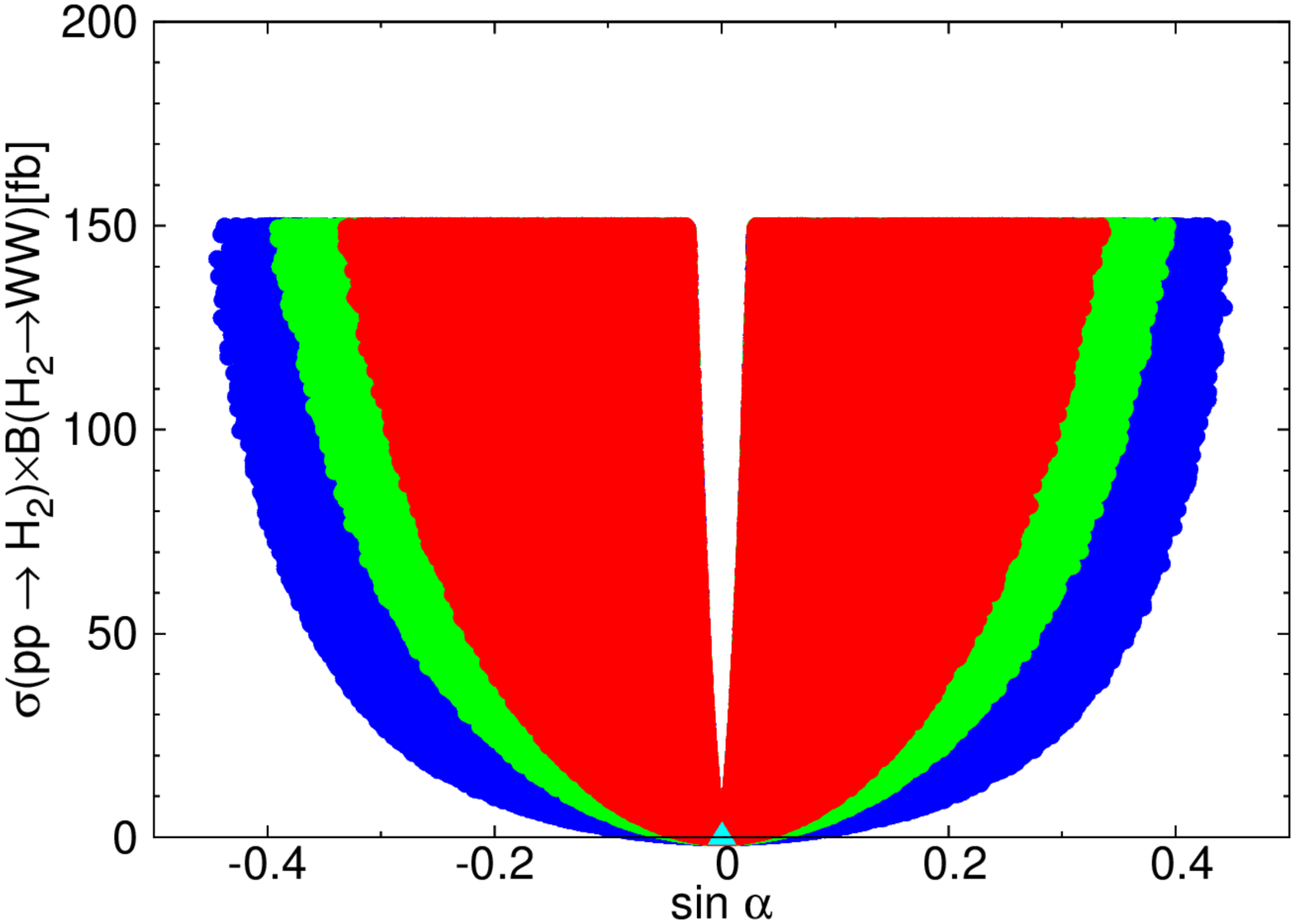}
\includegraphics[height=2.0in,angle=270]{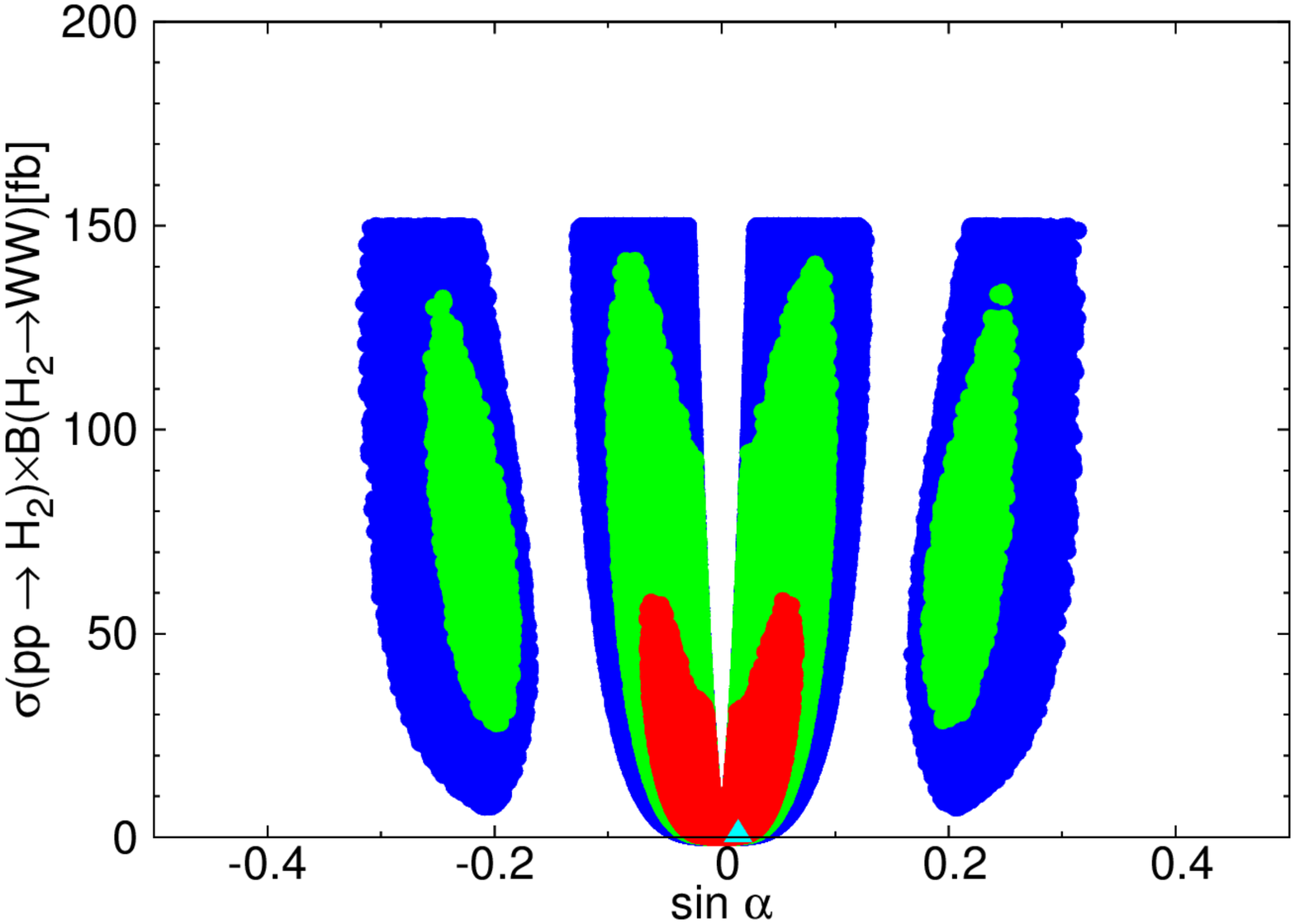}
\includegraphics[height=2.0in,angle=270]{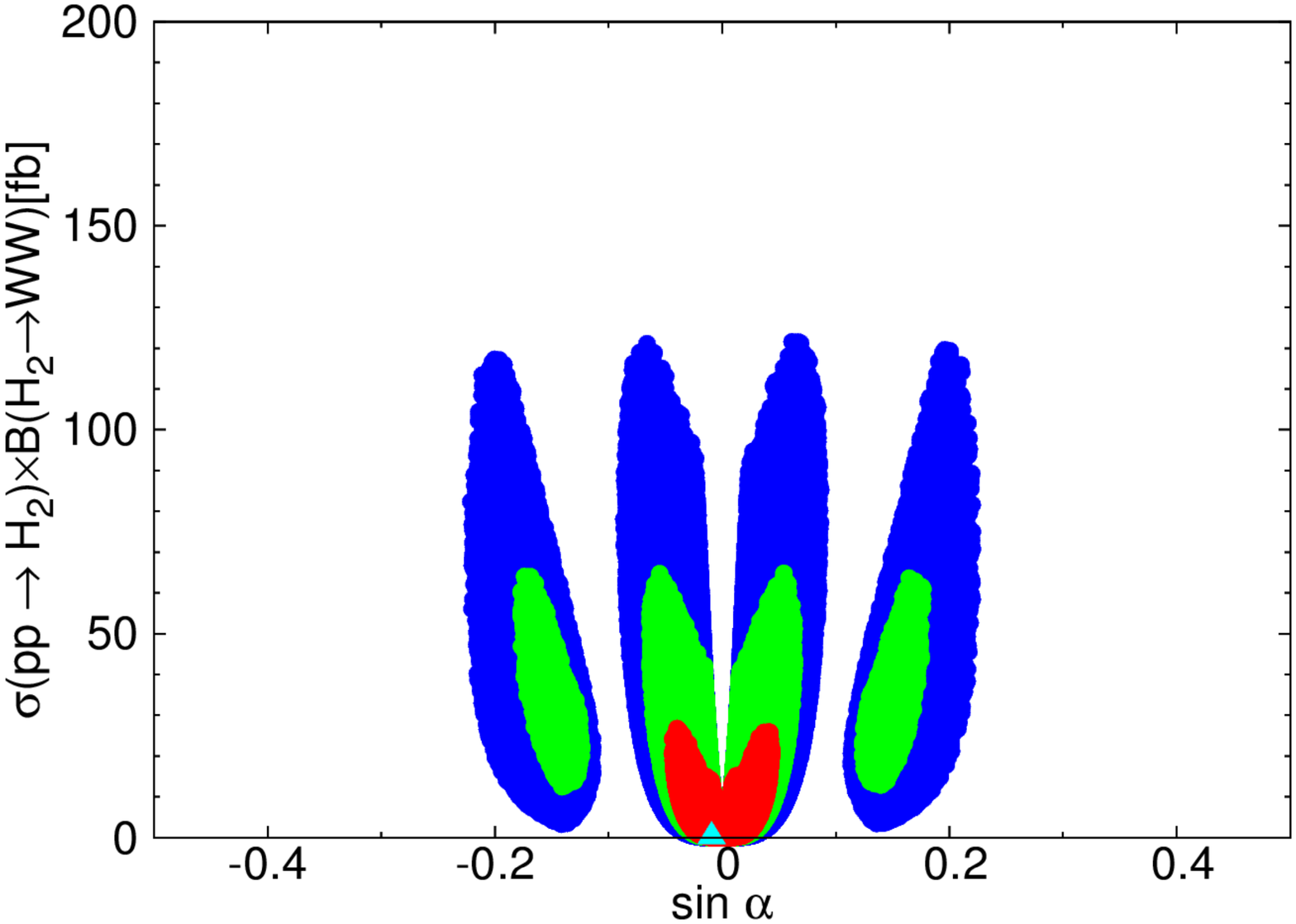}
\caption{\small \label{F4:sin_ww}
{\bf{F4 fits}}:
The same as in FIG.~\ref{F4:sin_sg} but for the CL regions in the
$(\sin\alpha,\sigma(gg\to H_2)\times B(H_2\to WW)$ plane.
}
\end{figure}

Figure~\ref{F4:sin_ww}  shows the CL regions in the 
$(\sin\alpha,\sigma(gg\to H_2)\times B(H_2\to WW))$ plane.
As $\sin\alpha$ deviates from $0$,
the cross section can be as large as $150$ fb,
limited by the current $H_2$ diboson constraint.

\begin{figure}[th!]
\centering
\includegraphics[height=2.0in,angle=270]{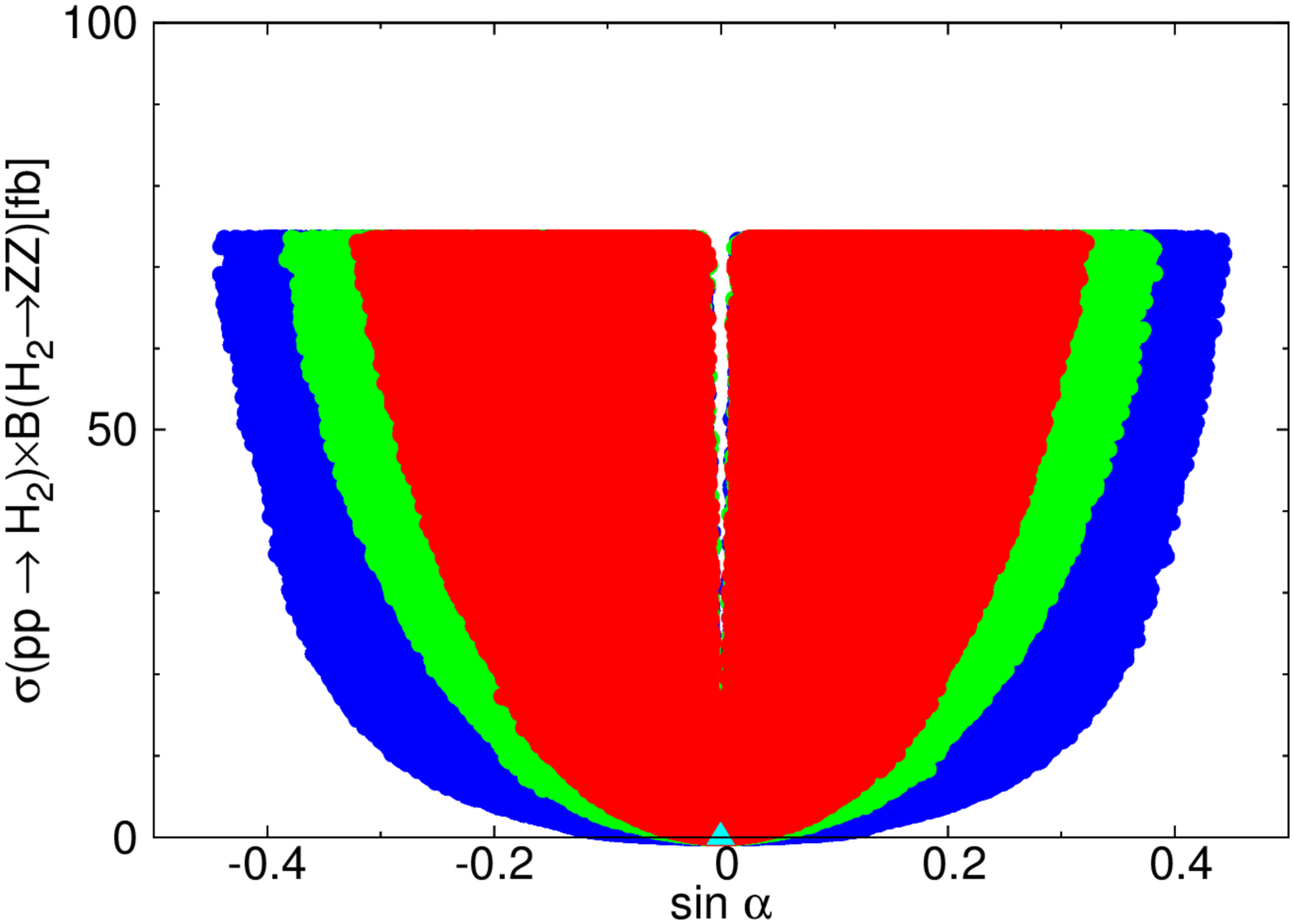}
\includegraphics[height=2.0in,angle=270]{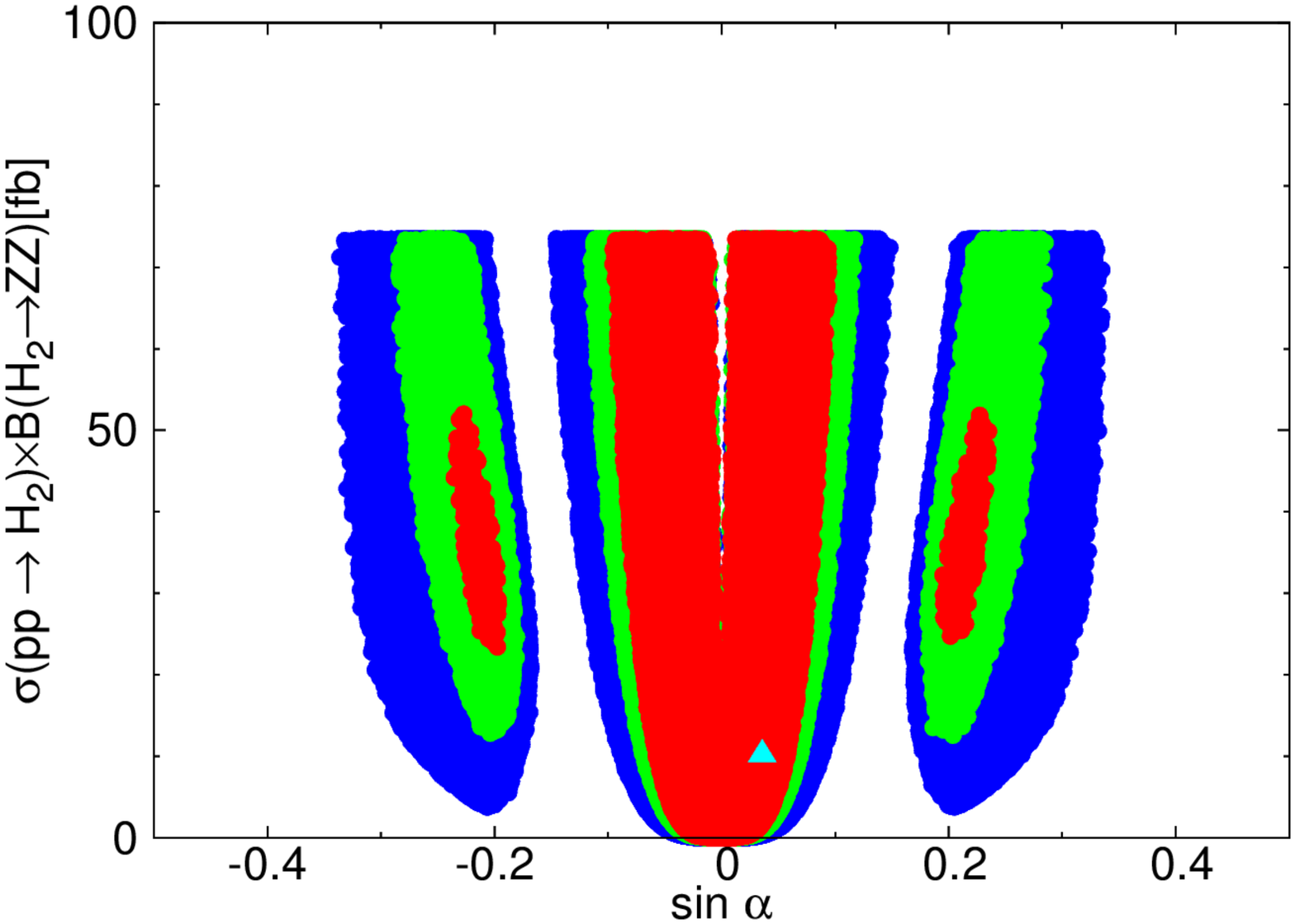}
\includegraphics[height=2.0in,angle=270]{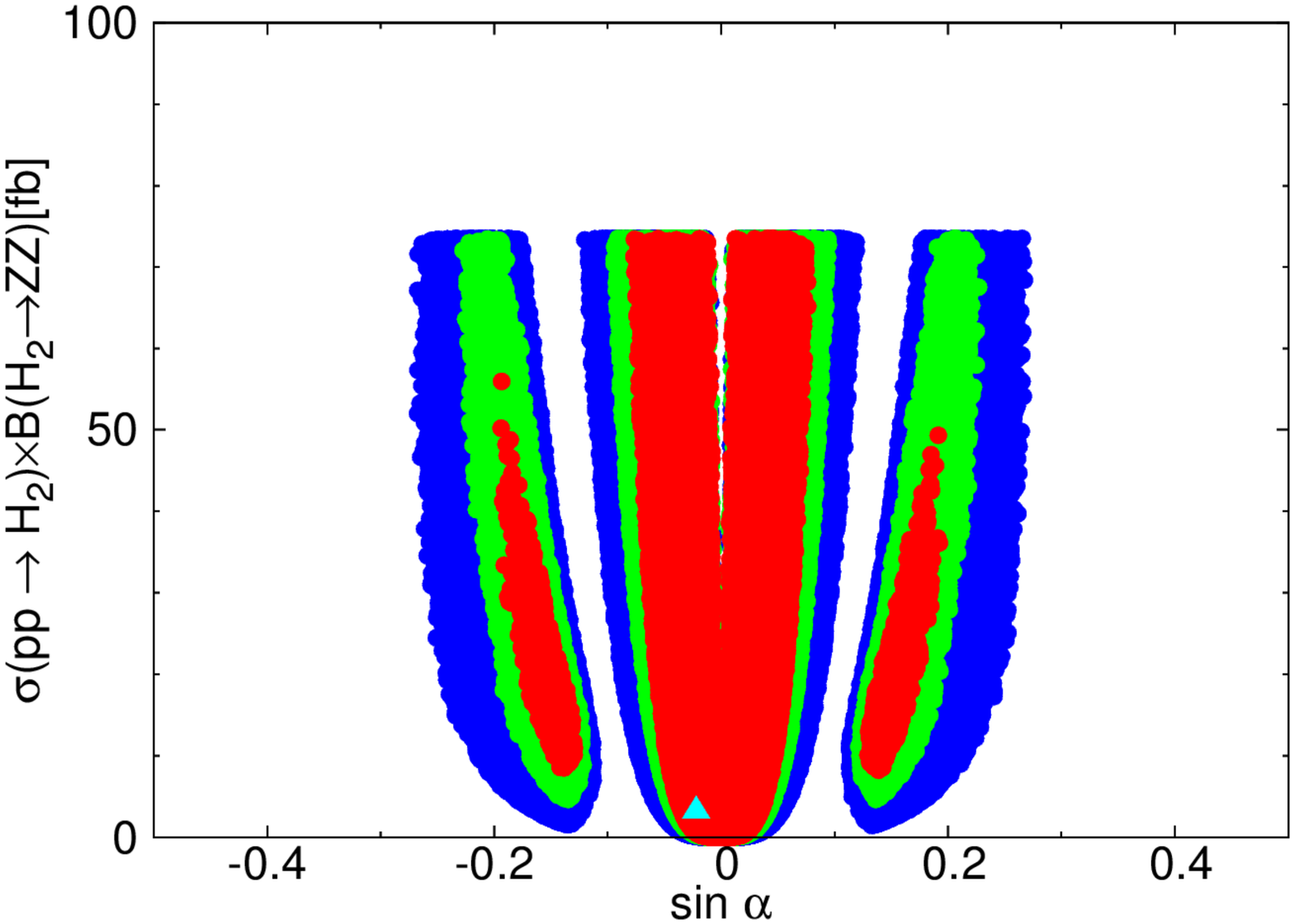}
\includegraphics[height=2.0in,angle=270]{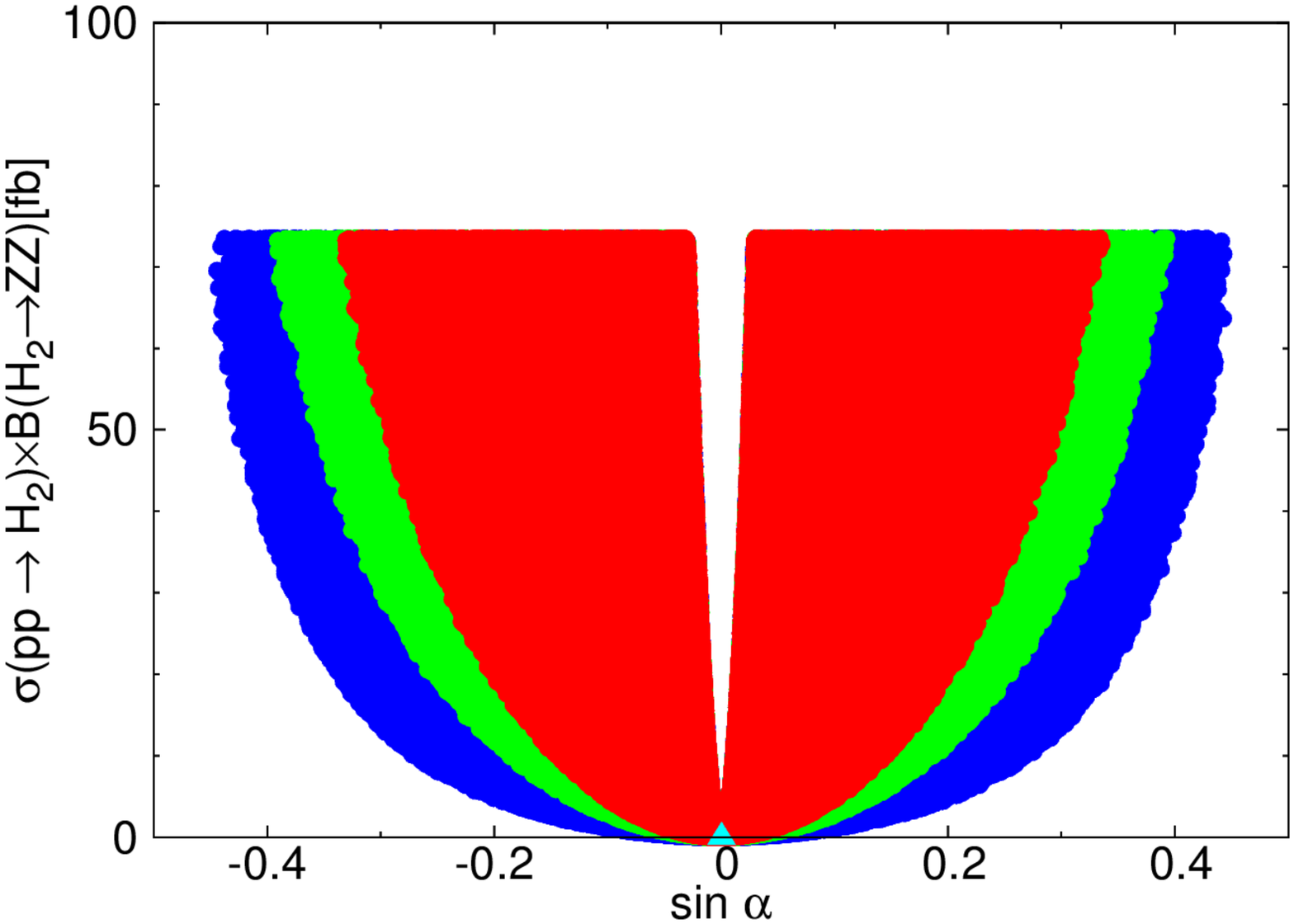}
\includegraphics[height=2.0in,angle=270]{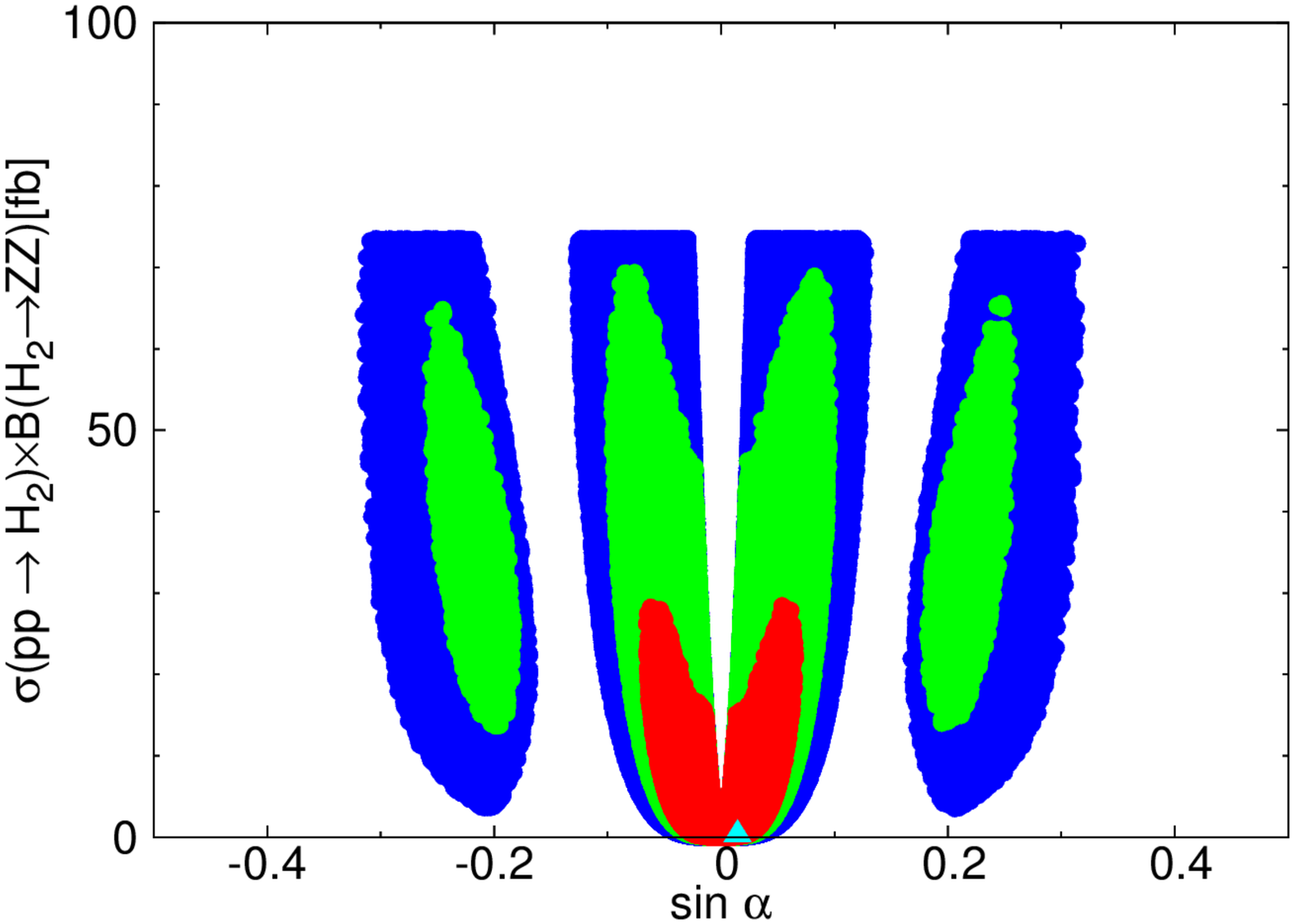}
\includegraphics[height=2.0in,angle=270]{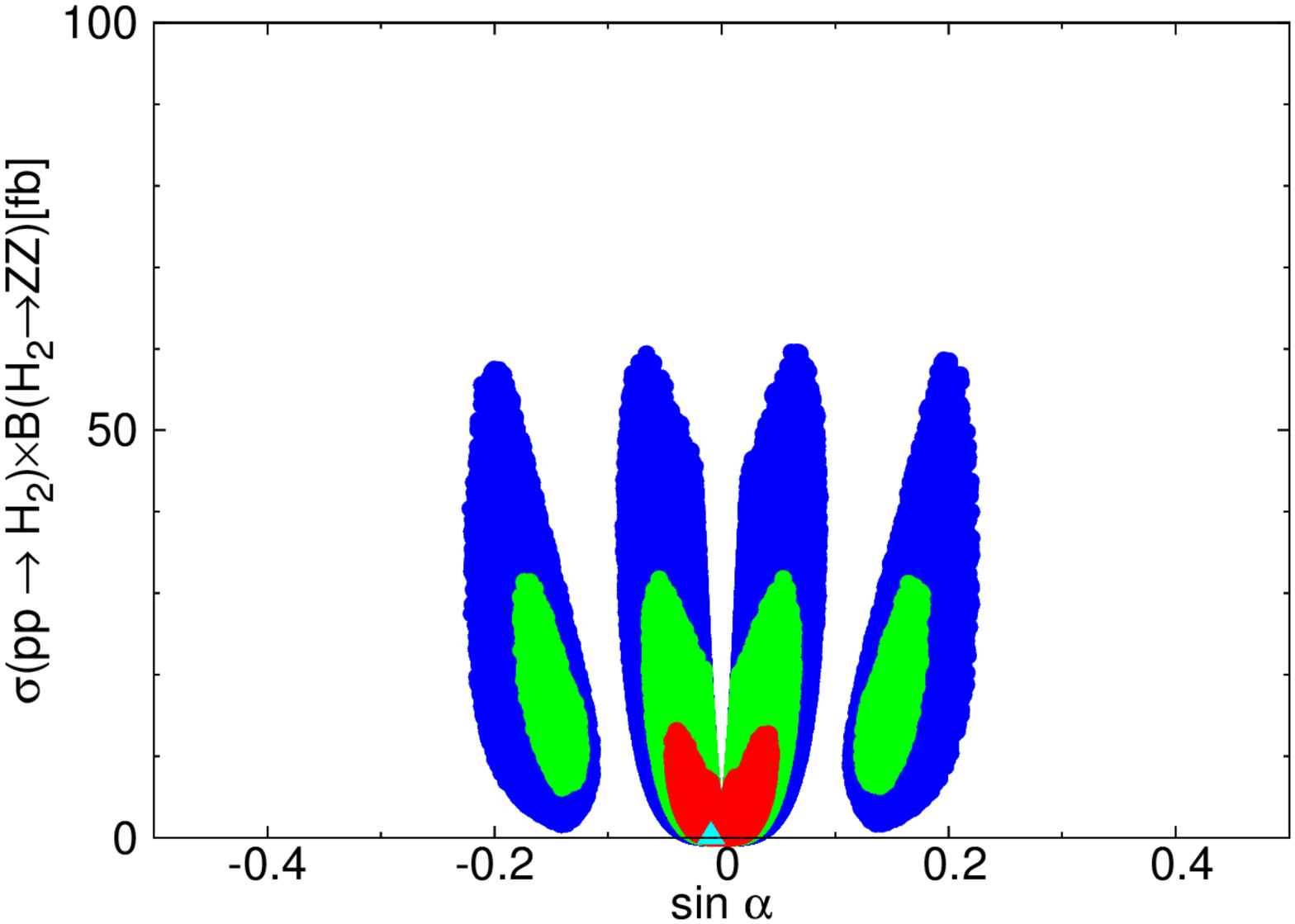}
\caption{\small \label{F4:sin_zz}
{\bf{F4 fits}}:
The same as in FIG.~\ref{F4:sin_sg} but for the CL regions in the
$(\sin\alpha,\sigma(gg\to H_2)\times B(H_2\to ZZ)$ plane.
}
\end{figure}

\begin{figure}[th!]
\centering
\includegraphics[height=2.0in,angle=270]{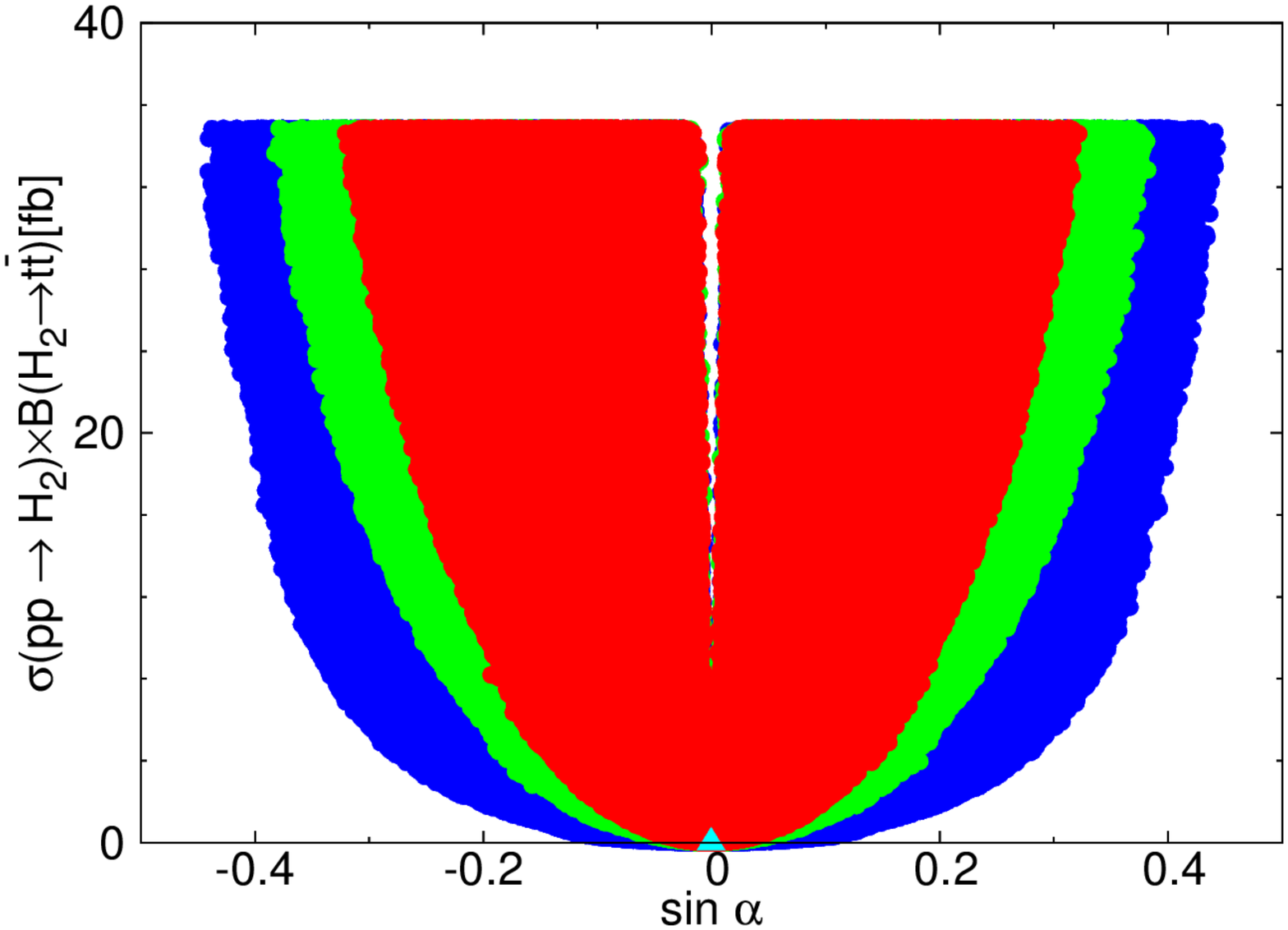}
\includegraphics[height=2.0in,angle=270]{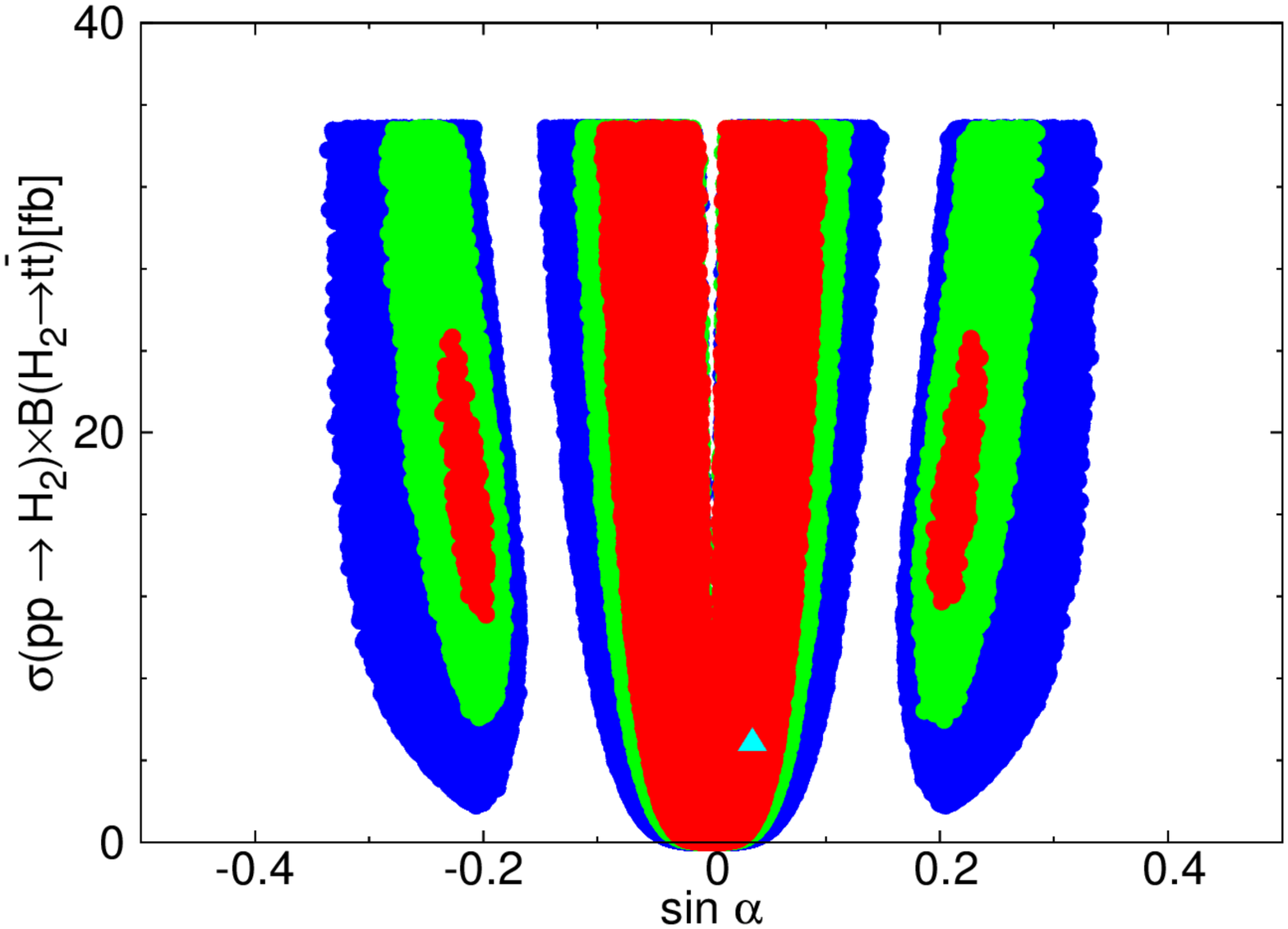}
\includegraphics[height=2.0in,angle=270]{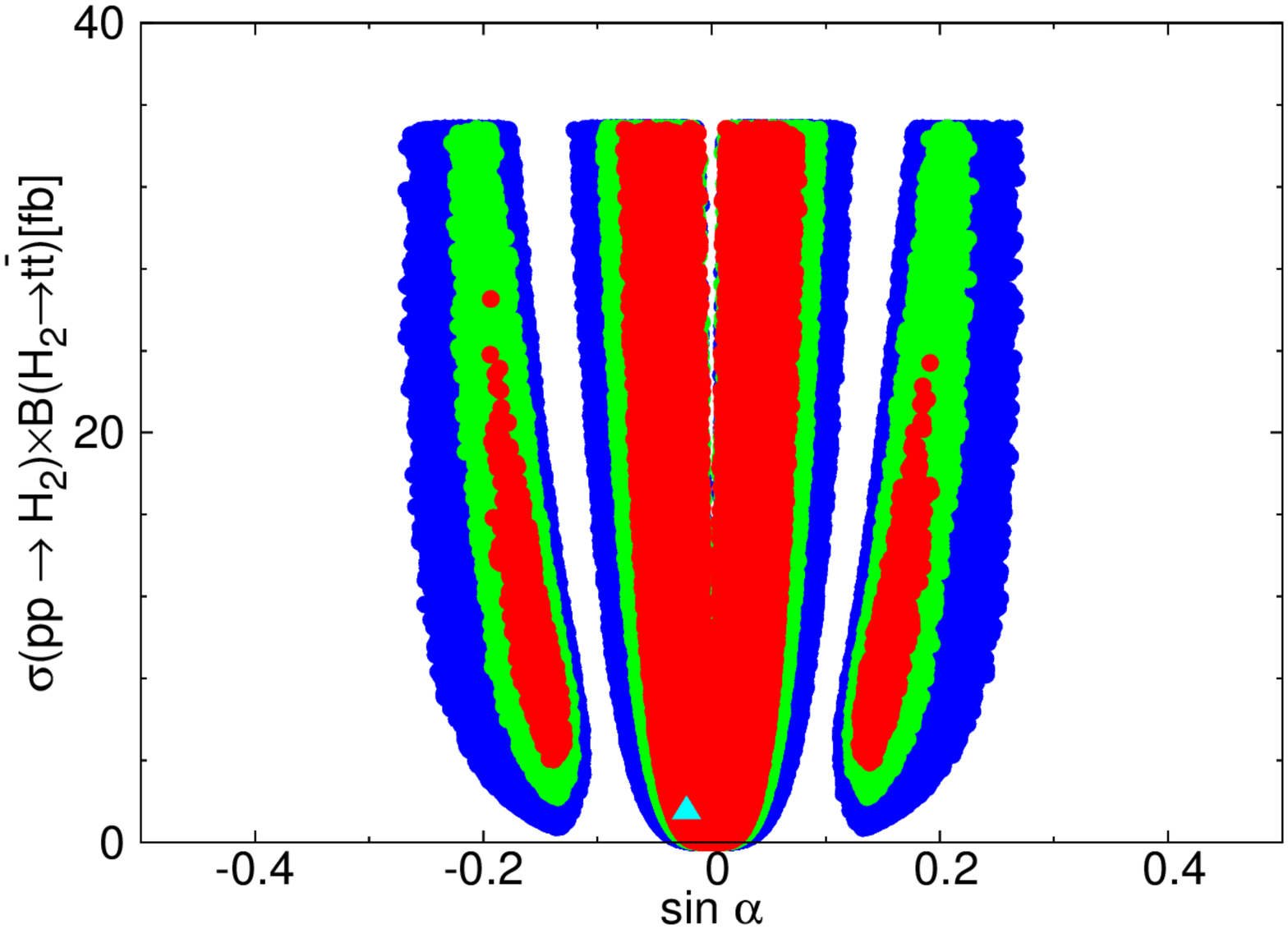}
\includegraphics[height=2.0in,angle=270]{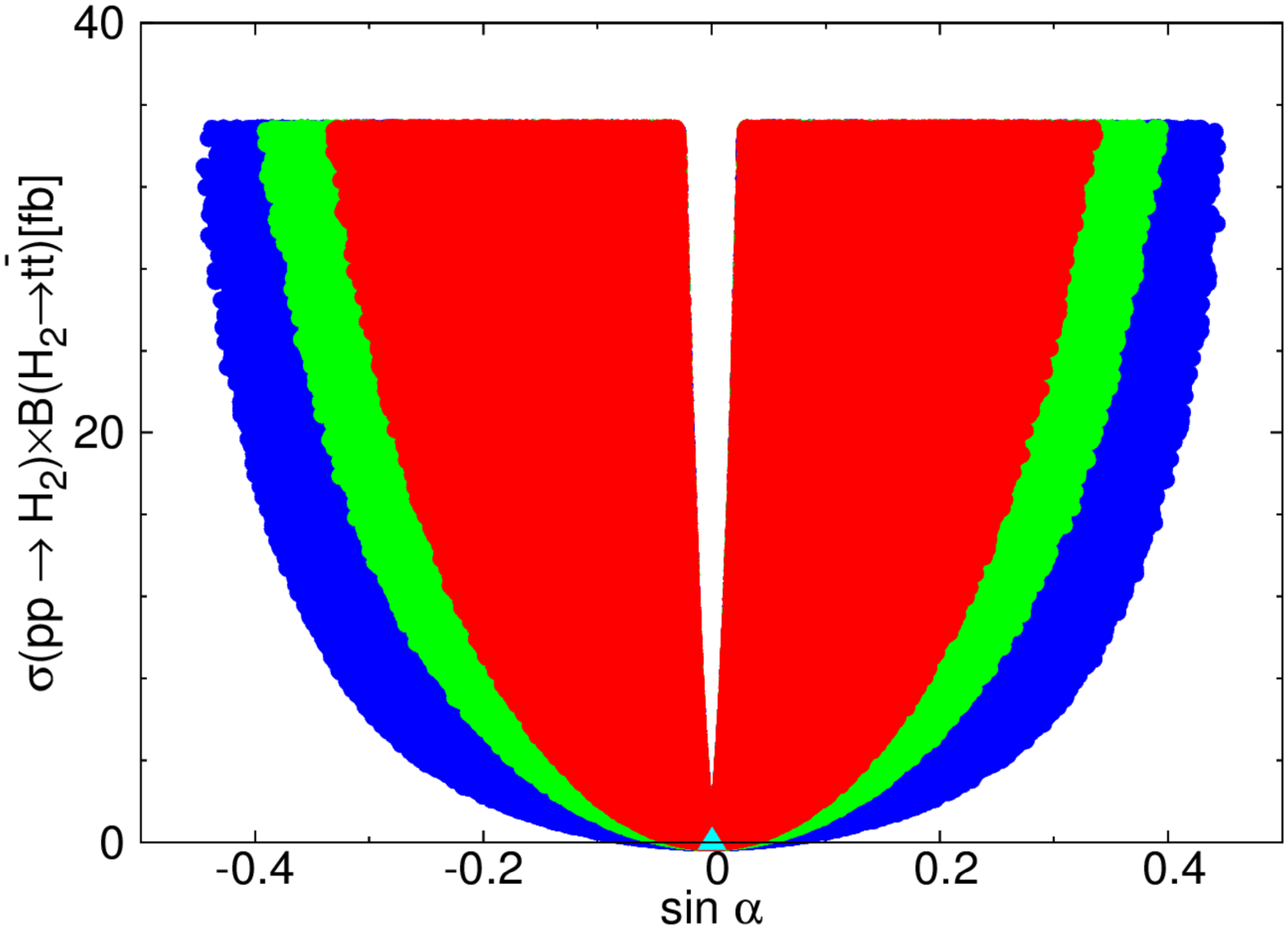}
\includegraphics[height=2.0in,angle=270]{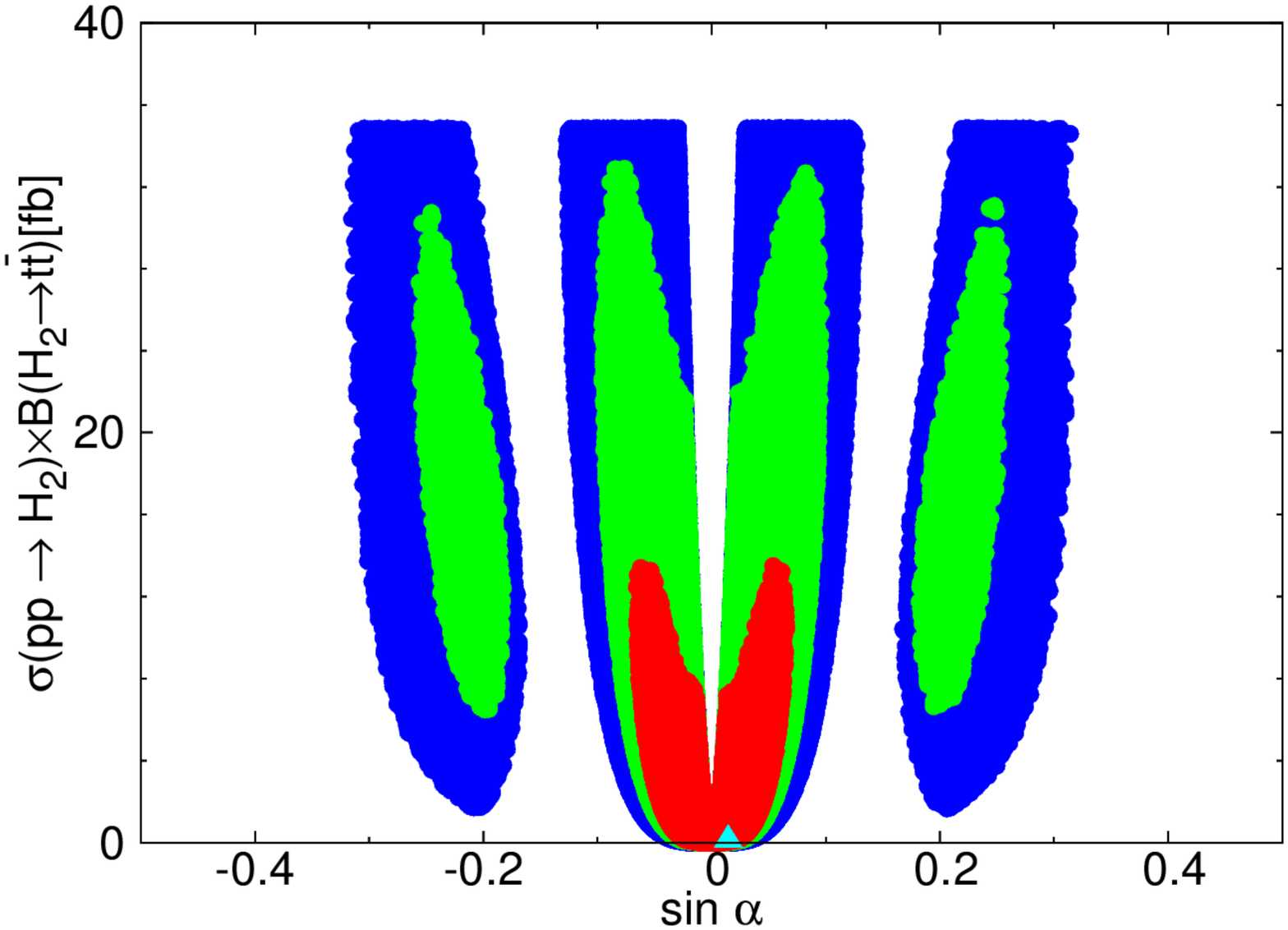}
\includegraphics[height=2.0in,angle=270]{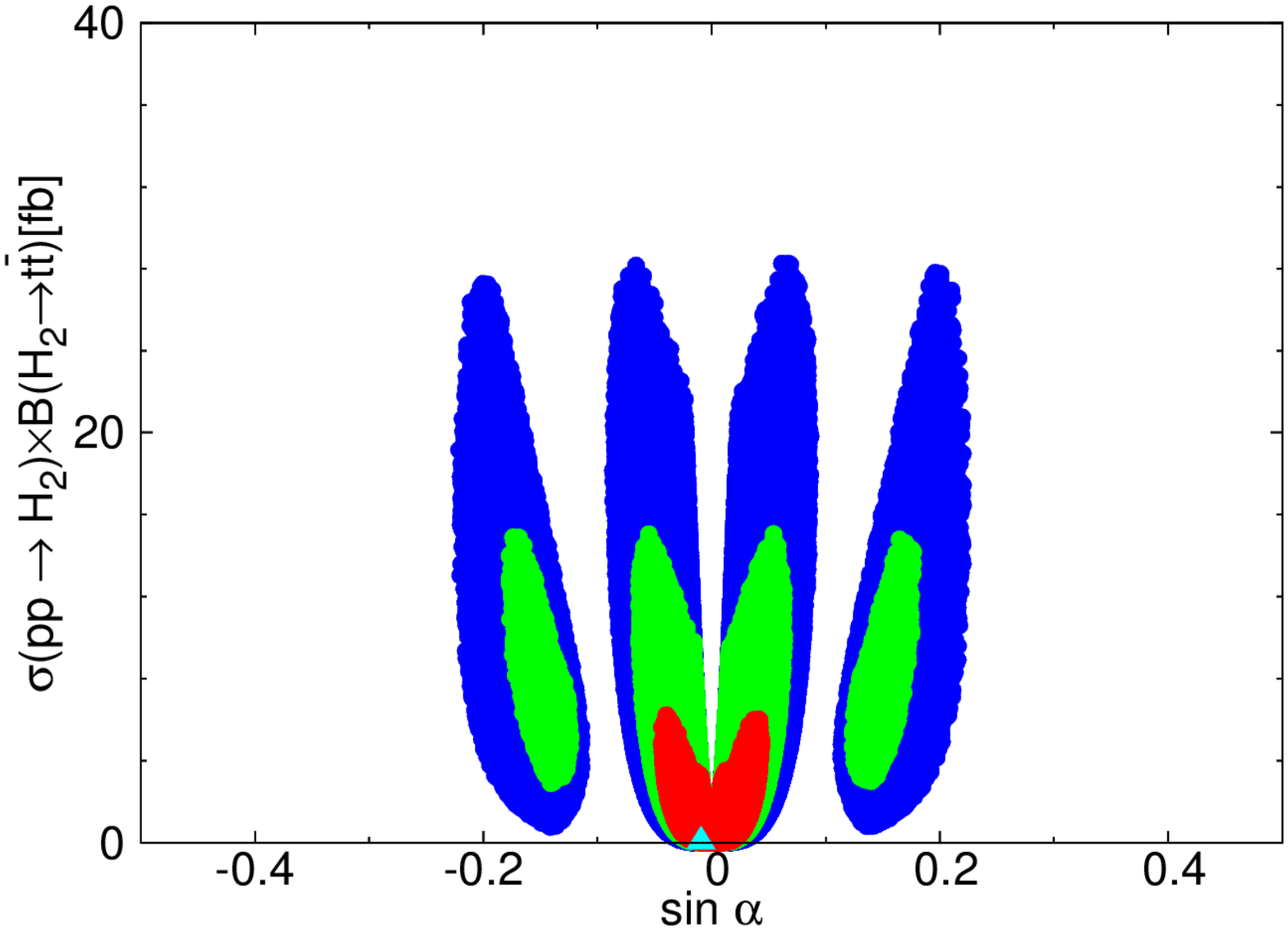}
\caption{\small \label{F4:sin_tt}
{\bf{F4 fits}}:
The same as in FIG.~\ref{F4:sin_sg} but for the CL regions in the
$(\sin\alpha,\sigma(gg\to H_2)\times B(H_2\to t\bar{t})$ plane.
}
\end{figure}

Compared to $\sigma(gg\to H_2)\times B(H_2\to WW)$,
the cross sections
$\sigma(gg\to H_2)\times B(H_2\to ZZ)$ and
$\sigma(gg\to H_2)\times B(H_2\to t\bar{t})$ 
can be as large as $70$ fb and $36$ fb, respectively,
suppressed by the factors
$B(H_2\to ZZ)/B(H_2\to WW)\sim 1/2$ and
$B(H_2\to t\bar{t})/B(H_2\to WW) \sim 1/5$:
see FIGs.~\ref{F4:sin_zz} and ~\ref{F4:sin_tt}.
Otherwise, their patterns are similar.

\begin{figure}[th!]
\centering
\includegraphics[height=2.0in,angle=270]{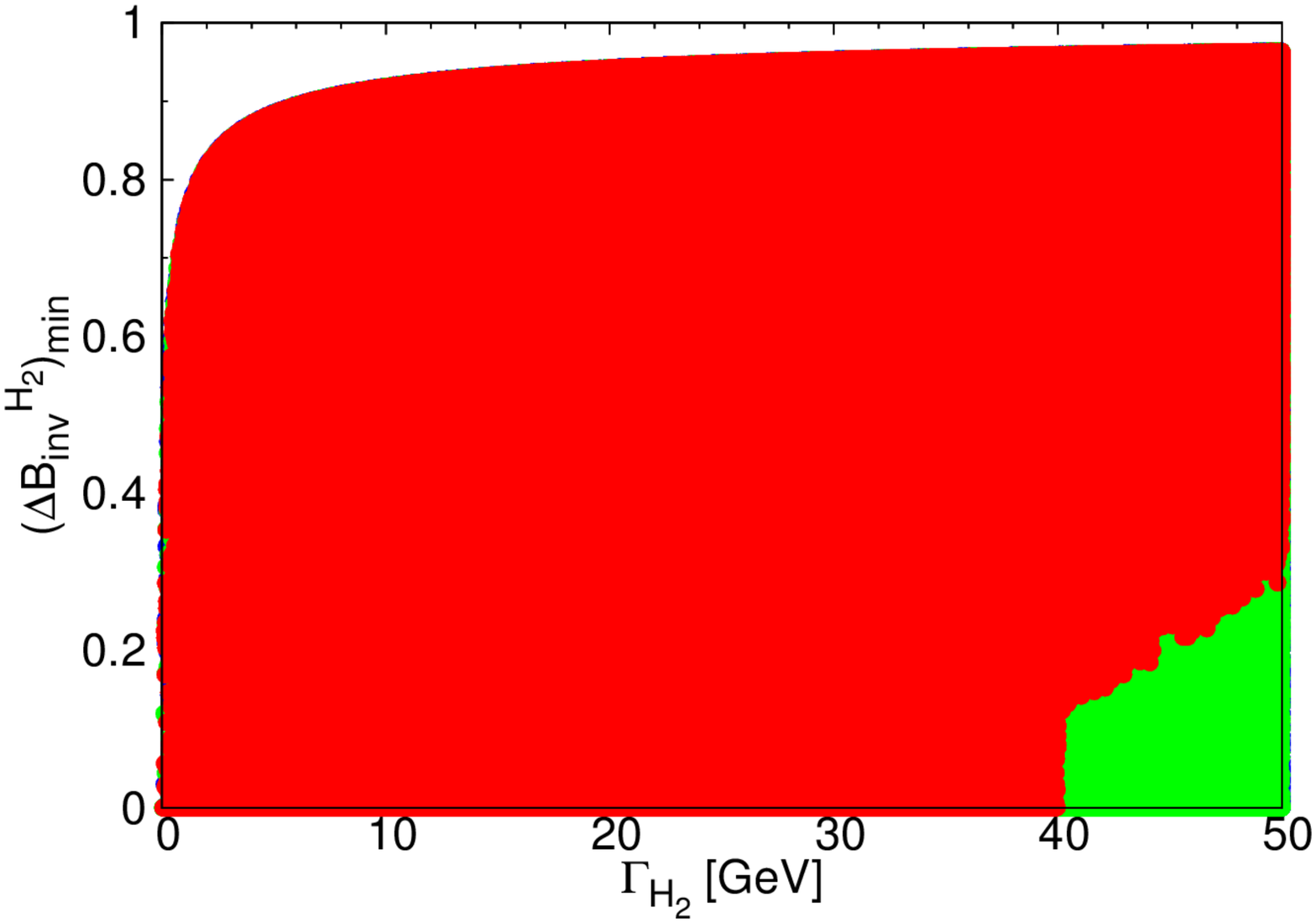}
\includegraphics[height=2.0in,angle=270]{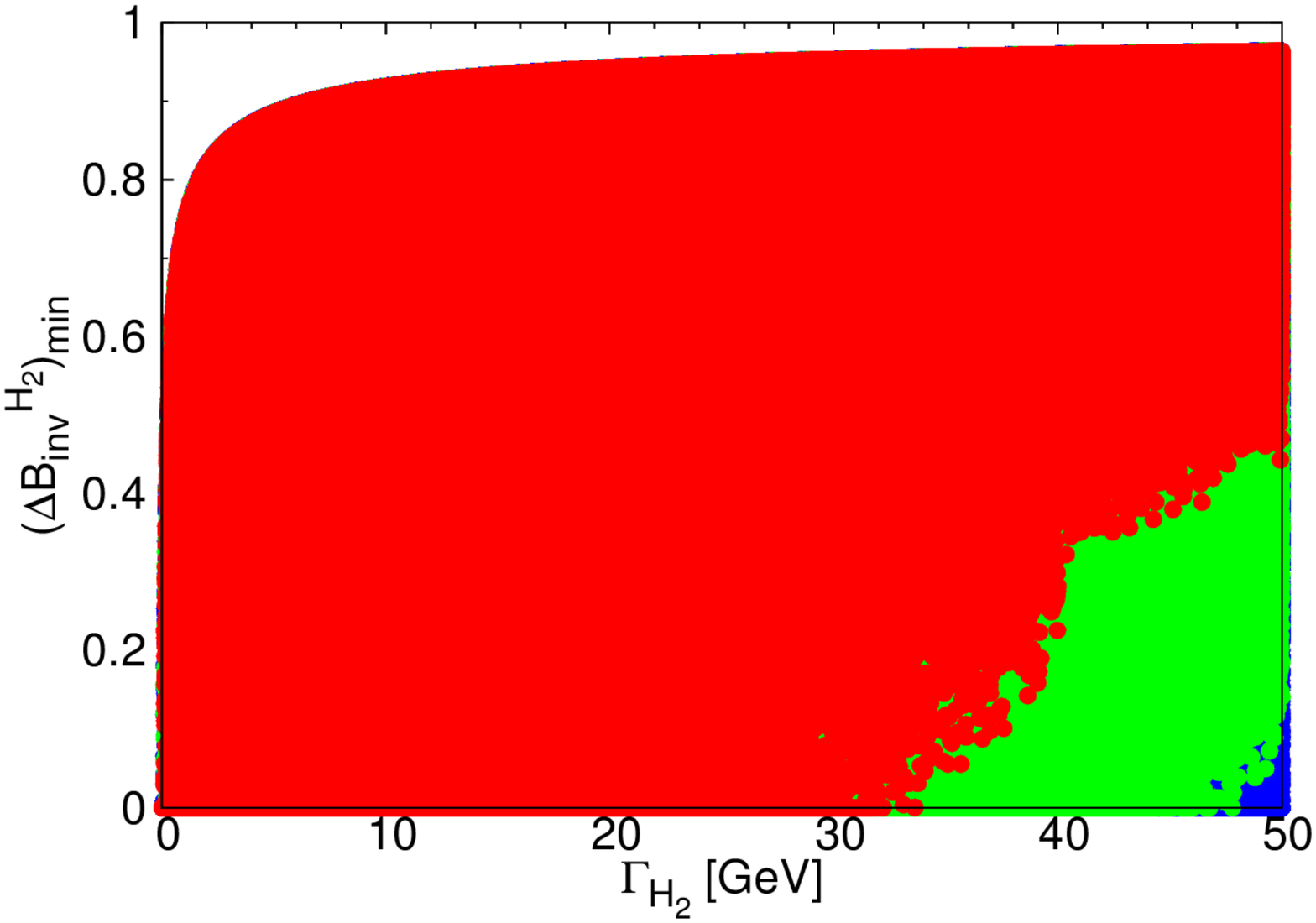}
\includegraphics[height=2.0in,angle=270]{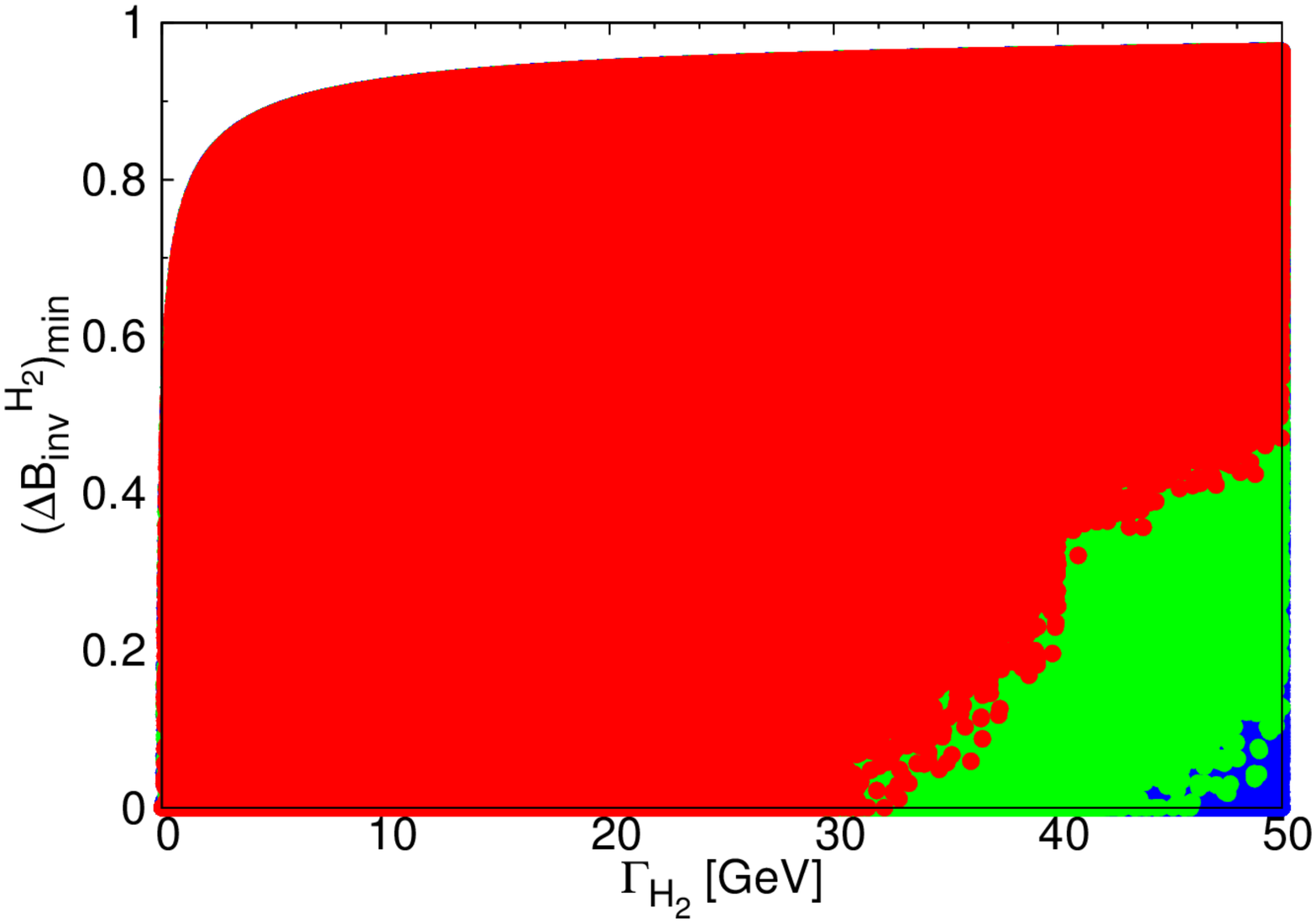}
\caption{\small \label{F4:db_gam}
{\bf{F4 fits}}: 
The CL regions for the {\bf F4-1} (left), {\bf F4-2} (middle),
and {\bf F4-3} (right) fits in the
$\left(\Gamma_{H_2},\left(\Delta B_{\rm inv}^{H_2}\right)_{\rm min}\right)$ plane.
The colors are the same as in FIG.~\ref{F4:sin_sg}.
See text for the definition of $\left(\Delta B_{\rm inv}^{H_2}\right)_{\rm min}$.
}
\end{figure}

Finally, in FIG.~\ref{F4:db_gam}, we show the CL regions in the 
$\left(\Gamma_{H_2},\left(\Delta B_{\rm inv}^{H_2}\right)_{\rm min}\right)$ plane
where
$$
\left(\Delta B^{H_2}_{\rm inv}\right)_{\rm min} \equiv \frac{
\left(\Delta\Gamma^{H_2}_{\rm inv}\right)_{\rm min}}{\Gamma_{H_2}}
$$
denoting the minimum value of the $H_2$ branching ratio into invisible particles.
The minimum invisible decay width is obtained by requiring
the decay width $\Gamma(H_2\to H_1H_1)$ to saturate the current upper limit on 
$\sigma(gg\to H_2)\times B(H_2\to H_1 H_2)$.  More explicitly, we have
\begin{equation}
\left(\Delta\Gamma^{H_2}_{\rm inv}\right)_{\rm min} =
\Gamma_{H_2}-\left[
\sin^2\alpha\Gamma^{\rm SM}_{H_2} +
\Delta\Gamma_{\rm vis}^{H_2\to\gamma\gamma} +
\Delta\Gamma_{\rm vis}^{H_2\to gg} +
\Gamma(H_2\to H_1H_1)_{\rm max}\right] 
\end{equation}
with
\begin{equation}
\Gamma(H_2\to H_1H_1)_{\rm max} =  15\,{\rm GeV}\,
\left(\frac{150\,{\rm fb}}{\left.\sigma(gg\to H_2)\right|_{\sqrt{s}=8\,{\rm TeV}}}
\right)\,
\left(\frac{\Gamma_{H_2}}{50\,{\rm GeV}}\right)
\end{equation}
where we again take $\left.\sigma(gg\to H_2)\right|_{\sqrt{s}=8\,{\rm TeV}}
=\left.\sigma(gg\to H_2)\right|_{\sqrt{s}=13\,{\rm TeV}}/4.7$,
see Eq.~(\ref{eq:h2h1h1}).
We observe that, at 68\% CL, 
$\left(\Delta B^{H_2}_{\rm inv}\right)_{\rm min} \sim 0$ can accommodate 
the situation with 
$\Gamma_{H_2}\lsim 40\,(32\,,32)$ GeV for {\bf F4-1} ({\bf F4-2, F4-3}).
But it should be larger than $\sim 0.2$ in order to accommodate the value
 $\Gamma_{H_2}\gsim 40$
GeV. Especially, when $\Gamma_{H_2}\sim 50$ GeV, 
the invisible branching ratio should be larger than
$0.3$, $0.45$, and $0.45$
for {\bf F4-1}, {\bf F4-2}, and {\bf F4-3}, respectively, at
68\% CL.

\section{$H_2\to WW\,,ZZ\,,Z\gamma$}
\label{sec:vlq}
In this section, we consider the more general case 
in which there exist interactions between VLQs and $W/Z$ bosons.
Then, even in the limit of $\sin\alpha=0$, 
$H_2$ can decay into $WW$ and $ZZ$ via VLQ loops and, more importantly, 
into $Z\gamma$.

Note that the couplings of VLQs to $W/Z$ bosons are highly model dependent 
 on the weak isospin and the $U(1)_Y$ hypercharges. 
In order to be specific but without much loss of generality, 
we introduce $N_d$ copies of VLQ doublets 
$Q_d=(U,D)^T$ and $N_s$ copies of VLQ singlets $Q_s$ which couples to
the SM gauge bosons as follows:
\begin{eqnarray}
-{\cal L}_{\rm QCD} &=&  g_s 
\left( \bar U \gamma^\mu T^a U
     + \bar D \gamma^\mu T^a D
     + \bar Q_s \gamma^\mu T^a Q_s \right )\, G^a_\mu\,, \nonumber \\[3mm]
-{\cal L}_{\rm EW} &=& 
\overline{Q_d}  \gamma^\mu \left ( 
     g \frac{\tau^a}{2} W^a_\mu +  g' \frac{Y_d}{2} B_\mu \right ) \, Q_d
 + \overline{Q_s}  \gamma^\mu \left(  g' \frac{Y_s}{2} B_\mu
  \right ) Q_s\,,
\end{eqnarray}
where $g_s$ denotes the $SU(3)$ gauge couping,
$g=e/s_W$ and $g'=e/c_W=gt_W$ with 
$s_W\equiv\sin\theta_W$, $c_W\equiv\cos\theta_W$, and $t_W=s_W/c_W$, and
$T^a$ and $\tau^a/2$  are generators of $SU(3)$ and $SU(2)$ groups.
And $Y_d$ and $Y_s$ denote the $U(1)_Y$ hypercharges of 
doublet $Q_d$ and singlet $Q_s$, respectively.  They 
are related with the  electric charges of VLQs by:
\begin{equation}
Q_{Q_s}=Y_s/2\,, \ \ \
Q_{U}=1/2+Y_d/2\,, \ \ \
Q_{D}=-1/2+Y_d/2\,. 
\end{equation}
Note that $Q_U-Q_D=1$ independently of $Y_d$.
After rotating $(W_\mu^3,B_\mu)^T$ into $(Z_\mu,A_\mu)^T$ as usual
or replacing $W_\mu^3$ and $B_\mu$ with
$c_W Z_\mu +s_W A_\mu$ and $-s_W Z_\mu +c_W A_\mu$, respectively,
one may have
\begin{eqnarray}
\label{eq:ew}
-{\cal L}_{\rm EW} &=& 
 e \left[ Q_U \bar U \gamma^\mu U + Q_D \bar D \gamma^\mu D +
     Q_{Q_s} \bar Q_s \gamma^\mu Q_s \right ]   \, A_\mu \nonumber \\[3mm]
 &+& \frac{g}{c_W} \left[
   \bar U \gamma^\mu U \left(\frac{1}{2} - s_W^2 Q_U \right )
 + \bar D \gamma^\mu D \left(- \frac{1}{2} - s_W^2 Q_D \right )
 + \bar Q_s \gamma^\mu Q_s \left( - s_W^2 Q_{Q_s} \right )  \right ] \, Z_\mu
  \nonumber \\[3mm]
 &+& \frac{g}{\sqrt{2}} ( \bar U \gamma^\mu D\, W^+_\mu +
                             \bar D \gamma^\mu U\, W^-_\mu  )\,.
\end{eqnarray}
We note the couplings to the $Z$ boson are purely vector-like and 
proportional to the factors $\pm 1/2 -s_W^2 Q_{U,D}$ which are different 
from the SM case where only the left-handed quarks
are participating in the $SU(2)$ interaction. 
Further we note that the couplings to the $Z$ boson become the same as those to
photons taking $(I_3^F-s_W^2 Q_F)/s_Wc_W  \,\to\,  Q_F$ with
$I_3^U=-I_3^D=1/2$ and $I_3^{Q_s}=0$.
Incidentally, the Yukawa couplings of VLQs to the singlet $s$ are given by
\begin{equation}
-{\cal L}_Y=g^S_{s\bar{Q}_dQ_d}s\bar{Q}_dQ_d + 
g^S_{s\bar{Q}_sQ_s}s\bar{Q}_sQ_s\,.
\end{equation}

With all these couplings given, one can calculate the VLQ-loop
contributions to the $H_2$ couplings to
$gg$, $\gamma\gamma$, $Z\gamma$, $ZZ$, and $WW$,
which are proportional to $\cos\alpha$.
For the $H_2$ couplings to two gluons and two photons,
adopting the same notations as in Eqs.~(\ref{eq:hgg}) and (\ref{eq:hpp}),
we have
\begin{eqnarray}
S^{g(Q)}_{H_2}      &=&
\sum_{Q_d} g^S_{s\bar{Q}_dQ_d}\left[
\frac{v}{m_{U}}F_{sf}(\tau_{2U})+\frac{v}{m_{D}}F_{sf}(\tau_{2D})
\right]+
\sum_{Q_s} g^S_{s\bar{Q}_sQ_s}\frac{v}{m_{Q_s}}F_{sf}(\tau_{2Q_s})\;,
\\[3mm]
S^{\gamma(F)}_{H_2} &=&
2\sum_{Q_d} N_Cg^S_{s\bar{Q}_dQ_d}\left[
Q^2_U \frac{v}{m_{U}} F_{sf}(\tau_{2U})+
Q^2_D \frac{v}{m_{D}} F_{sf}(\tau_{2D})
\right]+
2\sum_{Q_s} N_C g^S_{s\bar{Q}_sQ_s} Q^2_{Q_s}\frac{v}{m_{Q_s}}
F_{sf}(\tau_{2Q_s})\;. \nonumber
\end{eqnarray}
On the other hand, for the $H_2$ coupling to $Z$ and $\gamma$,
following the convention of Eq.~(\ref{eq:hzp}), we have
\begin{eqnarray}
\label{eq:SZgamma}
S^{Z\gamma(F)}_{H_2}&=&
2\sum_{Q_d} N_C g^S_{s\bar{Q}_dQ_d} \left\{
 Q_U \frac{(1/2-s^2_WQ_U) }{s_Wc_W} \frac{v}{m_{U}} 
 \left[2m_U^2F^{(0)}_{f}(M_{H_2}^2,m_U^2)\right] \right.  \nonumber \\[3mm]
&& \hspace{2.5cm} \left.
+Q_D \frac{(-1/2-s^2_WQ_D)}{s_Wc_W} \frac{v}{m_{D}} 
 \left[2m_D^2F^{(0)}_{f}(M_{H_2}^2,m_D^2)\right]
\right\} \nonumber \\[3mm]
&+& 2\sum_{Q_s}  N_C g^S_{s\bar{Q}_s{Q_s}} \left\{
 Q_{Q_s} \frac{(-s^2_WQ_{Q_s})}{s_Wc_W} \frac{v}{m_{Q_s}}
 \left[2m_{Q_s}^2 F^{(0)}_{f}(M_{H_2}^2,m_{Q_s}^2)\right] \right\}\;.
\end{eqnarray}
Note that, in the limit of $M_Z=0$,
we have $2m_X^2F^{(0)}_{f}(M_{H_2}^2,m_X^2)=F_{sf}(\tau_{2X})$
leading to
$S^{Z\gamma(F)}_{H_2} = S^{\gamma(F)}_{H_2}$ after replacing
$(I_3^F-s_W^2 Q_F)/s_Wc_W$ with $Q_F$ 
as noted following Eq.~(\ref{eq:ew}).

For the decay processes $H_2\to VV$ with $V=W/Z$, the amplitude is given by
\begin{equation}
{\cal M}_{VVH_2}=\sin\alpha{\cal M}_{VVH_2}^{(0)}
                +\cos\alpha{\cal M}_{VVH_2}^{(1)}
\end{equation}
and, in the leading order neglecting the SM one-loop contributions to the
$hWW$ vertex,
the tree-level and one-loop amplitudes are 
\begin{eqnarray}
\label{eq:ampVV}
{\cal M}^{(0)}_{WWH_2} &=& -gM_W \epsilon_1^*\cdot\epsilon_2^*\;,
\nonumber \\[3mm]
{\cal M}^{(0)}_{ZZH_2} &=& -\frac{gM_W}{c^2_W}
\epsilon_1^*\cdot\epsilon_2^*\;,
\nonumber \\[3mm]
{\cal M}^{(1)}_{VVH_2} &=&
-\frac{\alpha}{2\pi v} S^{VV(F)}_{H_2}\left[ k_1\cdot
k_2\,\epsilon_1^*\cdot\epsilon_2^*
-k_1\cdot\epsilon_2^*\,k_2\cdot\epsilon_1^* \right]\;,
\end{eqnarray}
where $k_{1,2}$ are the momenta of the two massive
vector bosons with $2k_1\cdot k_2 = M_{H_2}^2-2M_V^2$
and
$\epsilon_{1,2}$ are their polarization vectors.
Note that there exists a tree-level 
contribution to the amplitude when $\sin\alpha\neq 0$
which has different vertex structure from the loop-induced one.

The form factor $S_{H_2}^{ZZ(F)}$ can be cast into the form
\begin{eqnarray}
\label{eq:SZZ}
S^{ZZ(F)}_{H_2}&=& 2\sum_{Q_d}N_C g^S_{s\bar{Q}_dQ_d} \left\{
 \frac{(1/2-s_W^2Q_U)^2}{s_W^2c_W^2} \frac{v}{m_U}
 \left[2m_U^2F_f^{(1)}(M_{H_2}^2,m_U^2)\right] \right.
\nonumber \\
&& \hspace{2.5cm} \left.
+\frac{(-1/2-s_W^2Q_D)^2}{s_W^2c_W^2} \frac{v}{m_D}
 \left[2m_D^2F_f^{(1)}(M_{H_2}^2,m_D^2)\right]
\right\}
\nonumber \\[3mm]
&+& 2\sum_{Q_s}N_C g^S_{s\bar{Q}_sQ_s} 
 \frac{(-s_W^2Q_{Q_s})^2}{s_W^2c_W^2} \frac{v}{m_{Q_s}}
 \left[2m_{Q_s}^2F_f^{(1)}(M_{H_2}^2,m_{Q_s}^2)\right] \,.
\end{eqnarray}
Note that, in the limit of $M_Z=0$,
we have $2m_X^2F^{(1)}_{f}(M_{H_2}^2,m_X^2)=F_{sf}(\tau_{2X})$
leading to
$S^{ZZ(F)}_{H_2} = S^{\gamma(F)}_{H_2}$ after replacing
$(I_3^F-s_W^2 Q_F)/s_Wc_W$ with $Q_F$, see Eq.~(\ref{eq:ew}).

Similarly, $S_{H_2}^{WW(F)}$ may take the form
\begin{eqnarray}
\label{eq:SWW}
S^{WW(F)}_{H_2}      &=& 
2\sum_{Q_d} N_C g^S_{s\bar{Q}_dQ_d} \frac{1}{2s^2_W} 
\frac{v}{m_{Q_d}}
\left[2m^2_{Q_d} F^{(2)}_f(M_{H_2}^2,m_{Q_d}^2)\right]\;,
\end{eqnarray}
in the limit of $m_U=m_D=m_{Q_d}$.
Also note that, in the limit of $M_W=0$, 
$2m_{Q_d}^2F^{(2)}_{f}(M_{H_2}^2,m_{Q_d}^2)=F_{sf}(\tau_{2{Q_d}})$
and 
$S^{WW(F)}_{H_2}$ becomes the same as the singlet 
contribution to $S^{\gamma(F)}_{H_2}$ after replacing
$1/\sqrt{2}s_W$ with $Q_{Q_s}$  and, subsequently, $Q_d$ with $Q_s$.

In the previous section, we are taking
$S_{H_2}^{\gamma (F)}$ and $S_{H_2}^{g(Q)}$ as 
our independent fitting parameters.
In general, the form factors
$S^{Z\gamma(F)}_{H_2}$, $S^{ZZ(F)}_{H_2}$, and $S^{WW(F)}_{H_2}$
are independent of $S_{H_2}^{\gamma (F)}$ and $S_{H_2}^{g(Q)}$ and
they should be treated as independent parameters.
But we find that they can be expressed in terms of
$S_{H_2}^{\gamma (F)}$ and $S_{H_2}^{g(Q)}$
when
\begin{equation}
\label{eq:limit1}
\frac{M_{Z,W}}{M_{H_2}} \rightarrow  0\,, \ \ \
g^S_{s\bar{Q}_dQ_d}=g^S_{s\bar{Q}_sQ_s}\,, \ \ \
m_U=m_D=m_{Q_s}\,.
\end{equation}
In the above limit, we have
\begin{eqnarray}
\label{eq:limit2}
S^{Z\gamma(F)}_{H_2}&=& 
-t_W S^{\gamma(F)}_{H_2}+\left( \frac{N_d}{2N_d+N_s} \right)
\frac{N_C}{s_Wc_W} S^{g(Q)}_{H_2} \;,\nonumber \\[3mm]
S^{ZZ(F)}_{H_2}&=& 
t^2_W S^{\gamma(F)}_{H_2}+\left( \frac{N_d}{2N_d+N_s} \right)
\frac{1-2s^2_W}{s^2_Wc^2_W} N_C S^{g(Q)}_{H_2} \;,\nonumber \\[3mm]
S^{WW(F)}_{H_2}&=& \left( \frac{N_d}{2N_d+N_s} \right)  
\frac{N_C}{s^2_W} S^{g(Q)}_{H_2}\;,
\end{eqnarray}
where we use $Q_U-Q_D=1$.
Note that the form factors
$S^{Z\gamma(F)}_{H_2}$, $S^{ZZ(F)}_{H_2}$, and $S^{WW(F)}_{H_2}$ are all fixed
once  $S_{H_2}^{\gamma (F)}$, $S_{H_2}^{g(Q)}$, $N_d$, and $N_s$
are given and, accordingly, one can calculate the decay widths of 
$H_2$ into $Z\gamma$, $ZZ$, and $WW$, see Appendix A.  

In Fig.~\ref{fig:sin_ww_vlq}, we shows the CL regions in the 
$(\sin\alpha,\sigma(gg\to H_2)\times B(H_2\to WW))$ plane
including the VLQ-loop induced contributions to $B(H_2\to WW)$
in the presence of interactions between VLQs and $W/Z$ bosons.
We are taking the limits suggested in Eq.~(\ref{eq:limit1}) and $N_d=N_s$.
Compared to Figure~\ref{F4:sin_ww}, we observe that there are non-vanishing
VLF contributions to $\sigma(gg\to H_2)\times B(H_2\to WW)$
when $\sin\alpha\to 0$ because:
\begin{equation}
\left.  \Gamma(H_2\to WW)  \right|_{\sin\alpha\to 0} \ \simeq \
\frac{M_{H_2}^3\alpha^2}{128\pi^3 v^2} \left(S^{WW(F)}_{H_2}\right)^2
\ = \
\frac{M_{H_2}^3\alpha^2}{128\pi^3 v^2} 
\frac{\left(S^{g(Q)}_{H_2}\right)^2}{s_W^4}\,.
\end{equation}

Figure~\ref{fig:sin_zz_vlq} shows the CL regions in the
$(\sin\alpha,\sigma(gg\to H_2)\times B(H_2\to ZZ))$ plane
in the same context.
Compared to Figure~\ref{F4:sin_zz}, we observe that there are 
also non-vanishing VLF contributions 
when $\sin\alpha\to 0$ because:
\begin{equation}
\left.  \Gamma(H_2\to ZZ)  \right|_{\sin\alpha\to 0} \ \simeq \
\frac{M_{H_2}^3\alpha^2}{256\pi^3 v^2} \left(S^{ZZ(F)}_{H_2}\right)^2
\ = \
\frac{M_{H_2}^3\alpha^2}{256\pi^3 v^2}
\left(t_W^2 S^{\gamma(F)}_{H_2}+\frac{1-2s_W^2}{s_W^2c_W^2}S^{g(Q)}_{H_2}
\right)^2\,.
\end{equation}

Figure~\ref{fig:sin_za_vlq} shows the CL regions in the
$(\sin\alpha,\sigma(gg\to H_2)\times B(H_2\to Z\gamma))$ plane.
In the limits suggested in Eq.~(\ref{eq:limit1}) with $N_d=N_s$,
we have
\begin{equation}
\left.  \Gamma(H_2\to Z\gamma)  \right|_{\sin\alpha\to 0} \ \simeq \
\frac{M_{H_2}^3\alpha^2}{128\pi^3 v^2} \left(S^{Z\gamma(F)}_{H_2}\right)^2
\ = \
\frac{M_{H_2}^3\alpha^2}{128\pi^3 v^2}
\left(-t_W S^{\gamma(F)}_{H_2}+\frac{S^{g(Q)}_{H_2}}{s_Wc_W}
\right)^2\,.
\end{equation}
We observe $\sigma(gg\to H_2)\times B(H_2\to Z\gamma)$ can be as large as
about $8$ fb around $\sin\alpha =0$ at 68 \% CL.

In Figs.~\ref{fig:aa_za_S001} and \ref{fig:aa_za_G001}, we show 
the correlations between
$\sigma(gg\to H_2)\times B(H_2\to \gamma\gamma)$ and
$\sigma(gg\to H_2)\times B(H_2\to Z\gamma)$ taking $|\sin\alpha|<0.1$
and  $|\sin\alpha|\geq 0.1$, respectively. In the limits suggested in
Eq.~(\ref{eq:limit1}) with $N_d=N_s$, we have
\begin{equation}
\label{eq:za_aa}
\left.  \frac{\Gamma(H_2\to Z\gamma)}{\Gamma(H_2\to \gamma\gamma)}
\right|_{\sin\alpha\to 0} \ \simeq \ 2\,
\left(-t_W +\frac{1}{s_Wc_W}\,
\frac{S^{g(Q)}_{H_2}}{S^{\gamma(F)}_{H_2}} \right)^2\,.
\end{equation}
When $|\sin\alpha|< 0.1$, we observe 
{\small
\begin{equation}
\frac{1}{5}\sigma(gg\to H_2)\times B(H_2\to \gamma\gamma)
\lsim
\sigma(gg\to H_2)\times B(H_2\to Z\gamma)
\lsim \frac{4}{5}\,\sigma(gg\to H_2)\times B(H_2\to \gamma\gamma)
\ + \ 2\,{\rm fb}\,. 
\end{equation}
}
We find the correlation is strong when $|\sin\alpha|\geq 0.1$ 
and $|S^{g(Q)}_{H_2}/S^{\gamma(F)}_{H_2}|$ is small, see 
Figs.~\ref{F4:sin_sg} and \ref{F4:sin_sa}.
In this case, we find
\begin{equation}
\frac{
\sigma(gg\to H_2)\times B(H_2\to Z\gamma)}{
\sigma(gg\to H_2)\times B(H_2\to \gamma\gamma)} \simeq 2t_W^2 \simeq 2/3
\end{equation}
as shown in Fig.~\ref{fig:aa_za_G001}
and indicated by Eq.~(\ref{eq:za_aa}).

\section{Discussion}

We have performed a ``Double Higgcision'' -- Higgs precision study on
both the 125 GeV Higgs boson and 
a potential diphoton resonance that may appear in the
near future data.  The recent 750 GeV diphoton resonance serves as
a concrete example and we borrow the diphoton resonance data
collected in 2015 in our analysis.
We have used 
all the available Higgs boson data from 7 \& 8 TeV runs as well as 
the diphoton cross sections of the 750 GeV boson from the 13 TeV run 
in 2015.

The important findings and a few comments are summarized as follows: 

\begin{enumerate}
\item
We have divided the analysis into two cases: (i) the width is varied freely
and (ii) a broad-width defined by $\Gamma_{H_2} > 40 $ GeV is enforced.
In the former case, a narrow width is always preferred and the width is of
order $1-3$ GeV.  On the other hand, in the broad-width case the width is 
around 45 GeV.  Note that the minimal $\chi^2$ for these two cases 
only differ by a small amount, which is statistically not significant.

\item 
As we have shown that $S^{g(Q)}_{H_1}$ and $S^{g(Q)}_{H_2}$ and similarly
$S^{\gamma(F)}_{H_1}$ and $S^{\gamma(F)}_{H_2}$ are independent 
parameters, but, however, they share the same form and with varying VLF mass
their ratios $\eta^{g(Q)}$ and $\eta^{\gamma(F)}$ range 
between $2/3$ and $1$ for VLF mass from $m_{H_2}$ to
infinity. We have shown the results of our analysis for these two 
representative values of $\eta$s in {\bf F4-2} and {\bf F4-3} fits,
which have similar features.

\item 
We have also demonstrated the extreme case of $\eta$s equal to zero, i.e.,
the VLFs do not affect the gluon-fusion production of $H_1$ and the
decays of $H_1$ into photons and gluons. In such a scenario, the effect
of $H_2$ on $H_1$ Higgs boson data is only via the mixing angle $\cos\alpha$.
Also, in the case the decays of $H_2$ into other modes such as $WW$, $ZZ$,
and $t\bar t$ are only via the mixing with $H_1$. The best-fit shown allows
a very tiny mixing angle $|\sin\alpha| \sim 10^{-4}$ such that the decays 
of $H_2 \to WW, ZZ,t\bar t$ are negligible. 
If this is the case in the future data, there arises  one immediate question why
the mixing angle $\alpha$ is so tiny.  This should be accommodated in any models
for the 750 GeV diphoton excess. 

\item 
In the fits of {\bf F4-2} and {\bf F4-3}, the mixing angles are not too small
and of order $10^{-2}$. The cross sections of $H_2$ into $WW,ZZ,t\bar t$ are
not negligible and demonstrate the cross sections in the ratio of 
$4:2:1$, because of they are all proportional to the square of the 
SM couplings and $\sin\alpha$. 

\item 
Both narrow and broad-width options in all three {\bf F4} fits, the 
$\Gamma^{\rm non-SM}_{H_2}$ dominates the total width $\Gamma_{H_2}$ of $H_2$,
especially in the broad-width case the non-SM decay accounts for more than
99\% branching ratio.

\item Should the $WW$, $ZZ$, $Z\gamma$ or $t\bar t$ modes of the $H_2$ be 
observed in near future, they would be extremely useful to tell 
the information on the VLFs. 

\item If we assume that VLQs are weak isospin singlets and/or doublets,
we can make more explicit and specific
predictions on $H_2 \rightarrow WW, ZZ , Z\gamma$, which
are shown in Figs. 11-15.  Discovery or upper bounds on the branching ratios of
750 GeV boson into these channels would shed more light on the nature of the
VLQs in the loop. This could be complementary to the direct search for VLQs
at the LHC through QCD interactions, keeping in mind that the decays of VLQs
would be more model dependent.

\item Our procedure can be applied in the future discovery of a new resonance in
the loop-induced diphoton and/or $Z\gamma$ channels, in particular,
taking into account a possible mixing with the SM Higgs boson.  
Through this work, we demonstrate in detail how to carry out the 
relevant analysis in a proper way.

\end{enumerate}

\section*{Acknowledgments}
This work is supported in part by National Research Foundation of Korea 
(NRF) Research Grant NRF-2015R1A2A1A05001869, and by the NRF grant 
funded by the Korea government (MSIP) (No. 2009-0083526) through Korea 
Neutrino Research Center at Seoul National University (PK).
The work of K.C. was supported by the MoST of Taiwan under Grants 
No. NSC 102-2112-M-007-015-MY3.

%
%
\def\theequation{\Alph{section}.\arabic{equation}}
\begin{appendix}
\setcounter{equation}{0}
\section{Decay widths of $H_2$ into $Z\gamma$, $ZZ$, and $WW$}
In  this appendix, we present the explicit forms for the decay widths of $H_2$ into 
$Z\gamma$, $ZZ$, and $WW$ 
including the mixing between the SM Higgs boson and the singlet scalar.

The amplitude for the decay process $H_2 \to
Z(k_1,\epsilon_1)\
\gamma(k_2,\epsilon_2)$ can be written as
\begin{equation}
{\cal M}_{Z\gamma H_2} = -\,\frac{\alpha}{2\pi v}\,
S^{Z\gamma}_{H_2}\,
\left[ k_1\cdot k_2\,\epsilon_1^*\cdot\epsilon_2^*
-k_1\cdot\epsilon_2^*\,k_2\cdot\epsilon_1^* \right] 
\end{equation}
where $k_{1,2}$ are the $4-$momenta of the $Z$ boson 
and the photon, respectively, and 
$\epsilon_{1,2}$ are their polarization vectors.
We note that $2k_1\cdot k_2 = M_{H_2}^2-M_Z^2$.
The form factor $S^{Z\gamma}_{H_2}$ is given by the sum
$$
S^{Z\gamma}_{H_2} = 
\sin{\alpha}S^{Z\gamma({\rm SM})}_{H_2} +\cos{\alpha}S^{Z\gamma(F)}_{H_2}
$$
with $S^{Z\gamma({\rm SM})}_{H_2}\simeq-0.0771-1.805i$.
The VLF contribution $S^{Z\gamma(F)}_{H_2}$ is model dependant and
it is given by Eq.~(\ref{eq:SZgamma}) 
in the context of $SU(2)_L$-doublet and singlet 
VLFs discussed in Section~\ref{sec:vlq}.
Then, the decay width of $H_2$ into $Z\gamma$ is given by
\begin{equation}
\Gamma(H_2\to Z\gamma) =
\frac{M^3_{H_2}\alpha^2}{128\pi^3v^2}\left( 1-\frac{M^2_Z}{M^2_{H_2}} \right)^3
\left|S^{Z\gamma}_{H_2}\right|^2 \;,
\end{equation}
with
\begin{equation}
\left|S^{Z\gamma}_{H_2}\right|^2=
\sin^2\alpha\,\left|S^{Z\gamma({\rm SM})}_{H_2}\right|^2
+2\cos\alpha\sin\alpha\,S^{Z\gamma(F)}_{H_2}\,
\real\left({S^{Z\gamma({\rm SM})}_{H_2}}\right)
+\cos^2\alpha\left(S^{Z\gamma(F)}_{H_2}\right)^2  
\end{equation}
assuming that the VLF mass satisfies $m_F>M_{H_2}/2$ 
so that $S^{Z\gamma(F)}_{H_2}$ becomes real.

The amplitude for the decay process $H_2 \to
V(k_1,\epsilon_1)\ V(k_2,\epsilon_2)$ can be written as
$$
{\cal M}_{VVH_2}=\sin\alpha{\cal M}_{VVH_2}^{(0)}
                +\cos\alpha{\cal M}_{VVH_2}^{(1)}\,.
$$
In the leading order neglecting the SM one-loop contributions to the
$hWW$  vertex,
the tree-level and one-loop amplitudes are given by Eq.~(\ref{eq:ampVV}):
\begin{eqnarray}
{\cal M}^{(0)}_{WWH_2} &=& -gM_W \epsilon_1^*\cdot\epsilon_2^*\;, \ \ \ 
{\cal M}^{(0)}_{ZZH_2} = -\frac{gM_W}{c^2_W}
\epsilon_1^*\cdot\epsilon_2^*\;;
\nonumber \\[3mm]
{\cal M}^{(1)}_{VVH_2} &=&
-\frac{\alpha}{2\pi v} S^{VV(F)}_{H_2}\left[ k_1\cdot
k_2\,\epsilon_1^*\cdot\epsilon_2^*
-k_1\cdot\epsilon_2^*\,k_2\cdot\epsilon_1^* \right]\;. \nonumber
\end{eqnarray}
The VLF contributions $S^{ZZ(F)}_{H_2}$  and $S^{WW(F)}_{H_2}$
are model dependent and  they are given by Eqs.~(\ref{eq:SZZ}) and (\ref{eq:SWW}), 
respectively,  in the context discussed in Section~\ref{sec:vlq}.
Finally, the decay width of $H_2$ into $VV$ is given by
\begin{equation}
\Gamma(H_2 \to VV) = \frac{\delta_V}{32\pi M_{H_2}}
\sum|{\cal M}_{VVH_2}|^2\sqrt{1-\frac{4M^2_V}{M^2_{H_2}}}\;
\end{equation}
with $\delta_W=2$ and $\delta_Z=1$. For the amplitude squared,
explicitly, we obtain
\begin{eqnarray}
\sum|{\cal M}_{WWH_2}|^2 &=&
\sin^2\alpha\,g^2M^2_W \left[
3-\frac{M^2_{H_2}}{M^2_W}+\frac{M^4_{H_2}}{4M^4_W} \right] \nonumber \\
&+& \sin\alpha\,\cos\alpha \frac{gM_W\alpha}{\pi \,v} \left[
-3M^2_W+\frac{3M^2_{H_2}}{2} \right] S^{WW(F)}_{H_2}  \nonumber \\
&+& \cos^2\alpha \frac{\alpha^2}{4\pi^2\, v^2} \left[
3M^4_W-2M^2_WM^2_{H_2}+\frac{M^4_{H_2}}{2} \right]
 \left(S^{WW(F)}_{H_2} \right)^2\;; \\
 \sum|{\cal M}_{ZZH_2}|^2 &=&
\sin^2\alpha\,\frac{g^2M^2_W}{c^4_W} \left[
3-\frac{M^2_{H_2}}{M^2_Z}+\frac{M^4_{H_2}}{4M^4_Z} \right] \nonumber \\
&+& \sin\alpha\,\cos\alpha \frac{gM_W\alpha}{c^2_W \pi \,v} \left[
-3M^2_Z+\frac{3M^2_{H_2}}{2} \right] S^{ZZ(F)}_{H_2}  \nonumber \\
&+& \cos^2\alpha \frac{\alpha^2}{4\pi^2\, v^2} \left[
3M^4_Z-2M^2_ZM^2_{H_2}+\frac{M^4_{H_2}}{2} \right]
 \left(S^{ZZ(F)}_{H_2} \right)^2\;. 
\end{eqnarray}

\end{appendix}


\newpage


\begin{figure}[th!]
\centering
\includegraphics[height=1.5in,angle=0]{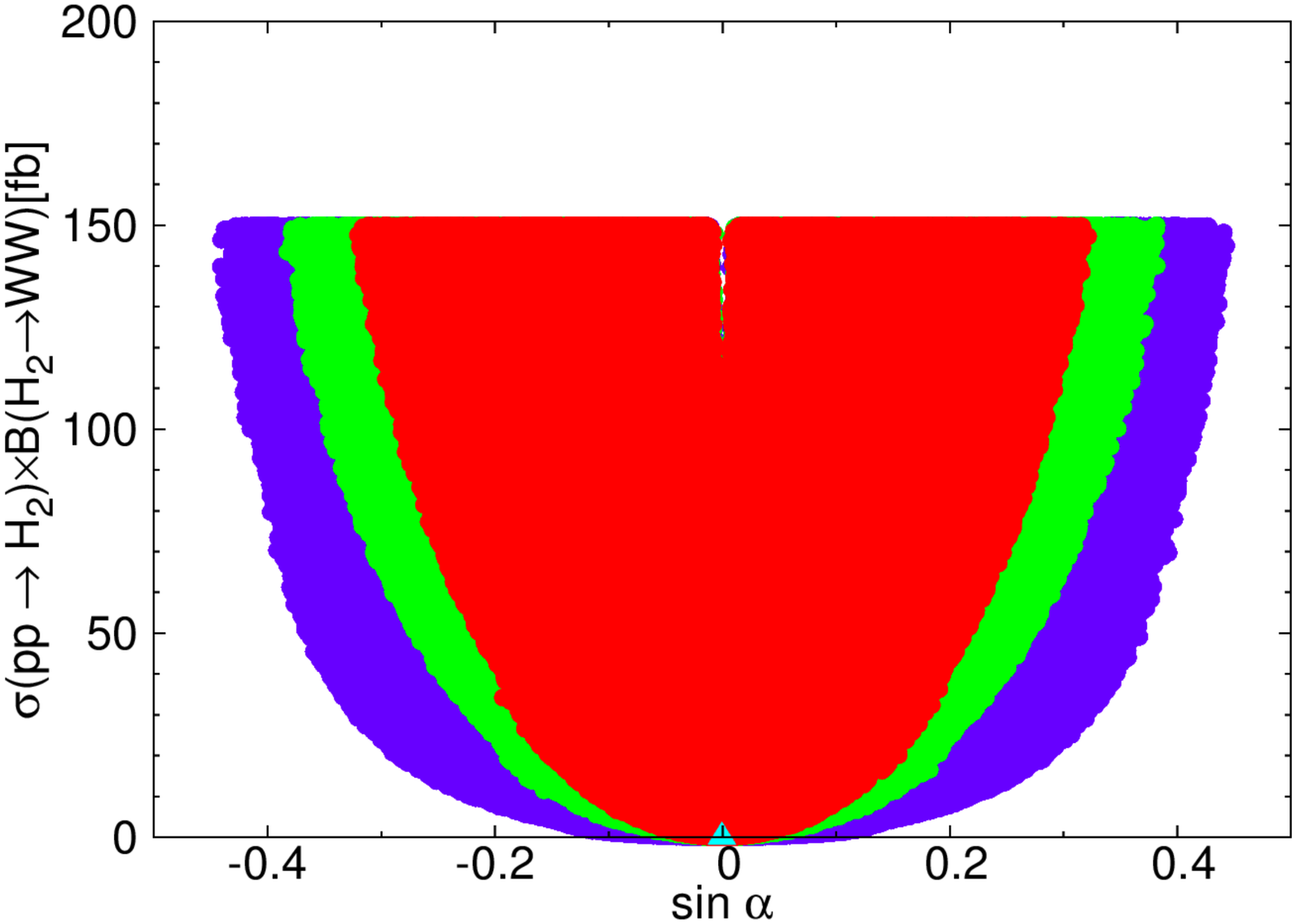}
\includegraphics[height=1.5in,angle=0]{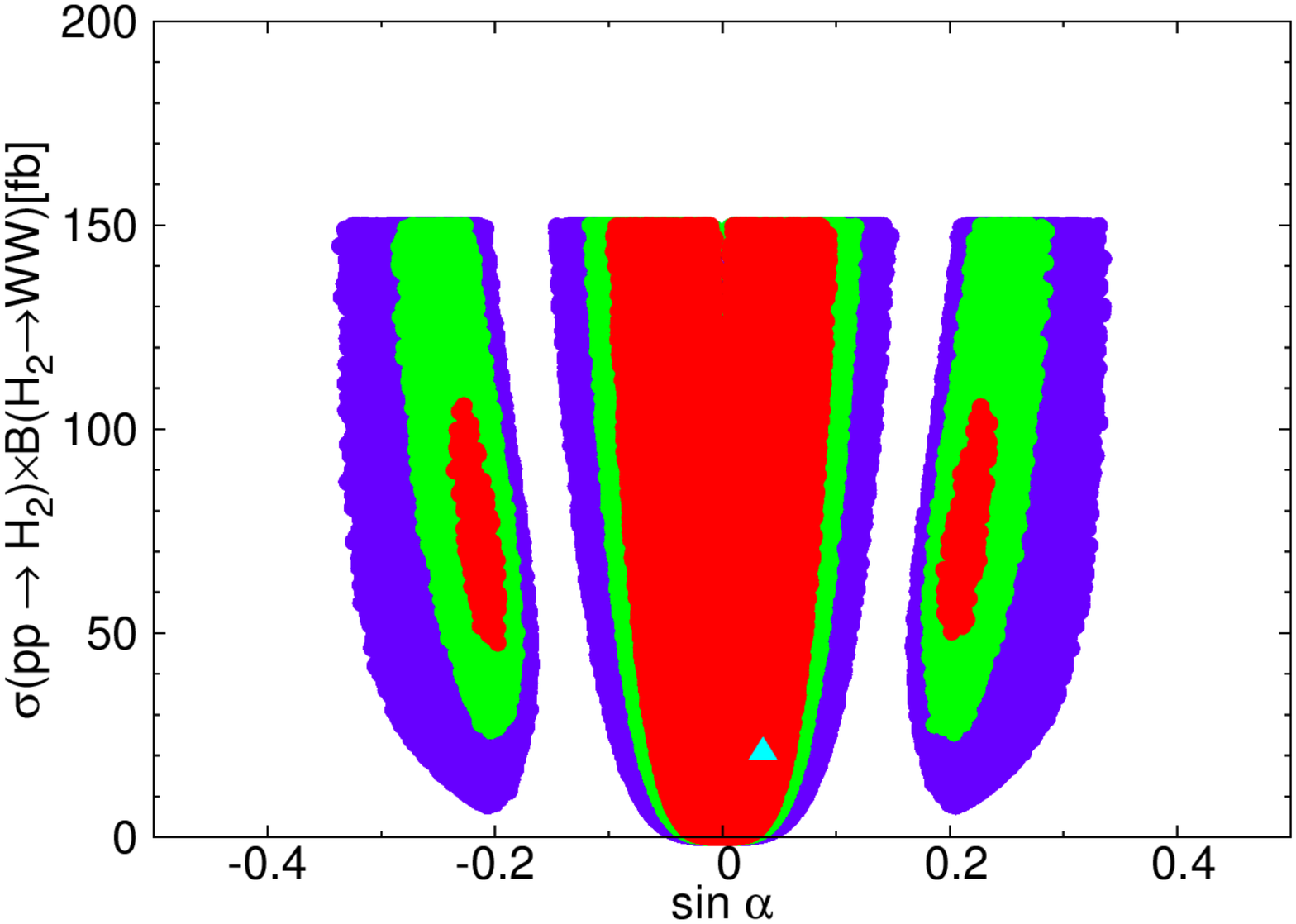}
\includegraphics[height=1.5in,angle=0]{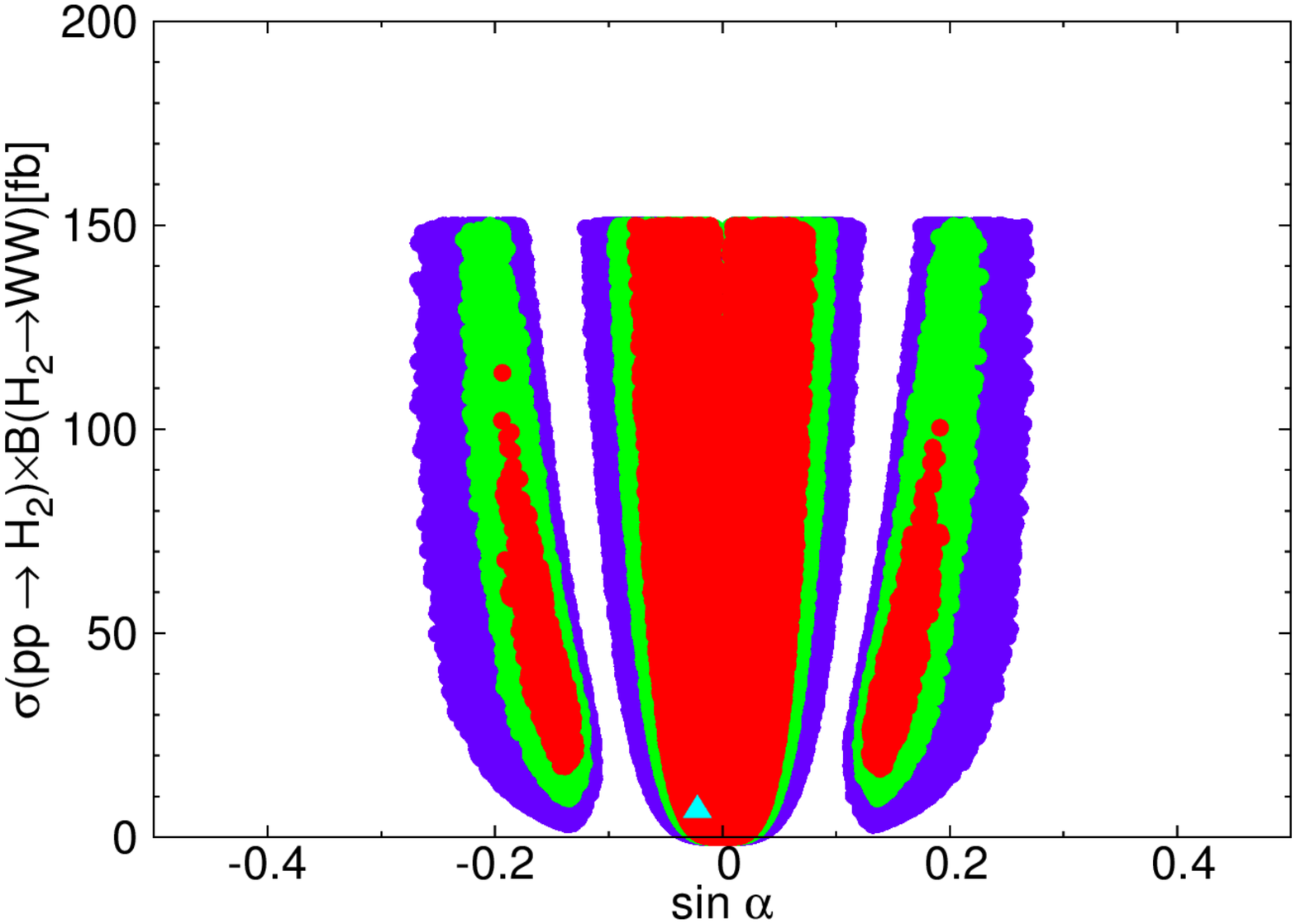}
\includegraphics[height=1.5in,angle=0]{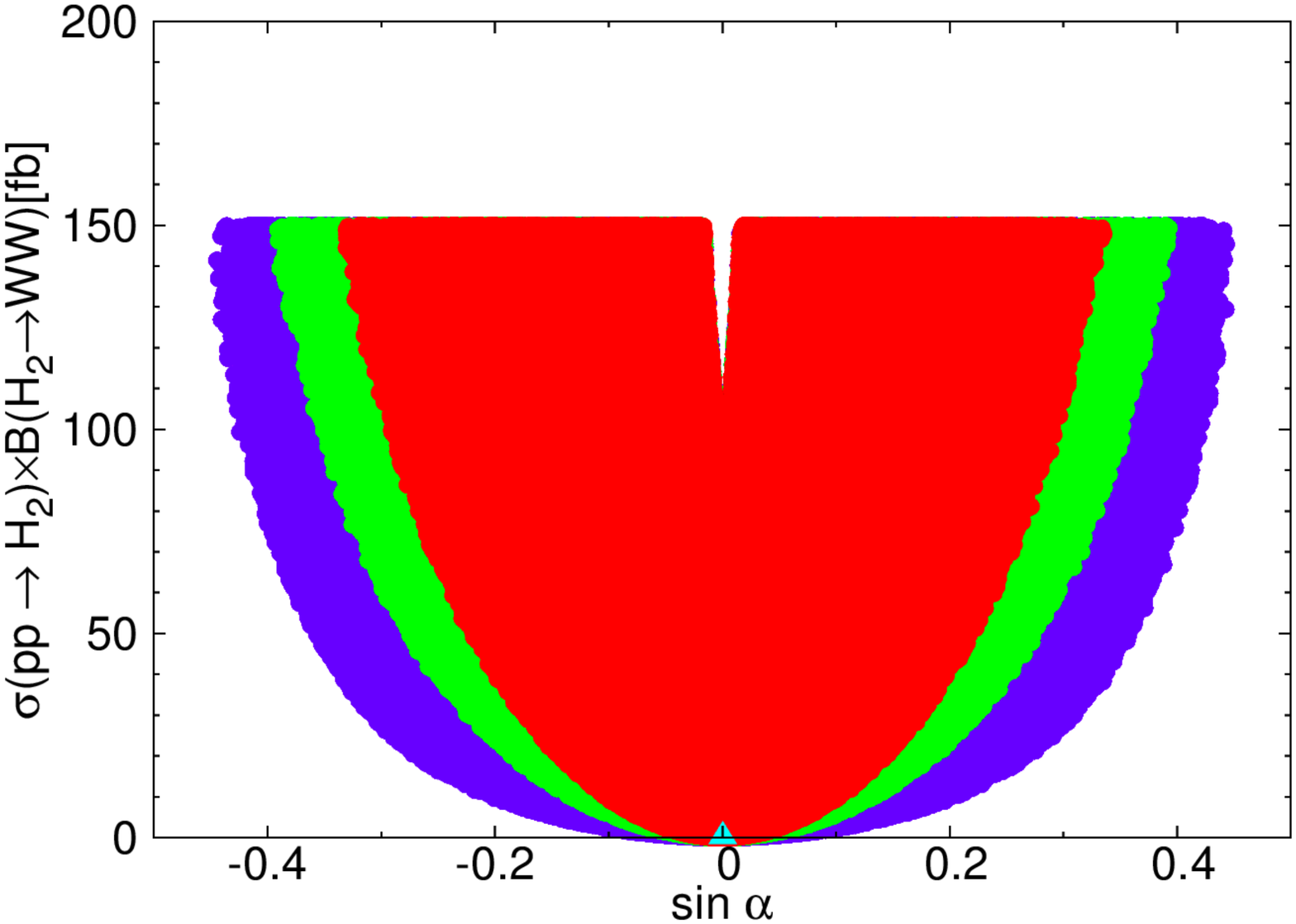}
\includegraphics[height=1.5in,angle=0]{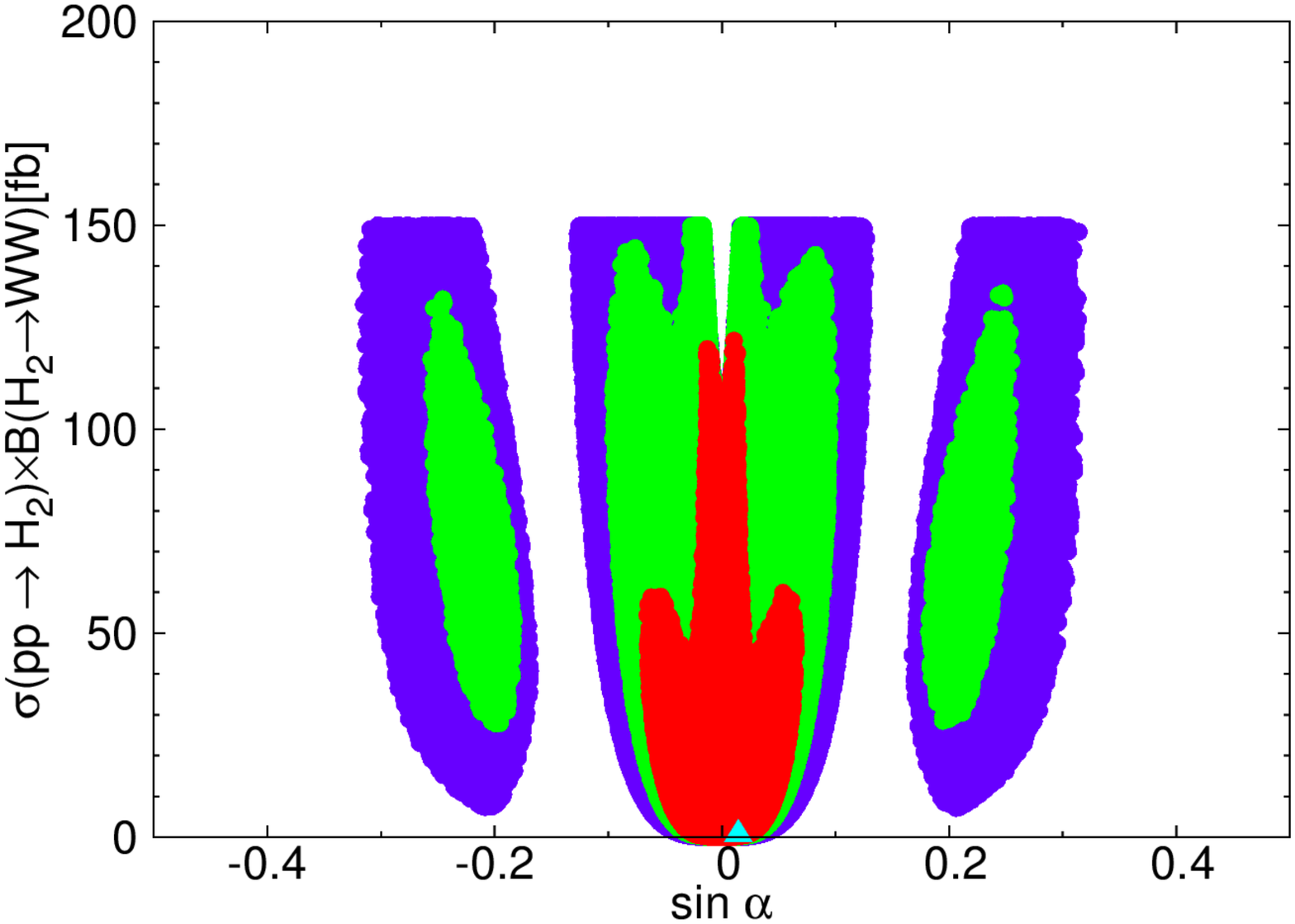}
\includegraphics[height=1.5in,angle=0]{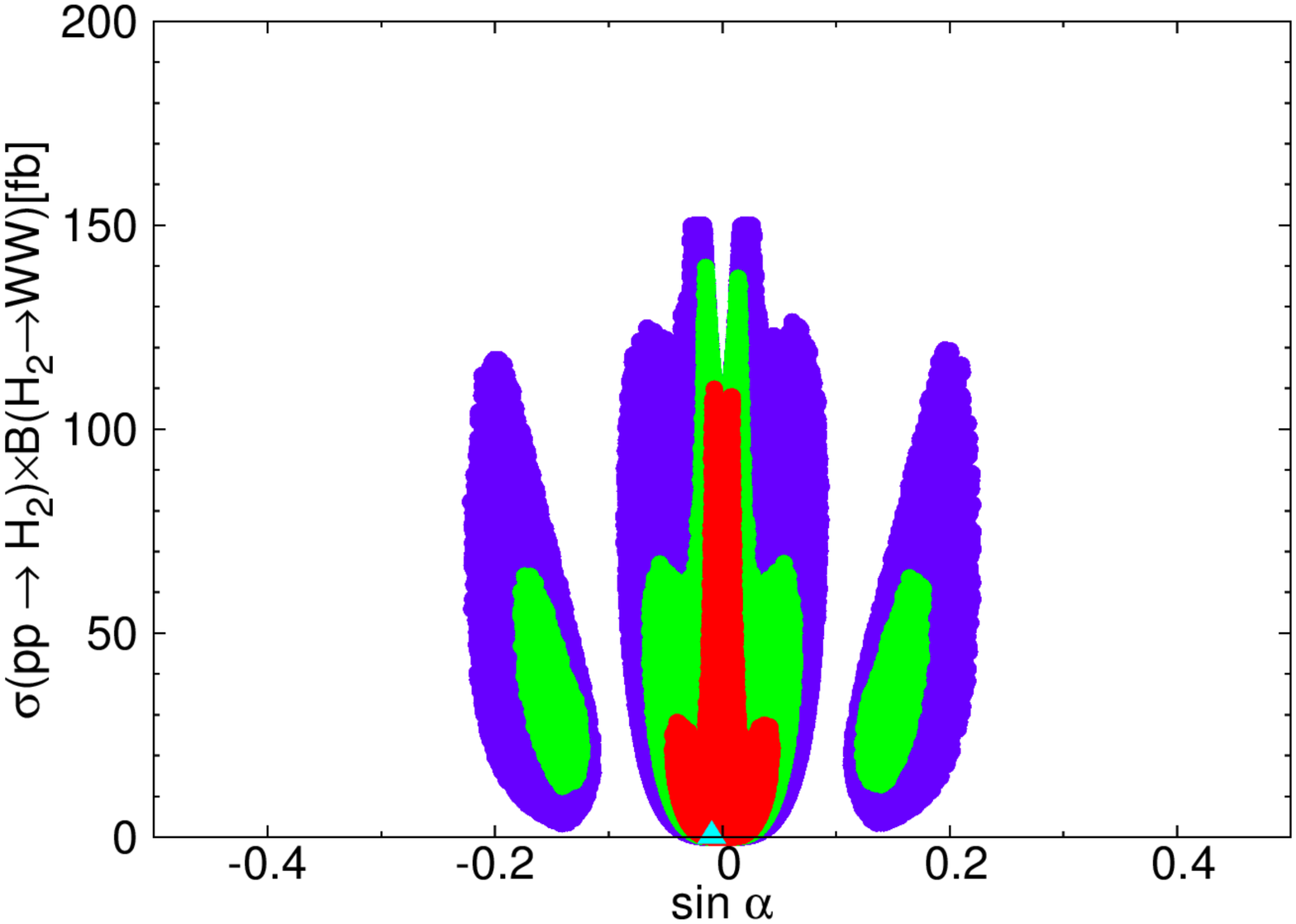}
\caption{\small \label{fig:sin_ww_vlq}
{\bf{VLQ-W/Z}}:
The same as in FIG.~\ref{F4:sin_ww} but including
the VLQ-loop induced contributions to $B(H_2\to WW)$
in the presence of interactions between VLQs and $W/Z$ bosons
discussed in Section~\ref{sec:vlq}.
We are taking the limits in Eq.~(\ref{eq:limit1}) and $N_d=N_s$.
}
\end{figure}

\begin{figure}[th!]
\centering
\includegraphics[height=1.5in,angle=0]{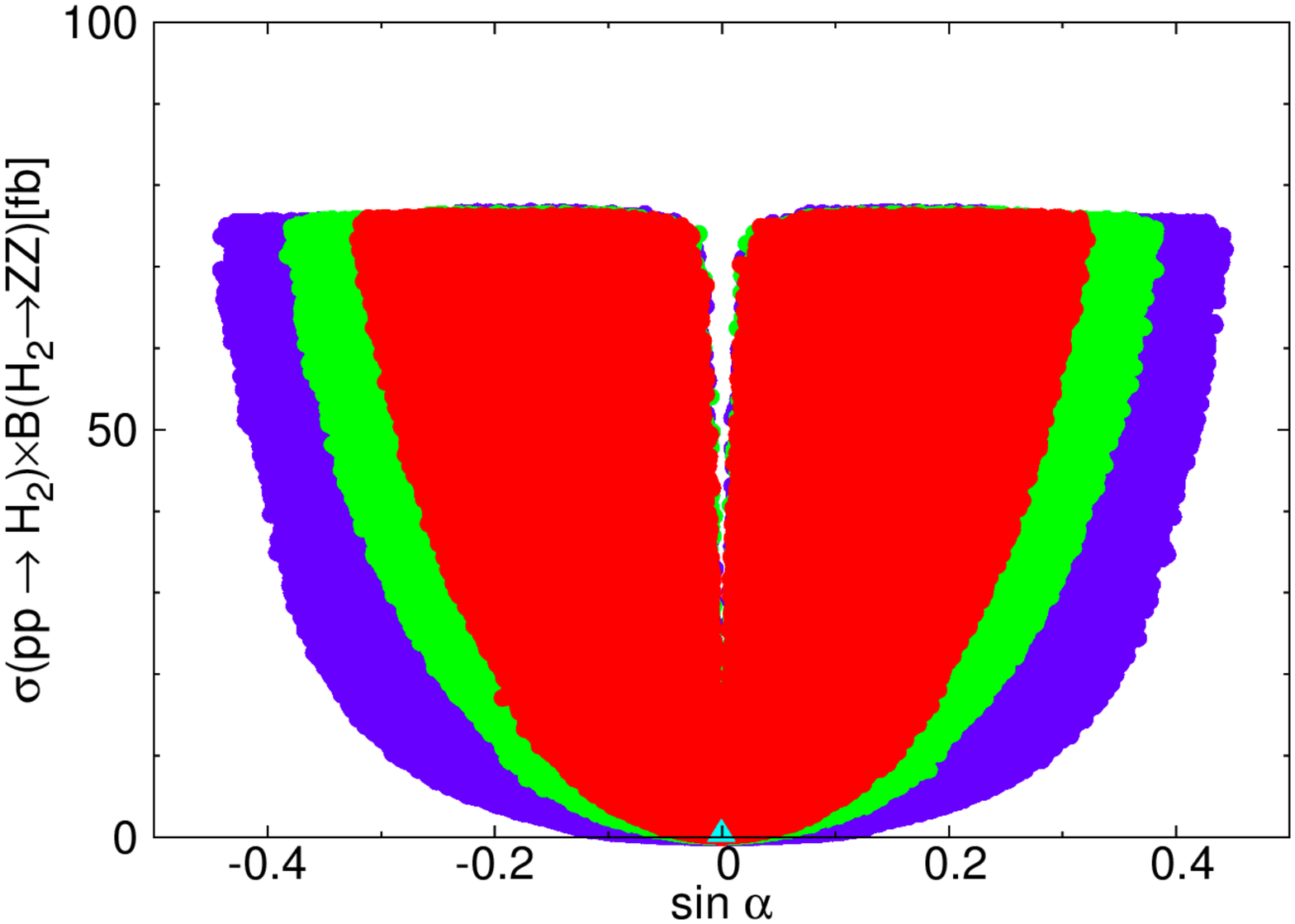}
\includegraphics[height=1.5in,angle=0]{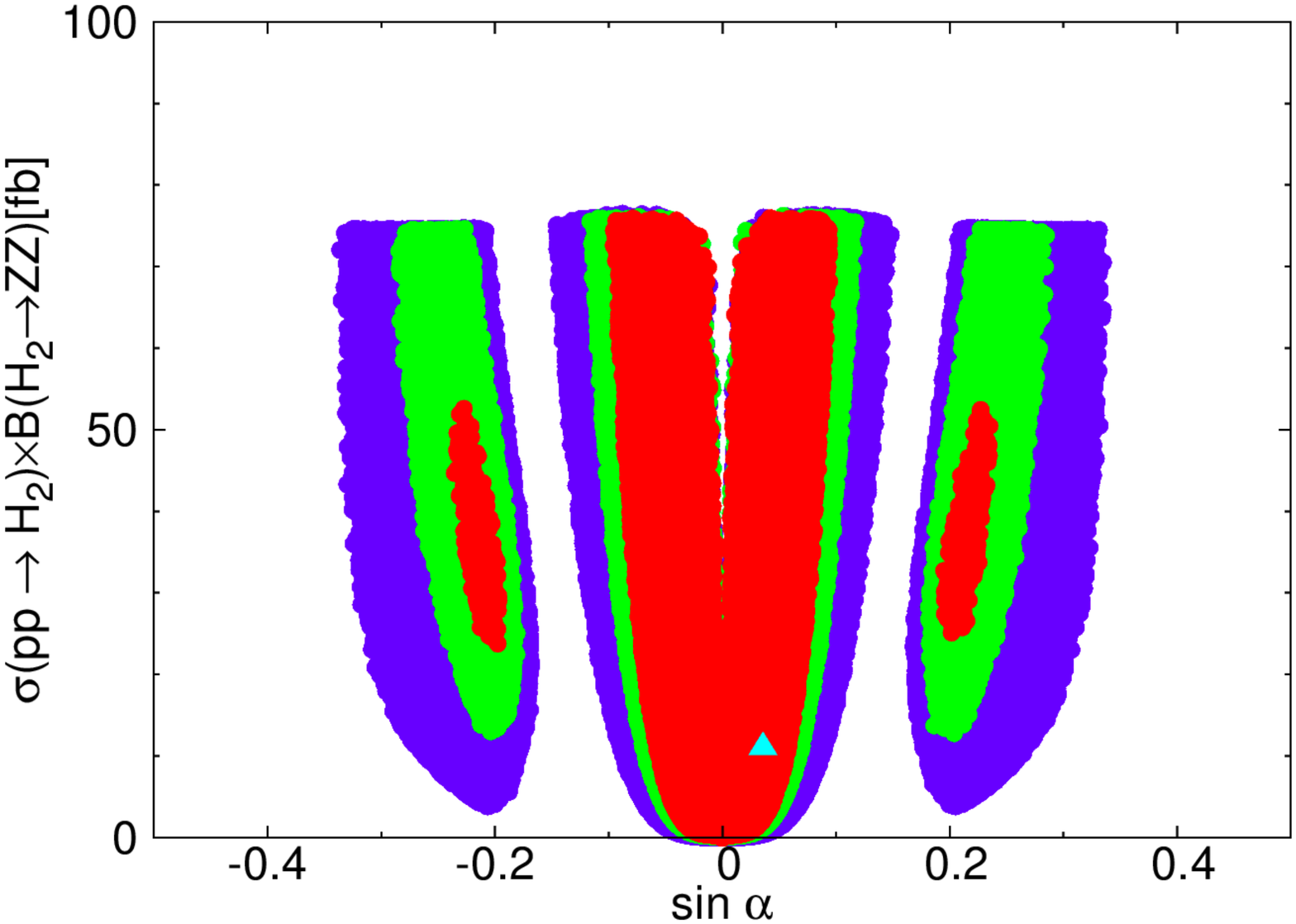}
\includegraphics[height=1.5in,angle=0]{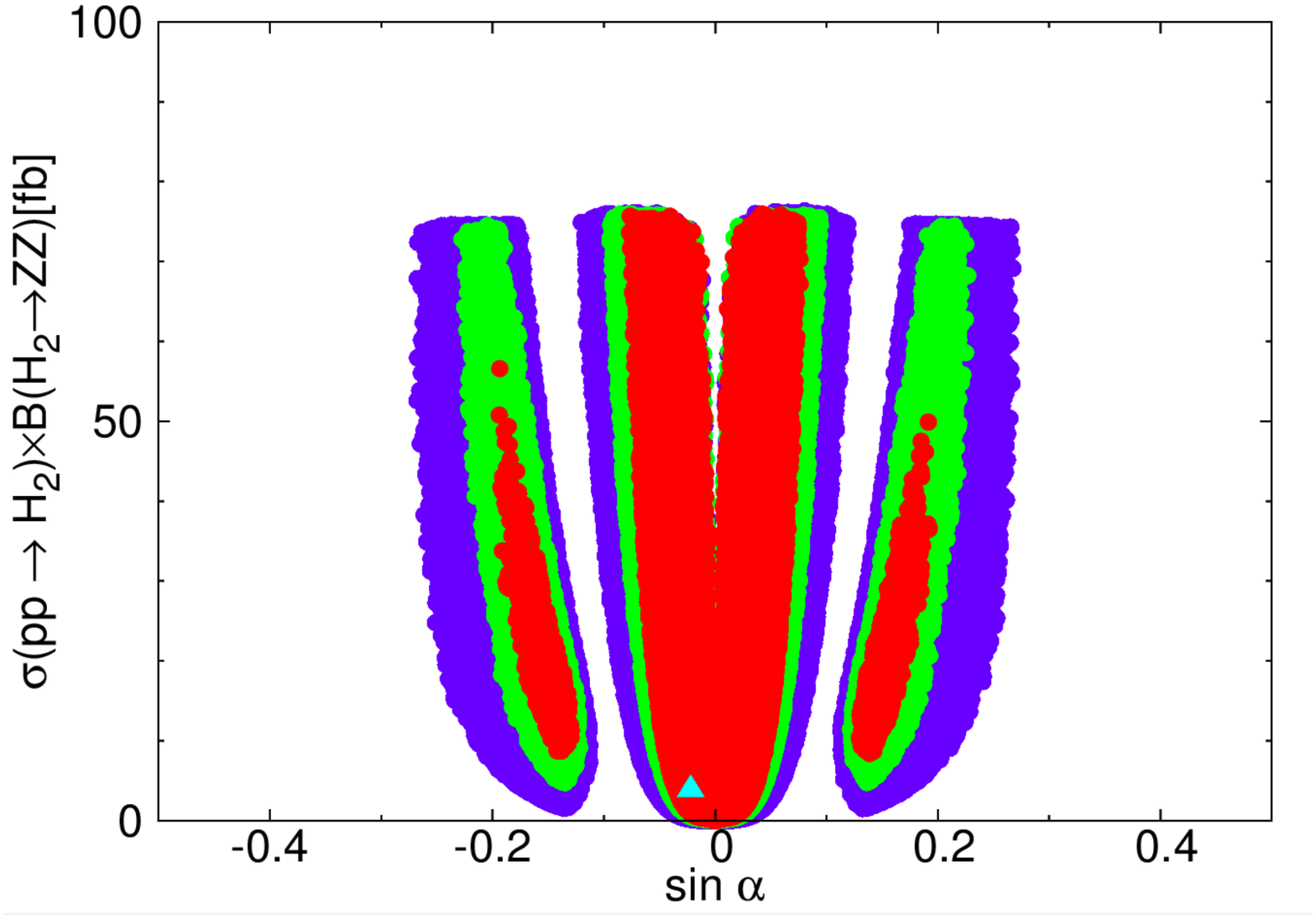}
\includegraphics[height=1.5in,angle=0]{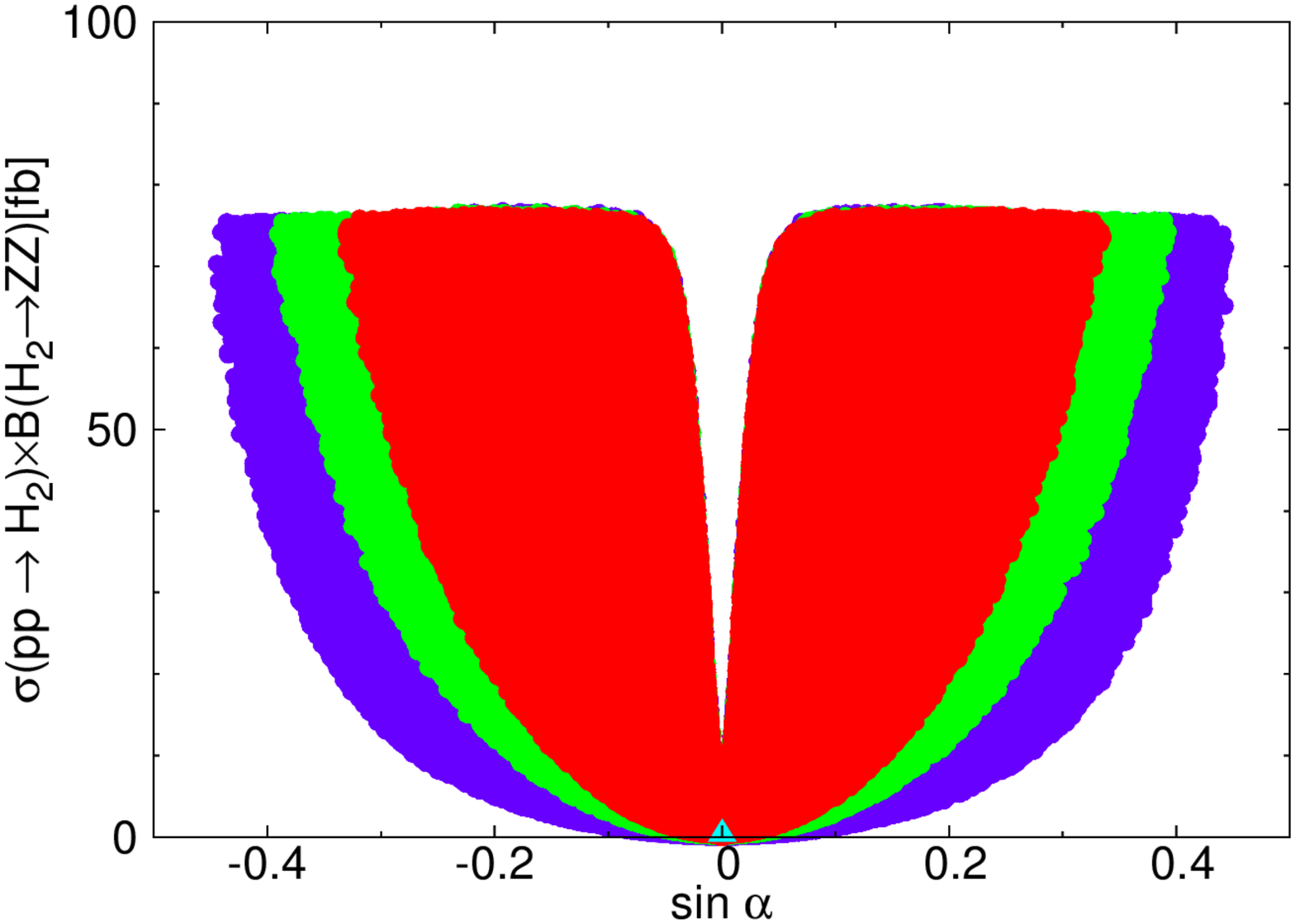}
\includegraphics[height=1.5in,angle=0]{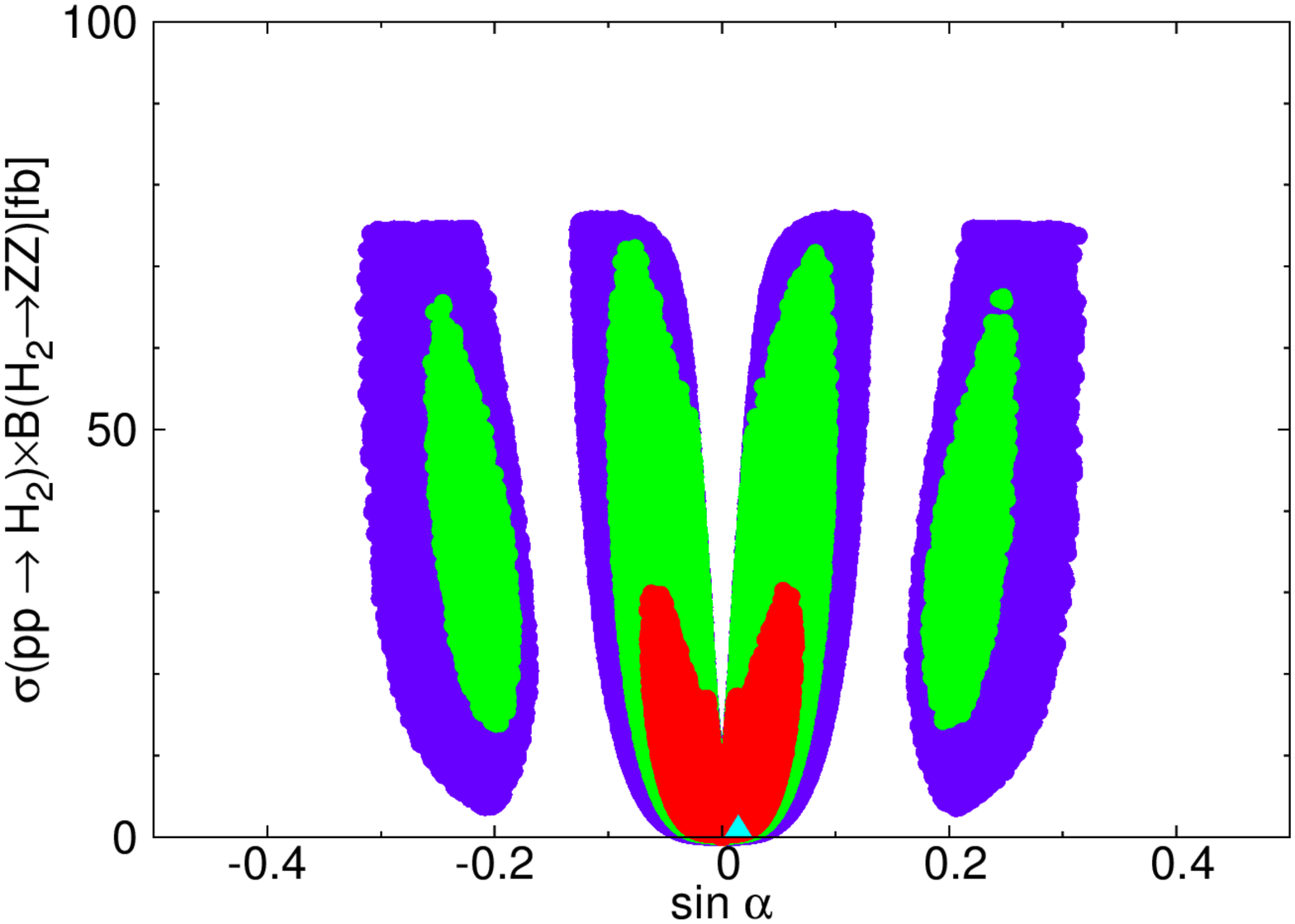}
\includegraphics[height=1.5in,angle=0]{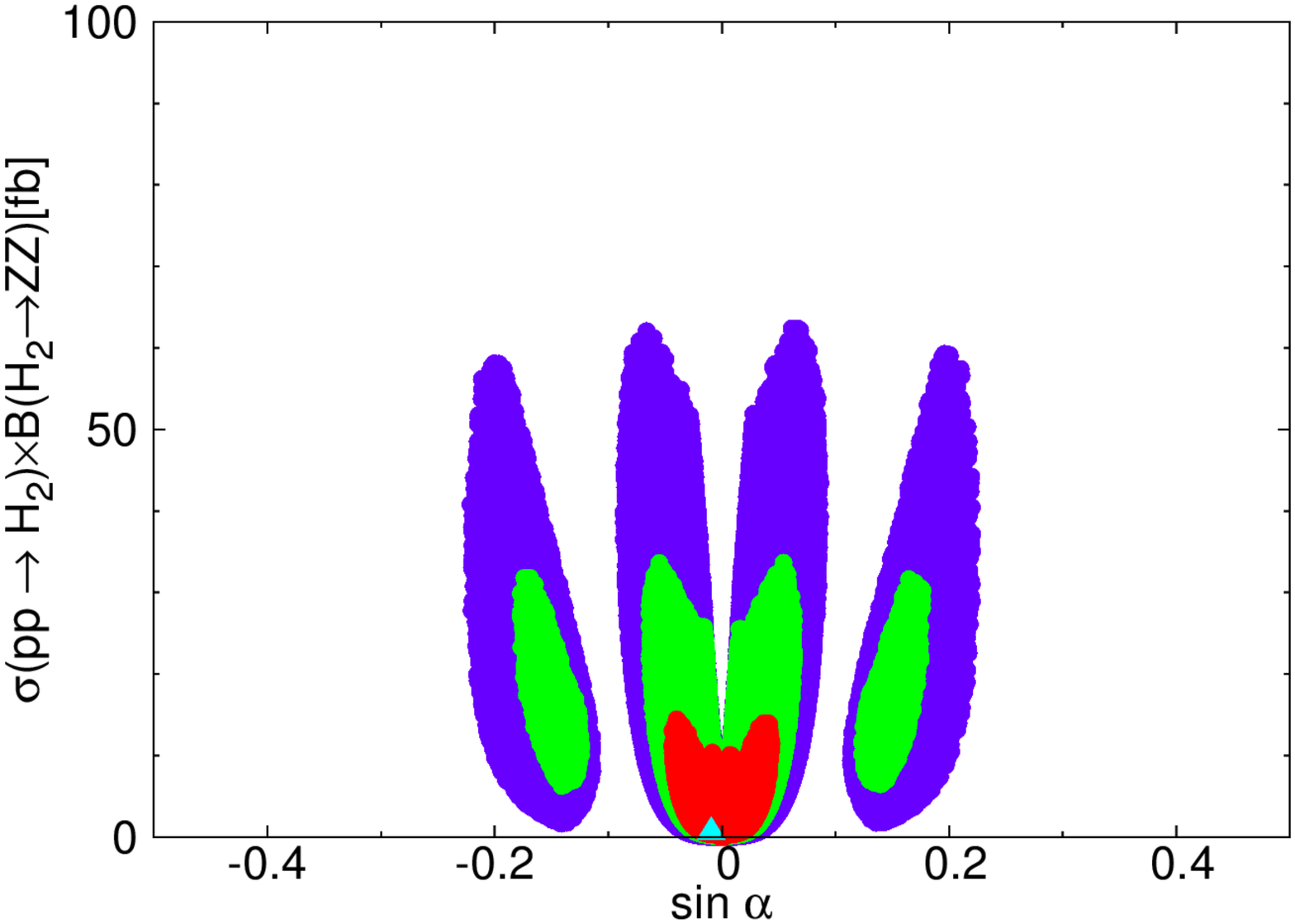}
\caption{\small \label{fig:sin_zz_vlq}
{\bf{VLQ-W/Z}}:
The same as in FIG.~\ref{F4:sin_zz} but including
the VLQ-loop induced contributions to $B(H_2\to ZZ)$
in the presence of interactions between VLQs and $W/Z$ bosons
discussed in Section~\ref{sec:vlq}.
We are taking the limits in Eq.~(\ref{eq:limit1}) and $N_d=N_s$.
}
\end{figure}

\begin{figure}[th!]
\centering
\includegraphics[height=1.5in,angle=0]{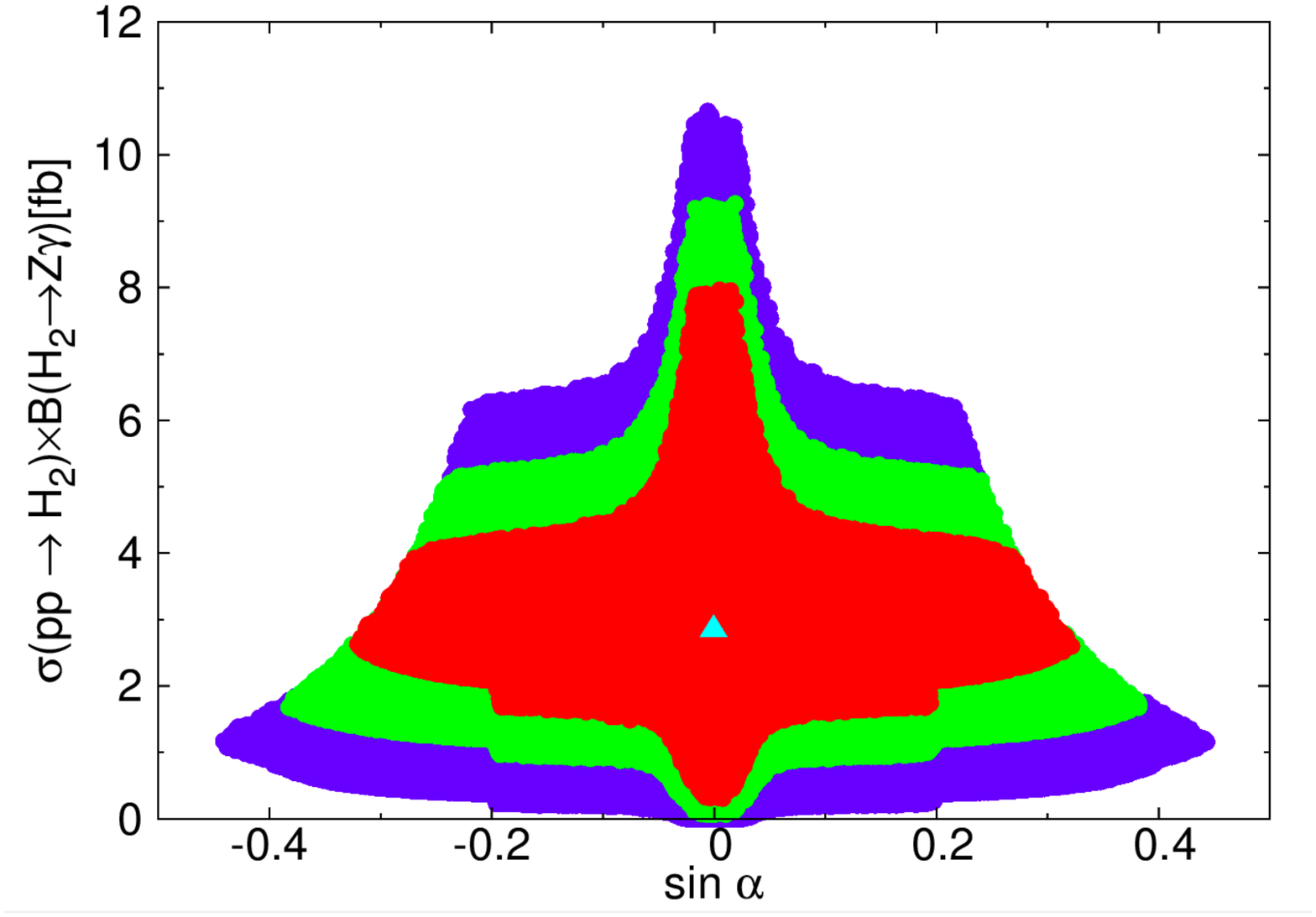}
\includegraphics[height=1.5in,angle=0]{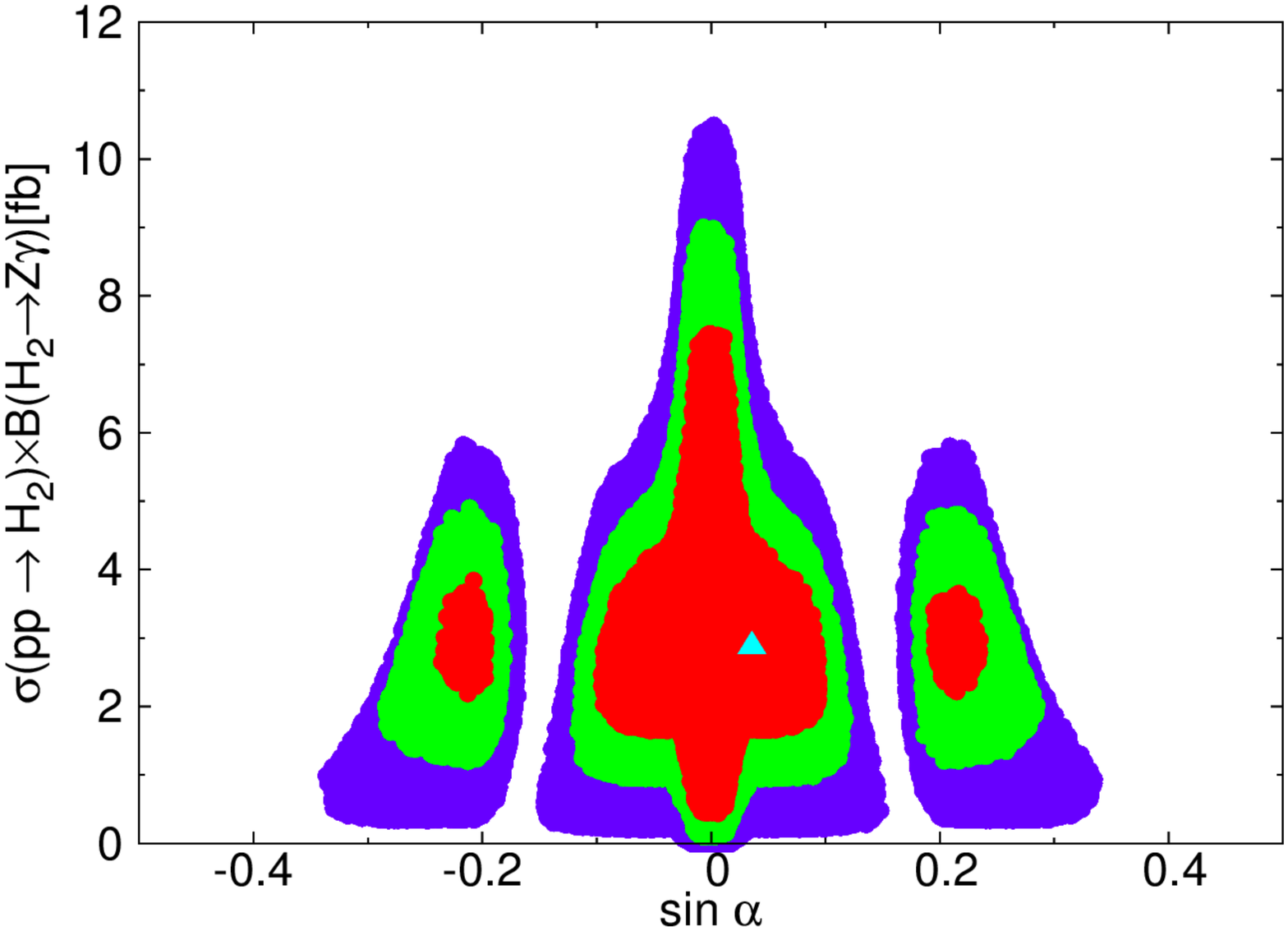}
\includegraphics[height=1.5in,angle=0]{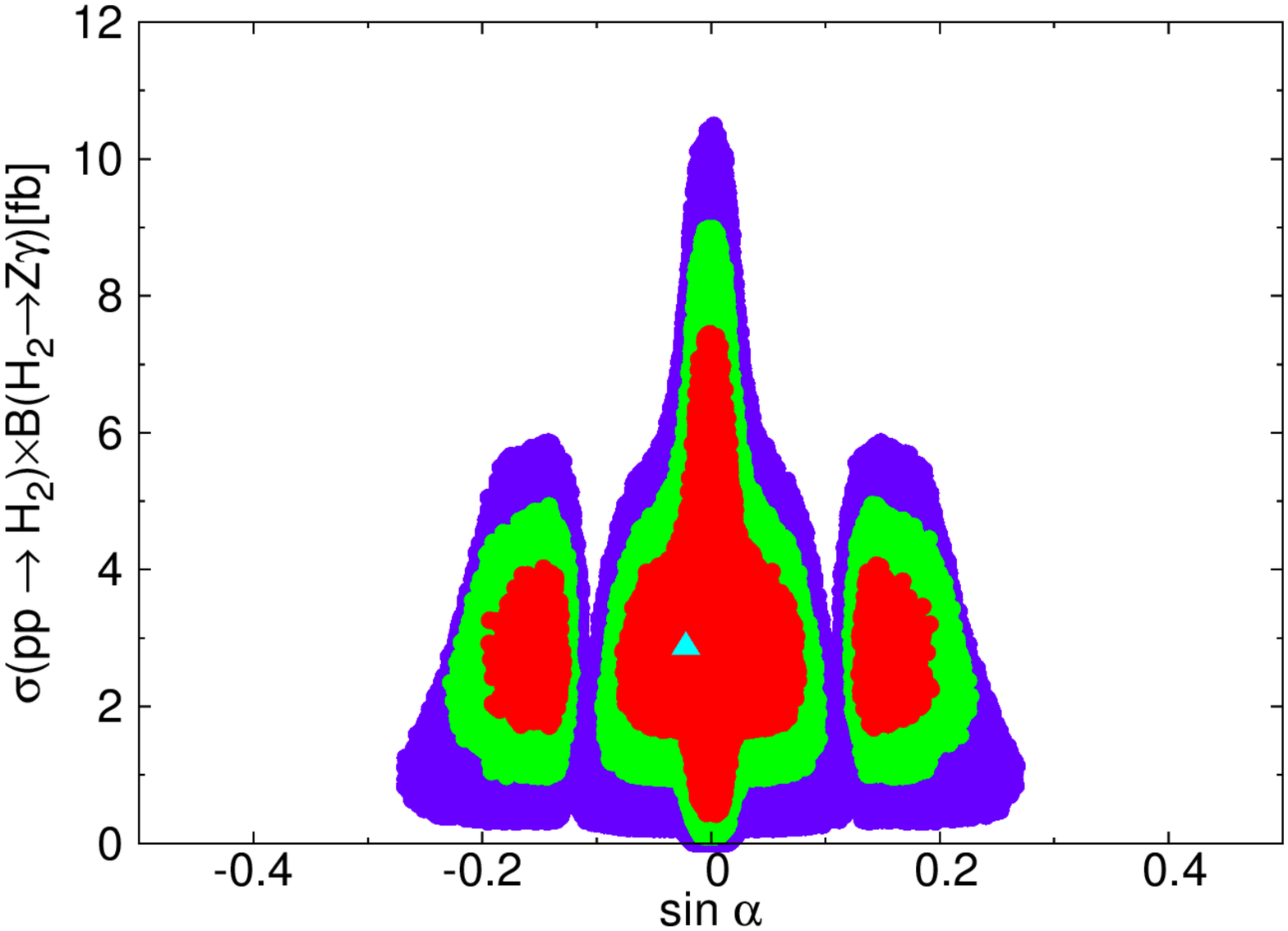}
\includegraphics[height=1.5in,angle=0]{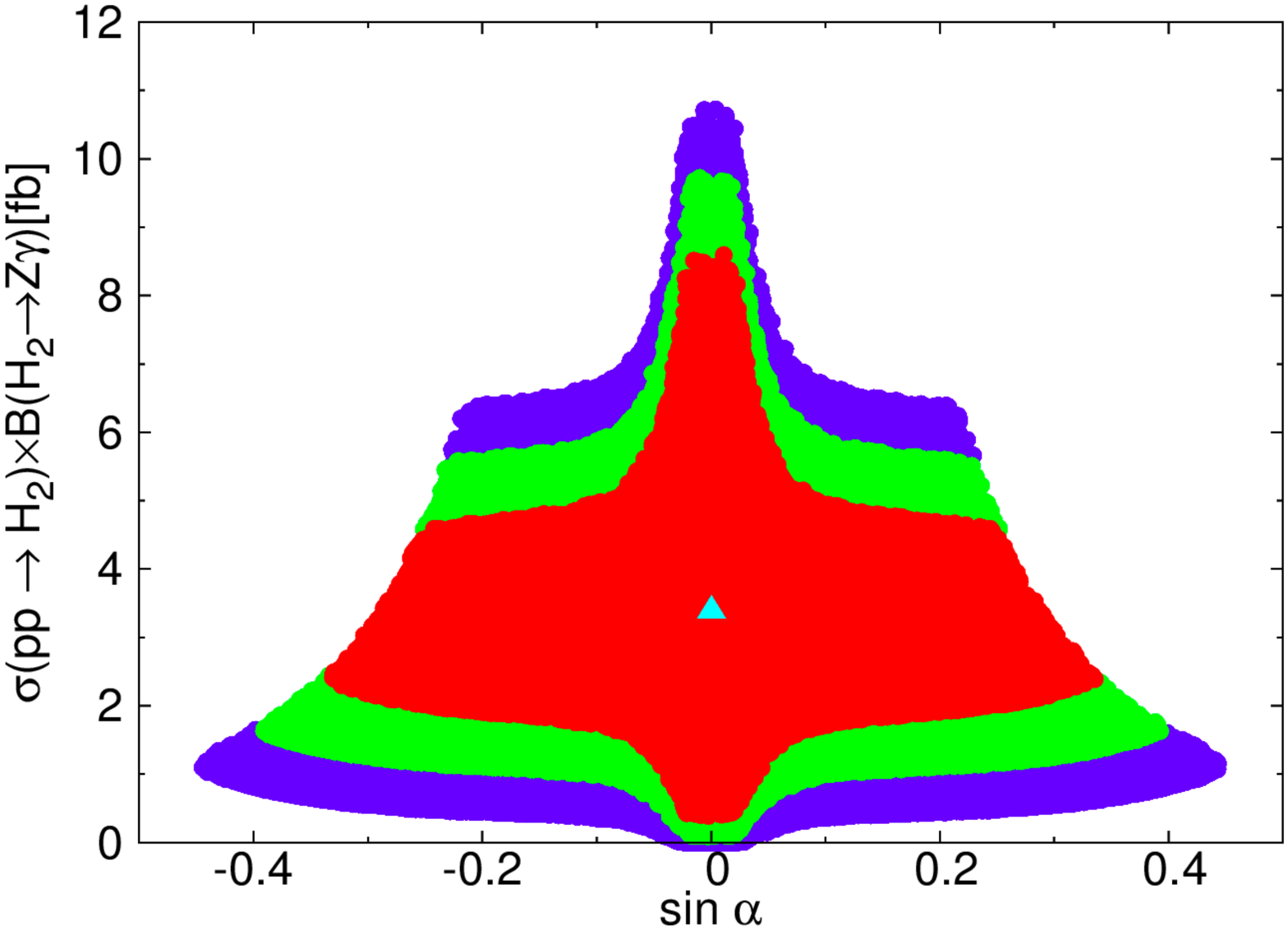}
\includegraphics[height=1.5in,angle=0]{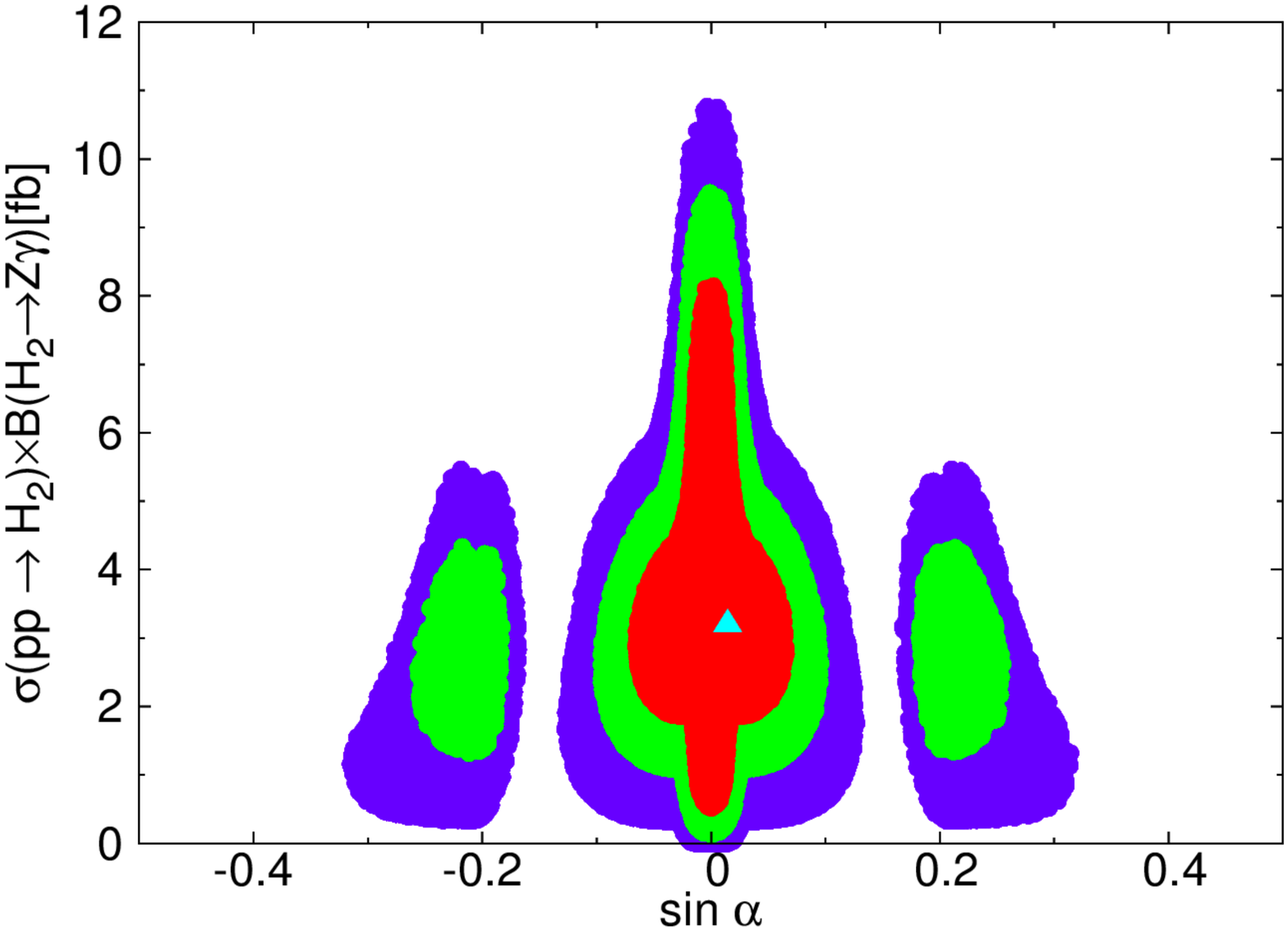}
\includegraphics[height=1.5in,angle=0]{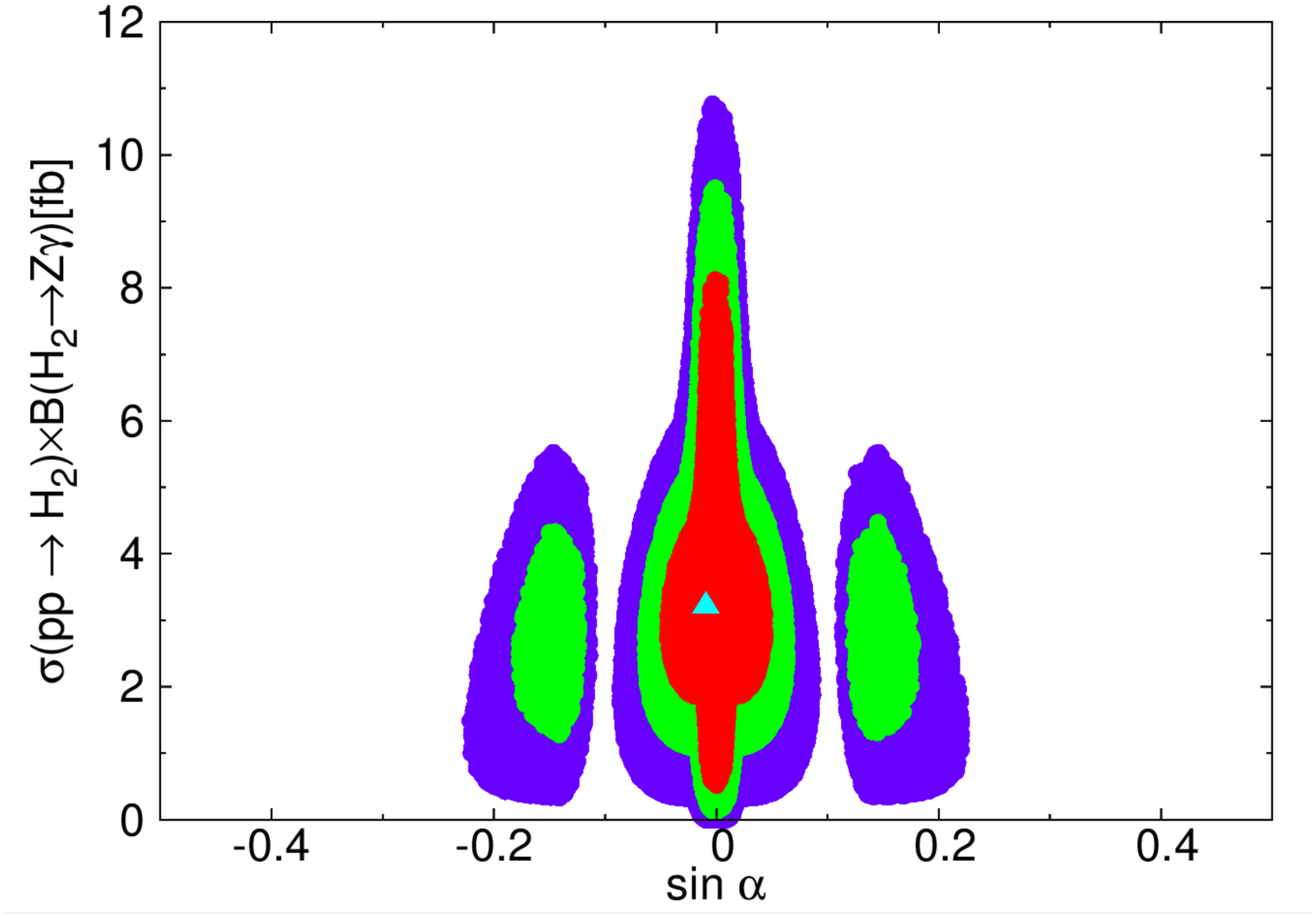}
\caption{\small \label{fig:sin_za_vlq}
{\bf{VLQ-W/Z}}:
The same as in FIG.~\ref{F4:sin_sg} but for the CL regions in the
$(\sin\alpha,\sigma(gg\to H_2)\times B(H_2\to Z\gamma))$ plane,
including the VLQ-loop induced contributions
in the presence of interactions between VLQs and $W/Z$ bosons
discussed in Section~\ref{sec:vlq}.
We are taking the limits in Eq.~(\ref{eq:limit1}) and $N_d=N_s$.
}
\end{figure}

\begin{figure}[th!]
\centering
\includegraphics[height=1.5in,angle=0]{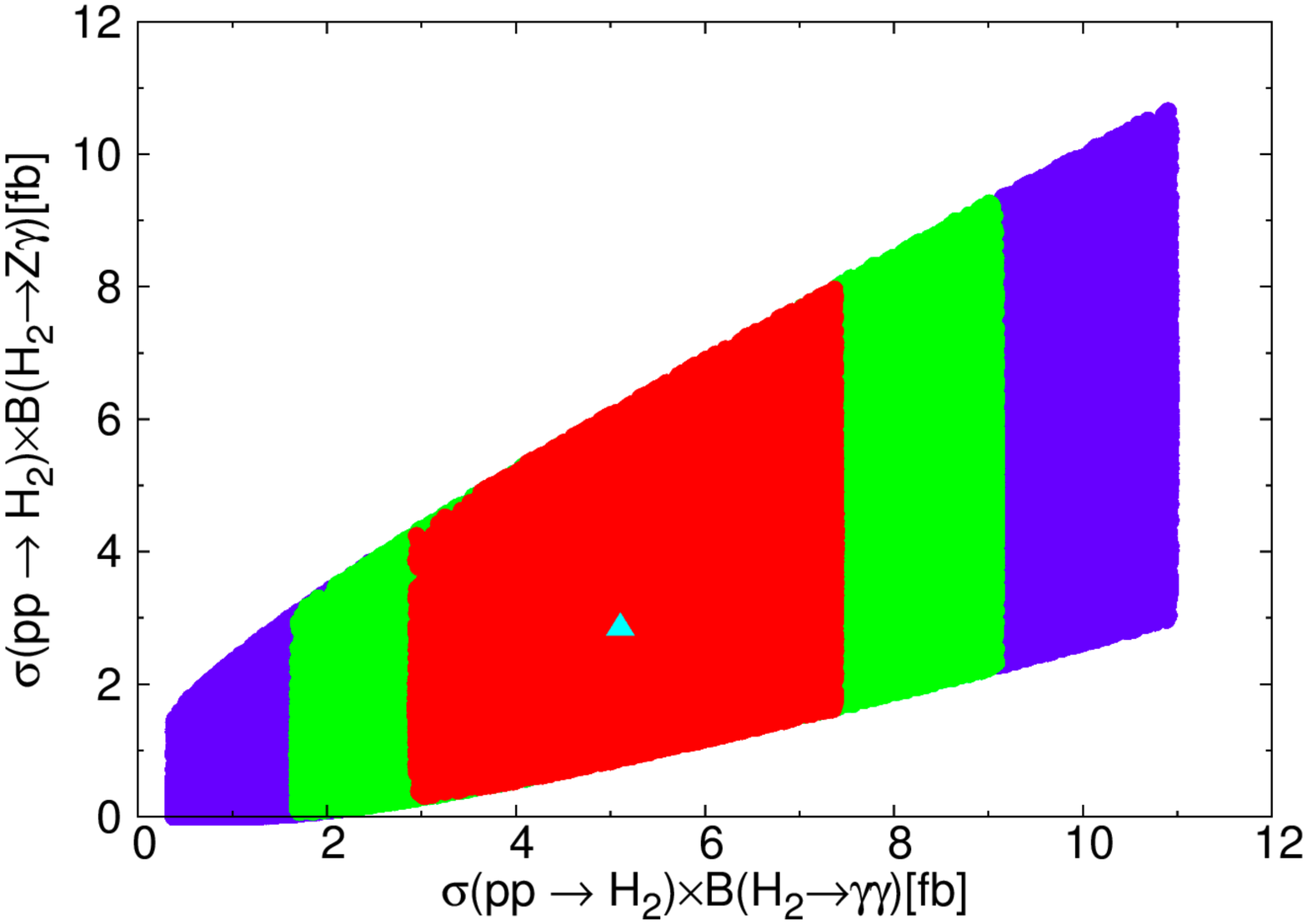}
\includegraphics[height=1.5in,angle=0]{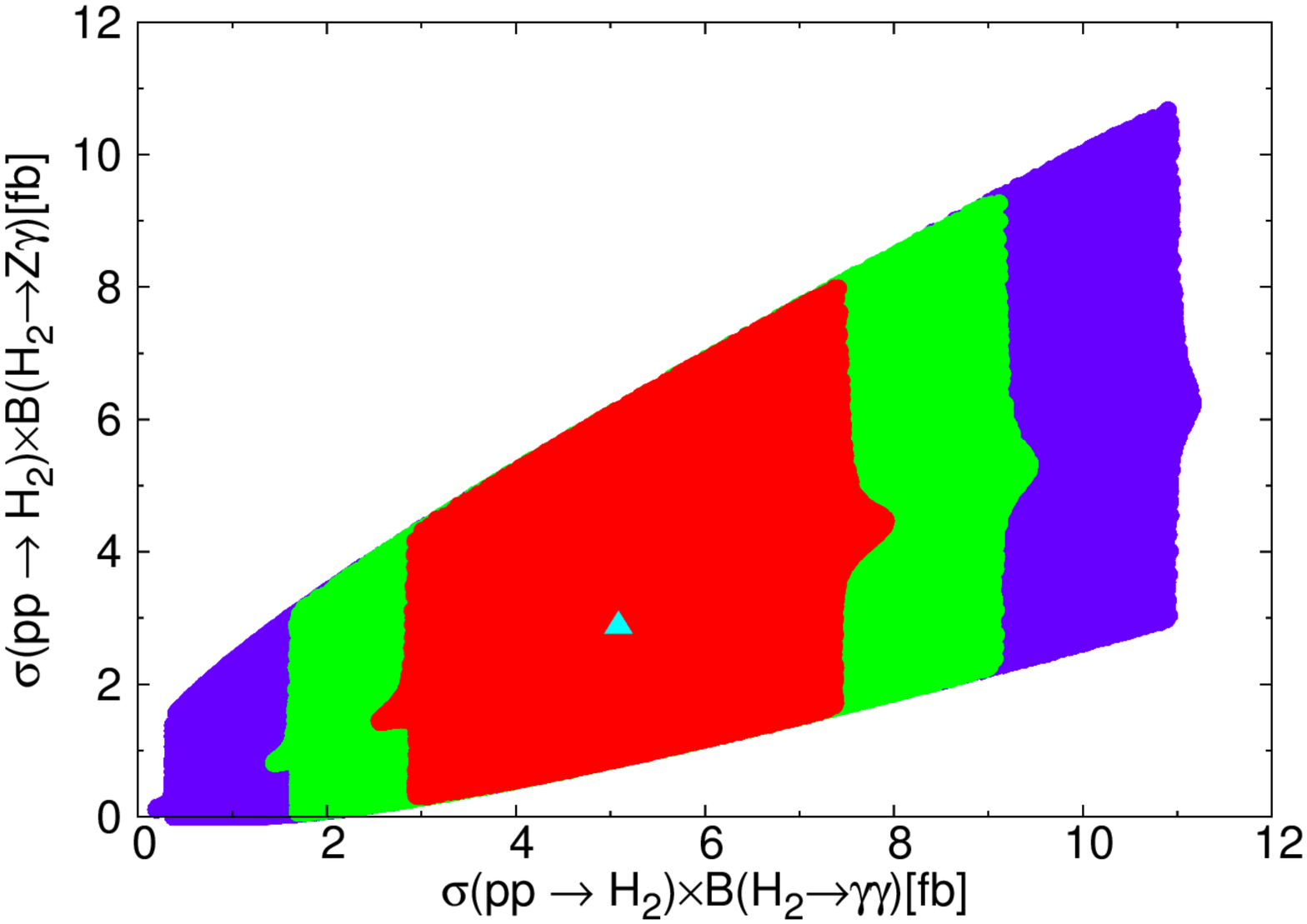}
\includegraphics[height=1.5in,angle=0]{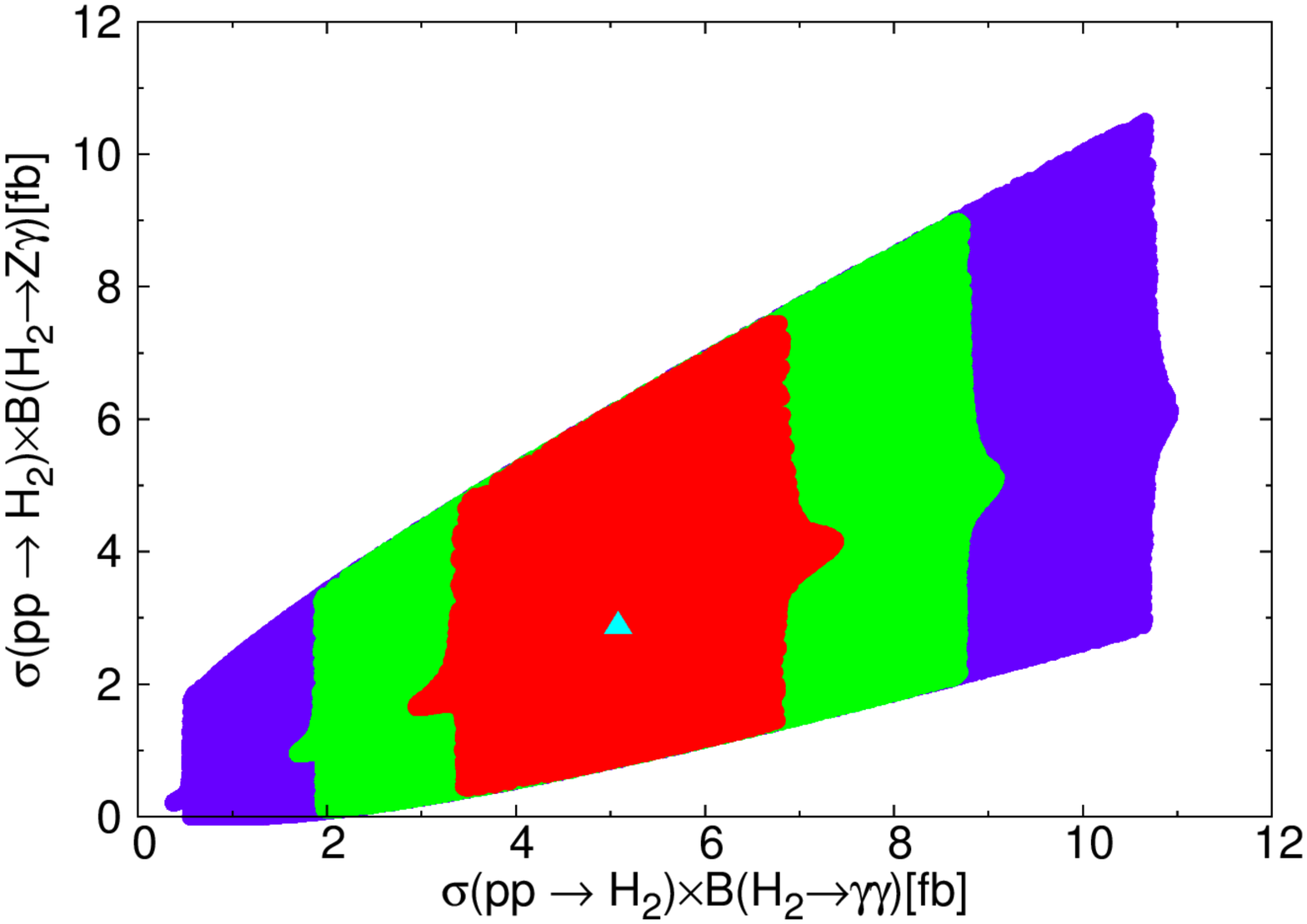}
\includegraphics[height=1.5in,angle=0]{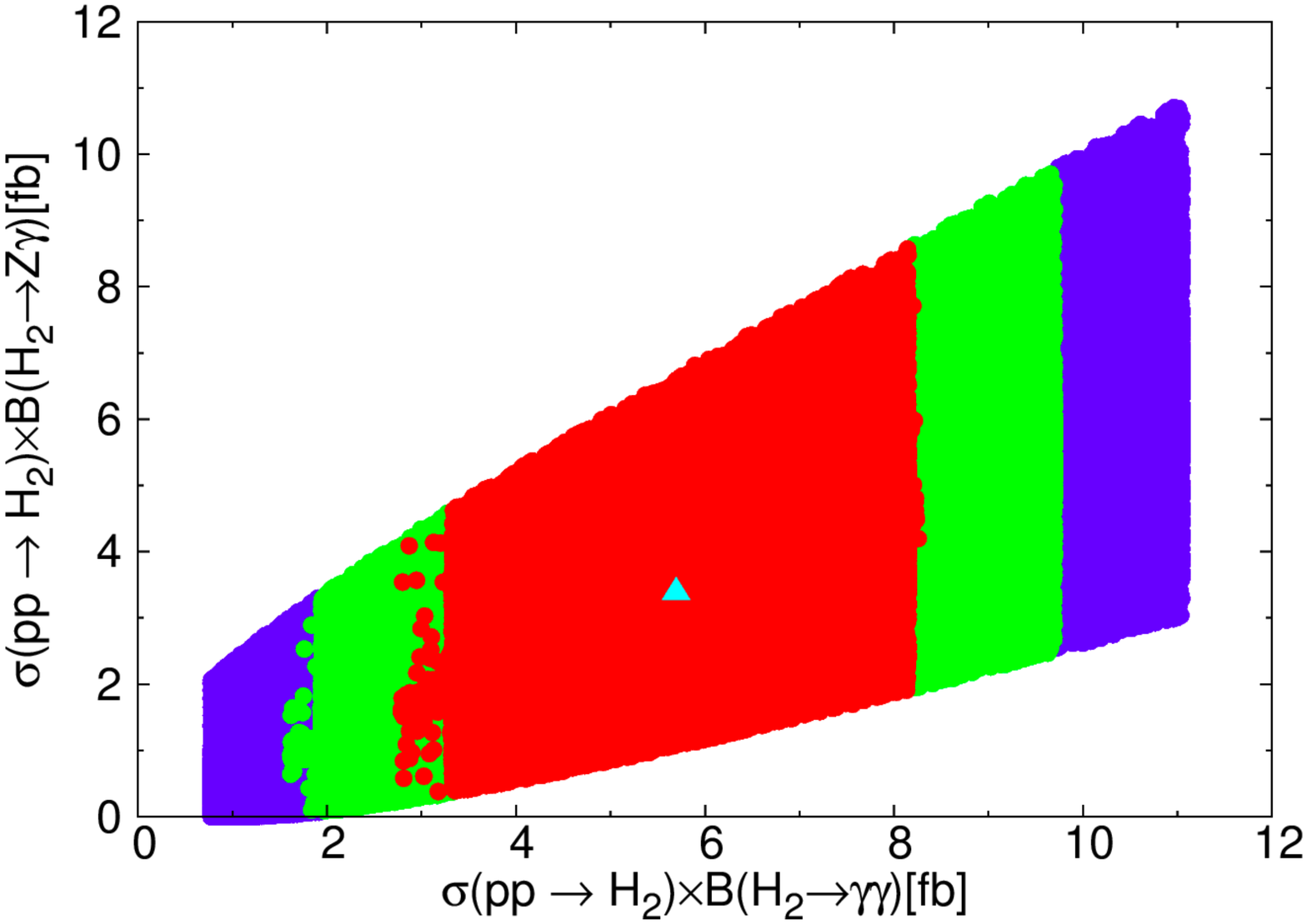}
\includegraphics[height=1.5in,angle=0]{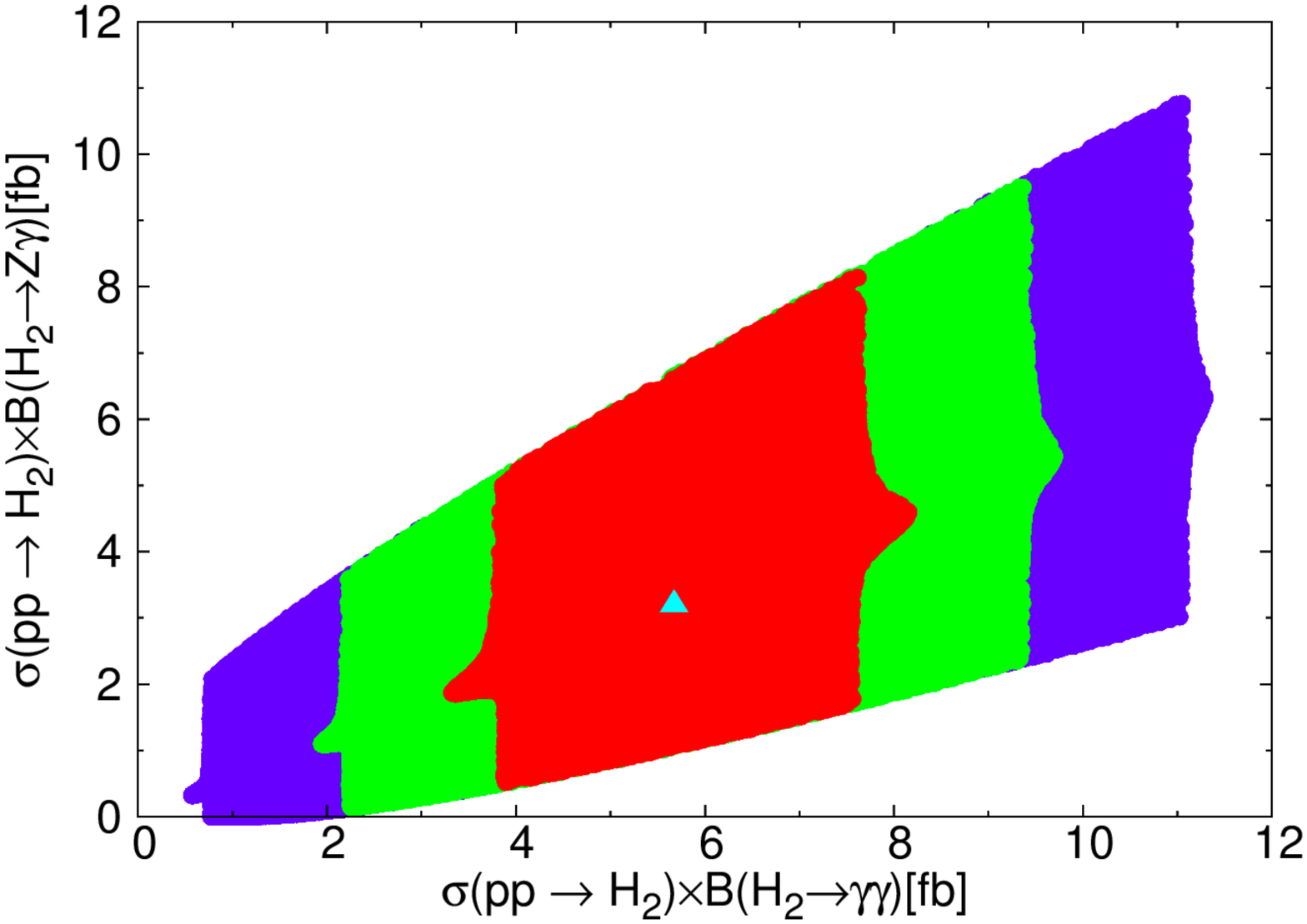}
\includegraphics[height=1.5in,angle=0]{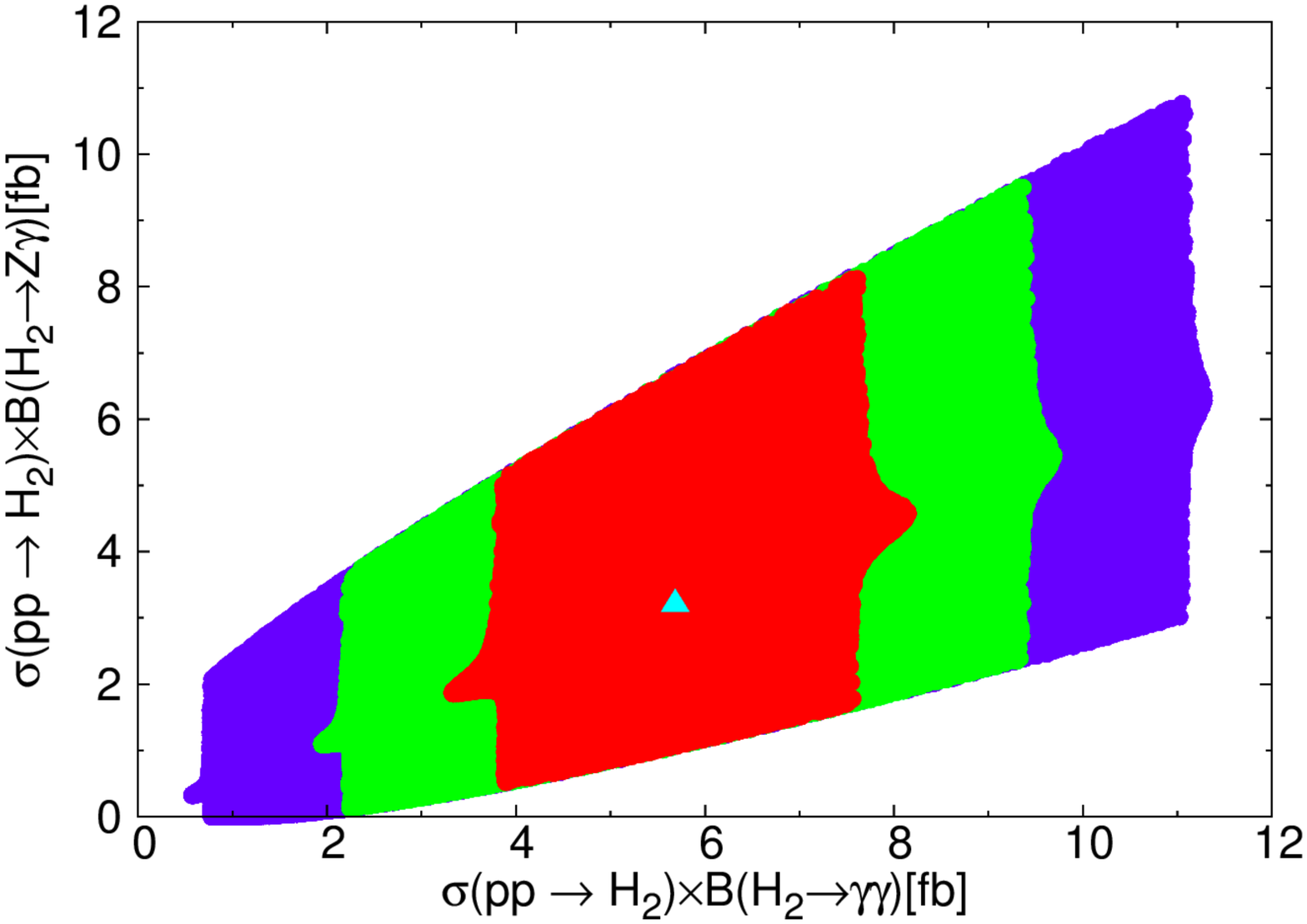}
\caption{\small \label{fig:aa_za_S001}
{\bf{VLQ-W/Z}}:
The CL regions for the correlations between
$\sigma(gg\to H_2)\times B(H_2\to \gamma\gamma)$ and
$\sigma(gg\to H_2)\times B(H_2\to Z\gamma)$ when $|\sin\alpha|<0.1$,
including the VLQ-loop induced contributions
in the presence of interactions between VLQs and $W/Z$ bosons
discussed in Section~\ref{sec:vlq}.
We are taking the limits in Eq.~(\ref{eq:limit1}) and $N_d=N_s$.
The colors are the same as in FIG.~\ref{F4:sin_sg}.
}
\end{figure}

\begin{figure}[th!]
\centering
\includegraphics[height=1.5in,angle=0]{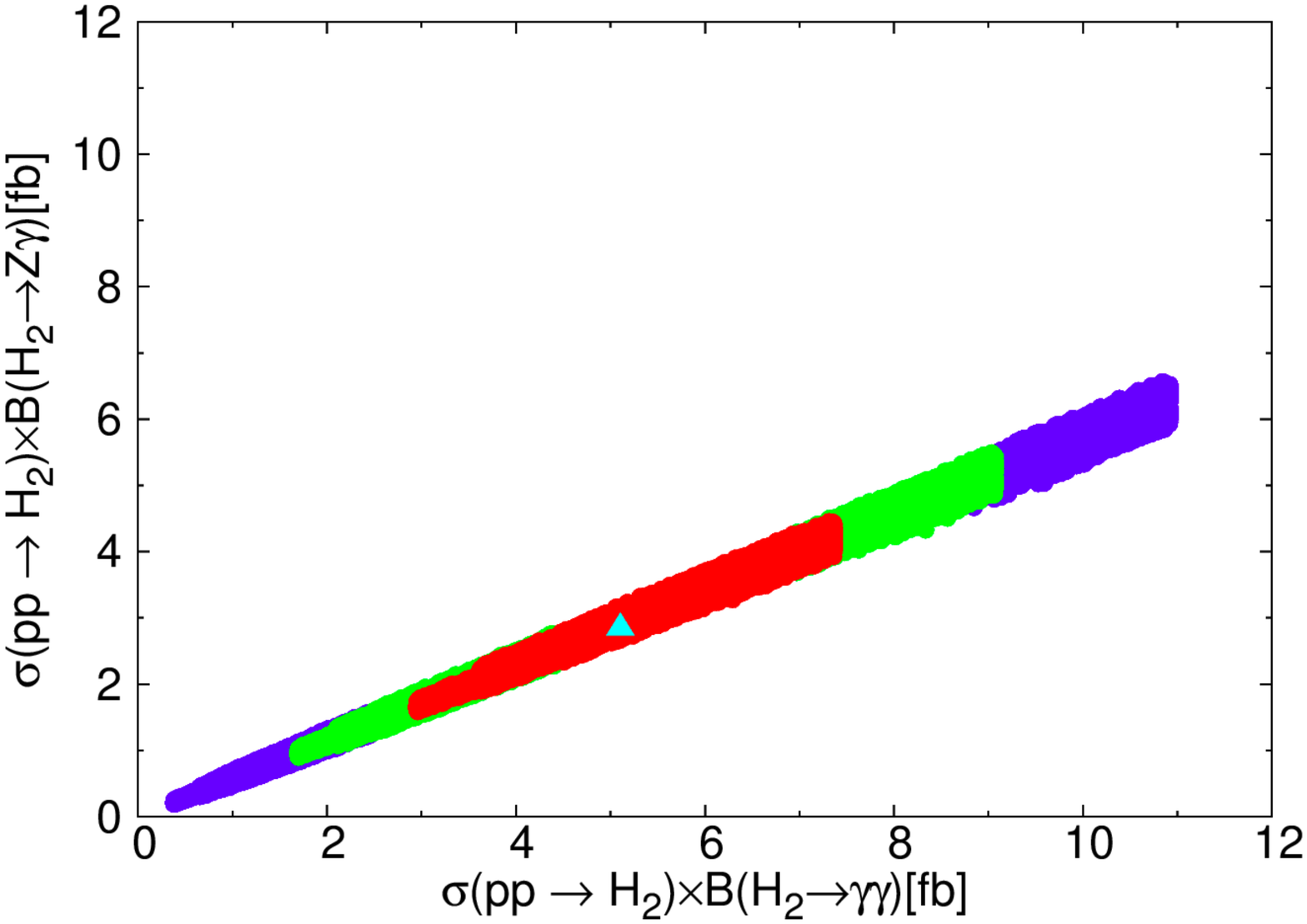}
\includegraphics[height=1.5in,angle=0]{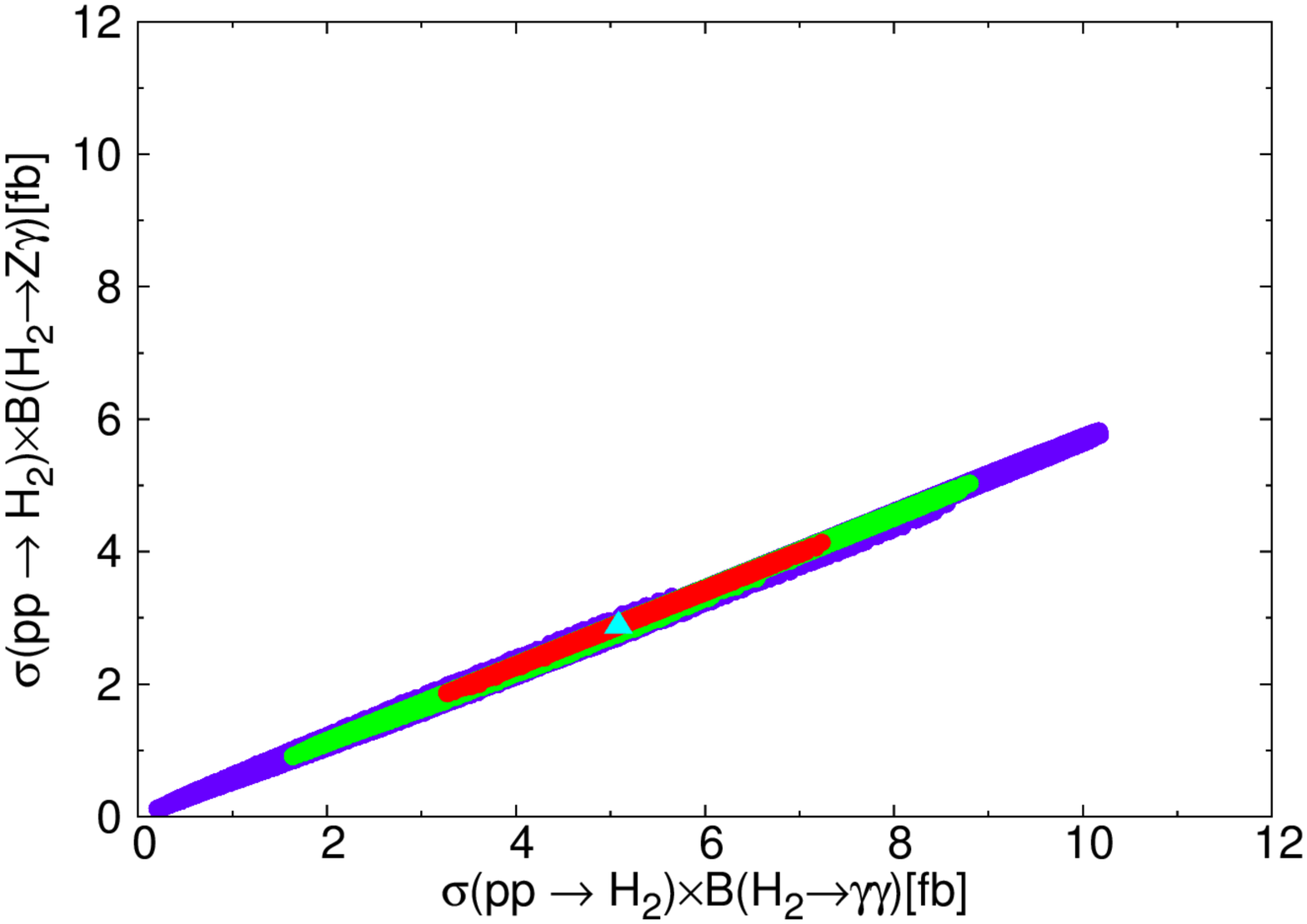}
\includegraphics[height=1.5in,angle=0]{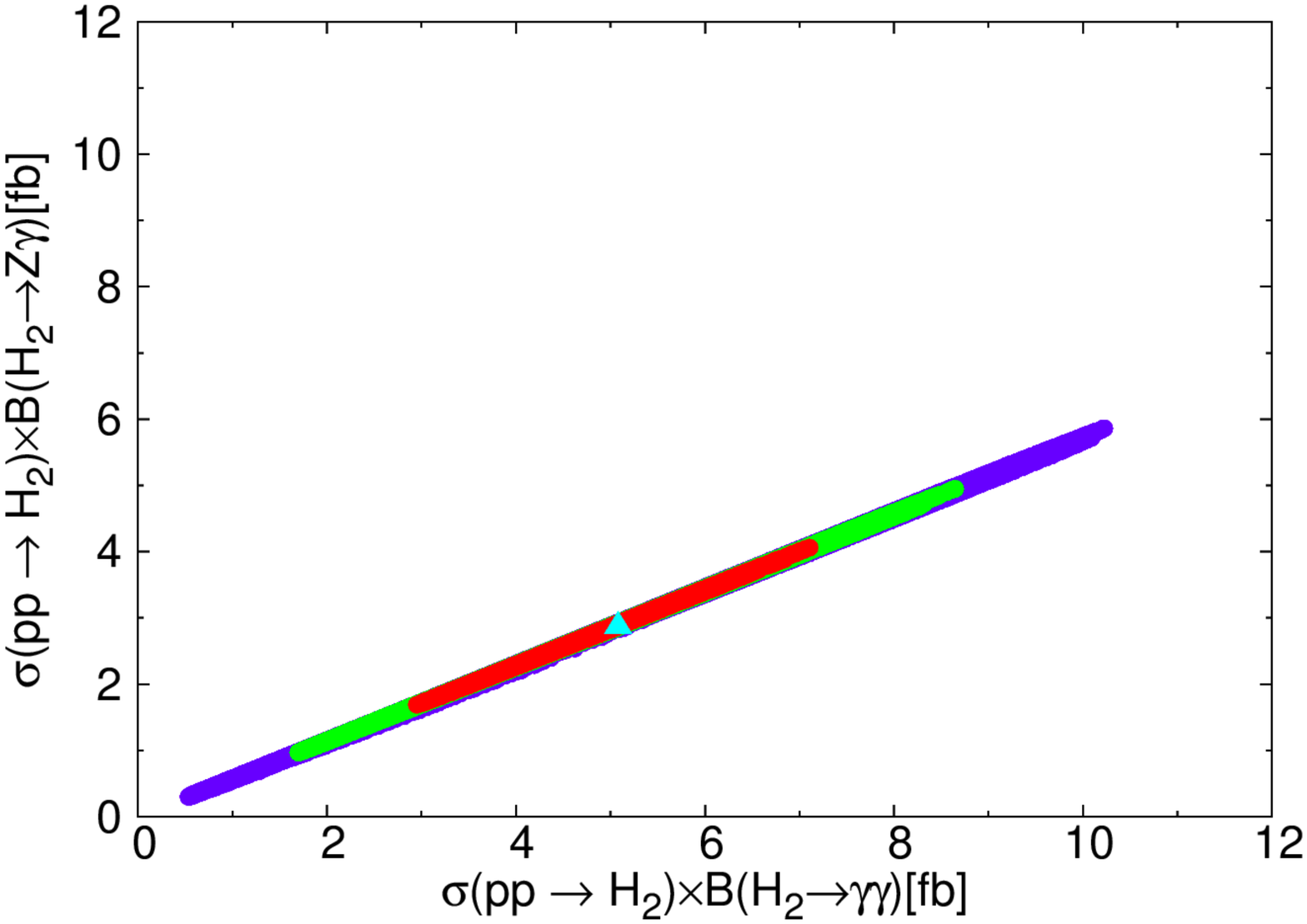}
\includegraphics[height=1.5in,angle=0]{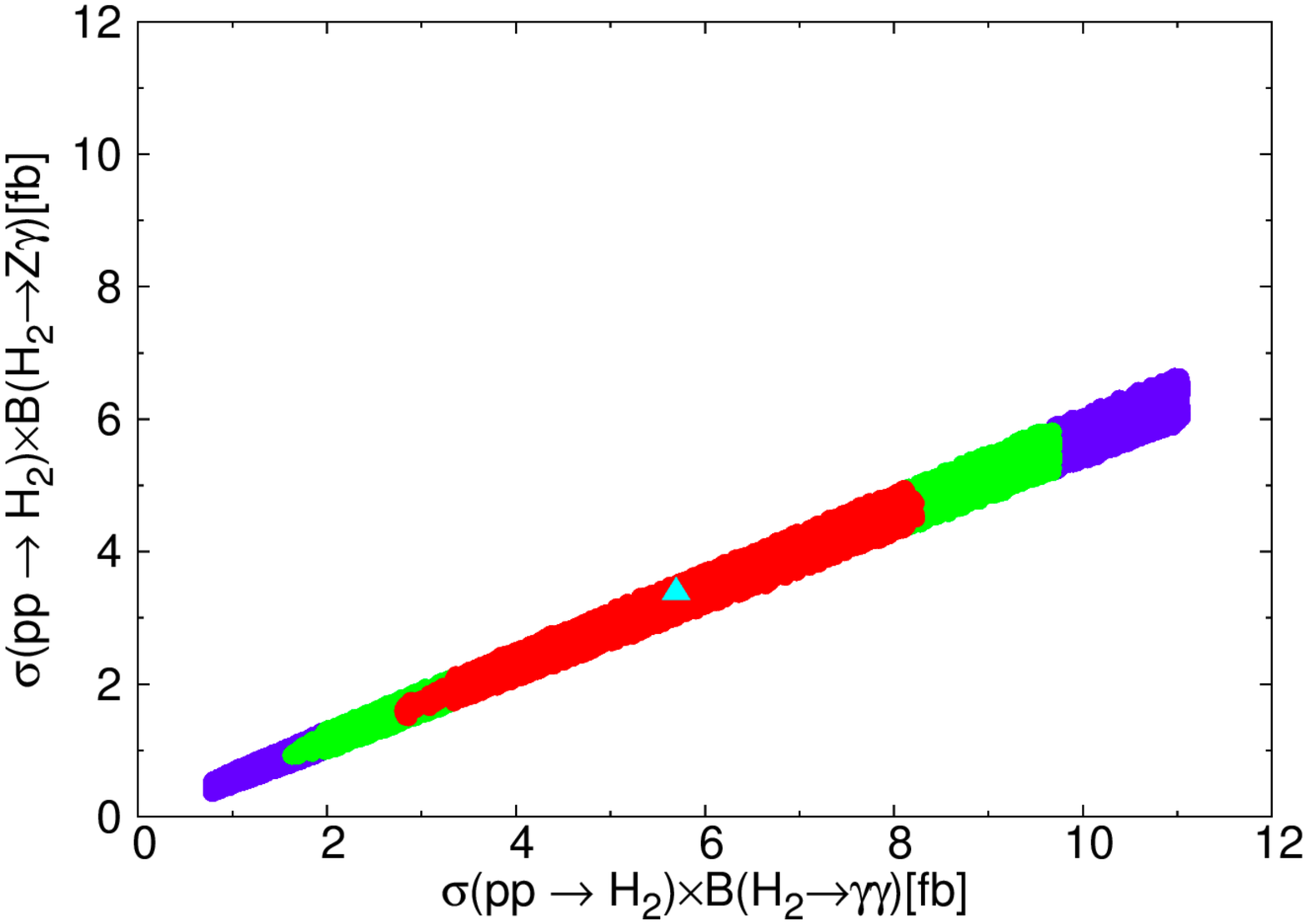}
\includegraphics[height=1.5in,angle=0]{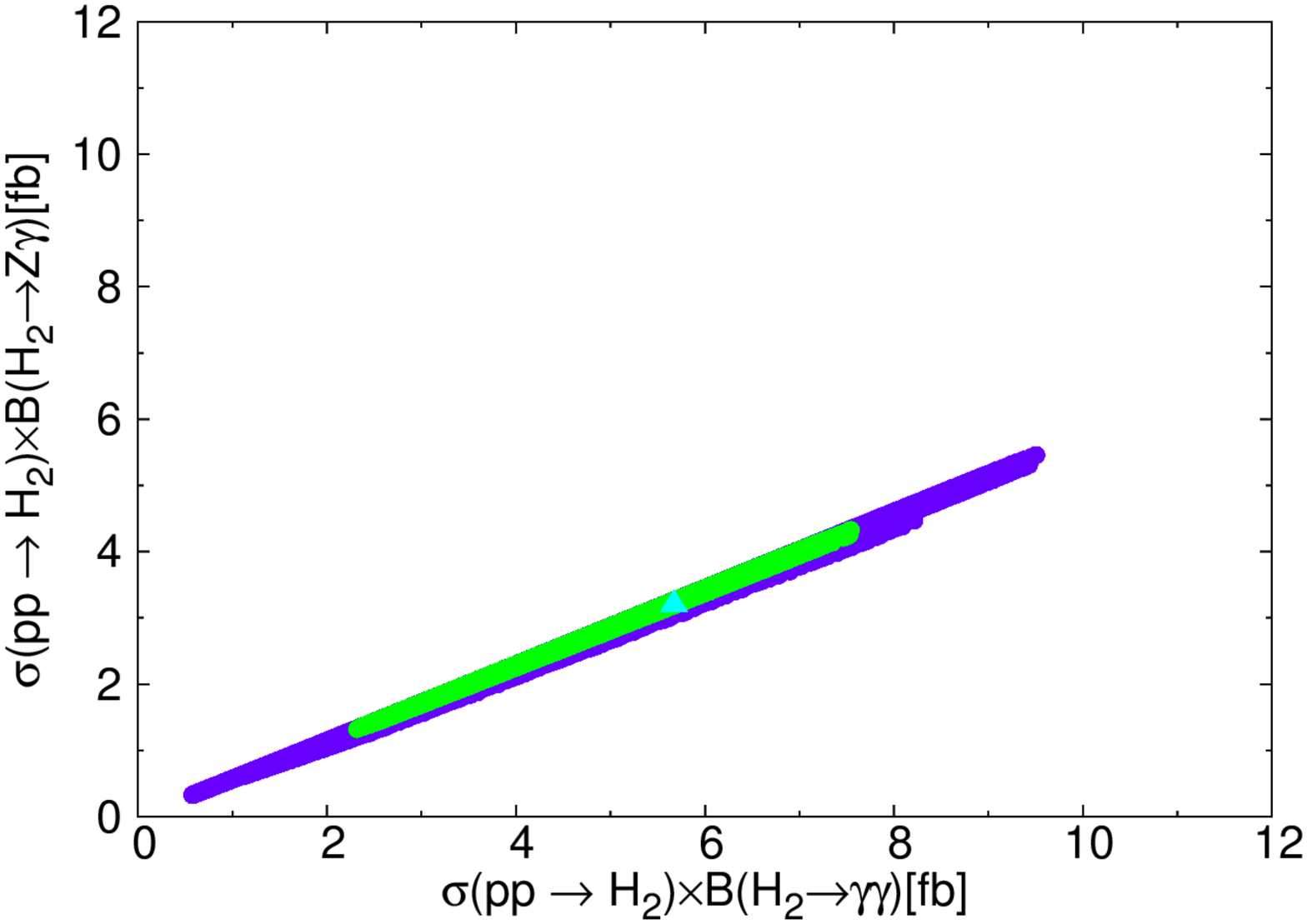}
\includegraphics[height=1.5in,angle=0]{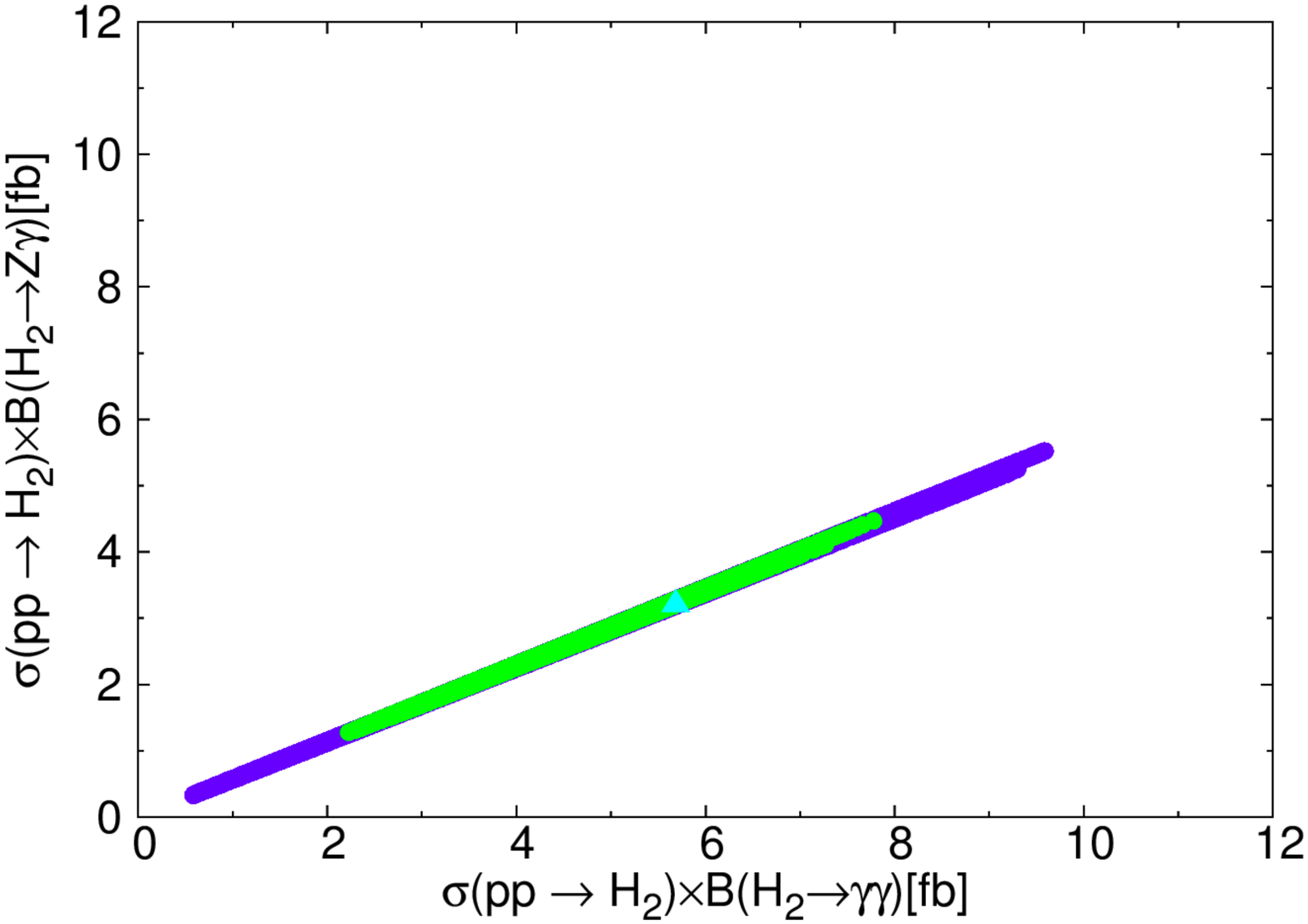}
\caption{\small \label{fig:aa_za_G001}
{\bf{VLQ-W/Z}}: 
The same as in FIG.~\ref{fig:aa_za_S001} but for 
$|\sin\alpha|\geq 0.1$.
}
\end{figure}

\end{document}